\documentclass[twocolumn,tighten, twocolappendix]{aastex631}

\usepackage{amsmath,amstext,eqnarray}
\usepackage[T1]{fontenc}
\usepackage{apjfonts} 
\usepackage[figure,figure*]{hypcap}
\usepackage{commath}
\usepackage{CJKutf8}
\usepackage{tabularx}
\usepackage{graphicx,array,bm,booktabs}
\usepackage{float}
\usepackage{longtable}
\usepackage{rotating}
\usepackage{color} 
\usepackage{xcolor}
\usepackage{CJKutf8}

\usepackage[none]{hyphenat}
\usepackage{array}
\newcolumntype{+}{>{\global\let\currentrowstyle\relax}}
\newcolumntype{^}{>{\currentrowstyle}}

\newcommand\clearrow{\global\let\rowmac\relax}
\clearrow

\shorttitle{SMA Taurus Survey at 200--400 GHz}
\shortauthors{Chung et al.}
\graphicspath{{./}{figures/}}

\begin{document}
\begin{CJK}{UTF8}{bsmi}

\title{SMA 200--400 GHz Survey for Dust Properties in the Icy Class II Disks in the Taurus Molecular Cloud}

\correspondingauthor{Hauyu Baobab Liu}
\email{baobabyoo@gmail.com}

\author[0009-0007-3677-8040]{Chia-Ying Chung}
\affiliation{Department of Physics, National Sun Yat-Sen University, No. 70, Lien-Hai Road, Kaohsiung City 80424, Taiwan, R.O.C.}
\affil{Department of Physics, National Taiwan University, 10617 Taipei, Taiwan}
\affiliation{Academia Sinica Institute of Astronomy and Astrophysics (ASIAA), No. 1, Section 4, Roosevelt Road, Taipei 10617, Taiwan}

\author[0000-0003-2253-2270]{Sean M. Andrews}
\affiliation{Center for Astrophysics \textbar\, Harvard \& Smithsonian, 60 Garden St., Cambridge, MA 02138, USA}

\author[0000-0003-0685-3621]{Mark A. Gurwell}
\affiliation{Center for Astrophysics \textbar\, Harvard \& Smithsonian, 60 Garden St., Cambridge, MA 02138, USA}

\author[0000-0002-9154-2440]{Melvyn Wright}
\affiliation{Department of Astronomy, University of California, Berkeley, 501 Campbell Hall, Berkeley, CA 94720-3441, USA}

\author[0000-0002-7607-719X]{Feng Long}
\affiliation{Center for Astrophysics \textbar\, Harvard \& Smithsonian, 60 Garden St., Cambridge, MA 02138, USA}
\affiliation{Lunar and Planetary Laboratory, University of Arizona, Tucson, AZ 85721, USA}
\altaffiliation{NASA Hubble Fellowship Program Sagan Fellow}

\author[0000-0002-9408-2857]{Wenrui Xu}
\affiliation{Center for Computational Astrophysics, Flatiron Institute, 162 5th Ave, New York, NY10010, USA}

\author[0000-0003-2300-2626]{Hauyu Baobab Liu}
\affiliation{Department of Physics, National Sun Yat-Sen University, No. 70, Lien-Hai Road, Kaohsiung City 80424, Taiwan, R.O.C.}
\affiliation{Center of Astronomy and Gravitation, National Taiwan Normal University, Taipei 116, Taiwan}



\begin{abstract}
We present a new SMA survey of 47 Class II sources in the Taurus-Auriga region.
Our observations made 12 independent samples of flux densities over the 200--400 GHz frequency range. 
We tightly constrained the spectral indices of most sources to a narrow range of $2.0\pm0.2$; only a handful of spatially resolved (e.g., diameter $>$250 au) disks present larger spectral indices.
The simplest interpretation for this result is that the (sub)millimeter luminosities of all of the observed target sources are dominated by very optically thick (e.g., $\tau\gtrsim$5) dust thermal emission.
Some previous works that were based on the optically thin assumption thus might have underestimated optical depths by at least one order of magnitude.
Assuming DSHARP dust opacities, this corresponds to underestimates of dust masses by a similar factor. 
For our specific selected sample, the lower limits of dust masses implied by the optically thick interpretation are 1--3 times higher than those previous estimates that were made based on the optically thin assumption.  
Moreover, some population synthesis models show that to explain the observed, narrowly distributed spectral indices, the disks in our selected sample need to have very similar dust temperatures ($T_{\mbox{\scriptsize dust}}$).
Given a specific assumption of median $T_{\mbox{\scriptsize dust}}$, the maximum grain sizes ($a_{\mbox{\scriptsize max}}$) can also be constrained, which is a few times smaller than 0.1 mm for $T_{\mbox{\scriptsize dust}}\sim$100 K and a few mm for $T_{\mbox{\scriptsize dust}}\sim$24 K.
The results may indicate that dust grain growth outside the water snowline is limited by the bouncing/fragmentation barriers. 
This is consistent with the recent laboratory experiments, which indicated that the coagulation of water-ice coated dust is not efficient and the water-ice free dust is stickier and thus can coagulate more efficiently.
In the Class II disks, the dust mass budget outside of the water snowline may be largely retained instead of being mostly consumed by planet formation. 
While Class II disks still possess sufficient dust masses to feed planet formation at a later time, it is unknown whether or not dust coagulation and planet formation can be efficient or natural outside of the water snowline.
\end{abstract}

\keywords{Circumstellar dust (236) --- Protoplanetary disks (1300) --- Pre-main sequence (1289) --- Planet formation (1241)}


\section{Introduction} \label{sec:intro}

\begin{deluxetable*}{ccccccccc}
\tabletypesize{\footnotesize}
\tablecolumns{9}
\tablewidth{0pt}
\tablecaption{SMA Observation Summary  \label{tab:obs_summary}}
\tablehead{ 
\colhead{Track ID} & \colhead{UTC Date} & LSB Freq. & USB Freq. & \colhead{Array Config.} & \colhead{uv--range} & \colhead{Flux Calib.} & \colhead{Passband Calib.} & \colhead{Gain Calib.} \\
\colhead{} & \colhead{(YYYY MM DD)} & \colhead{(GHz)} & \colhead{(GHz)} & \colhead{} & \colhead{(k$\lambda$)} & \colhead{} & \colhead{} & \colhead{} }
\startdata 
230 GHz--1 & 2021 10 18 & 193--205 & 213--225 & {SUB--COM} & 5--60 & {Uranus} & 3C84 & J0510+180, J0418+380 \\
 & & 222--234 & 242--254 & & & & & \\
230 GHz--2 & 2021 11 26 & 193--205 & 213--225 & {COM} & 15--60 & 3C84 & 3C84 & J0510+180, J0418+380 \\
 & & 222--234 & 242--254 & & & & & \\
270 GHz--3 & 2021 11 20 & 254--266 & 274--286 & {COM} & 18--80 & {Uranus} & {BL Lac} & J0510+180, J0418+380 \\
 & & 286--298 & 306--318 & & & & & \\
400 GHz--4 & 2021 09 03 & 331--343 & 351--363 & {SUB} & 10--90 & {Neptune} & 3C84 & J0510+180, J0418+380, 3C84 \\
 & & 391.5--403.5 & 411.5--423.5 & & & & & \\
400 GHz--5 & 2021 09 25 & 331--343 & 351--363 & {SUB} & 10--90 & {Uranus} & 3C84 & J0510+180, J0418+380, 3C84 \\
 & & 391.5--403.5 & 411.5--423.5 & & & & & \\
400 GHz--6 & 2021 11 22 & 331--343 & 351--363 & {COM} & 24--106 & {Uranus} & 3C84 & J0510+180, J0418+380, 3C84 \\
 & & 391.5--403.5 & 411.5--423.5 & & & & & \\
\enddata
\end{deluxetable*}

Observations have unveiled that rocky planets ubiquitously form around young stellar objects (YSOs; e.g., \citealt{Batalha2013ApJS..204...24B,Barclay2018ApJS..239....2B}). Over the evolution of YSOs, some dust grains in the natal protoplenatary disks must grow from (sub-)micron sizes to form millimeter-sized chondrules, centimeter-sized pebbles and, eventually, kilometer-sized planetesimals (for a review see \citealt{Testi2014prpl.conf..339T}).
Although it appears that dust growth occurs naturally in the protoplanetary disks, the underlying physical mechanisms remain puzzling and have been actively debated (for a review see \citealt{Birnstiel2016SSRv..205...41B}).


Firstly, in the protoplanetary disk, the 0.1--1 mm sized dust grains will decouple from gas and then undergo inward migration until they reach a gas pressure maximum \citep{Weidenschilling1977MNRAS.180...57W}.
The dust inward migration is predicted to be more efficient than dust growth.
Without trapping dust at pressure maximum, which may appear as substructures in the resolved, (sub)millimeter images of protoplanetary disks (for a review see \citealt{Andrews2020ARA&A..58..483A}), the maximum dust grain size may be limited to $\sim$100 $\mu$m.
This is so-called the {\it inward migration barrier} for dust growth.

In addition, the 0.1--1 mm sized dust grains trapped at pressure maximum may not coagulate upon collision.
Instead, they may bounce off or even fragment to smaller sizes. 
This is so called the {\it bouncing/fragmentation barrier} (\citealt{Zsom2010A&A...513A..57Z}). 
Depending on the strength of turbulence in the gaseous disk, this
also can limit $a_{\mbox{\scriptsize max}}$ to $\sim$0.1--1 mm sizes 
(for a recent simulation see \citealt{Jiang2023arXiv231107775J}). 

The outcome of dust collision is dictated by the surface energy of the dust grains which depend on the morphology and chemical composition of dust grains. 
For dust grains to grow larger than centimeter sizes, the effects of the bouncing/fragmentation barrier need to be alleviated by other physical mechanisms. 
For example, the dynamical instability, streaming instability (\citealt{Johansen2007ApJ...662..627J,Youdin2007ApJ...662..613Y}), can help suppress the collisional velocities between dust grains and thereby promote dust coagulation.
Other micro-physics, such as enhancing the grain porosity (\citealt{Okuzumi2012ApJ...752..106O}) or enhancing the dust stickiness by coating certain species onto the grain surfaces (e.g., \citealt{Wada2009ApJ...702.1490W}) may also alleviate the effect of bouncing/fragmentation barrier\footnote{Note that the effect of ice-coating is very subtle and may conversely make the dust grains more fragile (\citealt{Okuzumi2019ApJ...878..132O}).}.
Whether or not these physical mechanisms indeed help alleviate or bypass the inward migration and/or bouncing/fragmentation barriers in certain evolutionary stages or at certain regions in the protoplanetary disks needs to be tested by observationally measuring $a_{\mbox{\scriptsize max}}$.

In most of the previous observational studies, the constraints on $a_{\mbox{\scriptsize max}}$ were made by comparing the observed (sub)millimeter spectral energy distributions (SEDs) with the modified black body formulation $S_\nu = \Omega B_\nu(T_{\mbox{\tiny dust}})(1-e^{-\tau_{\nu}})$, 
where at frequency $\nu$ the optical depth $\tau_{\nu}$ is the product of dust mass surface density $\Sigma$ and dust absorption opacity $\kappa_\nu^{\mbox{\tiny abs}}$ (for a review see \citealt{Hildebrand1983QJRAS..24..267H}), $B_\nu(T_{\mbox{\tiny dust}})$ is the Planck function at dust temperature $T_{\mbox{\tiny dust}}$, and $\Omega$ is the solid angle. 
At (sub)millimeter wavelength, $\kappa_\nu^{\mbox{\tiny abs}}$ is approximately proportional to $\nu^\beta$, where the quantity $\beta$ is known as the dust opacity spectral index.
Under the optically thin assumption and Rayleigh-Jeans (i.e., high temperature or long wavelength) limit, $\beta$ is numerically related to the observed spectral index $\alpha$ by $\alpha\sim\beta+2$.
As compared with the black body emission, the role of $\beta$ can be qualitatively understood since dust grains absorb/emit more efficiently at wavelengths that are comparable or shorter than the grain sizes.
In the case of the diffuse interstellar medium (ISM), observations show steep (sub)millimeter SED with spectral index $\alpha \approx 4$ ($\beta \approx 2$) (\citealt{Erickson1981ApJ...245..148E,Draine1984ApJ...285...89D,Li2001ApJ...554..778L}). 
An increase in the dust grain size will result in better emission efficiency at longer wavelengths and, thereby, the shallower (sub)millimeter spectral indices. 

\begin{figure*}[ht!]
    \hspace{-0.35cm}
    \begin{tabular}{ll}
    \includegraphics[height=7cm]{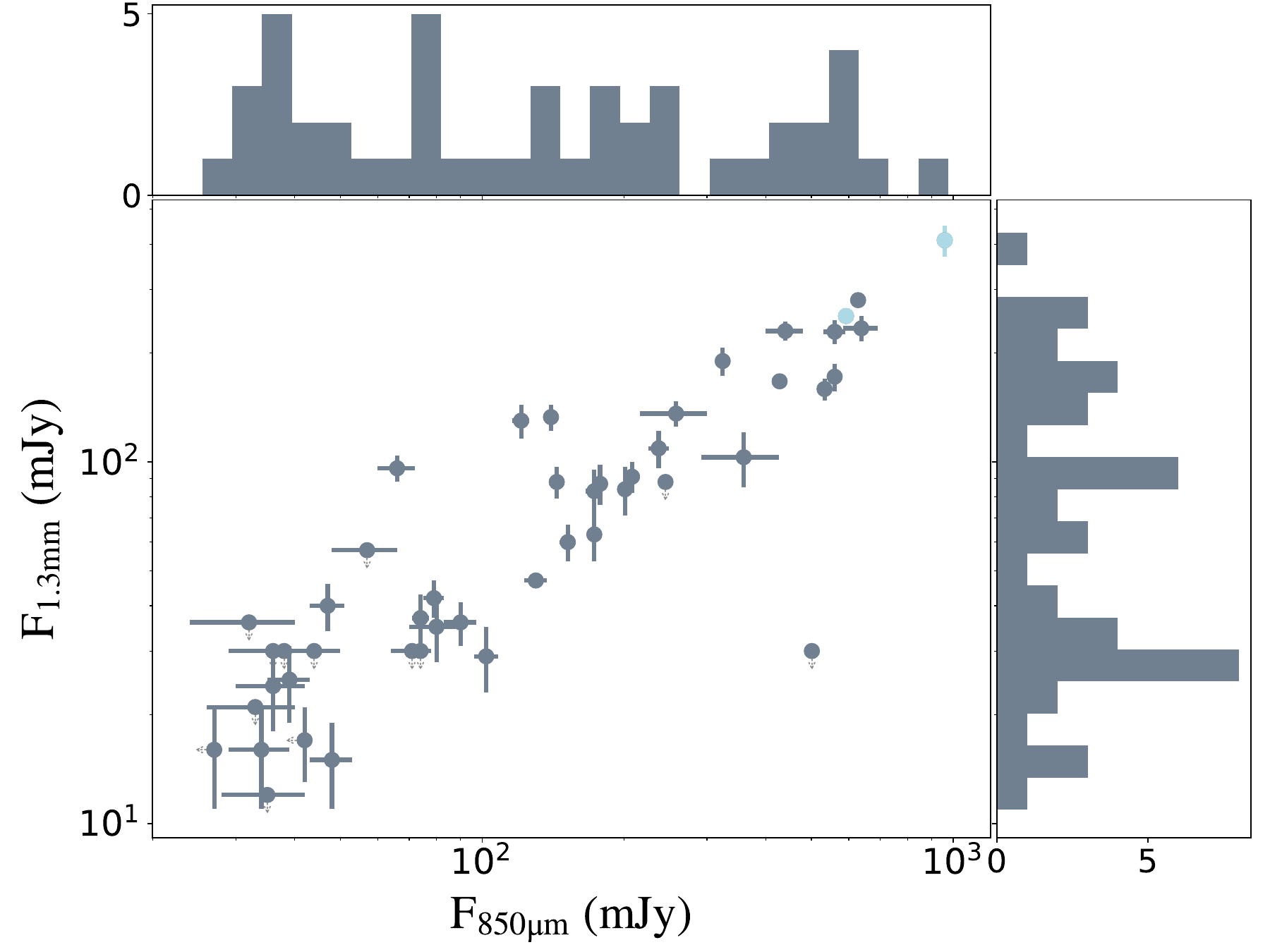}  &
    \includegraphics[height=7cm]{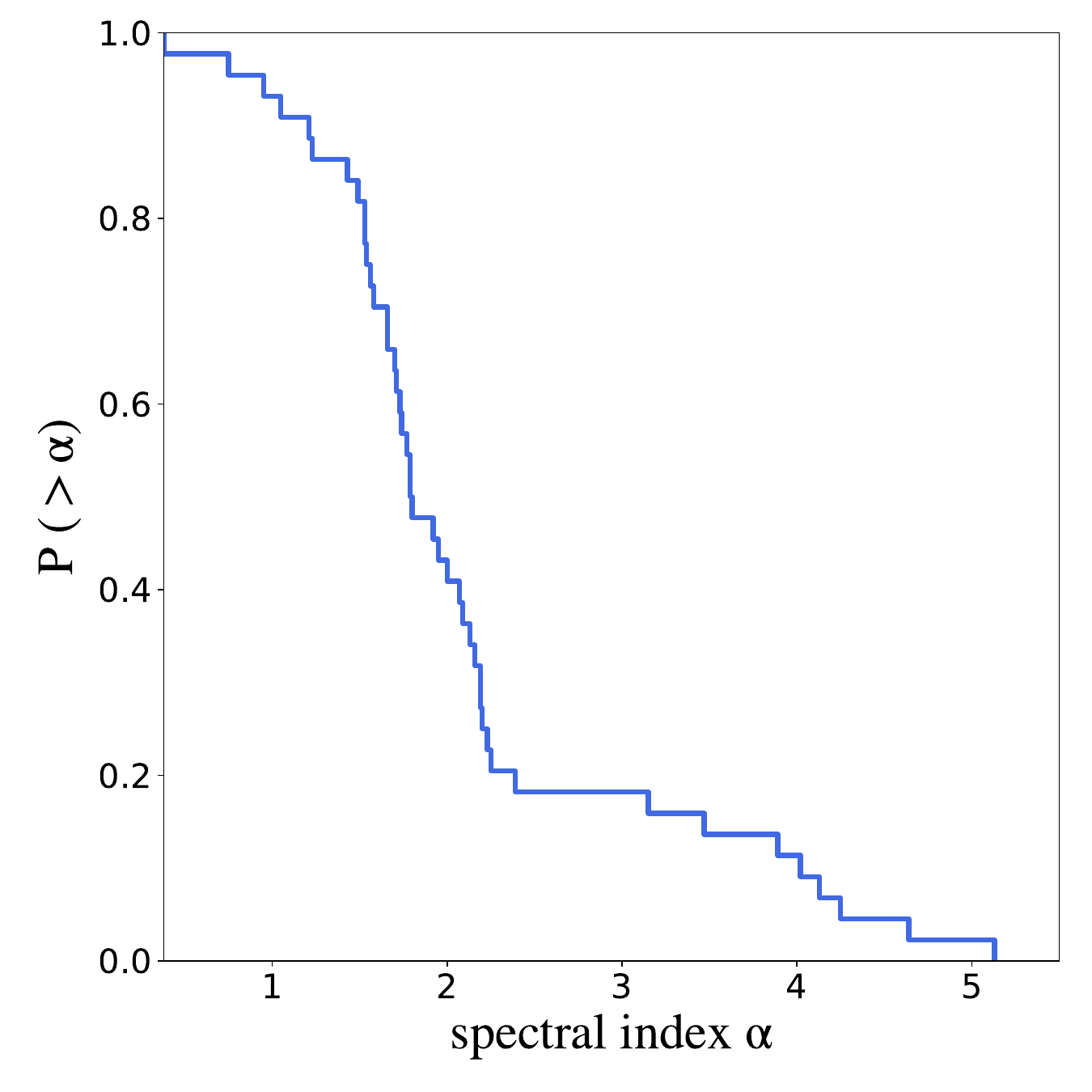}  \\
    \end{tabular}
    \caption{
    Properties of the selected 47 Class II sources in the Taurus-Auriga region (quoted from \citealt{Andrews_Williams_2005}). 
    {\it Left :} The 1.3 mm flux densities of the target sources plotted against 850 $\mu$m flux densities. Blue dots label the two sources for which the 850 $\mu$m flux densities were estimated from the 1.3 mm flux densities based on an $\alpha = 2.0$ assumption.
    {\it Right :} The cumulative distribution of (sub)millimeter spectral index $\alpha$ of the 44 sources in our sample.
    }
    \label{fig:target}
\end{figure*}

The previous (sub)millimeter interferometry surveys towards Class II protoplanetary disks ubiquitously detected low (2.0--3.0) (sub)millimeter spectral indices (e.g., \citealt{Beckwith1991ApJ...381..250B,Ricci2010A&A...512A..15R,Perez2012ApJ...760L..17P,Tsukagoshi2016ApJ...829L..35T,Tazzari2021MNRAS.506.2804T}).
This had been regarded as evidence of the presence of grown dust with $a_{\mbox{\scriptsize max}}\gg1$ mm. 
Limited by angular resolutions, most of the previous (sub)millimeter observations towards protoplanetary disks preferentially observed the regions outside of the 150--170 K water snowline (\citealt{Pollack1994ApJ...421..615P}). 
Therefore, these previous studies also implied that the coagulation of the water-ice-coated dust grains residing outside of the water snow line is not prohibited by the aforementioned bouncing/fragmentation barriers. 

At the same time, some spatially resolved observations on Class II protoplanetary disks (e.g., \citealt{Perez2012ApJ...760L..17P,Tazzari2021MNRAS.506.2804T,Tsukagoshi2016ApJ...829L..35T}) also suggested that outside of the inner few tens of AU regions, the dust brightness temperatures appear lower than the expected dust temperatures.
Their interpretation for the low dust brightness temperatures was that the (sub)millimeter dust emission beyond the inner few tens of AU region is optically thin, and the optically thin components dominate the total flux densities as the cumulative surface area increase rapidly with radius.
Applying the optically thin assumption, the previous demographic surveys at (a single or multiple) (sub)millimeter wavelength(s) estimated that the median dust mass in the Class II protoplanetary disks is $\sim$10 times the Earth mass ($M_{\oplus}$) and suggested that the dust masses dissipate rapidly during the Class II stage (e.g., \citealt{Andrews_Williams_2005,Andrews_Williams_2007,Carpenter2014ApJ...787...42C,Ansdell2016ApJ...828...46A,Barenfeld2016ApJ...827..142B,Testi2016A&A...593A.111T,Akeson_2019,Babaian2019AAS...23410503B,Cazzoletti2019A&A...626A..11C,Williams2019ApJ...875L...9W,Testi2022A&A...663A..98T}).
Such low dust masses appear insufficient to feed the ubiquitously observed super-Earths (c.f., \citealt{Hasen2012ApJ...751..158H,Liu2020A&A...638A..88L}).
Therefore, it has been proposed that icy dust grains grow very rapidly and thus planet-formation occurs predominately during the embedded, Class 0/I stages of the YSOs (e.g., \citealt{Greaves2010MNRAS.407.1981G,Najita2014MNRAS.445.3315N}).
The present paradigms of planet-formation have been largely motivated by these observational studies (e.g., \citealt{Drazkowska2017A&A...608A..92D,Lichtenberg2021Sci...371..365L}).

However, \citet{Liu2016ApJ...816L..29L} and \citet{,Liu2017A&A...602A..19L} (see their appendix) observed a sample of accretion outburst YSOs and found that based on the commonly adopted, optically thin modified black body assumption, the gas masses inferred from the (sub)millimeter dust continuum observations appeared too low and thus may have a tension with the observed protostellar accretion rates.
They suggested that the majority of dust masses may be hidden in the inner disks where the optically thin assumption breaks down (see also the similar measurements in \citealt{Cieza2018MNRAS.474.4347C, Kospal2021ApJS..256...30K}).
\citet{Hartmann1998ApJ...495..385H} made a more general theoretical argument that the previously proposed disk gas and dust masses may be considerably underestimated; otherwise, it is hard to explain both the protostellar accretion rates and planet formation.
Potentially, both dust and gas masses might have been underestimated in the observational studies (see also the related discussion in \citealt{Pascucci2023ApJ...953..183P}, \citealt{Andrews2020ARA&A..58..483A} and references therein).

Moreover, the observations of polarized dust self-scattered light commonly indicated that $a_{\mbox{\scriptsize max}}$ is only $\sim$100 $\mu$m, which is 2--3 orders of magnitude smaller than the derivations based on the aforementioned SED analyses (e.g., \citealt{Kataoka2016ApJ...820...54K,Kataoka2016ApJ...831L..12K,Stephens2017ApJ...851...55S,Bacciotti2018ApJ...865L..12B,Hull2018ApJ...860...82H,Dent2019MNRAS.482L..29D}).
Such a large discrepancy in the estimates of $a_{\mbox{\scriptsize max}}$ indicates that the previous analyses of dust spectral energy distributions (SEDs) to derive dust masses and $a_{\mbox{\scriptsize max}}$ may be biased due to quoting incomplete physical formulation.
Alternatively, \citet{Lin2023MNRAS.520.1210L} argued that mm-sized dust with irregular grain morphology makes this a non-issue. 

\citet{Liu2019ApJ...877L..22L} realized that the inner HD~163296 disk (\citealt{Dent2019MNRAS.482L..29D}) presents anomalously low ($<$2.0) (sub)millimeter spectral indices, which is hard to explain with the modified black body formulation.
The anomalously low spectral indices may have been detected in other (sub)millimeter observations (e.g., \citealt{Andrews_Williams_2005,Andrews_Williams_2007}; see Figure \ref{fig:target}). 
Another evolved disk, MP Mus, was also reported spectral index $<$2.0 in the inner 30 AU in the ALMA observations (\citealt{Ribas2023A&A...673A..77R}).
Some of these anomalously low (sub)millimeter spectral indices had been overlooked or had been incorrectly attributed to data calibration errors and/or imaging artifacts.
In the multi-wavelength study on the TW~Hya disk, \citet{Macias2021A&A...648A..33M} interpreted the anomalously low (sub)millimeter spectral index by the confusion of free-free emission, and suggested that the $a_{\mbox{\scriptsize max}}$ can be either $>$1 mm or $<$50 $\mu$m.
Alternatively, \citet{Liu2019ApJ...877L..22L} hypothesized that such anomalously low spectral indices may signify the effects of dust self-scattering in the optically thick disks if $a_{\mbox{\scriptsize max}}$ is around $\sim$100 $\mu$m, and suggested that this is the only probable interpretation for the 145--225 GHz observations in the innermost region of the TW~Hya disk. 
The interpretations for the HD~163296 and TW~Hya observations with the effects of dust self-scattering have been supported by follow-up observational studies (\citealt{Ueda2020ApJ...893..125U,Guidi2022A&A...664A.137G}).

The new interpretation incorporating the effects of dust self-scattering does not only resolve the tension between the derivations of $a_{\mbox{\scriptsize max}}$ based on analyzing polarized dust self-scattered light and dust SEDs.
In addition, in this case, the dust scattering attenuates the observed brightness temperature to considerably lower than the dust temperature, making a very optically thick disk look optically thin.
Some previous observational studies might incorrectly adopt an optically thin assumption and thereby were subject to a $\gtrsim$1 orders of magnitude underestimate of dust masses.
A comprehensive parameter survey of \citet{Zhu2019ApJ...877L..18Z} based on numerical radiative transfer simulations has arrived a similar conclusion.

\begin{figure}
\includegraphics[width=8.5cm]{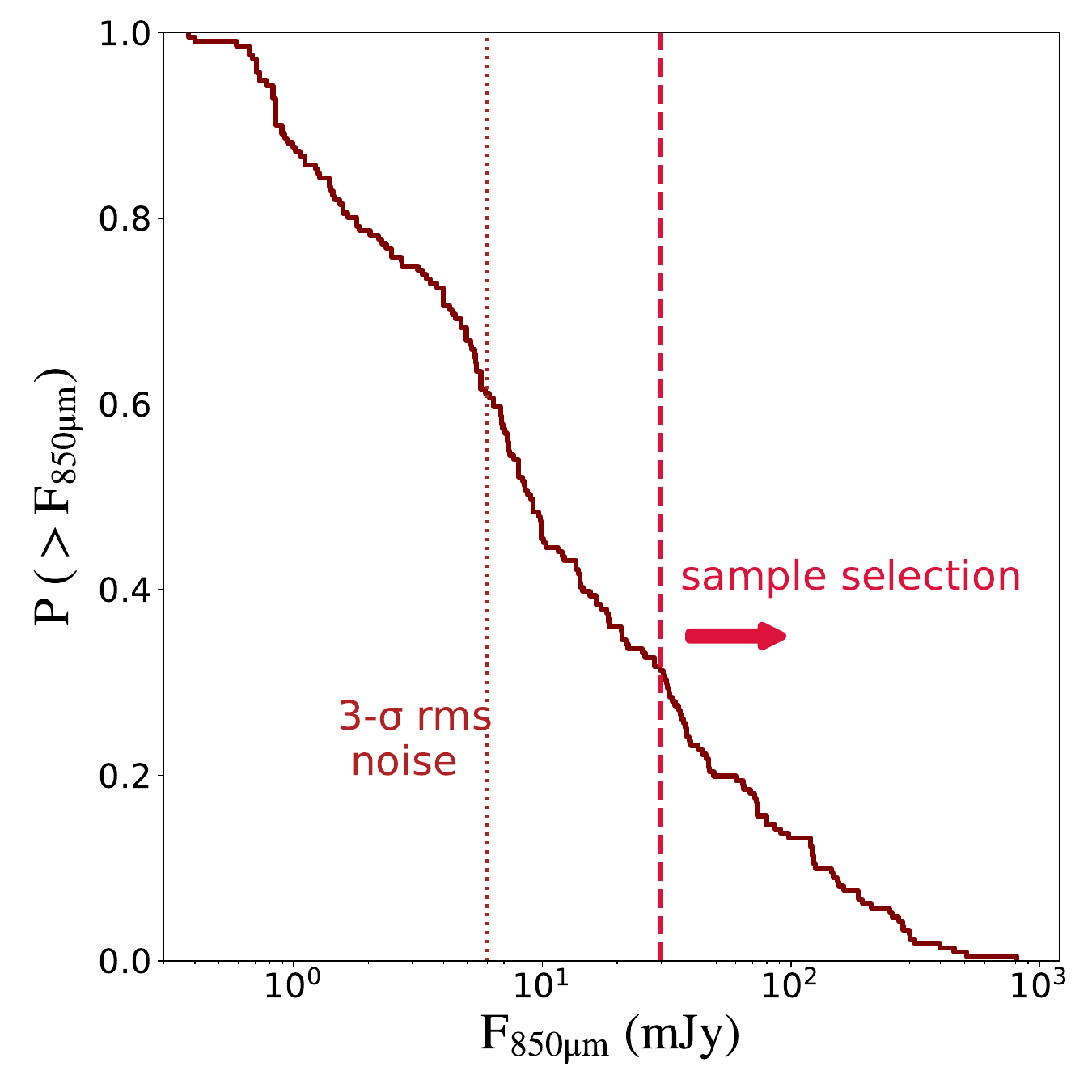}
\caption{The cumulative distribution of 850 $\mu$m flux density of 211 Class II disks in Taurus-Auriga region. The 850 $\mu$m flux densities were estimated from the 1.3 mm flux densities quoted from \citet{Akeson_2019} assuming $\alpha = 2.0$. The red dashed line labels our sample selection criterion (Section \ref{sub:source}). The brick-red dotted line labels the 3-$\sigma$ rms noise at 337 GHz over 12 GHz bandwidth achieved by our observations (Section \ref{sub:obs}). 
}
\label{fig:totalClassII}
\end{figure}

At this moment, how common the optically thick (at (sub)millimeter wavelengths) disks are and how common the anomalously low spectral indices are, have not yet been constrained by systematic surveys. 
This has been largely caused by the uncertainties in the previously measured spectral indices. 
In particular, the anomalously low spectral indices can only be detected with precise measurements of flux densities. 
Since dust self-scattering makes the spectral index a rather complicated function of frequency, measuring spectral indices with a sparse sampling on frequency domain (e.g. $\sim$100 GHz separations) will erase important spectral information through averaging the possible spectral index variation in the range of a few hundreds GHz (\citealt{Liu2019ApJ...877L..22L}). 
To correctly derive dust masses and $a_{\mbox{\scriptsize max}}$, we also need to avoid the frequency-smearing of spectral indices that is caused by coarsely sampling flux densities over the frequency domain.  

\begin{deluxetable*}{cccccccc}
\tabletypesize{\footnotesize}
\tablecolumns{9}
\tablecaption{
Properties of the Host Protostars and the (Sub)millimeter Disk
\label{tab:targ_properties}}
\tablehead{ 
\colhead{Source} & \colhead{2MASS name} & \colhead{Distance} & \colhead{Spectral Type} & \colhead{Multiplicity} & \colhead{Separation} & \colhead{$R_{\mbox{\tiny mm, disk}}$} & \colhead{Group} \\
\colhead{} & \colhead{} & \colhead{(pc)} & \colhead{} & \colhead{} & \colhead{($''$)} & \colhead{($''$)} & \colhead{} }
\colnumbers
\startdata 
04113+2758 A  & J04142626+2806032 & $134$ & M2.5 & B & 4.01 & 0.51\tablenotemark{\scriptsize a} & 1 \\
04113+2758 B  & J04142626+2806032 & $134$ & M2.5 & SB & 4.01 & 0.54\tablenotemark{\scriptsize a} & 1 \\
04278+2253  & J04305028+2300088 & $125$ & F1 &   &   &   & 4 \\
AA Tau  & J04345542+2428531 & $135$ & M0.6 & S &   & 0.69\tablenotemark{\scriptsize a} & 3 \\
AB Aur  & J04554582+3033043 & $156$ & A1.0 & S &   & 1.44\tablenotemark{\scriptsize b} & 6 \\
BP Tau  & J04191583+2906269 & $127$ & M0.5 & S &   & 0.321\tablenotemark{\scriptsize c}  & 2 \\
CIDA-7  & J04422101+2520343 & $141$ & M5.1 & S &   & 0.154\tablenotemark{\scriptsize d} & 5 \\
CIDA-9 A/B  & J05052286+2531312 & $175$ & M1.8 & B & 2.34 & 0.371\tablenotemark{\scriptsize c}  & 6 \\
CI Tau  & J04335200+2250301 & $160$ & K5.5 & S &   & 1.195\tablenotemark{\scriptsize c} & 4 \\
CW Tau  & J04141700+2810578 & $132$ & K3.0 & S &   & 0.366\tablenotemark{\scriptsize e} & 1 \\
CY Tau  & J04173372+2820468 & $126$ & M2.3 & S &   & 0.56\tablenotemark{\scriptsize a} & 1 \\
CoKu Tau 1  & J04185147+2820264 & $115$ & M0 &   &   &   & 1 \\
DD Tau AB  & J04183112+2816290 & $127$ & M4.8 & B & 0.55 &   & 1 \\
DE Tau  & J04215563+2755060 & $128$ & M2.3 & S &   &   & 2 \\
DH Tau A/B  & J04294155+2632582 & $133$ & M2.3 & B & 2.34 & 0.146\tablenotemark{\scriptsize c} & 3 \\
DK Tau AB  & J04304425+2601244 & $132$ & K8.5 & B & 2.36 & 0.117\tablenotemark{\scriptsize c} & 3 \\
DL Tau  & J04333906+2520382 & $160$ & K5.5 & S &   & 1.033\tablenotemark{\scriptsize c} & 3 \\
DM Tau  & J04334871+1810099 & $144$ & M3.0 & S &   & 1.227\tablenotemark{\scriptsize d} & 4 \\
DN Tau  & J04352737+2414589 & $129$ & M0.3 & S &   & 0.313\tablenotemark{\scriptsize c} & 5 \\
DO Tau  & J04382858+2610494 & $139$ & M0.3 & S &   & 0.263\tablenotemark{\scriptsize c} & 5 \\
DQ Tau  & J04465305+1700001 & $195$ & M0.6 & SB &   & 0.219\tablenotemark{\scriptsize c} & 4 \\
DR Tau  & J04470620+1658428 & $193$ & K6 & S &   & 0.276\tablenotemark{\scriptsize c} & 4 \\
DS Tau  & J04474859+2925112 & $158$ & M0.4 & S &   & 0.446\tablenotemark{\scriptsize c} & 5,6 \\
FM Tau  & J04141358+2812492 & $132$ & M4.5 &   &   & 0.55\tablenotemark{\scriptsize a} & 1 \\
FT Tau  & J04233919+2456141 & $130$ & M2.8 & S &   & 0.357\tablenotemark{\scriptsize c} & 2 \\
FV Tau AB  & J04265352+2606543 & $136$ & M0.0 & B & 0.7 &   & 2 \\
FY Tau  & J04323058+2419572 & $130$ & M0.1 &   &   &   & 3 \\
GI Tau  & J04333405+2421170 & $130$ & M0.4 & B & 13.2 & 0.190\tablenotemark{\scriptsize c} & 3 \\
GM Aur  & J04551098+3021595 & $158$ & K6.0 & S &   & 1.114\tablenotemark{\scriptsize f} & 6 \\
GO Tau  & J04430309+2520187 & $142$ & M2.3 & S &   & 1.187\tablenotemark{\scriptsize c} & 5 \\
HO Tau  & J04352020+2232146 & $164$ & M3.2 & S &   & 0.267\tablenotemark{\scriptsize c} & 5 \\
HV Tau C  & J04383528+2610386 & $138$ & M4.1 & T &   & 0.37\tablenotemark{\scriptsize a} & 5 \\
Haro 6-37 C/AB  & J04465897+1702381 & $195$ & K8.0 & T & 2.66 & 0.97\tablenotemark{\scriptsize a} & 4 \\
Haro 6-39  & J04520970+3037454 & $127$ &   &   &   &   & 6 \\
IC 2087 IR  & J04395574+2545020 & $145$ &   &   &   &   & 5,6 \\
IP Tau  & J04245708+2711565 & $129$ & M0.6 & S &   & 0.28\tablenotemark{\scriptsize c} & 2 \\
IQ Tau  & J04295156+2606448 & $132$ & M1.1 & S &   & 0.838\tablenotemark{\scriptsize c} & 3 \\
LkCa 15  & J04391779+2221034 & $157$ & K5.5 & S &   & 0.707\tablenotemark{\scriptsize g} & 4 \\
RW Aur AB  & J05074953+3024050 & $156$ & K0 & B & 1.42 & 0.132\tablenotemark{\scriptsize c} & 6 \\
RY Tau  & J04215740+2826355 & $138$ & G0 & S &   & 0.509\tablenotemark{\scriptsize c} & 2 \\
SU Aur  & J04555938+3034015 & $157$ & G4 & S &   & 0.49\tablenotemark{\scriptsize a} & 6 \\
T Tau N/Sab  & J04215943+1932063 & $145$ & K0 & T & 0.68 & 0.143\tablenotemark{\scriptsize c} & 4 \\
UX Tau A/C  & J04300399+1813493 & $142$ & K0.0 & Q & 2.67 & 0.402\tablenotemark{\scriptsize g} & 4 \\
UY Aur AB  & J04514737+3047134 & $152$ & K7.0 & B & 0.84 & 0.044\tablenotemark{\scriptsize c} & 6 \\
UZ Tau E/Wab  & J04324303+2552311 & $130$ & M1.9 & Q & 3.54 & 0.667\tablenotemark{\scriptsize c} & 3 \\
V710 Tau A/B  & J04315779+1821380 & $146$ & M1.7 & B & 3.22 & 0.317\tablenotemark{\scriptsize c} & 4 \\
V836 Tau  & J05030659+2523197 & $167$ & M0.8 & S &   & 0.188\tablenotemark{\scriptsize c} & 6 \\
V892 Tau  & J04184061+2819155 & $134$ & A0 & S &   & 0.46\tablenotemark{\scriptsize h} & 1 \\
\enddata
\tablecomments{The source name in Column 1 is adopted from the name in \citet{Andrews_Williams_2005}. 
The two or more components are labelled if they were resolved and detected by high-resolution studies (\citet{Akeson_2014}, \citet{Akeson_2019}, \citet{Long2019ApJ...882...49L}). 
The slash symbol which separates the multiplicity components means that their separation is large enough to be resolved, while in these systems we only detected the former component except for 04113+2758. 
The distance of the source in Column 4 are quoted from Gaia DR3 (\citealt{Gaia_2016,Gaia_2023}). 
The spectral type of the host protostar in Column 5 is quoted from \citet{Herczeg2014ApJ...786...97H}. 
The multiplicity of the source and separation of binary in Column 6, 7 are quoted from \citet{White2001ApJ...556..265W}, \citet{Itoh2005ApJ...620..984I}, \citet{Kraus2009ApJ...704..531K}, \citet{Kraus2011ApJ...731....8K}, \citet{Akeson_2014}, and \citet{Akeson_2019}. 
Multiplicity: S = Single, B = Binary, SB = spectroscopic binary, T = Tertiary, Q = Quaternary. 
The disk radius is the $R_{\mbox{\tiny 95\%}}$ or the $R_{\mbox{\tiny 90\%}}$. $^{a}R_{\mbox{\tiny 90\%}}$ measured from disk images in ALMA archival data. $^{b}R_{\mbox{\tiny 90\%}}$ quoted from \citet{Stapper2022}. $^{c}R_{\mbox{\tiny 95\%}}$ quoted from \citet{Long2019ApJ...882...49L}. $^{d}R_{\mbox{\tiny 90\%}}$ quoted from \citet{Long2022ApJ...931....6L}. $^{e}R_{\mbox{\tiny 95\%}}$ quoted from \citet{Ueda_2022}. $^{f}R_{\mbox{\tiny 95\%}}$ quoted from \citet{Sierra2021ApJS..257...14S}. $^{g}R_{\mbox{\tiny 90\%}}$ coverted from $R_{\mbox{\tiny 68\%}}$ in \citet{Parker2022MNRAS.511.2453P} by the $R_{\mbox{\tiny 90\%}}-R_{\mbox{\tiny 68\%}}$ relation given in \citet{Hendler2020ApJ...895..126H}. $^{h}R_{\mbox{\tiny 90\%}}$ quoted from \citet{Long2021ApJ...915..131L}. }
\end{deluxetable*}

In this paper, we present a new Submillimeter Array (SMA) survey with a densely sampled (at 10--40 GHz intervals) 200--400 GHz SED towards a sample of 47 Class II sources in the Taurus-Auriga region. 
With the more accurate measurements on the spectral indices, we aim to provide new insight on the globally-averaged $a_{\mbox{\scriptsize max}}$ and dust column density of Class II disks. 
This statistical study will help distinguish the optically thick picture from optically thin picture and address the aforementioned two crucial questions: (1) whether the dust mass budget is sufficient to feed planet-formation in Class II disks. (2) whether the dust growth can overcome the 0.1mm size barrier outside the water snow line. 
The details of our observations and data calibrations are provided in Section \ref{sec:observation} and Section \ref{sec:reduction}. 
The results are presented in Section \ref{sec:result}.
Our tentative interpretation of the observational results and the physical implication of our interpretation are discussed in Section \ref{sec:discussions}.
The conclusion is in Section \ref{sec:conclusion}.
Appendix \ref{appendix:popsynthesis} introduces how we perform population syntheses for dust SED based on simple parametric models of radial dust temperature and column density distributions.
Appendix \ref{appendix:dustmass} provides the details of how we estimated dust masses in the individual sources.
We adopted the DSHARP dust opacity table (\citealt{Birnstiel2018ApJ...869L..45B}) in the discussion about dust spectral indices and dust masses. 
Our rationale for adopting this opacity table is discussed in Appendix \ref{appendix:diana}.
We provide some necessary comments on some individual target sources in Appendix \ref{appendix:source}.
Our survey strategies are presented in \ref{appendix:group}.
Some specific minor issues of our observations are commented in Section \ref{appendix:obs_issue}.
Appendix \ref{appendix:selcal} provides examples of our gain-phase self-calibration solutions.
Appendix \ref{appendix:image} summarizes the qualities of our SMA images for individual target sources at certain representative observing frequencies.

\section{Observations}\label{sec:observation}
\subsection{Target selection}\label{sub:source}
\cite{Andrews_Williams_2005} reported the continuum survey on 153 young stellar objects in Taurus-Auriga star-forming region at $350 \mu$m, $450 \mu$m, $850 \mu$m, and $1.3$mm, from which we selected the brighter Class II disks with $F_{850 \mu m} > 30$ mJy.
For the sources that were only detected by \cite{Andrews_Williams_2005} at $\sim$1.3 mm but not at $850 \mu$m, we estimated the $F_{850\mu m}$ based on the $F_{\mbox{\scriptsize 1.3mm}}$ measurement assuming a spectral index of $\alpha = 2.5$. 
Our selection criteria are to allow 
$\gtrsim 10$-$\sigma$ significance in every 12 GHz bandwidth within realistic observing time.
As we will show in Section \ref{sec:result}, the spectral indices of the selected samples are very close to 2.0.
We excluded GG~Tau from our sample due to its multiplicity and yielded 47 selected targets.

The left panel of Figure \ref{fig:target} shows $F_{850 \mu m}$ and $F_{\mbox{\scriptsize 1.3mm}}$ of the selected sample; the right panel shows the cumulative distribution of the spectral indices reported in \cite{Andrews_Williams_2005} which were derived based on the flux density measurements $F_{350 \mu m}$, $F_{450 \mu m}$, $F_{850 \mu m}$, and $F_{\mbox{\scriptsize 1.3mm}}$.
The mean value of these previously reported spectral indices was 2.1; over 50\% of the sample shows lower than 2.0 spectral indices.
Among the 211 Class~II objects covered by the ALMA survey toward the Taurus-Auriga region (\citealt{Akeson_2019}), our selected sample represent the top 32 percentile $F_{850 \mu m}$ bright sources (Figure \ref{fig:totalClassII}).
Table \ref{tab:targ_properties} summarizes properties of the host stars and the disk properties that were constrained from the higher angular resolution ALMA observations (for more detail see the table note in Table \ref{tab:targ_properties}). 

\subsection{Technical details and calibration strategy}\label{sub:obs}

We carried out the SMA survey towards the selected sample in 2021, between September and November. 
The observing dates and some technical details are summarized in Table \ref{tab:obs_summary}. 

The observations were conducted in the subcompact or compact array configurations, resulting in the {\it uv} distance ranges of $\sim$5--105 k$\lambda$.
The corresponding angular resolutions are 1\farcs8--3\farcs7 ($\sim$252--518 AU).
Since most of the dusty disks in our sample have $< 250$ AU diameters, our observations are not subject to missing short-spacing.

Our observations employed the dual-receivers mode.
Each receiver observed a single polarization, and the obtained signal was mixed with that of an independent local oscillator (LO).
For each receiver, the SMA correlator capability allows simultaneously covering the 12 GHz intermediate frequency (IF) ranges in both the lower and upper sidebands (LSB, USB).
We tuned the two independent LOs for the receivers to different frequencies to achieve extended frequency coverages.

There were six tracks of observations in total: two of them employed the receiver tunings that the central frequency is around 230 GHz; one of them employed the tuning that the central frequency is around 270 GHz; and three of them employed the tuning that the central frequency is around 400 GHz (Table \ref{tab:obs_summary}).
Combining these observations yields a $\sim$200--420 GHz frequency coverage which avoids the regions of poor atmospheric transmission. 

We adaptively allocated on-source integration time
to ensure robust flux density measurements for all of them at all observing frequencies.
Specifically, the allocated on-source time allowed for a signal-to-noise ratio (SNR) of at least 10-$\sigma$ within every 12 GHz bandwidth.
For any source, the minimum on-source time was 5 minutes in case of any momentary technical issues that could result in data loss. 

We divided the 47 selected targets into six groups according to their proximity to one another and their angular separation from the two selected complex gain calibrators, J0510+180 and J0418+380 (see Appendix \ref{appendix:group}).
We additionally adopted 3C84 as an ancillary gain calibrator in case J0510+180 or J0418+380 were not bright enough at some observing frequencies (see Appendix \ref{appendix:group}).
The grouping was made according to the minimum spanning tree algorithm which at the same time allowed us to find the shortest slewing path.
In our target source loops, each target-source scan had a 12 minutes duration and observed only one of the six groups of sources; each target-source scan was sandwiched by two $\sim$3 minutes' scans on the complex gain calibrator. 
In other words, the duration of our calibration duty cycle is 15 minutes, slightly shorter than the typical SMA observations ($\sim$20 minutes). 
We observed the absolute flux (Solar system planets) and passband calibrators (bright quasars) at the beginning and/or end of each track of observations.
Using this survey strategy, over a night of observations (4--6 hours in the target source loop), each of the 6 target sources groups could be visited at least 3 times.
The visits were well separated in time, which helps achieve a good {\it uv} coverage.
Spreading the scans on every target source over a night also helps lower the risk of entirely losing the observations on certain target sources at certain frequencies due to tentative weather or instrumental issues.
Visiting each target source more frequently will lead to a high slewing time, which will result in higher thermal noise.

\section{Data Reduction}\label{sec:reduction}

\subsection{Routine Calibration}\label{sub:basic}

We performed the basic SMA data calibration using the MIR IDL software package (\citealt{Qi2003cdsf.conf..393Q}).
Below we outline some specific issues of the observations and our specific treatments to achieve accurate absolute flux calibrations.
 
The T$_{\mbox{\scriptsize sys}}$ were taken from the wide-band measurements that have been averaged over the spectral domain. 
In the observations of tracks 230 GHz-2, 270 GHz-3 and 400 GHz-6 (Table \ref{tab:obs_summary}), the baseline-based T$_{\mbox{\scriptsize sys}}$ information for Antenna 1 was corrupted due to malfunction of the instrument. 
To permit re-scaling the electronic amplitude signal of Antenna 1 to Jansky (Jy) units, we adopted the Tsys measurement of Antenna 2 for Antenna 1 as the two are closely spaced during these observations. 

In track 400 GHz-4, 400 GHz-5 and 400 GHz-6, the 391.5--395.5 GHz and 419.5--423.5 GHz passbands exhibited high phase and amplitude noises which were likely due to the poor receiver response or the low atmospheric transition. 
In light of this, we discarded these passbands during the post processing.

We derived the passband calibration solutions based on the observations on 3C84 except for the track 270 GHz--3.
To accurately determine the inter-band flux density ratios, we derived and applied the passband amplitude solutions for small ranges of elevation.
This approach allows us to take into consideration the elevation-dependent effects of the frequency-dependent atmosphere transmission (e.g., due to the atmospheric absorption lines).
This strategy may not be necessary in the future, when the spectrally resolved  T$_{\mbox{\scriptsize sys}}$ becomes available to the general SMA users. 

In track 270 GHz-3, we observed over-corrected passband amplitudes in the high and low frequency edges of the 254--258 GHz, 282--290 GHz, and 314--318 GHz passbands.
Due to the thermal noise, we could not identify the exact causes of the amplitude over-corrections.
Therefore, we discarded these passbands in the subsequent data calibrations and did not use these passbands in any scientific analyses.

\begin{figure}[]
    \hspace{-0.8cm}
    {\renewcommand{\arraystretch}{2.5} 
    \begin{tabular}{c}
    \includegraphics[width=9.3cm]{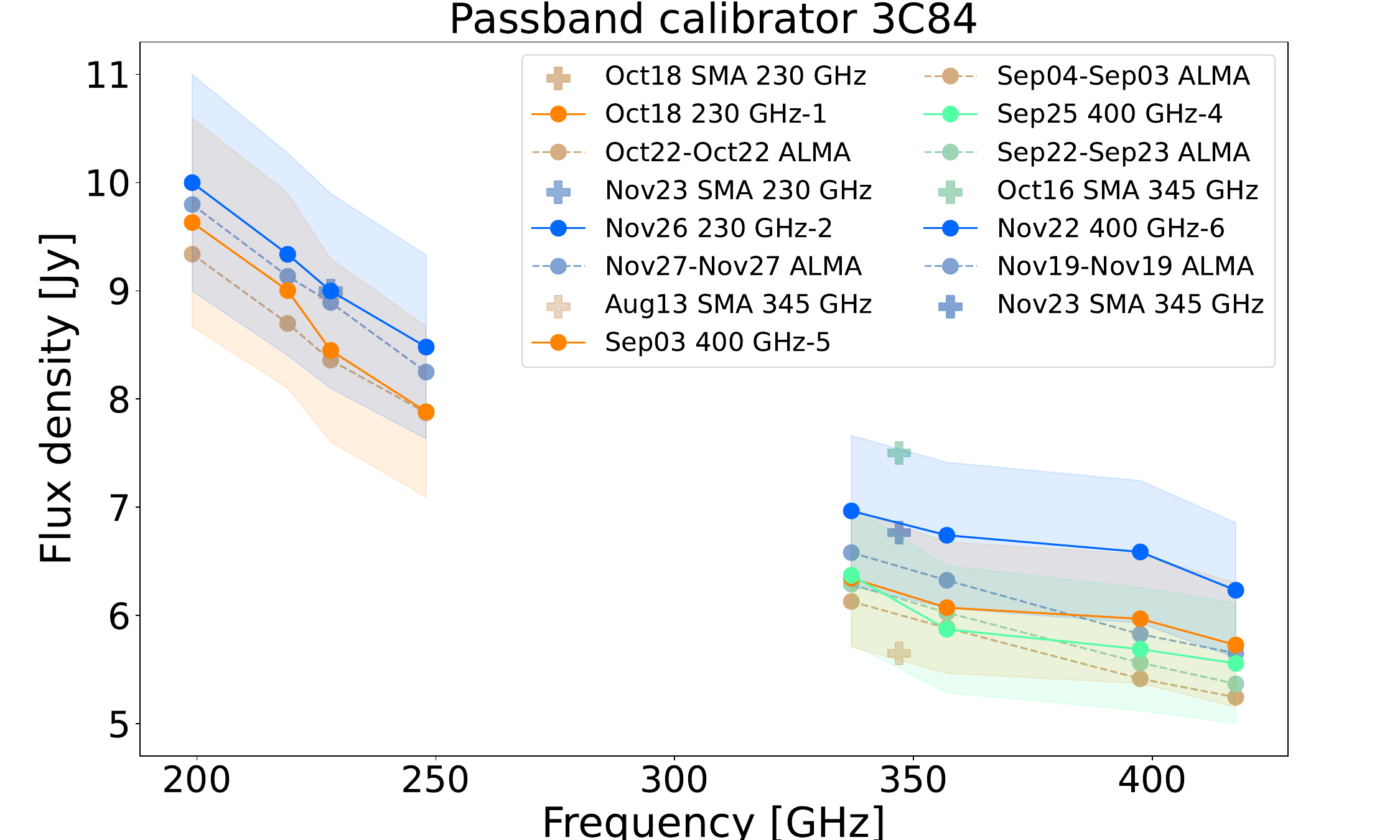} \\
    \includegraphics[width=9.3cm]{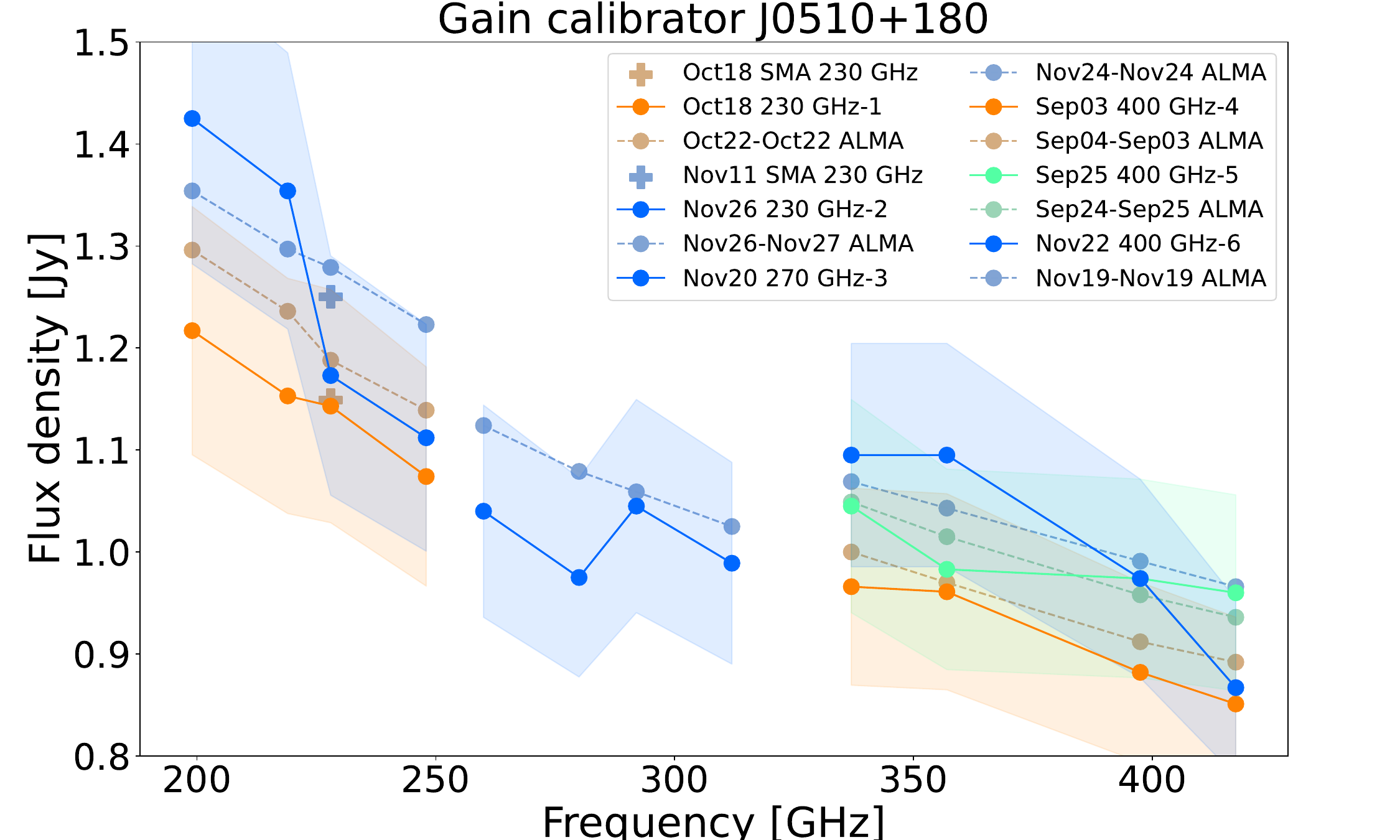} \\
    \includegraphics[width=9.3cm]{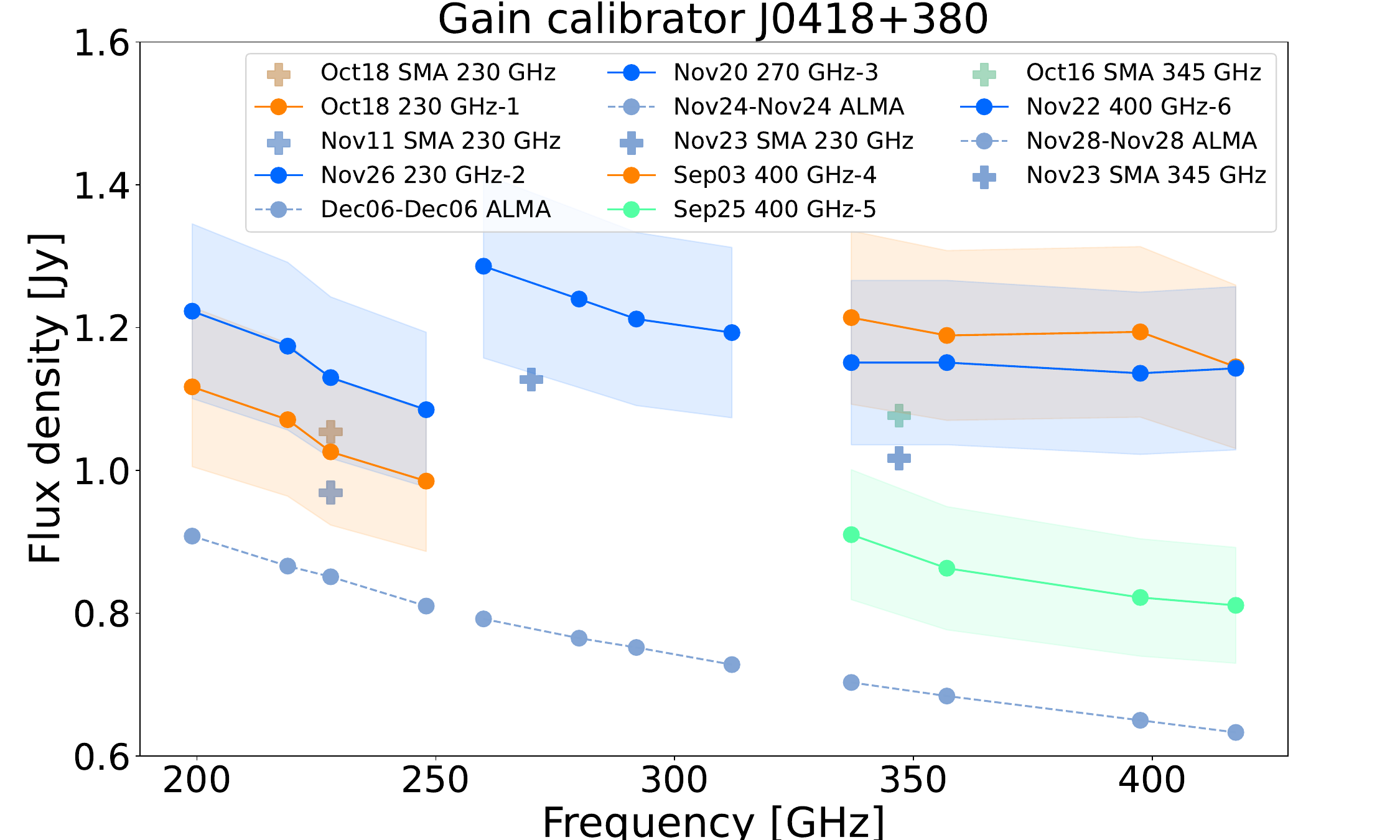}
    \end{tabular}
    }
    \caption{
    The measurements of the flux densities of our passband calibrator and gain calibrators.
    We include the measurements from our own data calibration procedures (Section \ref{sub:basic}) and the records quoted from the SMA Calibrator List and the ALMA calibrator database for comparisons. 
    The circular dots and straight lines show our measurements, and the shaded region shows the 10\% uncertainty; the dashed line are the calibrator flux densities interpolated from the ALMA calibrator database; the plus symbols are the measurements quoted from the SMA Calibrator List. 
    The observing dates are color-coded.
    The months and dates (UTC) are labeled in the legends.
    The ALMA measurements are labeled with two dates that are separated by a hyphen: the former and the latter are the dates of the Band 3 and 7 measurements, respectively.
    }
    \label{fig:calibrator_flux_3c84} 
\end{figure}

\begin{figure*}[]
    \centering
    \includegraphics[width=15cm]{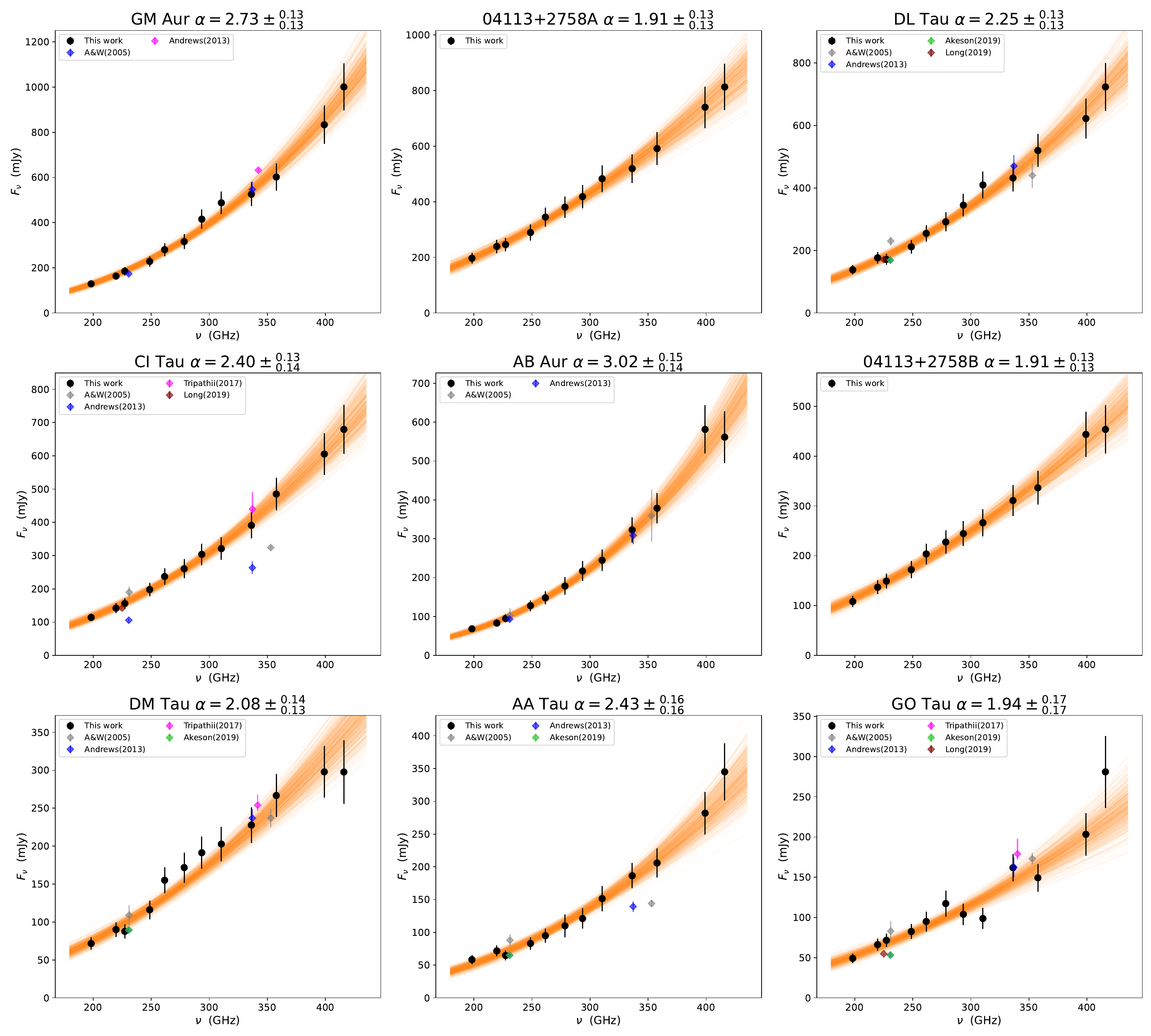}
    \caption{
    The 200--420 GHz SEDs of the 7 spatially resolved disks and the resolved binary. Note that the 415.5 GHz flux density of these 9 specific sources have been discarded (Section \ref{sub:flux}). Black points show the flux densities measured from our observations (Section \ref{sub:flux}). 
    Vertical error bars ($\pm$1-$\sigma$) on our measurements are calculated by error propagating the $\pm$1-$\sigma$ errors returned from our image or visibility domain fittings (Section \ref{sub:flux}) and the 10\% nominal absolute flux calibration errors. 
    Colored symbols are the flux densities quoted from literature (\citealt{Andrews_Williams_2005,Andrews_2013,Akeson_2014,Tripathi_2017,Akeson_2019,Long2019ApJ...882...49L}). 
    For binary sources, we quoted the summed flux densities of the two components if our observations did not resolve the binarity. 
    The derived $\alpha_{\mbox{\scriptsize 200-420 GHz}}$ for each source is labeled on the top of each panel. 
    The orange lines in each panel are 50 random draws of the fitting parameters extracted from the MCMC samplings. }
    \label{fig:SED_0510_resolved}
\end{figure*}

For each track of observations, we first derived the absolute flux calibration solutions for each sideband  based on the scans on Uranus or Neptune.
We have visually inspected the spectra of Uranus or Neptune to ensure that there were no strong emission lines.
After applying these absolute flux calibration solutions, we looked up the tabulated flux densities of our selected passband and complex gain calibrators in the SMA Calibrator List\footnote{http://sma1.sma.hawaii.edu/callist/callist.html} which were measured on the dates that were adjacent to our observations.
We also looked up the ALMA's grid source survey\footnote{https://almascience.eso.org/alma-data/calibrator-catalogue} and considered only the records where the Bands 3 and 7 measurements were taken close to our observing dates. 
We used the {\tt getALMAFlux} task in the CASA analysis utilities package to interpolate to the observing frequencies and the observation date assuming a constant quasar spectral index between the Bands 3 and 7 measurements. 

Figure \ref{fig:calibrator_flux_3c84} summarizes the comparisons of the bootstrapped flux densities of the calibrators from our own SMA observations with those tabulated flux densities. 
We note that our own SMA observations were single-polarization observations, while those tabulated flux densities were measured with dual linear polarizations.
The intrinsic polarization of the calibrators led to the apparent amplitude jumps in our bootstrapped flux densities but not in the tabulated SMA and ALMA measurements (Figure \ref{fig:calibrator_flux_3c84}), which does not impact our absolute flux accuracy.
In most cases, our bootstrapped flux densities of the calibrators were different from those tabulated values by less than 10\%, which can be attributed to the typical inherent absolute flux errors in both the SMA and ALMA observations. 
However, we visually identified that there might be significant antenna pointing errors in track 230 GHz-2 (Table \ref{tab:obs_summary}) during the observations on Uranus and BL~Lac but not during the observations on the other calibrators and target sources. 
Therefore, the absolute flux calibration solutions in track 230 GHz-2 cannot be applied to the target sources reliably.
For track 230-2, we set the flux densities of 3C84 to the tabulated SMA measurements on November 23 and derived the absolute flux calibration solutions based on the scans on 3C84.
We confirmed that in this way, the flux densities of our target sources derived from track 230 GHz-1 and track 230 GHz-2 are barely distinguishable. 

\begin{figure*}[ht!]
    \centering
    \includegraphics[width=15cm]{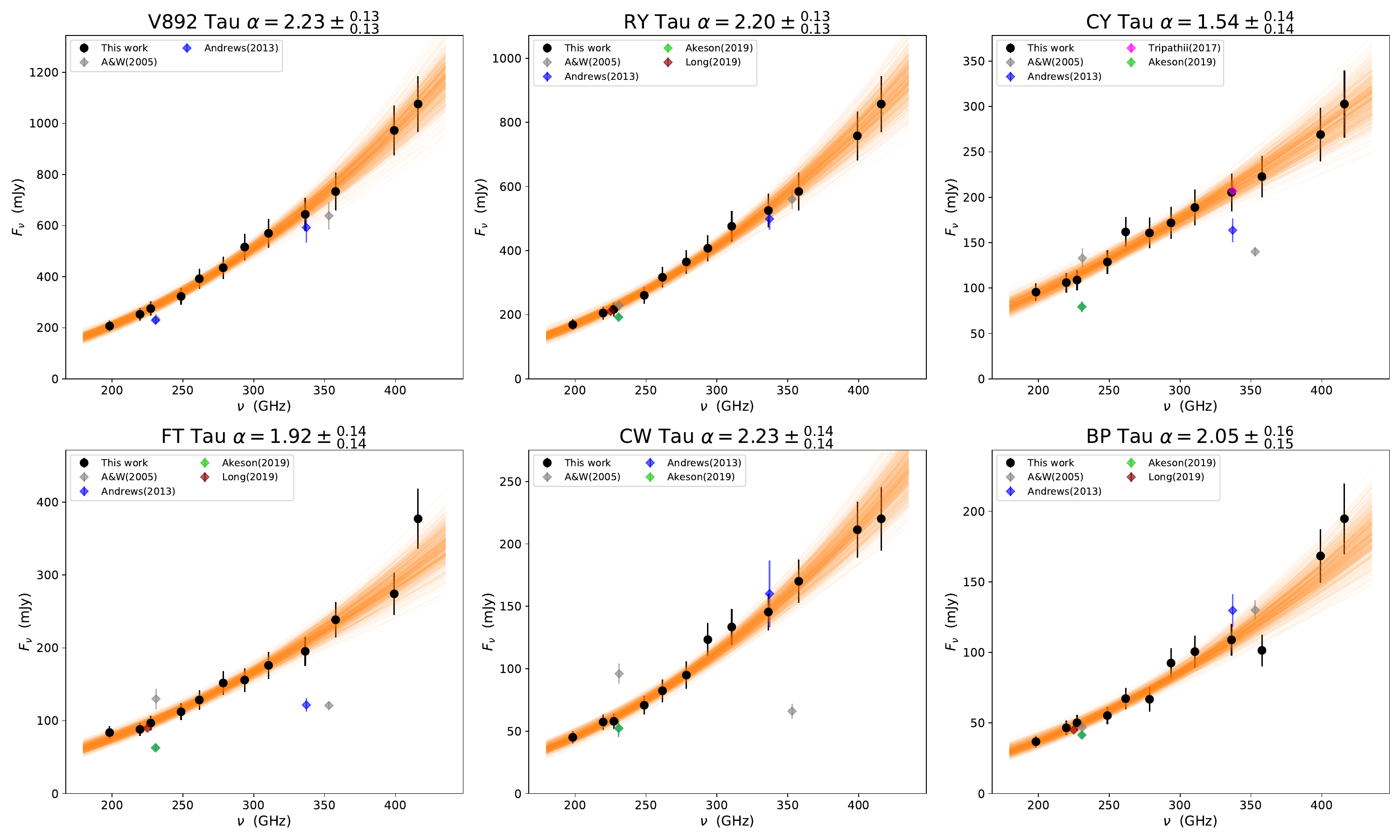} 
    \includegraphics[width=15cm]{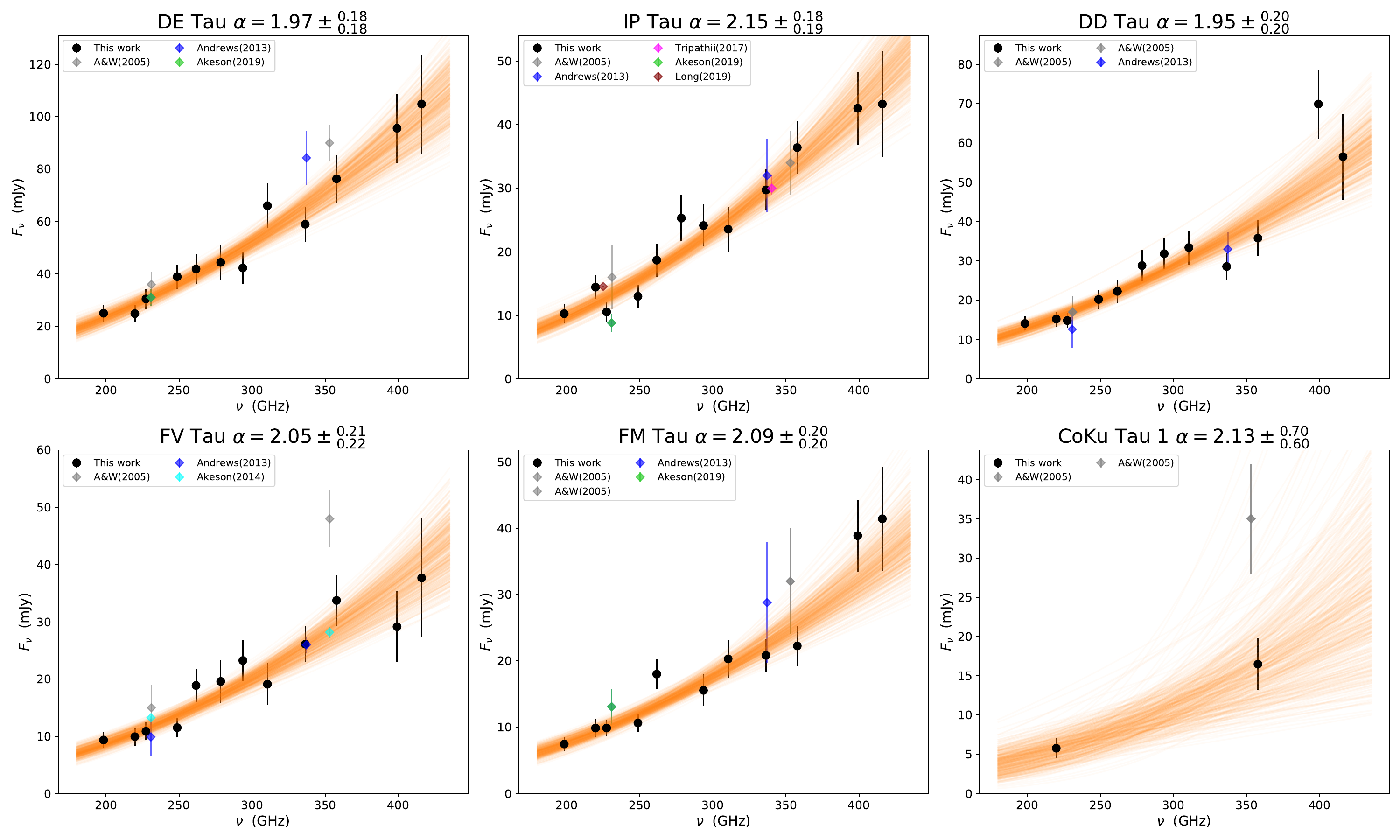}
    \caption{
    The 200--420 GHz SEDs of the 12 spatially unresolved sources that were calibrated by the gain calibrator J418+380. The style of the presentation is the same as that of Figure \ref{fig:SED_0510_resolved}. The sources are arranged in the descending order of their $F_{\mbox{\scriptsize 337 GHz}}$.
    }
    \label{fig:SED_0418}
\end{figure*}

\begin{figure*}[]
    \centering
    \includegraphics[width=15cm]{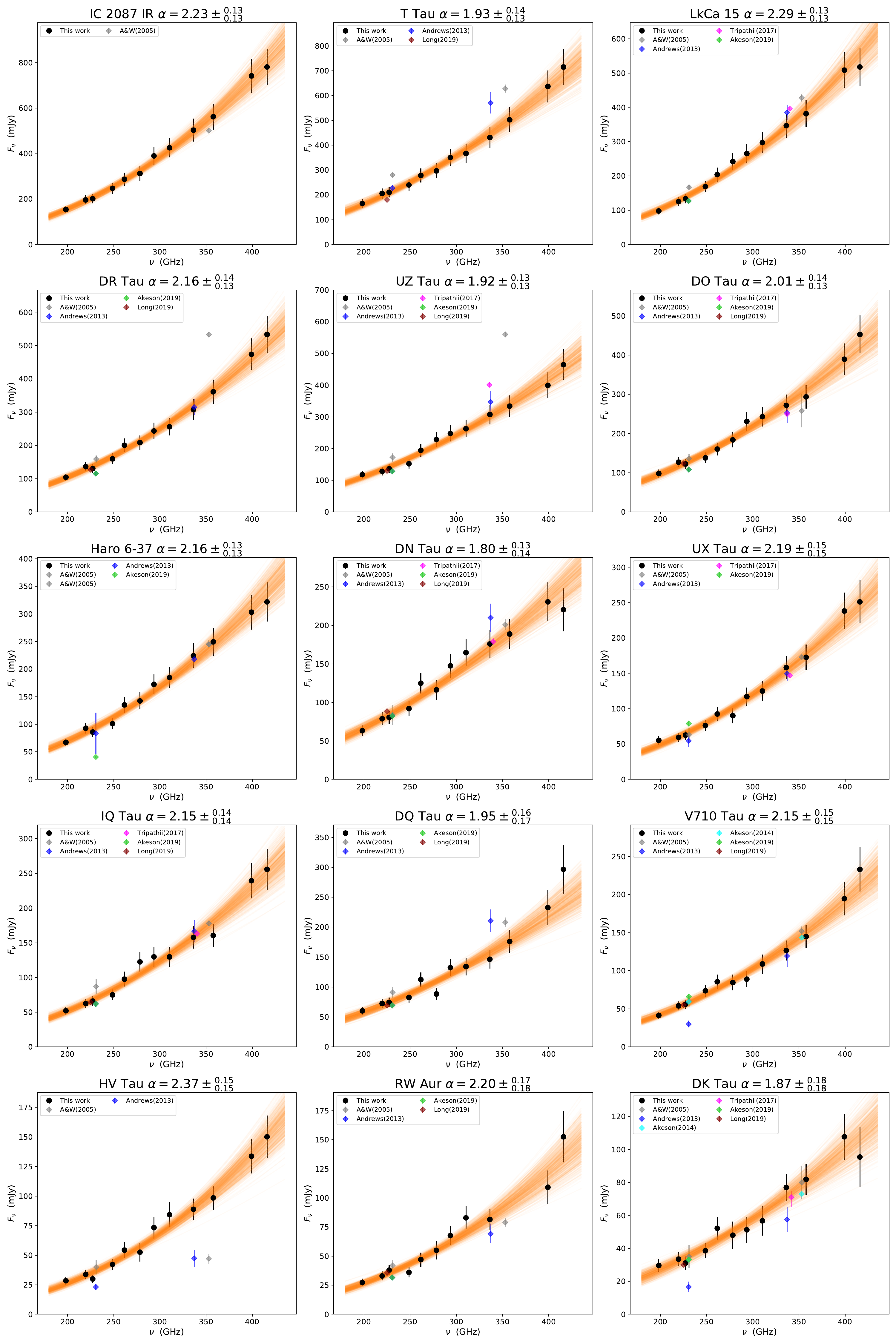} 
    \caption{
    The 200--420 GHz SEDs of the 15 spatially unresolved sources that were calibrated by the gain calibrator J0510+180. The style of the presentation is the same as that of Figure \ref{fig:SED_0510_resolved}. The sources are arranged in the descending order of their $F_{\mbox{\scriptsize 337 GHz}}$. 
    }
    \label{fig:SED_0510_0}
\end{figure*}

We derived the gain phase and amplitude calibration solutions primarily based on the observations on the quasars J0510+180 and J0418+380.
However, since these two gain calibrators were subject to poor signal-to-noise ratios (SNRs) in the 400-GHz tuning and phase incoherence, we used the brighter quasar 3C84 as an auxiliary gain calibrator. 
We removed the residual phase errors based on the gain phase self-calibrations (for more details see Section \ref{sub:selcal}).

\begin{figure*}
    \centering
    \includegraphics[width=15cm]{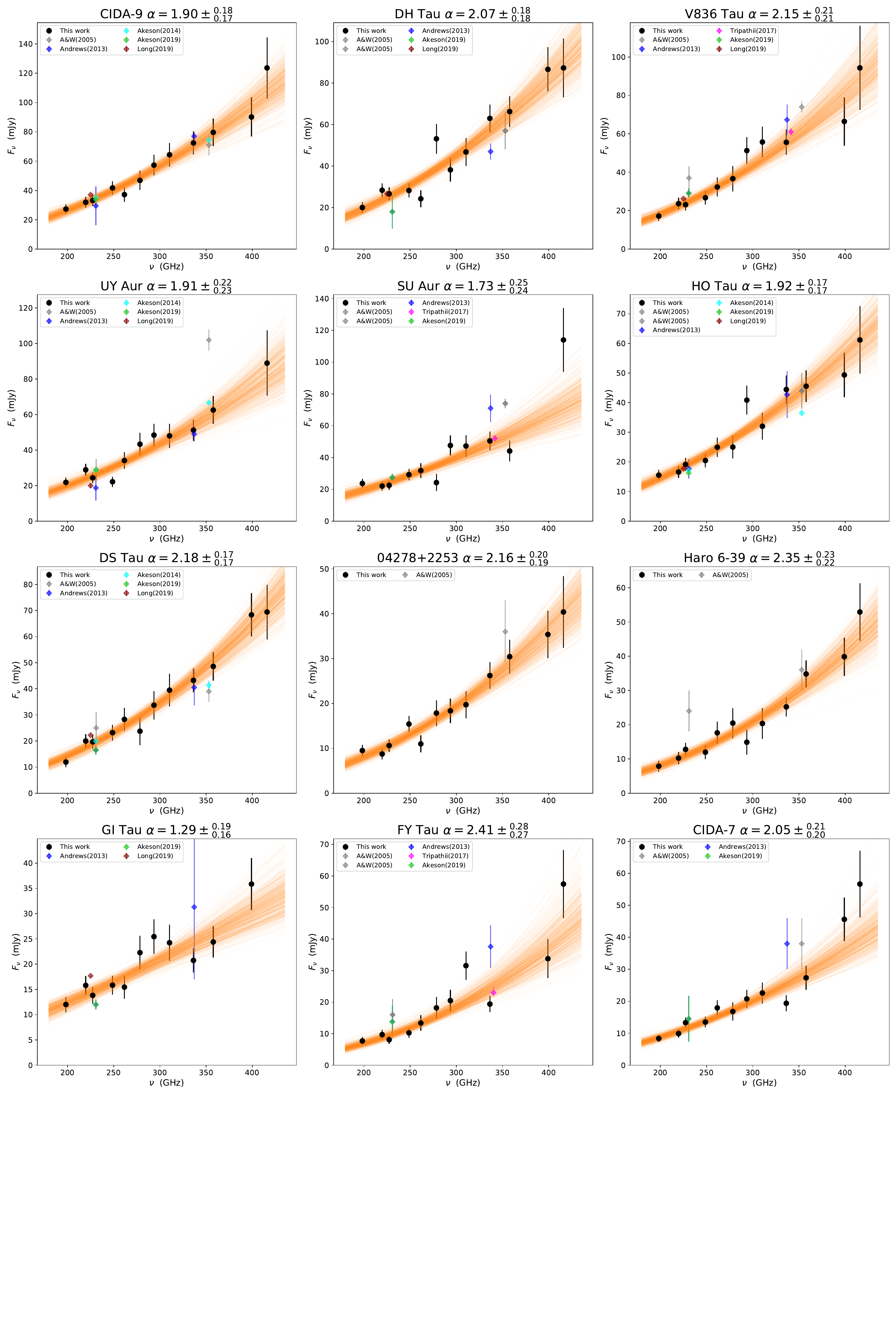}
    \caption{
    Continuation of Figure \ref{fig:SED_0510_0} to show the SEDs of another 12 spatially unresolved sources that were calibrated by the gain calibrator J0510+180.
    }
    \label{fig:SED_0510_1}
\end{figure*}

\subsection{Gain Phase Self-calibration}\label{sub:selcal}

After applying the routine, gain phase calibration solutions, the scans on science targets can still be subject to residual phase errors due to (i) the phase differences between the directions on the targets and the gain calibrators, (ii) the time variations of phase between the observations on the gain calibrators and target sources, and (iii) intrinsic errors in the gain phase solutions due to the limited SNRs. 
These are the most serious for the observations at $\sim$400 GHz.
Nevertheless, at this (and other) frequency, there are some bright target sources for which the SNRs achieved with 12 GHz continuum bandwidth are high enough for gain phase self-calibration.
The scans on these bright target sources are interleaving the observations on the remaining target sources.
To minimize the residual phase errors, after the routine calibrations (Section \ref{sub:basic}), we derived the gain phase self-calibration solutions based on the scans on these bright sources, following an iterative procedure.
First, we derived the gain phase self-calibration solutions for each of the 47 selected targets.
After inspecting the solutions (See Appendix \ref{appendix:selcal}), we identified 15 self-calibratable bright sources that had sufficient SNRs ($>$ 5 in $\sim$ 2.25 minutes of integration at 415.5 GHz) and the gain phase self-calibration solutions appear smooth in time.
We then re-derived the multi-pointing gain phase self-calibration solutions based on the observations only on these 15 bright target sources.
We fit the multi-pointing gain phase solutions by polynomial functions of time.
Based on the polynomial fittings, we interpolated and then applied the solutions to the scans on the other 32 fainter sources.

\begin{figure}
    \centering
    \includegraphics[width=8.5cm]{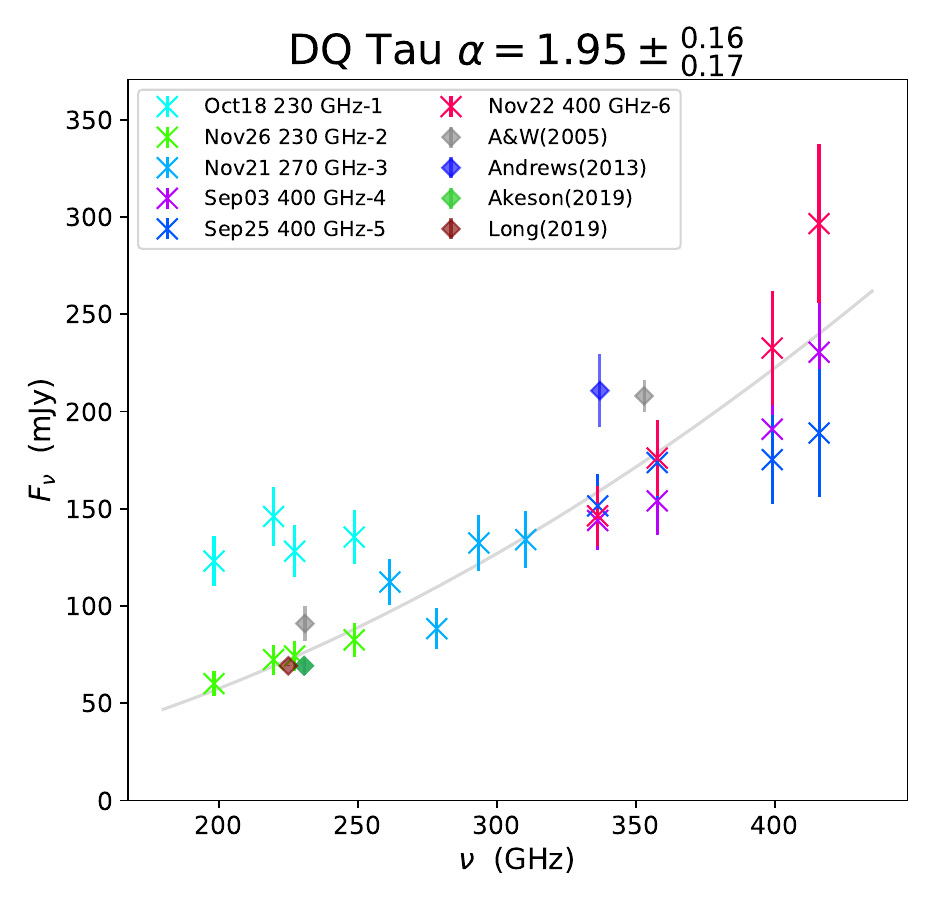}
    \caption{
    The 200--420 GHz SED of DQ Tau. The style of
the presentation is the same as that of Figure \ref{fig:SED_0510_resolved} despite that different epochs of measurements are separately presented with rainbow color code (see Table \ref{tab:DQ_Tau_flux}). The observing dates are labeled in the legend. The gray line shows the best fit parameters in SED fitting. }
    \label{fig:SED_DQtau}
\end{figure}

We performed the gain phase self-calibration using the \textsc{Miriad} {\tt selcal} task.
The "mosaic" option of the {\tt selcal} task allows combining multiple scans on different pointing positions when deriving one time-avaraged phase solution.
For AB Aur, RY Tau, FT Tau, UZ Tau, DL Tau, T Tau, DR Tau, and DO Tau, we adopted the previous high angular resolution ALMA images (\citealt{Tang_2017,Long2018ApJ...869...17L,Long2019ApJ...882...49L}) as the source models.
We corrected the phase centers of those previous images using the proper motions reported by Gaia DR3 (\citealt{Gaia_2016,Gaia_2023}). 
For the remaining self-calibratable sources, we used the {\tt clean} model (Section \ref{sub:imaging}) at 337 GHz as the source models.

The gain phase self-calibration solutions of the scans that were gain-calibrated by J0510+180 and J0418+380 were derived and applied separately.
Due to the calibration cycle time of our observations (Section \ref{sub:obs}), the gain phase self-calibration solutions are populated in time sectors with 12 minutes duration.
We fit the gain phase self-calibration solutions within each of the 12 minutes' time sectors with a polynomial and then interpolated the solutions in time according to the results of the polynomial fittings. 
For a time sector that includes $m$ valid gain phase self-calibration solutions, the order of the polynomial was chosen to be the integer nearest to $m/2$ when it is less than 3; otherwise, the order of the polynomial was chosen to be 3 to avoid potential over-fitting. 

To check the validity of interpolating the self-calibration solutions (e.g., to see whether or not applying the time-interpolated solutions indeed improves the image quality), we prepared another version of gain-phase self-calibration solutions from which the self-calibratable source, CI~Tau, was excluded when deriving the multi-pointing solutions.
We interpolated and applied these gain-phase self-calibration solutions onto the scans on CI~Tau, and then measured the flux densities of CI~Tau.
We found that these flux densities are consistent with the flux densities measured from the self-calibrated CI~Tau data to within 11\% at all frequency bands and in all the tracks, except for the 294 GHz flux densities measured from track 270 GHz-3.
In track 270 GHz-3, at 294 GHz, the flux density measured from the self-calibrated CI~Tau data is higher than the measurement from the CI~Tau data that were applied with the time-interpolated multi-pointing self-calibration solutions by $\sim$20\%.
This may indicate that the residual phase errors remain non-negligible at this frequency.

We applied the standard (i.e., not the polynomial fit) multi-pointing gain phase self-calibration solutions to the 15 bright self-calibratable sources, and applied the time-interpolated solutions made from polynomial fits to the remaining 32 fainter sources. 
The flux densities measured from the data that were applied with the self-calibration solutions are higher than those measured from the non-self-calibrated CI~Tau data by $\sim8$\%.
We assessed that the data of the self-calibratable bright sources can be accurately calibrated; for the fainter sources, the errors of flux densities ($\sim$10\% in most cases) induced by residual phase errors are compatible with the thermal noise.

In the visual inspection, we noticed that the real part of the visibilities of the faint sources, FM~Tau, V836~Tau, UY~Aur, SU~Aur, GI~Tau and FY~Tau, at 280, 357, 400 or 415 GHz showed negative values at some {\it uv}-ranges. 
This is caused by the residual phase errors in the calibrations at certain frequency bands which may be partly due to thermal noise.
The residual phase errors also appear as the scattered or vanished signal in their clean images (Section \ref{sub:imaging}) at these frequency bands. 
To avoid the potential biases to the visibility fitting (Section \ref{sub:flux}) induced by the phase errors, we excluded the flux measurements of these faint sources at these frequencies in the subsequent analysis (Appendix \ref{appendix:source}). 

\subsection{Imaging}\label{sub:imaging}

We performed imaging using the \textsc{Miriad} software package (\citealt{Sault1995ASPC...77..433S}).
Specifically, to achieve the best signal-to-noise ratio (SNR), we used the {\tt invert} task with natural weighting to perform the inverse Fourier transform for the calibrated visibilities.
We then used the {\tt clean} task to deconvolve the dirty images that were produced from inverse Fourier transform.
Finally, we used the {\tt restor} task to produce the clean images that have Gaussian synthesized beams.

\begin{figure*}[]
    \centering
    \includegraphics[width=15cm]{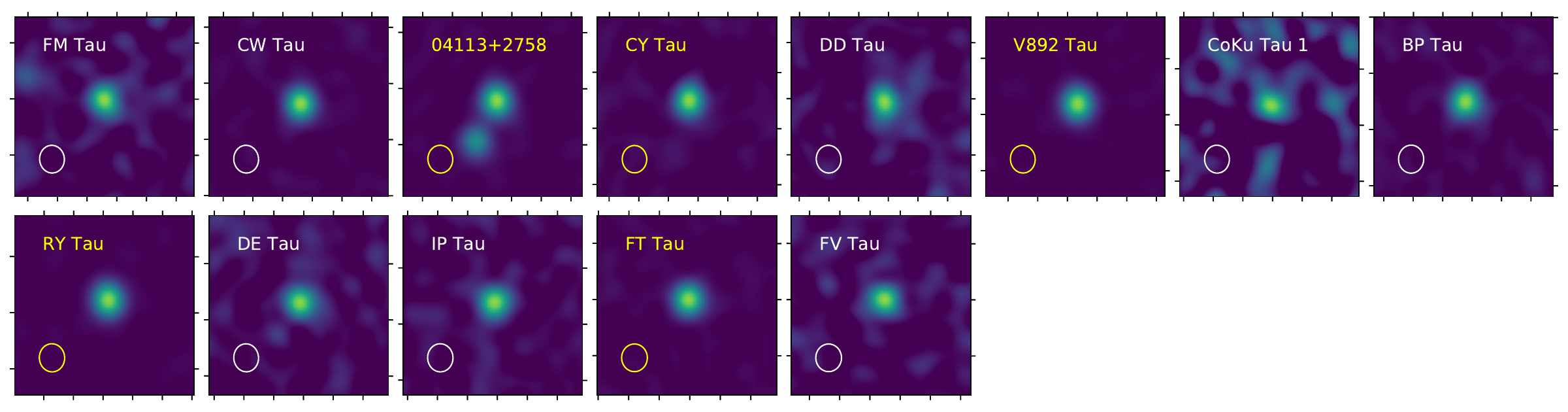}
    \includegraphics[width=15cm]{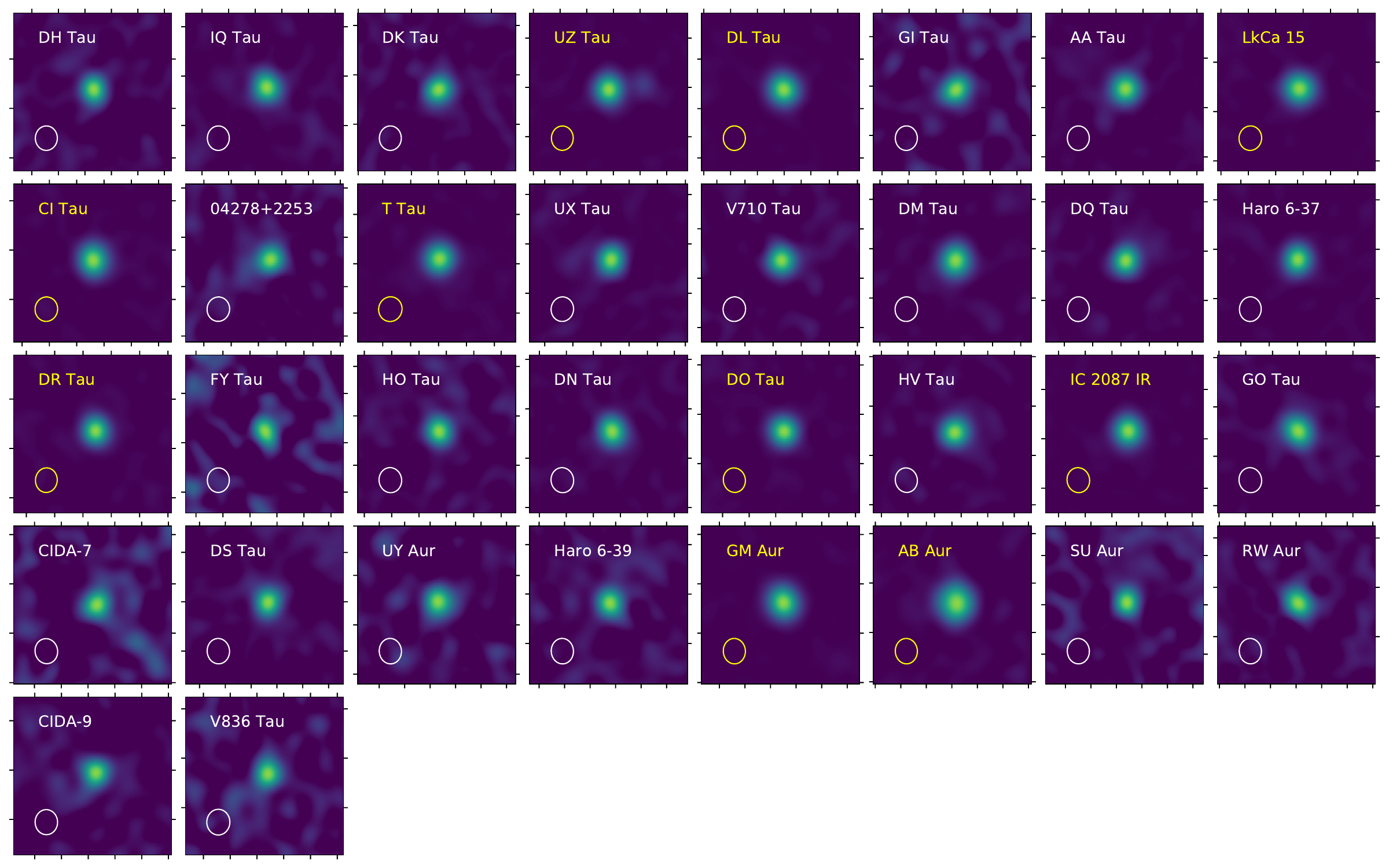}
    \caption{
    The 347 GHz images of the 47 Class II sources in the survey. The sources are ordered according to the sequence of our observations. Each panel is $16''×16''$ with source peak in the center. The upper right corner labels the source name, and the synthesized beam is shown in the lower left corner. The upper two rows are 13 sources calibrated by gain calibrator J0418+380; the lower five rows are 34 sources calibrated by J0510+180. Yellow text labels the 15 sources that are used to derive gain phase self-calibration solutions.
    }
    \label{fig:347GHz_image}
\end{figure*}

Prior to the imaging processes, for the observational data of each receiver in each track, we first binned the line-free spectral channels in each sideband to a single continuum channel using the \textsc{Miriad} task {\tt uvlin} and  {\tt uvaver} tasks. 
We note that the frequency range of the continuum observations did not cover the CO (J=3--2) line. 
We also excluded the CO (J=2--1) line as it was located at the edge channels of the spectral window in the tuning. 
Except for these two strong lines, we did not find any other line emission in all the spectral coverages. 

We produced the images for our scientific analyses using the following three steps.
First, for each sideband of a receiver tuning, we jointly imaged the observations of all tracks without using a {\it clean box} to produce rough images.
From the rough images, we assessed the root-mean-square (RMS) noise levels for each of the 47 targets.
We found that the SNRs of the rough 337 GHz images are better than those of the observations taken at other frequencies.
Therefore, we defined the {\it clean boxes} that will be used in the subsequent imaging steps based on the rough 337 GHz images.
We located the source centers based on the 2-dimensional (2D) Gaussian fitting (see Section \ref{sub:flux}) and defined the clean boxes to be the square regions where the side lengths are eight times the major axes of the synthesized beams. 
We then performed boxed imaging for each track of observations separately and then, at all observing frequencies, compared the flux densities measured from different dates to assess the consistency of absolute flux calibration and to pick out the truly variable sources. 
Finally, for each sideband covered by a receiver tuning, we jointly imaged the data taken from all tracks of observations to produce the images for our scientific measurements.
Our typically achieved synthesized beam full width at half maximum (FWHM) are $\sim$3\farcs0, $\sim$2\farcs8, and $\sim$2\farcs0 at 230 GHz, 270 GHz, and 400 GHz, respectively. 
The achieved synthesized beam sizes and RMS noise levels are summarized in Table \ref{tab:beam_size_rms}.

\begin{figure}
    \centering
    \includegraphics[width=8.5cm]{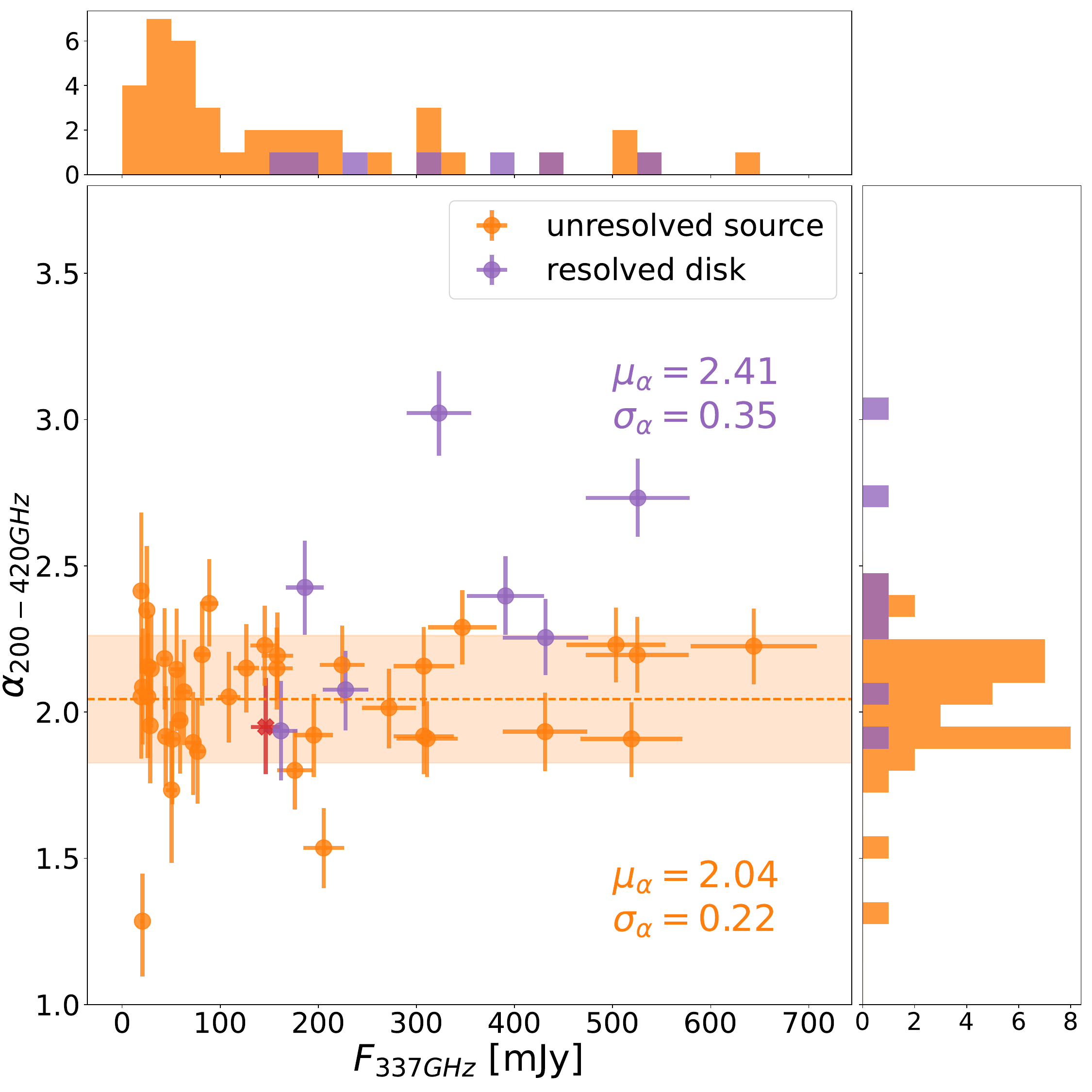}\\
    \caption{
    The 200--420 GHz spectral index versus 337 GHz flux density of the 47 selected sources. The dashed line and the shaded area are the mean and standard deviation of the spectral indices of unresolved sources, respectively. The red cross labels DQ Tau, which may be confused by free-free and/or synchrotron emission flare and thus was excluded from the statistical analysis. The orange histograms were made from the 39 unresolved sample, and the purple histograms were made from the 7 resolved disks.
    }
    \label{fig:337GHz_flux_spidx}
\end{figure}

\subsection{Measuring Flux Densities}\label{sub:flux}

In our observations, the disks are either spatially unresolved or marginally resolved.
By comparing the synthesized beam sizes of our observations with the disk radii quoted from higher angular resolution ALMA 1.3 mm observations (\citealt{Long2018ApJ...869...17L, Long2019ApJ...882...49L}), we separated the seven marginally resolved disks, DL Tau, CI Tau, GM Aur, AB Aur, DM Tau, AA Tau, and GO Tau, and the binary source 04113+2758, from the rest of the unresolved sources.
For the unresolved sources, we derived the flux densities in each sideband based on fitting the complex visibilities as point sources; for the resolved disks, we fit the complex visibilities assuming Gaussian profiles for the disks.
Before the fittings for each source, we visually inspected the image produced from each epoch of observations at each frequency and omitted the data that were subject to large residual phase errors.

The visibility fittings were carried out using the Markov chain Monte Carlo (MCMC) method, and the MCMC samplings were realized using the publicly available python package {\tt emcee} (\citealt{emcee}).
For unresolved sources, we assumed a point-source model that has three free parameters: flux density (flux), x-offset (dx) and y-offset (dy). 
For resolved disks, we assumed a circular Gaussian disk model that has the same three free parameters: flux density (flux), x-offset (dx), y-offset (dy), and the fourth free parameter: full width at half maximum (FWHM). 
The prior function of flux is uniform within the range of 0--2 Jy. 
The prior functions of dx and dy are Gaussian functions with $\sigma = 2\farcs0$ centering at the pixel of the source center which we determined in the imaging process.
The $\sigma$ value was chosen to be slightly smaller than our synthesized beam sizes (Table \ref{tab:beam_size_rms}).
In the MCMC fittings, the initial values of the parameters dx and dy were set to the pixel of the source center.
The prior function of FWHM is the product of a Gaussian function centering at $1\farcs0$ with $\sigma = 2\farcs0$ and a step function that is 0.0 at FWHM$\le$0 and is 1.0 at FWHM$>$0.
In this case, FWHM is always positive.


In our observations, 04113+2758 was resolved into two point-like sources, 04113+2758A and 04113+2758B. 
We assumed two point-source components in visibility fitting, which yields six parameters: flux density (flux1), x-offset (dx1) and y-offset (dy1) for the first component, and flux density (flux2), x-offset (dx2) and y-offset (dy2) for the second component. 
For the first component, we adopted the same prior functions for the three free parameters as that used in fitting point-source model on unresolved sources. 
For the second component, we set the prior function to be uniform within 0--$[\mbox{flux1}]$ Jy by inspecting that the secondary source is fainter than the primary source. 
To allow the offset of the secondary source, we chose $\sigma = 5\farcs0$ for the Gaussian prior functions of dx2 and dy2 and gave a slight offset from dx1 and dy1 in their initial values.

The hierarchical quadruple system, UZ~Tau was resolved into the two components, UZ~Tau~E and UZ~Tau~W in the 400 GHz tracks, while it was only marginally resolved in the 230 GHz and 270 GHz tracks. 
The achieved SNRs in their $\leq 310$ GHz clean images were not sufficient for a robust two-Gaussians fitting. 
Therefore, we treated UZ~Tau as an unresolved source in this work and bear in mind that the flux density is dominated by UZ~Tau~E (for more discussion of the flux density measurements of this source see Appendix \ref{appendix:uztau}).

Besides fitting complex visibilities, we tried image domain fitting on all the sources and compared the derived flux densities using the two fitting methods. 
In the image domain fitting, we obtained their integrated flux densities by fitting 2D Gaussian profiles on the clean images. 
For 04113+2758, we fit two 2-dimension Gaussian components on their clean images and recorded the peak intensities for each source.
The Gaussian fittings were implemented using the curve\_fit function in the scipy software package (\citealt{2020SciPy-NMeth}). 
When generating the clean images of resolved disks for this purpose, we limited the {\it uv} distances of the observations at all frequencies to the range of 20--60 k$\lambda$ to minimize the systematic biases in the derived spectral indices due to resolving different disk structures at different frequencies.
This results in a compromised SNRs of these seven disks at 415.5 GHz, which in turn impacted the reliability of the image (and visibility) domain fittings of their flux densities.
When preparing the images of 04113+2758 for these two-Gaussians fitting, we did not limit the {\it uv} distance ranges of the observations.
This is because achieving a higher angular resolution helps robustly separate the two binary components; and a better SNRs is essential to robustly fit the two Gaussians that are spatially close to each other. 
Nevertheless, the flux density measurements of point-like sources are unlikely to be significantly biased by the selection of {\it uv} distance range.

\begin{figure}
\hspace{-1cm}
\begin{tabular}{ c }
\includegraphics[width=9cm]{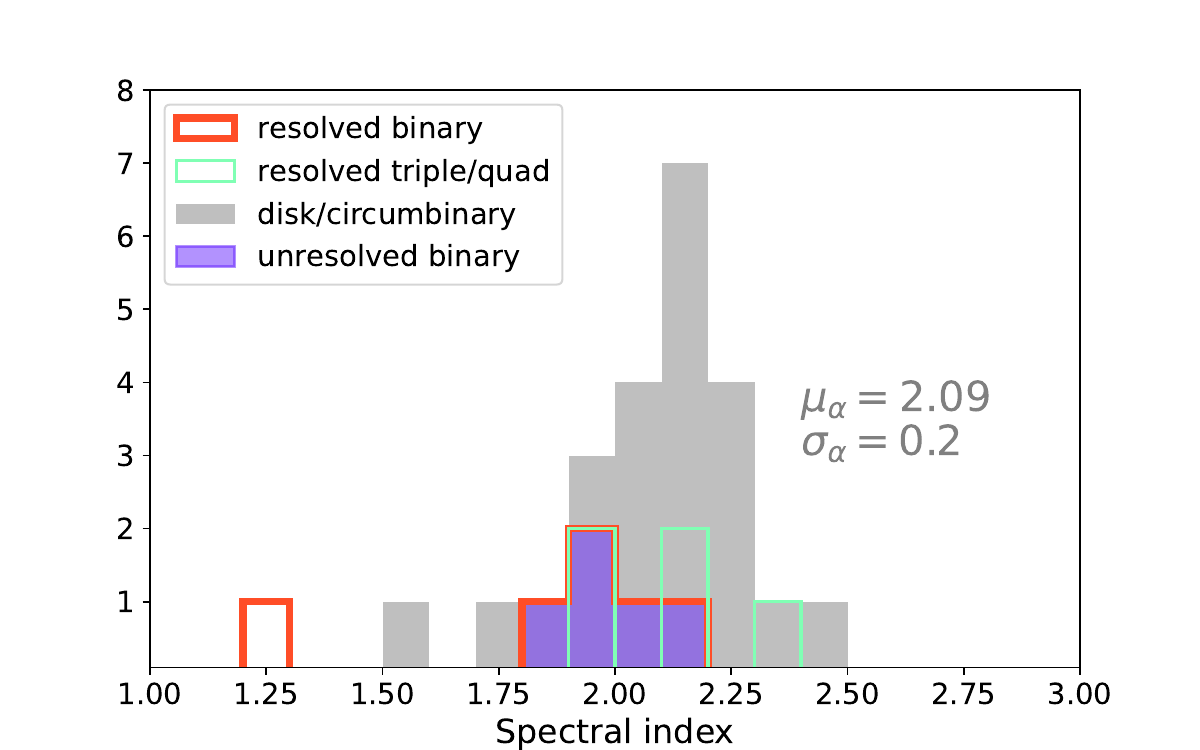}
\end{tabular}
\caption{
A comparison of the spectral index ($\alpha_{\mbox{\scriptsize 200-420 GHz}}$) distribution of the 23 unresolved disks/circumbinary disks with those of the 5 unresolved binaries, 6 resolved components in binary systems, and 5 resolved components in triple or quadruple systems.
}
\label{fig:alpha_multiple}
\end{figure}

Except for the observations taken with the 400 GHz receivers or the observations that have poor SNRs, the flux density measurements made from imaging and visibiilty fitting methods for all the unresolved sources are consistent within 20\%.
The SEDs yielded by the visibility fitting method show less inter-band flux density jumps and smaller uncertainties, especially for the resolved disks.
Therefore, we assessed that the visibility fitting method provides more robust measurements, in particular, at $\sim$400 GHz.

We quoted the flux densities at $\sim$230 GHz and $\sim$345 GHz from the previous SMA and ALMA surveys (\citealt{Andrews_Williams_2005,Andrews_2013,Akeson_2014,Tripathi_2017,Akeson_2019,Long2019ApJ...882...49L}) to check the consistency of our measurements with the previous observations (see Figure \ref{fig:SED_0510_resolved}--\ref{fig:SED_DQtau}). 

\subsection{SED fitting}\label{sub:SED}

Our observations yielded 12 independent samples of the 200--420 GHz SEDs for individual disks within our sample (Table \ref{tab:obs_summary}).
We fit the observed SED of each target source by a single power law using the MCMC method.
We extracted the spectral index $\alpha_{\mbox{\scriptsize 200-420 GHz}}$ for each source from the results of the SED fittings.
We used the Python package {\tt emcee} (\citealt{emcee}) to generate the MCMC samplings of the three free parameters: flux density at 1.3 mm wavelength ($F_{\mbox{\scriptsize 1.3mm}}$), spectral index ($\alpha$), and $\mbox{ln(f)}$, where $\mbox{ln(f)}$ is a nuisance parameter added in the Log-likelihood function to account for the additional, not well-understood errors in the data that cannot be simply accounted by thermal noise. 
The prior functions of the three parameters, $F_{\mbox{\scriptsize 1.3mm}}$, $\alpha$, and $\mbox{ln(f)}$, are uniform in the range of 0--2 Jy, 1.0--5.0, and ($-$10)--1, respectively. 
In the fitting process, the initial values of $F_{\mbox{\scriptsize 1.3mm}}$, $\alpha$, and $\mbox{ln(f)}$ were set to be 0.2 Jy, 2.0, and $-$5, respectively. 



\section{Results}\label{sec:result}

\subsection{Individual sources}\label{sub:individual}
Figure \ref{fig:347GHz_image} shows the 347 GHz continuum images of the 47 Class II sources.
The images at the other frequency bands appear similar (Appendix \ref{appendix:image}). 
These continuum images were produced using the 24 GHz bandwidth obtained from combining the LSB and USB data.
In addition, these images were produced without limiting the {\it uv} distance range to yield the best achievable SNRs.
The obtained source centers, synthesized beam sizes and RMS noise levels are listed in Table \ref{tab:beam_size_rms}.

We spatially resolved the binary components of 04113+2758, although the individual sources in each binary component are spatially unresolved.
The other detected sources appear spatially compact in spite that some of them were marginally resolved and the visibility amplitudes are decreasing with {\it uv} distances.
We did not detect CoKu~Tau~1 at all except for 219 and 357 GHz frequencies (over a 12 GHz bandwidth). 
We did not detect GK~Tau at all frequencies over 12 GHz bandwidth.
Instead, we detected its companion star, GI Tau, with a 13$''$ offset from the phase center. 
For multiple systems, CIDA-9, DH Tau, Haro 6-37, T Tau, UX Tau, V710 Tau, their flux densities were mainly contributed by their primary stars \citep{Akeson_2019}. 
For multiple systems DD Tau, DK Tau, FV Tau, RW Aur, UY Aur, UZ Tau, their flux densities are the sum of the primary and secondary components. 
During the observations of tracks 400 GHz-4,5,6, some faint sources such as FM~Tau, FV~Tau, UY~Aur, and SU~Aur were subject to high phase dispersion at 407.5 GHz frequency. 
Nevertheless, the $<$400 GHz data taken in these tracks have very good quality.
In fact, the 347 GHz images taken with these tracks have the best SNRs among the observations at all frequencies and thus yielded the best constraints on the source positions and radial extent of the emitting region.
The qualities of the 209 GHz and 270 GHz images are intermediate to those of the 407.5 GHz and 347 GHz images. 

Figure \ref{fig:SED_0510_resolved} shows the flux density measurements of the 7 spatially resolved disks and the resolved binary 04113+2758 A and B; Figure \ref{fig:SED_0418} shows the SED of 12 spatially unresolved sources that were calibrated by the gain calibrators J0418+380 and 3C84; and Figure \ref{fig:SED_0510_0} and \ref{fig:SED_0510_1} show the SED of 27 unresolved sources that were calibrated by the gain calibrators J0510+180 and 3C84 (for details of the measurements see Section \ref{sub:flux}). 
In these figures, we also present some quasi-random samples selected from our MCMC SED fittings (Section \ref{sub:SED}).
Table \ref{tab:flux} summarizes these flux density measurements. 

We found that DQ Tau had been flaring during our SMA observations and showed large variations in its  $F_{\mbox{\scriptsize 230 GHz}}$ (more discussion is provided in Appendix section \ref{appendix:dqtau}).
We separately plot the flux density measurements taken from individual epochs of observations on DQ~Tau in Figure \ref{fig:SED_DQtau}.
In the observations of tracks 230 GHz-1 (2021 October 18) and 230 GHz-2 (2021 November 26), the flux densities were different by $\sim$2 times.
The observations of tracks 400 GHz-4, 400 GHz-5, and 400 GHz-6, also showed noticeable flux density variations in spite of the large thermal noises.

We measured a very low spectral index from DQ~Tau when incorporating all measurements in SED fitting. 
In the observations of track 230 GHz-1, the spectral index of DQ~Tau appeared consistent with 0 (Figure \ref{fig:SED_DQtau}).
Such a low spectral index may be explained by a mixture of dust emission, free-free emission, and synchrotron emission (c.f., \citealt{Angelada1998AJ....116.2953A}), which might be due to the momentary flaring of the free-free and/or synchrotron emission.
This result is interesting on its own since (sub)millimeter free-free and/or synchrotron emission flares are rarely reported (e.g., \citealt{Bower2016ApJ...830..107B}) in the dedicated monitoring observations towards YSOs (e.g., \citealt{Herczeg2017ApJ...849...43H,Liu2018A&A...612A..54L,Johnstone2018ApJ...854...31J,Francis2019ApJ...871..149F,Lovell2024ApJ...962L..12L}).
We tried to derive the spectral index in the quiescent period using the data taken from track 230 GHz-2, 270 GHz-3 and 400 GHz-6, which were carried out on nearby dates in November. 
The obtained spectral index is consistent with 2. 
Because these observations might still be confused by free-free and/or synchrotron emission, we excluded DQ Tau from the sample in the subsequent statistical result and population synthesis. 
The previous JVLA survey on nearby ($d\sim$150 pc) Class II YSOs (e.g., \citealt{Liu2014ApJ...780..155L,Galvan2014A&A...570L...9G}) showed that the free-free and synchrotron emissions are generally $<$100 $\mu$Jy in these sources at $\gtrsim$10 GHz frequencies and thus are very unlikely to bias our studies about dust spectral indices in the remaining target sources.

The flux densities of the seven spatially resolved disks measured at $\sim$230 GHz and $\sim$345 GHz in this SMA survey are in good agreement with the previous measurements (Figure \ref{fig:SED_0510_resolved}). 
Except for GO Tau, these resolved disks consistently show $\alpha_{\mbox{\scriptsize 200-420 GHz}}>2.0$. 
In the statistical analysis and the related discussion (Section \ref{sub:statistical}, \ref{sec:discussions}), we will distinguish these 7 resolved disks from the other 39 unresolved sources. 

For the spatially unresolved sources that are brighter than BP Tau (Figure \ref{fig:SED_0418}) or RW Aur (Figure \ref{fig:SED_0510_0}, \ref{fig:SED_0510_1}), the flux densities measured at the 12 frequencies follow smooth power laws, as shown by the SED fitting.
These results indicate that our absolute flux calibrations were accurate between frequency bands (Section \ref{sub:basic}).
However, the fainter sources exhibit $>10 \%$ variations in flux densities between different sidebands, receivers and tuning, which is reflected by larger error bars in $\alpha_{\mbox{\scriptsize 200-420 GHz}}$.
This may be due to that the potential existing residual phase error in visibilities can bias the flux measurements when the SNR of the source is poor. 
The $\alpha_{\mbox{\scriptsize 200-420 GHz}}$ of the 39 unresolved sources are mostly around the value of 2.0. 
CY Tau, SU Aur and GI Tau show relatively low $\alpha_{\mbox{\scriptsize 200-420 GHz}}$ of 1.54, 1.73, and 1.29 respectively.

\begin{figure}
    \hspace{-1cm}
    \begin{tabular}{c}
    \includegraphics[width=9.2cm]{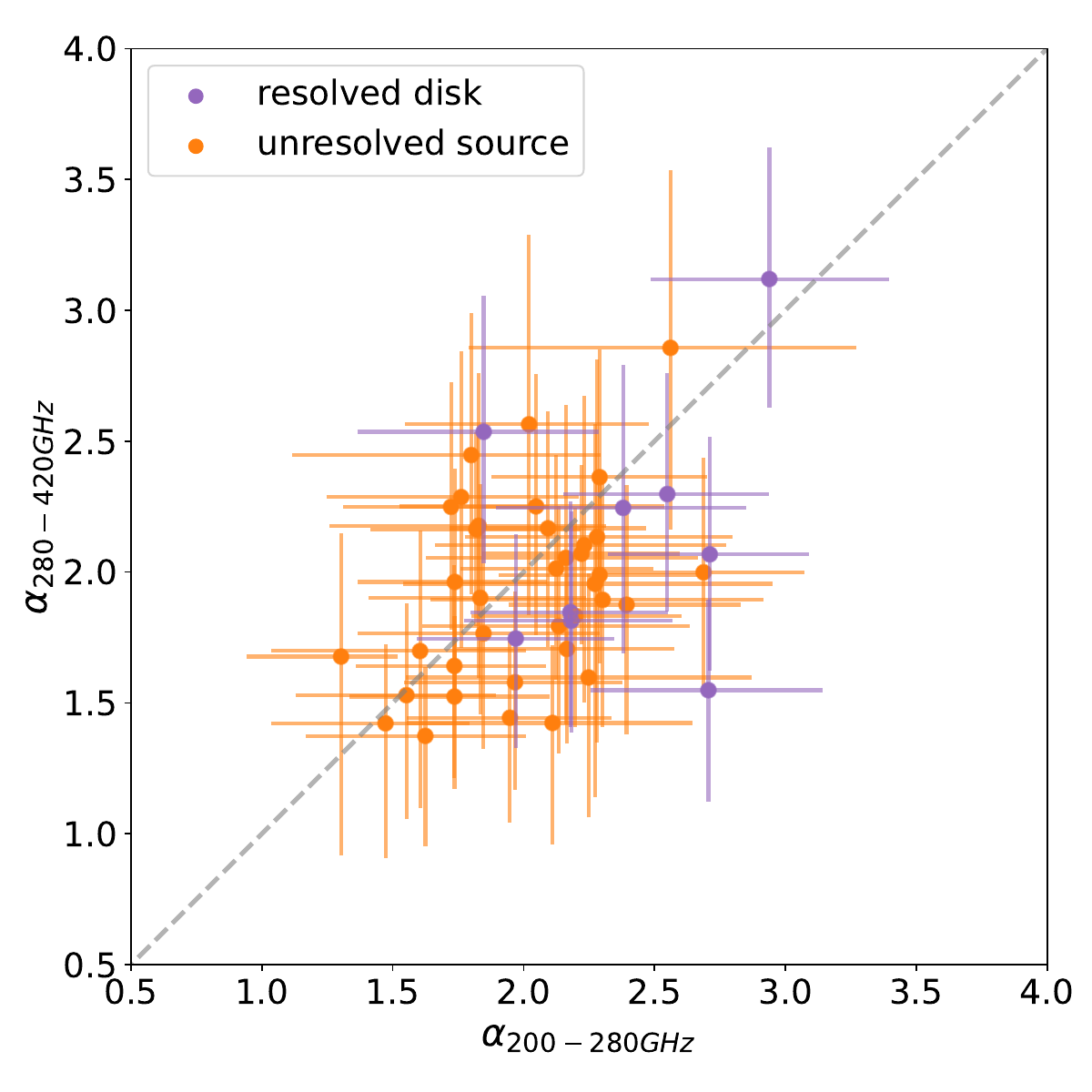}\\
    \includegraphics[width=8.8cm]{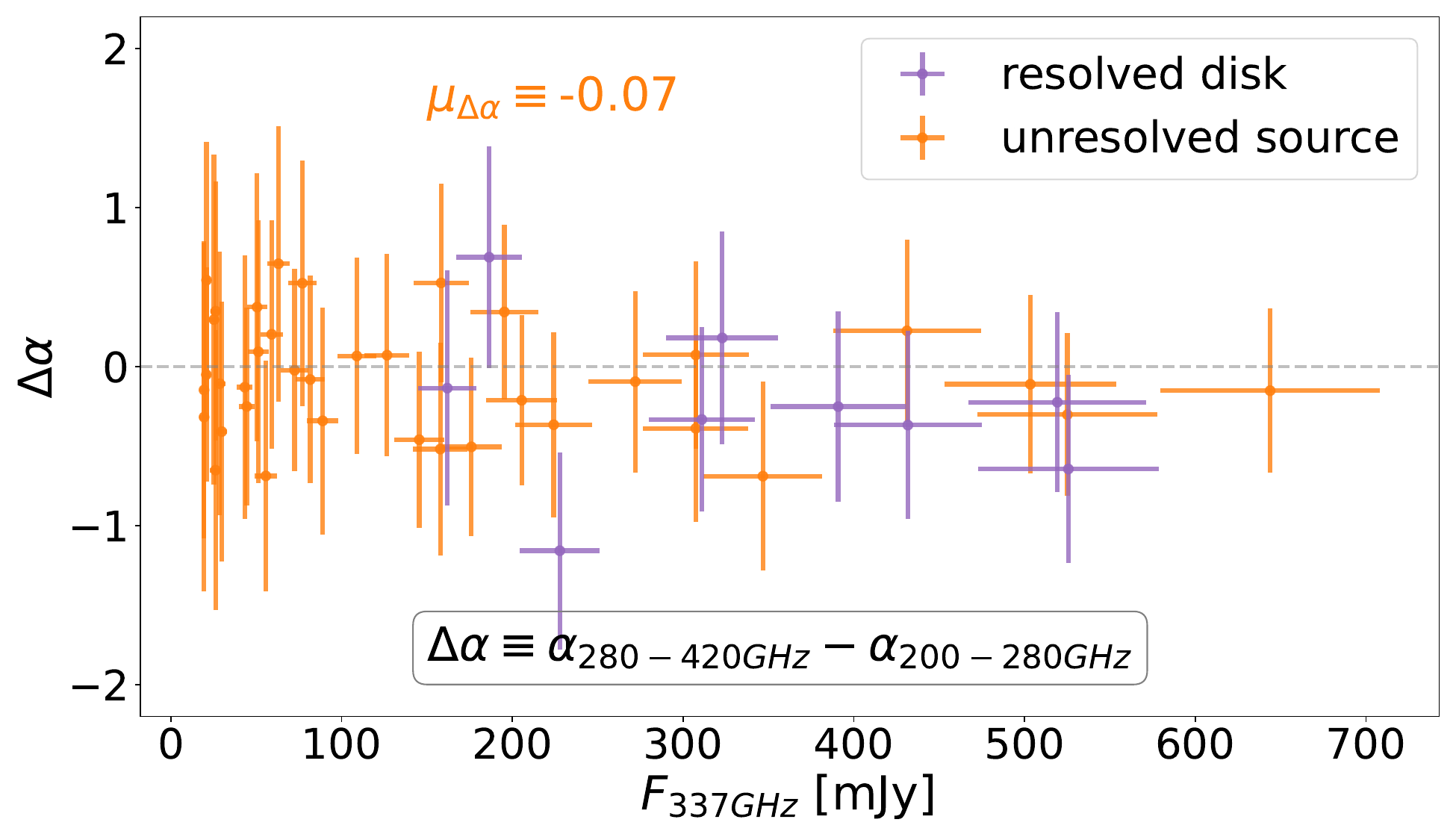}
    \end{tabular}
    \caption{ 
    A comparison of the observed spectral indices in two frequency sub-ranges, 200--280 GHz and 280--420 GHz.  
    The top panel shows the $\alpha_{\mbox{\scriptsize 280-420 GHz}}$ and $\alpha_{\mbox{\scriptsize 200-280 GHz}}$ of the detected, 46 protoplanetary disks. 
    The grey line shows where the two spectral indices are equal. 
    The bottom panel shows the 337 GHz flux densities and the differences between the 200--280 GHz and 280--420 GHz spectral indices ($\Delta \alpha$) of the 46 protoplanetary disks that were detected in our SMA survey. 
    The vertical error bars indicates $\pm$1-$\sigma$ uncertainties of $\Delta \alpha$, which were derived based on standard error propagation. }
    \label{fig:337GHz_flux_del_spidx}
\end{figure}

\begin{figure}
    \hspace{-0.5cm}
    \includegraphics[width=9cm]{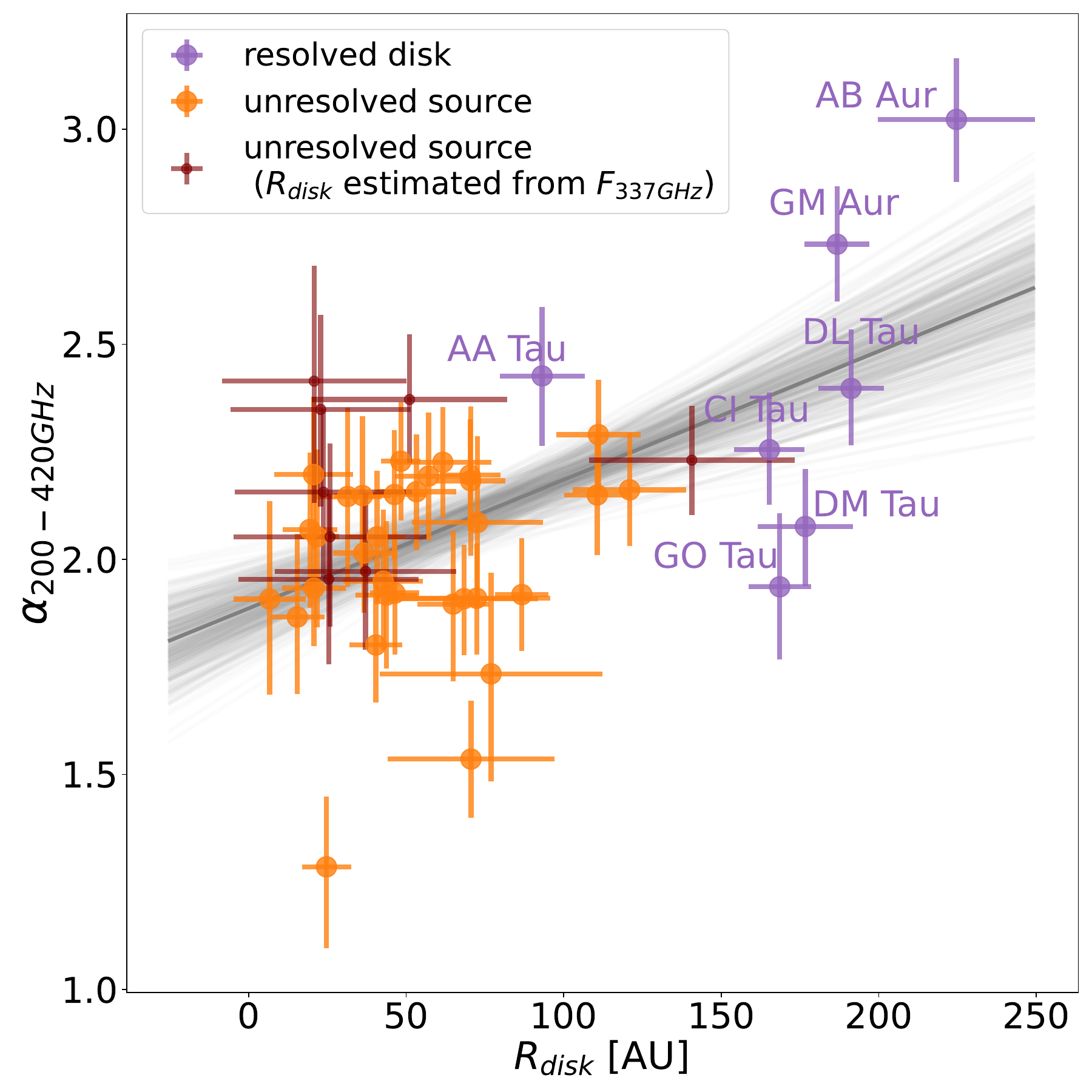}
    \caption{
    The 200--420 GHz spectral indices and the disk radius ($R_{\mbox{\scriptsize disk}}$) of the 46 protoplanetary disks that were constrained by the previous, high angular resolution 230 GHz observations  (see table note in Table \ref{tab:targ_properties}) or estimated from the $F_{\mbox{\scriptsize 337 GHz}}$. 
    Purple and orange symbols show the resolved and unresolved sources where the disk radius can be determined and are quoted from literature.
    The $R_{\mbox{\scriptsize disk}}$ of them are the $R_{\mbox{\scriptsize 95\%}}$ (The radius of a circle region that encloses 95\% of the total flux density), or $R_{\mbox{\scriptsize 90\%}}$ reported in previous observations or measured in ALMA archival images (Section \ref{sub:statistical}). 
    Brown symbols show the unresolved sources that the $R_{\mbox{\scriptsize 68\%}}$ were estimated based on the size--luminosity relation (\citealt{Tripathi_2017}) and then were converted to  $R_{\mbox{\scriptsize 90\%}}$ using a relation given by \citet{Hendler2020ApJ...895..126H}. Gray lines show the 25 random draw of the fitting parameters extracted from MCMC samplings in Bayesian linear regression fit.}
    \label{fig:disksize_spidx}
\end{figure}

\subsection{Statistical results}\label{sub:statistical}

Figure \ref{fig:337GHz_flux_spidx} shows the $F_{\mbox{\scriptsize 337 GHz}}$ and $\alpha_{\mbox{\scriptsize 200-420 GHz}}$ of the sources that were detected in our SMA observations.
We divided the 47 Class II sources sample into two subgroups, the resolved disks and the unresolved sources, based on the comparison between the angular resolution of our observations and the disk radii quoted from the ALMA 1.3 mm observations (Section \ref{sub:flux}).
The spatially resolved disks appear to have larger $\alpha_{\mbox{\scriptsize 200-420 GHz}}$ values than the unresolved sources. 
Except for AB~Aur and GM~Aur, all the disks have $\alpha_{\mbox{\scriptsize 200-420 GHz}}$ smaller than 2.5.
We performed the Kolmogorov–Smirnov (K-S) test on the $\alpha_{\mbox{\scriptsize 200-420 GHz}}$ distributions of unresolved sources and resolved disks. 
The p-value in the K-S test, $p = 0.013$, rejects the null hypothesis that the two samples are drawn from the same distribution with a 95\% confidence level. 

The $\alpha_{\mbox{\scriptsize 200-420 GHz}}$ of the spatially unresolved sources are narrowly populated in the range of $2.04 \pm 0.22$, which is indicated by the orange shaded area in Figure \ref{fig:337GHz_flux_spidx}. 
The two sources, CY~Tau and GI~Tau, seem to be outliers that have exceptionally low $\alpha_{\mbox{\scriptsize 200-420 GHz}}$ values.
If we exclude these two sources, the standard deviation of $\sigma_{\scriptsize \alpha}$ becomes 0.16.
The Pearson test p-value of their $\alpha_{\mbox{\scriptsize 200-420 GHz}}$ and $F_{\mbox{\scriptsize 337 GHz}}$ is 0.25, which is consistent with no obvious correlation between these two quantities.

There is no strong evidence that the $\alpha_{\mbox{\scriptsize 200-420 GHz}}$ of isolated disks and circumbinary disks are systematically different from those of the resolved/unresolved binary or multiple systems although our sample of resolved/unresolved binary or multiple systems is not representative in a statistical sense (Figure \ref{fig:alpha_multiple}).
We performed the K-S tests to compare the $\alpha_{\mbox{\scriptsize 200-420 GHz}}$ distribution of the isolated/circumbinary disks with those of the (1) unresolved binaries, (2) resolved binaries, and (3) multiple systems.
The derived p-values are 0.2, 0.08 and 0.9, respectively.

The densely-sampled 12 measurements on SED precisely constrain the $\alpha_{\mbox{\scriptsize 200-420 GHz}}$ for each source and yield lower uncertainty (mostly, the error bar is $\sim$0.2) than that of the studies which derived $\alpha$ from frequencies with $\sim$100 GHz separation (e.g., \citealt{Andrews_Williams_2005,Andrews_Williams_2007}, etc). 
In addition, a denser sampling on frequency domain allows us to constrain the spectral index in narrow frequency ranges, which mitigates the frequency smearing issue. 
The resolved distribution of $\alpha_{\mbox{\scriptsize 200-420 GHz}}$ appears narrower as compared with what was reported in \cite{Andrews_Williams_2005} (see Figure \ref{fig:target}, Right).
In particular, we confirm that some sources show spectral indices $\lesssim$2.0, which is inconsistent with optically thin emission (See Section \ref{subsub:thin}).

To examine the potential frequency variations of the spectral indices, we separately performed SED fittings using only the data in the two frequency ranges, 200--280 GHz and 280--420 GHz.
Based on these SED fittings, we obtained the spectral indices, $\alpha_{\mbox{\scriptsize 280-420 GHz}}$ and $\alpha_{\mbox{\scriptsize 200-280 GHz}}$, and their difference, $\Delta \alpha$.
Figure \ref{fig:337GHz_flux_del_spidx} shows the $F_{\mbox{\scriptsize 337 GHz}}$ $\Delta \alpha$ of our detected sources.
The $\Delta \alpha$ of unresolved sources have a mean value of $-0.1$ and are populated over the range of $-1.0$ to $1.0$. 
With the uncertainties of our measurements, the frequency dependence of spectral indices over the 200--400 GHz frequency range was not detected.

We examined how $\alpha_{\mbox{\scriptsize 200-420 GHz}}$ may depend on the sizes of the dusty disks.
The disk radius $R_{\mbox{\scriptsize disk}}$ of 39 of our sample were measured from the higher angular resolution ALMA observations: 
we quoted the $R_{\mbox{\scriptsize disk}}$ values from literature if available; 
Otherwise, we measured it by downloading the archival ALMA images.
A complete summary of where we quoted the disk radii of the 39 sources from can be found in the Table notes of Table \ref{tab:targ_properties}.
Specifically, the $R_{\mbox{\scriptsize disk}}$ of 25 sources are the $R_{\mbox{\scriptsize 95\%}}$ radii quoted from \citet{Long2019ApJ...882...49L}, \citet{Sierra2021ApJS..257...14S}, and \citet{Ueda_2022};
 the $R_{\mbox{\scriptsize disk}}$ of 4 sources are the $R_{\mbox{\scriptsize 90\%}}$ radii quoted from \citet{Long2021ApJ...915..131L}, \citet{Long2022ApJ...931....6L}, and \citet{Stapper2022};
 the $R_{\mbox{\scriptsize disk}}$ of 2 sources are the $R_{\mbox{\scriptsize 68\%}}$ radii quoted from \citet{Parker2022MNRAS.511.2453P}, which were converted 
 to $R_{\mbox{\scriptsize 90\%}}$ by the $R_{\mbox{\tiny 90\%}}-R_{\mbox{\tiny 68\%}}$ relation given in \citet{Hendler2020ApJ...895..126H};
 the $R_{\mbox{\scriptsize disk}}$ of 8 sources are the $R_{\mbox{\scriptsize 90\%}}$ radii we measured from the archival ALMA images. 
The $R_{\mbox{\scriptsize 95\%}}$ radius of a disk is the radius of the circular region that centers at the position of the host protostar and encloses 95 \% of the total flux density.
The definition of the $R_{\mbox{\scriptsize 90\%}}$ radius is similar.
There were 7 sources for which we could not find neither appropriate previous measurements of disk radii nor available high-resolution ALMA observations. For these 7 sources, we estimated their effective radius ($R_{\mbox{\scriptsize 68\%}}$) based on the size--luminosity relation presented in \citet{Tripathi_2017} using their $F_{\mbox{\scriptsize 337 GHz}}$ and converted to $R_{\mbox{\scriptsize 90\%}}$ by the $R_{\mbox{\tiny 90\%}}-R_{\mbox{\tiny 68\%}}$ relation given in \citet{Hendler2020ApJ...895..126H}. 

Figure \ref{fig:disksize_spidx} presents the $R_{\mbox{\scriptsize disk}}$ and $\alpha_{\mbox{\scriptsize 200-420 GHz}}$ of these 46 disks and shows a clear positive correlation. 
We used the Bayesian approach presented in \citet{Kelly2007ApJ...665.1489K} to perform linear regression. The best fit values for the linear relation are: intercept $\alpha_{\mbox{\scriptsize 200-420 GHz}} = 1.89 ^{+0.06} _{-0.06}$; slope $\gamma = 3.00 ^{+0.68} _{-0.69}\times10^{-3}$ AU$^{-1}$; correlation coefficient $\hat{\rho} = 0.67 ^{+0.12} _{-0.15}$. 
By performing the Pearson test, we obtained a Pearson correlation coefficient $r = 0.49$ and a p-value $p = 4.53\times10^{-4}$ for the null hypothesis.
The correlation coefficients suggest that the $\alpha_{\mbox{\scriptsize 200-420 GHz}}$ has moderate correlation with $R_{\mbox{\scriptsize disk}}$.

\section{Discussion}\label{sec:discussions}

\subsection{Qualitative interpretation}\label{sub:qualitative}

The narrowly distributed, close to 2.0 values of $\alpha_{\mbox{\scriptsize 200-420 GHz}}$ (Figure \ref{fig:337GHz_flux_spidx}) may indicate that certain physical properties of all the Class II disks that were spatially unresolved by our SMA observations are similar.
One possibility is that the dust properties in these unresolved Class II disks are the same. 

\subsubsection{Optically thin interpretation}\label{subsub:thin}
Assuming that the dust emission in all unresolved Class II disks is optically thin at $\sim$1 mm wavelength, the results of $\alpha_{\mbox{\scriptsize 200-420 GHz}}$ would require the maximum grain sizes $a_{\mbox{\scriptsize max}}$ to be commonly $\gg$1 cm in those unresolved Class II disks (c.f., \citealt{Testi2014prpl.conf..339T} and references therein).
In this case, both the correlation between $\alpha_{\mbox{\scriptsize 200-420 GHz}}$ and disk size (Figure \ref{fig:disksize_spidx}), and a potentially existing correlation between the millimeter continuum size and frequency (e.g., \citealt{Tripathi2018ApJ...861...64T}) could indicate that $a_{\mbox{\scriptsize max}}$ is larger at smaller radii.

We disfavor this optically thin and $a_{\mbox{\scriptsize max}}\gg1$ cm interpretation as a general interpretation for our SMA sample since (1) the multi-wavelength case studies on a few compact and extended Class II disks have demonstrated that they are optically thick ($\tau>$1) at 1 mm wavelength at least in the inner few tens au radii (e.g., \citealt{Liu2019ApJ...884...97L,Ueda2020ApJ...893..125U,Liu2021ApJ...923..270L,Guidi2022A&A...664A.137G,Hashimoto2022ApJ...941...66H,Liu2024arXiv240202900L}), and (2) the previous detections of polarized (sub)millimeter dust self-scattering are hard to interpret in this case (see the discussion in Section \ref{sec:intro}), and (3) the three sources CY~Tau, SU~Aur, and GI~Tau that present $<$2.0 spectral indices in the present SMA survey are hard to explain by the optically thin interpretation.
In particular, the previous detections of polarized (sub)millimeter dust self-scattering and the anomalously low (sub)millimeter that convincingly indicate $a_{\mbox{\scriptsize max}}\lesssim$100 $\mu$m in the innermost regions of the TW~Hya and HD~163296 disks  (\citealt{Dent2019MNRAS.482L..29D,Liu2019ApJ...877L..22L,Ueda2020ApJ...893..125U,Guidi2022A&A...664A.137G}), in contrast to the general idea of $a_{\mbox{\scriptsize max}}\gg1$ cm in the extended and optically thin disks.
How to comprehend the observations on TW~Hya and HD~163296 with the correlations mentioned above with a self-consistent physical model (of disk evolution, dust growth, and dust dynamics) is not clear to us. 

\subsubsection{Optically thick interpretation}\label{subsub:thick}
Alternatively, the observed $\alpha_{\mbox{\scriptsize 200-420 GHz}}$ distribution might be naturally interpreted if the (sub)millimeter luminosities of all observed Class II disks are dominated by optically thick dust emission (c.f. \citealt{Wu2024arXiv240304754W}). 
In particular, the Class II disks that are spatially unresolved in our SMA observations are ubiquitously optically thick at 200--420 GHz frequency; only the relatively spatially resolved ($\gtrsim$250 AU diameter) disks receive a nontrivial contribution from the low column density halos that are relatively optically thin ($\tau\lesssim$1) at 200--400 GHz.
This may be comprehended by the radially decreasing dust column density profiles of the Class II disks. 
In this case, the disk-averaged spectral indices of the resolved disks will be contributed by not only the compact, optically thick disk cores but also by the surrounding, optically thin halo part of the disks.
This makes the spectral indices of the resolved disks slightly higher than 2.
We may expect that new high-resolution, multi-frequency observations on the unresolved sources will reveal some small contribution from the optically thin emission. 
However, the globally-averaged spectral index would not change since the emission from the optically thin structures contributes less to total luminosity than that from the optically thick structures.

\begin{figure}
\hspace{-0.3cm}
\includegraphics[width=9cm]{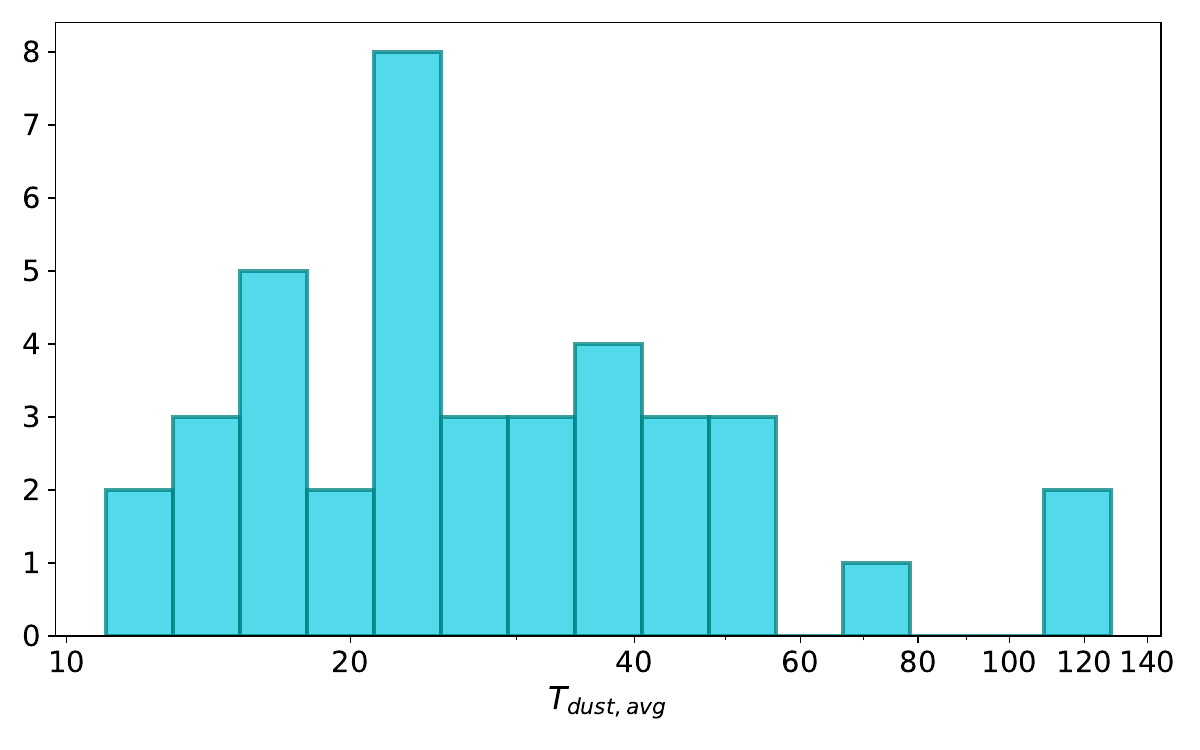}
\caption{
The estimated averaged dust temperature ($T_{\mbox{\scriptsize dust, avg}}$) distribution of the 39 unresolved sources, which were evaluated based on specific assumptions of radial temperature profiles and the inner and outer cutoff radii (see Section \ref{sub:Popsynthesis}). 
The assumed parameters are summarized in Table \ref{tab:Tdust}.
These temperatures are likely lower than the actual dust temperatures (see discussion in Section \ref{sub:Popcompar}).
}
\label{fig:Tdust_hist}
\end{figure}

\begin{deluxetable*}{ccccccc}
\tabletypesize{\footnotesize}
\tablecolumns{7}
\tablecaption{ 
Model parameters for the 46 sources
\label{tab:Tdust}}
\tablehead{ 
\colhead{Source} & \colhead{$L_{\scriptsize \mbox{bol}}$} & \colhead{$R_{\scriptsize \mbox{in}}$} & \colhead{$R_{\scriptsize \mbox{out}}$} & \colhead{$\mbox{T}_{\scriptsize \mbox{dust, 1AU}}$} & \colhead{$q$} & \colhead{$\mbox{T}_{\scriptsize \mbox{dust, avg}}$} \\
\colhead{} & \colhead{($L_{\scriptsize \mbox{$\odot$}}$)} & \colhead{(AU)} & \colhead{(AU)} & \colhead{(K)} & \colhead{} & \colhead{(K)}}
\startdata 
\hline
\multicolumn{7}{c}{Resolved Disks} \\
\hline
AA Tau & 0.75 & 0.06 & 93 & 129 & 0.56 & 19\\
AB Aur & 112.7 & 0.73 & 225 & 367 & 0.45 & 46\\
CI Tau & 1.65 & 0.09 & 191 & 152 & 0.56 & 15\\
DL Tau & 1.49 & 0.08 & 165 & 149 & 0.62 & 14\\
DM Tau & 0.24 & 0.03 & 177 & 111 & 0.51 & 14\\
GM Aur & 0.903 & 0.07 & 187 & 136 & 0.44 & 20\\
GO Tau & 0.27 & 0.04 & 169 & 90 & 0.62 & 11\\
\hline
\multicolumn{7}{c}{Unresolved Sources} \\
\hline
04113+2758A & 0.37 & 0.04 & 68 & 90\tablenotemark{\scriptsize b} & 0.58 & 15\\
04113+2758B & 0.45 & 0.05 & 72 & 95\tablenotemark{\scriptsize b} & 0.58 & 16\\
04278+2253 &  & 0.05 & 24\tablenotemark{\scriptsize a} & 100\tablenotemark{\scriptsize b} & 0.58 & 29\\
BP Tau & 0.98 & 0.07 & 41 & 117 & 0.64 & 23\\
CIDA-7 & 0.14 & 0.03 & 22 & 71\tablenotemark{\scriptsize b} & 0.58 & 22\\
CIDA-9 & 0.05 & 0.02 & 65 & 55\tablenotemark{\scriptsize b} & 0.58 & 11\\
CW Tau & 1.06 & 0.07 & 48 & 204 & 0.62 & 36\\
CY Tau & 0.36 & 0.04 & 71 & 90\tablenotemark{\scriptsize b} & 0.58 & 15\\
DD Tau & 0.7 & 0.06 & 25\tablenotemark{\scriptsize a} & 106\tablenotemark{\scriptsize b} & 0.58 & 29\\
DE Tau & 0.67 & 0.06 & 37\tablenotemark{\scriptsize a} & 130 & 0.55 & 30\\
DH Tau & 0.6 & 0.05 & 19 & 109 & 0.55 & 36\\
DK Tau & 2.16 & 0.1 & 15 & 175 & 0.7 & 56\\
DN Tau & 0.695 & 0.06 & 40 & 117 & 0.6 & 24\\
DO Tau & 1.24 & 0.08 & 37 & 193 & 0.52 & 47\\
DQ Tau & 1.55 & 0.09 & 43 & 143 & 0.6 & 28\\
DR Tau & 1.9 & 0.1 & 53 & 216 & 0.58 & 38\\
DS Tau & 1.04 & 0.07 & 70 & 97 & 0.67 & 15\\
FM Tau & 0.22 & 0.03 & 73 & 79\tablenotemark{\scriptsize b} & 0.58 & 14\\
FT Tau & 0.45 & 0.05 & 46 & 121 & 0.58 & 24\\
FV Tau & 0.56 & 0.05 & 26\tablenotemark{\scriptsize a} & 190 & 0.53 & 54\\
FY Tau & 1.136 & 0.07 & 21\tablenotemark{\scriptsize a} & 120\tablenotemark{\scriptsize b} & 0.58 & 36\\
GI Tau & 0.918 & 0.07 & 25 & 114\tablenotemark{\scriptsize b} & 0.58 & 31\\
HO Tau & 0.18 & 0.03 & 44 & 76\tablenotemark{\scriptsize b} & 0.58 & 16\\
HV Tau & 0.47 & 0.05 & 51 & 96\tablenotemark{\scriptsize b} & 0.58 & 19\\
Haro 6-37 & 6.28 & 0.17 & 121 & 184\tablenotemark{\scriptsize b} & 0.58 & 21\\
Haro 6-39 &  & 0.05 & 23\tablenotemark{\scriptsize a} & 100\tablenotemark{\scriptsize b} & 0.58 & 29\\
IC 2087 IR &  & 0.05 & 141\tablenotemark{\scriptsize a} & 100\tablenotemark{\scriptsize b} & 0.58 & 12\\
IP Tau & 0.52 & 0.05 & 36 & 107 & 0.63 & 23\\
IQ Tau & 0.86 & 0.06 & 111 & 121 & 0.6 & 15\\
LkCa 15 & 1.1 & 0.07 & 111 & 117 & 0.52 & 17\\
RW Aur & 3.12 & 0.12 & 21 & 154\tablenotemark{\scriptsize b} & 0.58 & 46\\
RY Tau & 23.4 & 0.33 & 70 & 342 & 0.66 & 42\\
SU Aur & 14.2 & 0.26 & 77 & 264 & 0.48 & 49\\
T Tau & 10.08 & 0.22 & 21 & 338 & 0.45 & 122\\
UX Tau & 3.34 & 0.13 & 57 & 132 & 0.41 & 34\\
UY Aur & 1.94 & 0.1 & 7 & 226 & 0.53 & 128\\
UZ Tau & 0.78 & 0.06 & 87 & 167 & 0.61 & 22\\
V710 Tau & 0.41 & 0.04 & 46 & 112 & 0.58 & 22\\
V836 Tau & 0.64 & 0.06 & 31 & 97 & 0.58 & 24\\
V892 Tau & 38 & 0.43 & 62 & 461 & 0.58 & 72\\
\enddata
\tablecomments{
$^a$ The $R_{\tiny \mbox{out}}$ estimated from $F_{\mbox{\tiny 337 GHz}}$ using the size-luminosity relation (\citealt{Tripathi_2017}). $^b$ The $T_{\mbox{\scriptsize dust, 1 AU}}$ estimated from the bolometric luminosities of the host stars (\citealt{Akeson_2019,Parker2022MNRAS.511.2453P,Manara2023ASPC..534..539M}).
}
\end{deluxetable*}

An important note to the optically thick interpretation is that it does not require most of the disk (by area, by radius, or even by mass) to be optically thick; instead, it only requires most of the flux to come from optically thick regions. 
There are at least three channels to produce such optically thick emission: The most trivial scenario is that the disk can be optically thick everywhere. 
Alternatively, dust in the disk may be organized into optically thick substructures (e.g., rings or clumps), which could leave the emission optically thick even when most of the disk area does not contain these substructures and are optically thin (e.g., \citealt{Tripathi2018ApJ...861...64T,Tazzari2021MNRAS.506.2804T}). Some continuum observations show that the dust rings can be slightly narrower than the pressure scale height when dust trapping is effective (\citealt{Ono2016ApJ...823...84O,Dullemond2018ApJ...869L..46D}). 
These highly concentrated rings are finitely geometrically narrow due to the turbulence effect and thereby define a finite fractional contribution to the overall flux density.
Finally, the disk can have a relatively steep radial profile. When $\Sigma_{\rm dust}(R) T_{\rm dust}(R)$ (which is proportional to the intensity when the emission is optically thick) scales steeper than $R^{-2}$, emission in the optically thin outer disk will become negligible compared to the emission from the optically thick inner disk because the flux per log radius peaks at the location where the emission is marginally optically thick, and the disk-integrated spectral index can be optically thick even when most of disk radius and/or area may belong to the optically thin outer disk. 
(When $\Sigma_{\rm dust}(R) T_{\rm dust}(R)$ is steeper than $R^{-2}$ but $\Sigma$ is shallower than $R^{-2}$, one could even have most disk mass being optically thin but most emission being optically thick.) 
This type of steep radial profile can be produced when the effective viscosity of the disk increases in radius, and has been used to explain low spectral indices observed in Class 0/I disks \citep{Xu2022ApJ...934..156X}. 
Each of these possibilities lead to different interpretations on disk evolution. For example, the trend that sub-mm flux decreases in time (e.g., \citealt{Carpenter2014ApJ...787...42C}) could be due to the decrease of disk size (when the disk is not very steep), the formation of more concentrated substructures, or simply a uniform decrease of surface density (when the disk is steep). 
It is therefore important to distinguish between these possibilities using future high-resolution observations.
In Section \ref{sub:slab}, we make a deeper exploration of the optically thick interpretation based on simple physical models for dust slabs. 

\begin{figure*}
    \hspace{-1.8cm} 
        \begin{tabular}{ lll }
        \includegraphics[width=7.0cm]{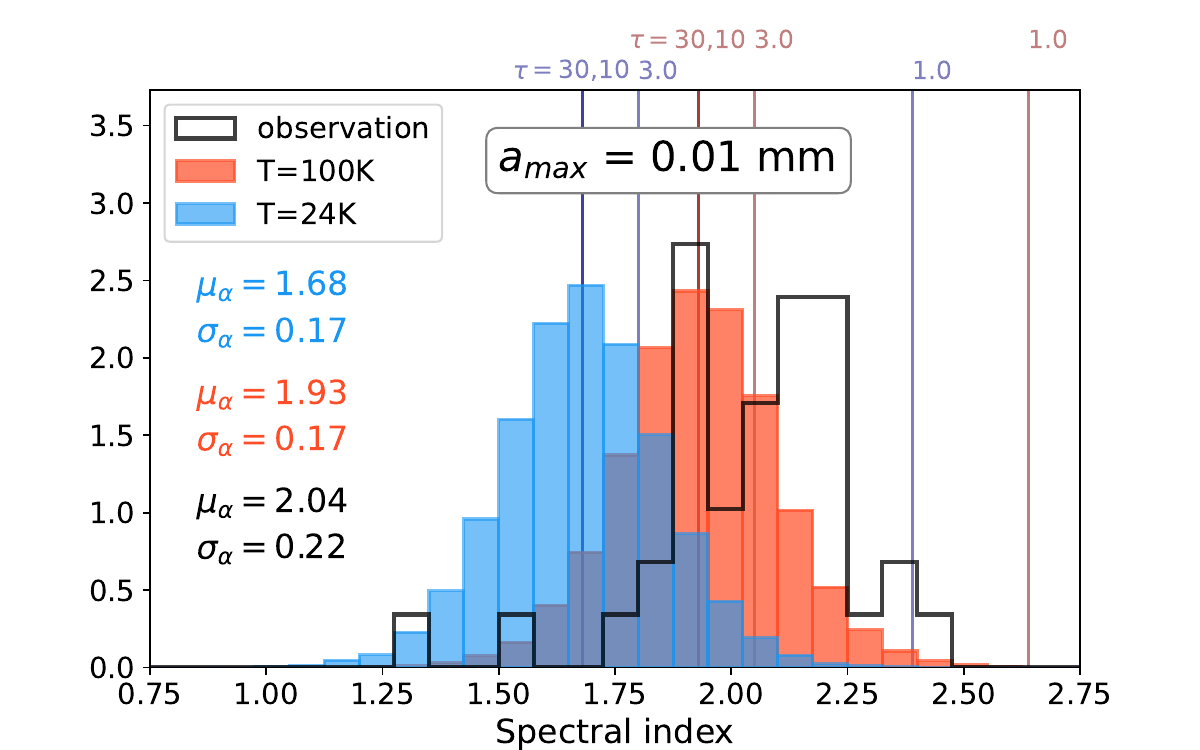}&
        \hspace{-1.1cm}
        \includegraphics[width=7.0cm]{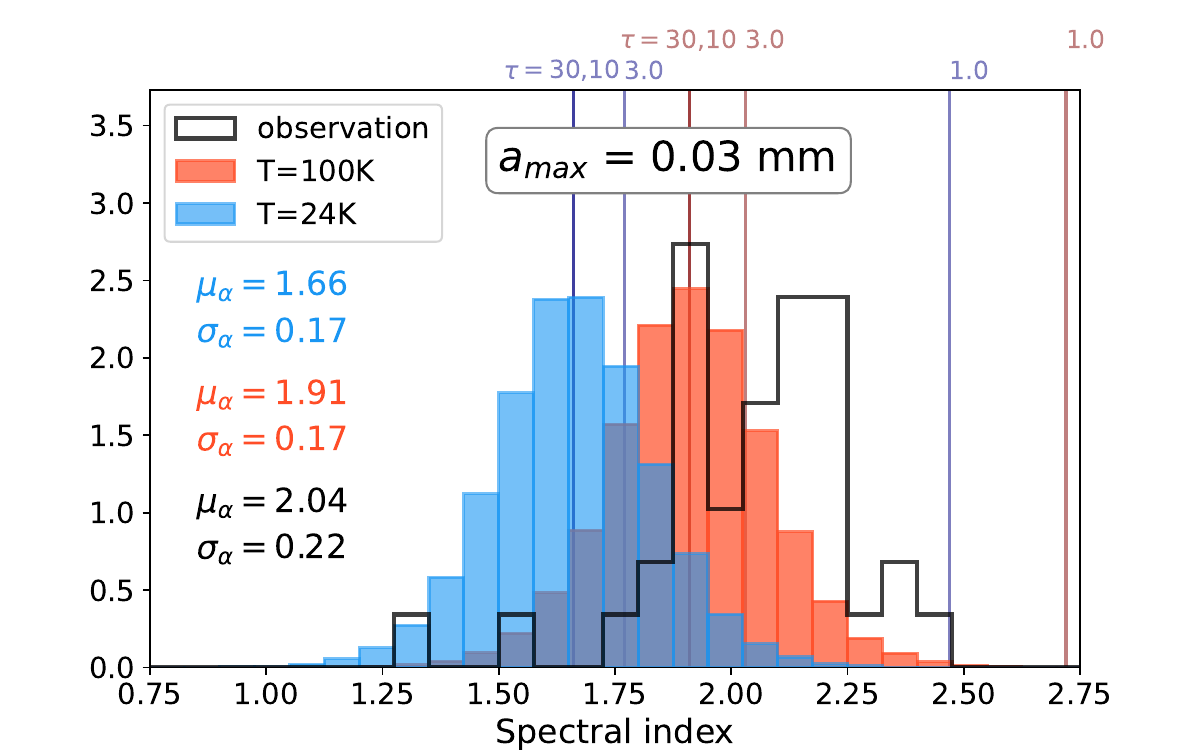}&
        \hspace{-1.1cm}
        \includegraphics[width=7.0cm]{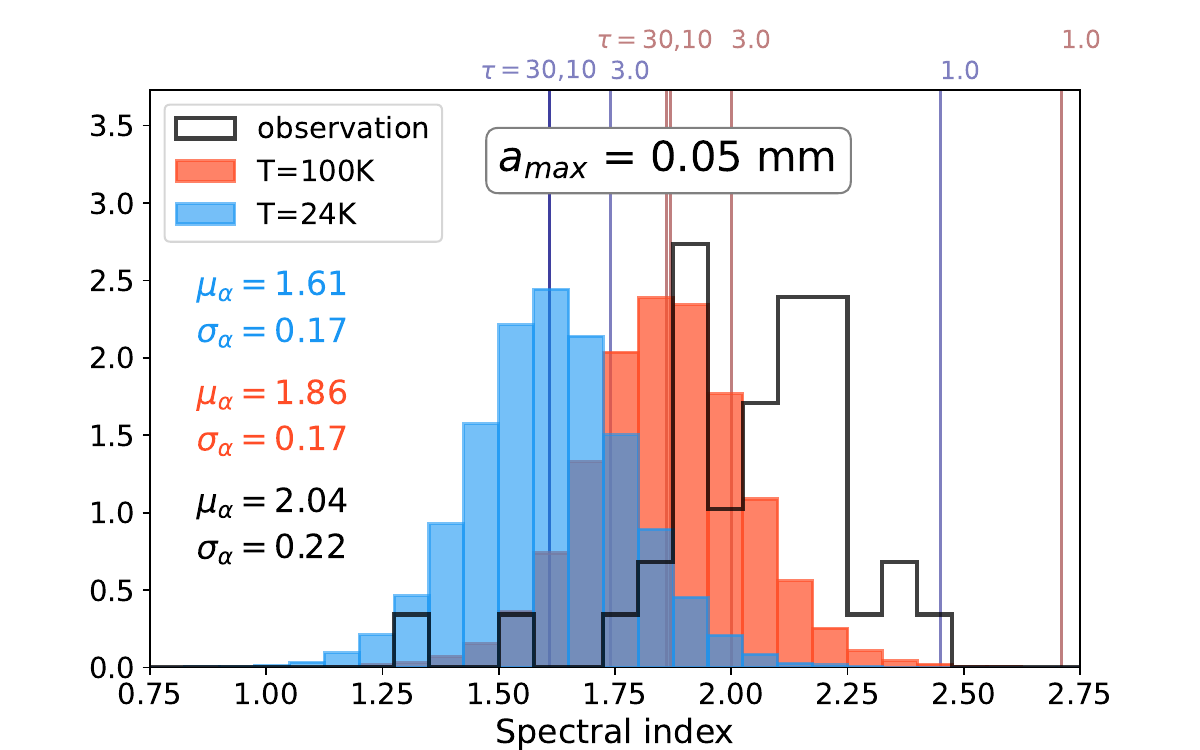}\\
        \includegraphics[width=7.0cm]{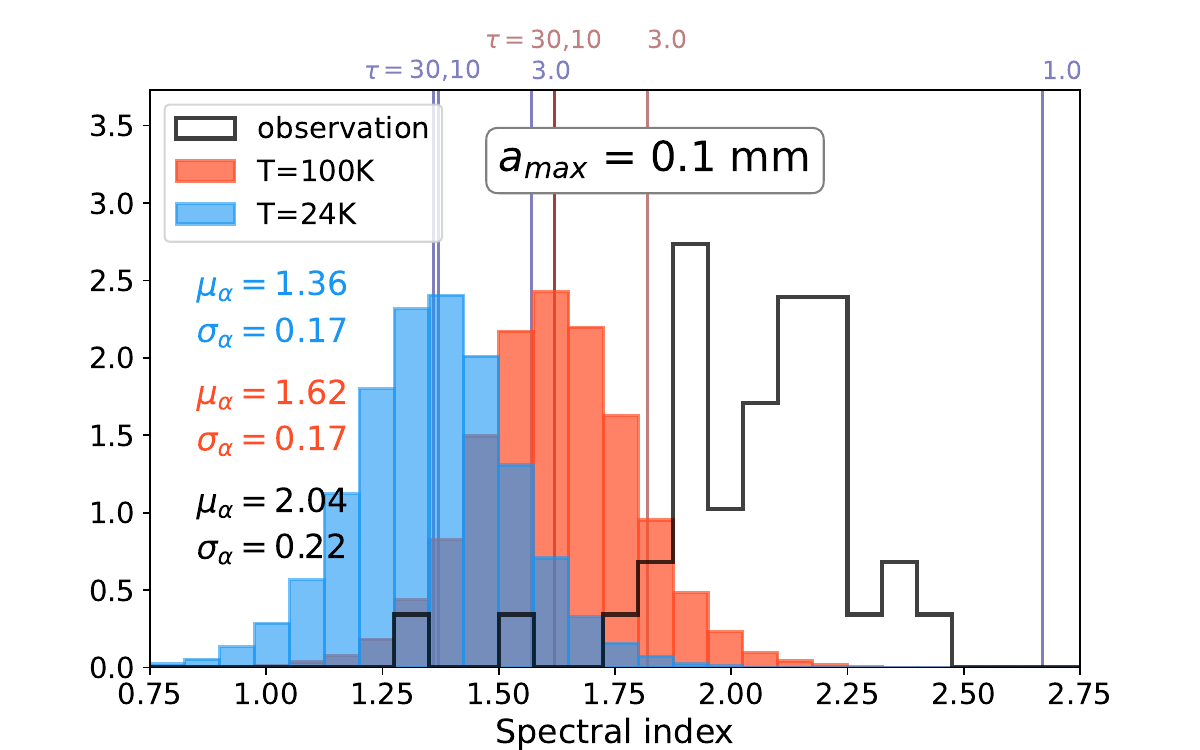}&
        \hspace{-1.1cm}
        \includegraphics[width=7.0cm]{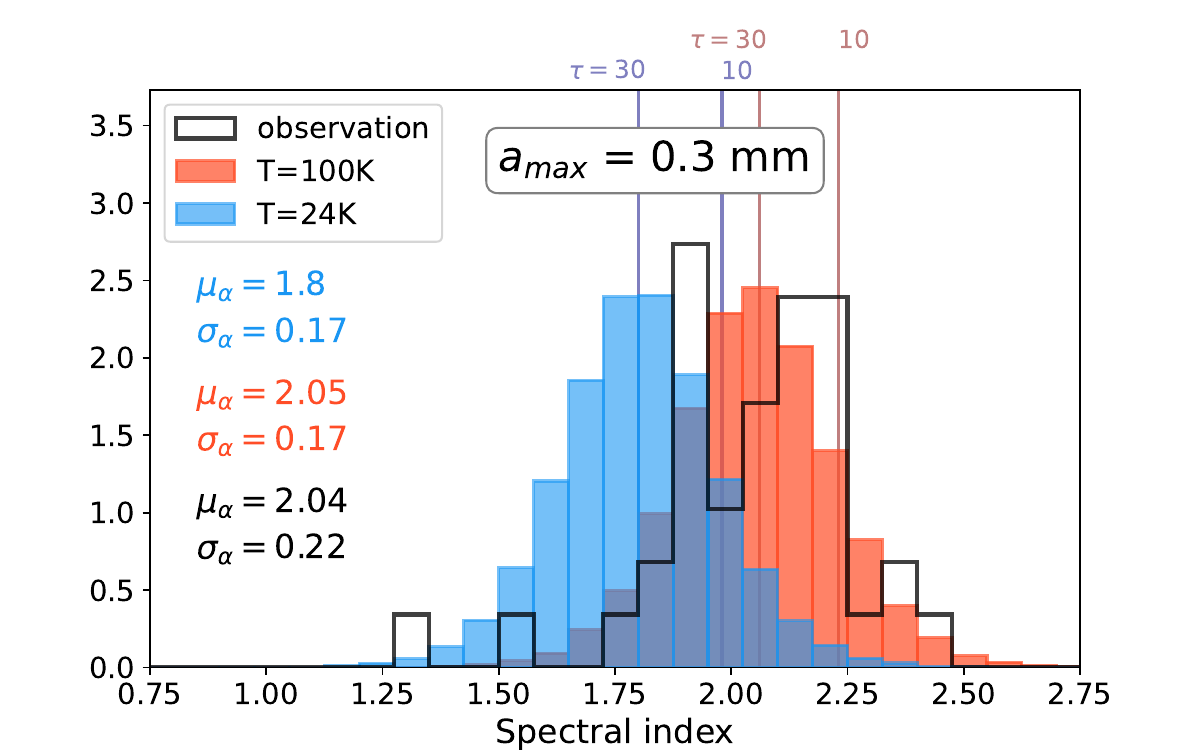}&
        \hspace{-1.1cm}
        \includegraphics[width=7.0cm]{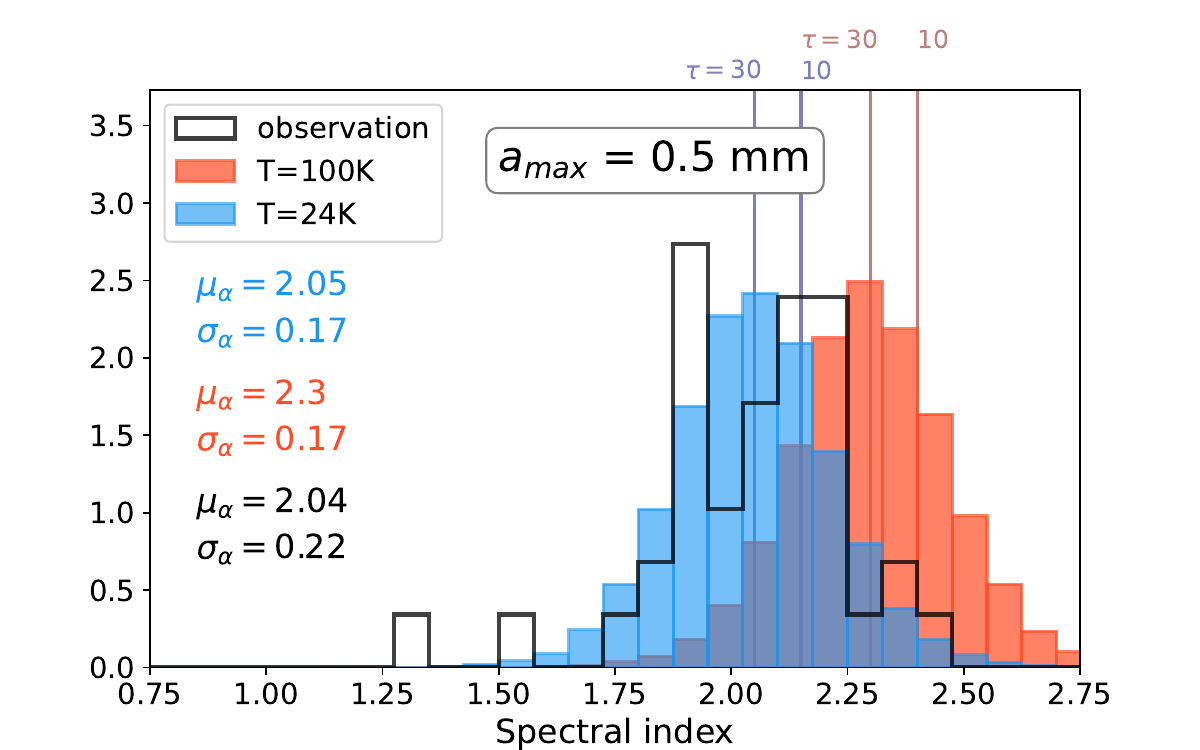}\\ 
        \includegraphics[width=7.0cm]{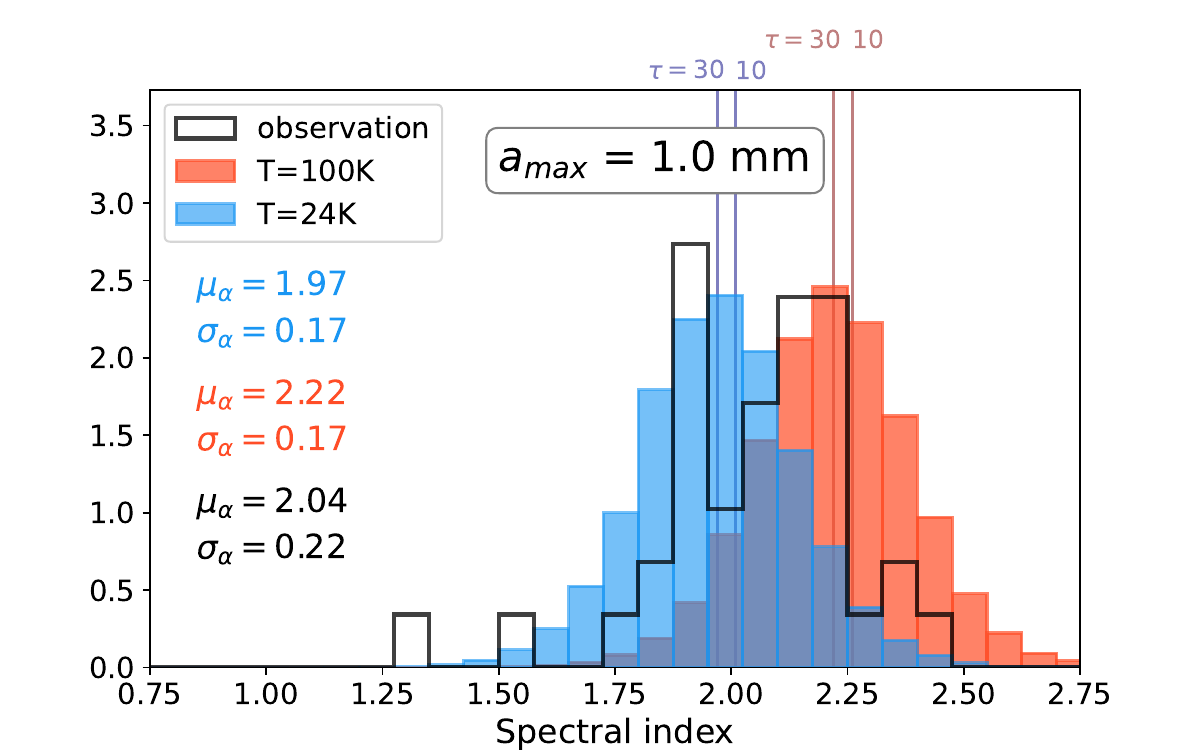}&
        \hspace{-1.1cm}
        \includegraphics[width=7.0cm]{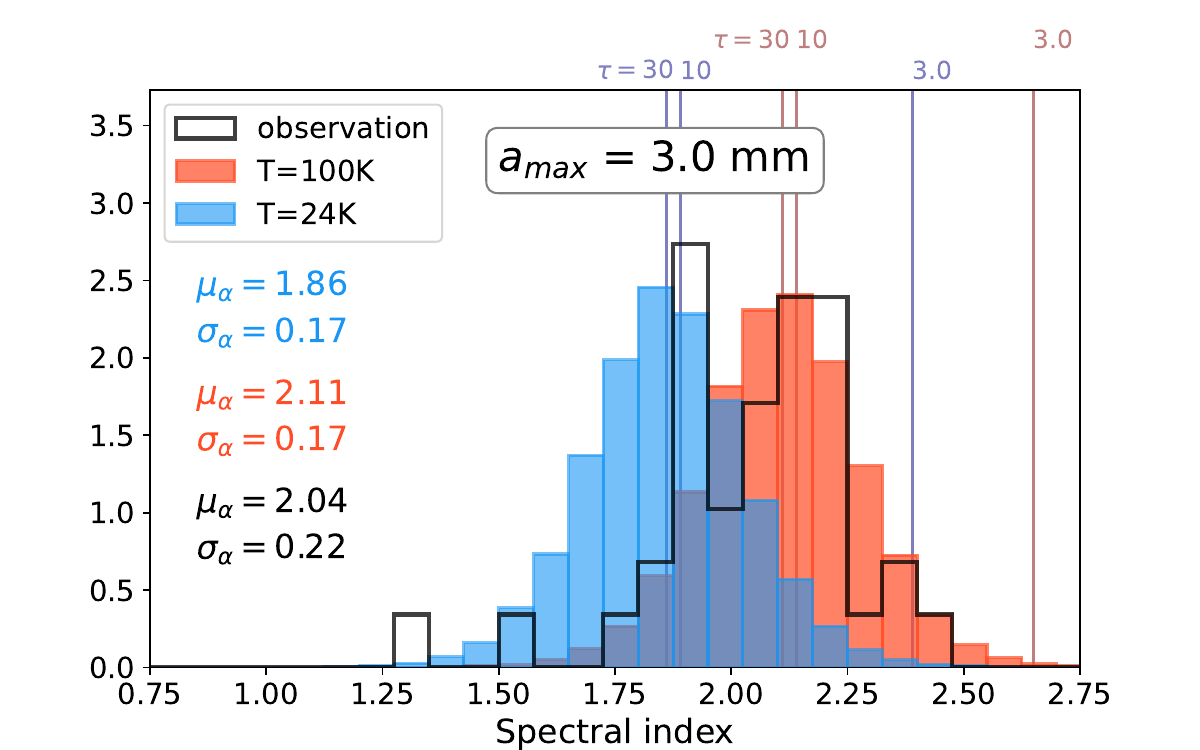}&
        \hspace{-1.1cm}
        \includegraphics[width=7.0cm]{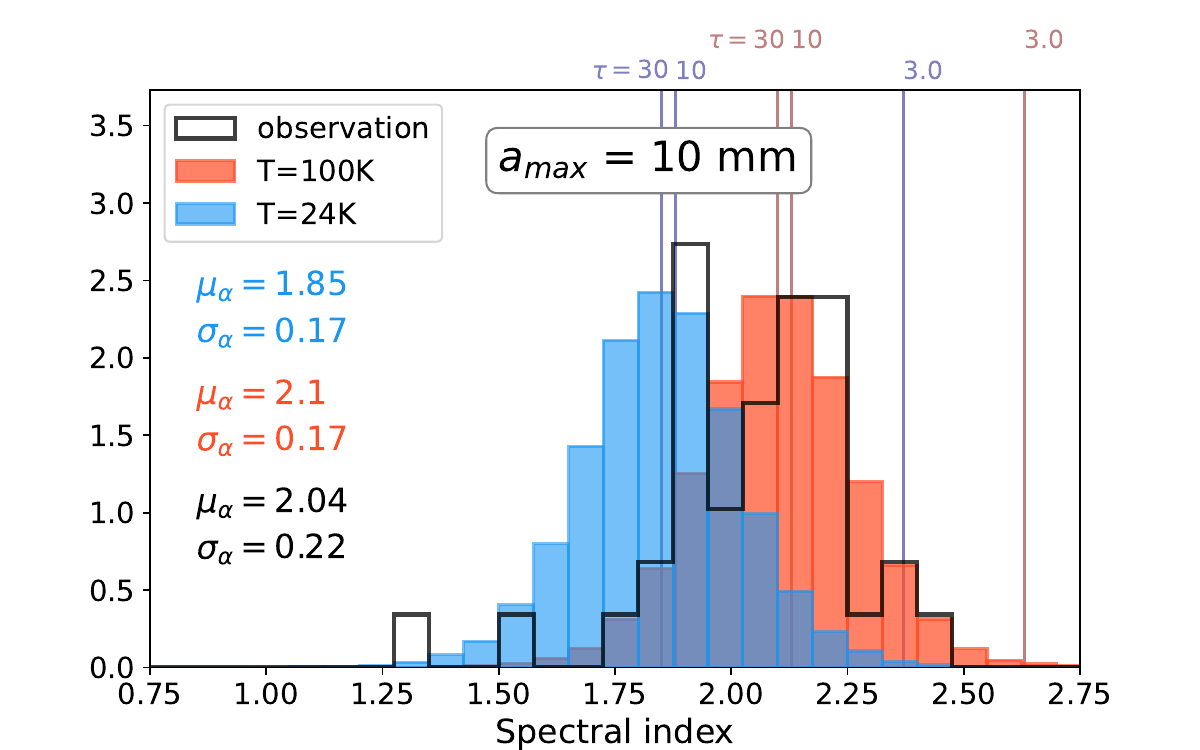}\\
        \end{tabular}
    \caption{Synthesized spectral index distribution ($\alpha_{\mbox{\scriptsize 200-420 GHz}}$) derived from the dust slab model overplotted with the spectral index distribution from the observation for the 39 unresolved sources. The top 3 panels show the synthesized distributions for $a_{\mbox{\scriptsize max}} = 0.01, 0.03, 0.05$ mm. The middle 3 panels show the synthesized distributions for $a_{\mbox{\scriptsize max}} = 0.1, 0.3, 0.5$ mm. The bottom 3 panels show the synthesized distributions for $a_{\mbox{\scriptsize max}} = 1.0, 3.0, 10$ mm. In each panel, the blue histogram shows the synthesized distribution generated by assuming the $T_{\mbox{\scriptsize dust,avg}}$ to be the median value of the distribution in Figure \ref{fig:Tdust_hist} (24 K). The red histogram shows the synthesized distribution generated by assuming $T_{\mbox{\scriptsize dust,avg}}=$100 K. The black histogram shows the spectral index distribution in the observational results. The derived $\mu_{\scriptsize \alpha}$ and $\sigma_{\scriptsize \alpha}$ of the synthesized distributions and observational results are shown in the left and have the same color with the histograms. The blue and brown vertical lines indicate the synthesized $\alpha_{\mbox{\scriptsize 200-420 GHz}}$ under different $\tau_{\mbox{\scriptsize 230 GHz}}$ assumptions for the dust slabs in $T_{\mbox{\scriptsize dust,avg}}=$24 K and 100 K cases, respectively. 
    }
    \label{fig:spid_hist}
\end{figure*}

\subsection{Simple physical models}\label{sub:slab}
\subsubsection{Population synthesis}\label{sub:Popsynthesis}
For a preliminary, quantitative understanding of the observed spectral index distribution (Figure \ref{fig:337GHz_flux_spidx}), we performed population synthesis based on simple dust-slab models. 
For simplicity, we approximated each observed Class II disk as a dust slab that has uniform dust column density, temperature, and $a_{\mbox{\scriptsize max}}$.
In addition, we assumed that the dust slabs are extremely optically thick and considered the dust self-scattering effect. 
We focused on the 39 spatially unresolved sources since the extremely optically thick assumption is unlikely a good approximation for all the spatially resolved disks. 
We assumed all unresolved sources to have the same $a_{\mbox{\scriptsize max}}$, and varied the value of $a_{\mbox{\scriptsize max}}$ to see which $a_{\mbox{\scriptsize max}}$ results in the spectral index distribution that matches the best to the observed one.
Our simple assumption were motivated by the narrowly distributed $\alpha_{\mbox{\scriptsize 200-420 GHz}}$ of the unresolved source. 

At first, we tried adopting the functional form $T_{\mbox{\scriptsize dust}}(R)=T_{\mbox{\scriptsize dust, 1au}} (R/\mbox{1 au})^{-q}$ for dust temperature profile, which was used by \citet{Andrews_Williams_2005} in modeling their observed Class~II disks.
We then evaluated the averaged dust temperature ($T_{\mbox{\scriptsize dust,avg}}$) in each disk based on
\begin{equation}
T_{\mbox{\scriptsize dust,avg}} = \frac{ \int_{R_{\mbox{\scriptsize in}}}^{R_{\mbox{\scriptsize out}}} T_{\mbox{\scriptsize dust}}(R) B_{\mbox{\scriptsize 337 GHz}}(T_{\mbox{\scriptsize dust}}(R)) 2\pi RdR }{ \int_{R_{\mbox{\scriptsize in}}}^{R_{\mbox{\scriptsize out}}} B_{\mbox{\scriptsize 337 GHz}}(T_{\mbox{\scriptsize dust}}(R)) 2\pi RdR }
\end{equation}
where $B_{\mbox{\scriptsize 337 GHz}}(T_{\mbox{\scriptsize dust}})$ is the Planck function at 337 GHz, $\mbox{R}_{\mbox{\scriptsize in}}$ and $\mbox{R}_{\mbox{\scriptsize out}}$ are the inner rim radius and the outer radius, respectively.
For the 23 unresolved sources that have been modeled in \citet{Andrews_Williams_2005}, we quote their best-fit values of $T_{\mbox{\scriptsize dust, 1au}}$ and $q$.
For the remaining 16 sources, we assumed $q = 0.58$, which is the median value of the quoted $q$. 
The $T_{\mbox{\scriptsize dust, 1au}}$ of these 16 sources were estimated from the bolometric luminosities of the host stars (\citealt{Akeson_2019,Parker2022MNRAS.511.2453P,Manara2023ASPC..534..539M}; Table \ref{tab:Tdust}) based on the flaring disk model (\citealt{Chiang1997ApJ...490..368C,Dullemond2001ApJ...560..957D}) that assumed that the disk is passively heated by protostellar irradiation. 
In this estimation, we assumed the disk flaring angle to be 0.015 which was calculated from the flaring power index obtained in \citet{Woitke2016A&A...586A.103W}. 
Based on the adopted temperature profiles, we estimated inner rim radius ($\mbox{R}_{\mbox{\scriptsize in}}$) to be at where $T_{\mbox{\scriptsize dust}}(R)$ is 1500 K, which is assumed to be the dust sublimation temperature  \citep{Dullemond2001ApJ...560..957D}. 
For the 32 sources which we quoted the disk radius in Table \ref{tab:targ_properties}, we used the quoted disk radius as the outer disk radius (${\mbox{R}}_{\mbox{\scriptsize out}}$). 
For the other 7 sources, we estimated their $\mbox{R}_{\mbox{\scriptsize out}}$ from the $\mbox{F}_{\mbox{\scriptsize 337 GHz}}$ using the same method in Section \ref{sub:statistical}.
The evaluated $\mbox{T}_{\mbox{\scriptsize dust, avg}}$ are listed in Table \ref{tab:Tdust}. 
The distribution of them is shown in Figure \ref{fig:Tdust_hist}. 

For many of the estimated $\mbox{T}_{\mbox{\scriptsize dust, avg}}$, our 200--400 GHz observing frequencies were already not in the Rayleigh-Jeans tail of the Planck function.
Therefore, for a specific assumption of $a_{\mbox{\scriptsize max}}$ value, the synthesized spectral index of an isothermal dust slab will depend on the value of $\mbox{T}_{\mbox{\scriptsize dust, avg}}$: a lower $\mbox{T}_{\mbox{\scriptsize dust, avg}}$ yields a lower spectral index.
The broad $\mbox{T}_{\mbox{\scriptsize dust, avg}}$ distribution in our estimates (Figure \ref{fig:Tdust_hist}) thereby resulted in a standard deviation ($\sigma_{\scriptsize \alpha}$) in the synthesized spectral index distribution that is higher than that of the observed distribution (See Appendix \ref{appendix:popsynthesis}).
To suppress $\sigma_{\scriptsize \alpha}$ in the synthesized spectral index distribution, one way is to consider that the dispersion of $\mbox{T}_{\mbox{\scriptsize dust, avg}}$ in Figure \ref{fig:Tdust_hist} was over-estimated.
In other words, the actual $\mbox{T}_{\mbox{\scriptsize dust, avg}}$ in our sample may be distributed over a much narrower range.
To explore this possibility, we re-ran the population synthesis by adopting the median value of the evaluated dust temperature, $\mbox{T}_{\mbox{\scriptsize dust, avg}}=24$ K for all 39 unresolved sources.
This assumption of $\mbox{T}_{\mbox{\scriptsize dust, avg}}$ is not particularly favorable to us since the narrowly distributed $\mbox{T}_{\mbox{\scriptsize dust, avg}}$ seems to require a coincidence. 
Another possibility to resolve the too large $\sigma_{\scriptsize \alpha}$ issue is to consider that many $\mbox{T}_{\mbox{\scriptsize dust, avg}}$ in Figure \ref{fig:Tdust_hist} were underestimated (e.g., due to unrealistic consideration of disk geometry or missing the consideration of viscous heating).
If the actual $\mbox{T}_{\mbox{\scriptsize dust, avg}}$ of most of the selected sources can make the 200--400 GHz observing frequencies in the Rayleigh-Jeans limit, the synthesized spectral index will not be sensitive to the value of $\mbox{T}_{\mbox{\scriptsize dust, avg}}$.
In this case, having a distribution of $\mbox{T}_{\mbox{\scriptsize dust, avg}}$ will not lead to broadened spectral index distribution.
In light of this, we produced another version of population synthesis that assumes $T_{\mbox{\scriptsize dust,avg}}=$ 100 K for all 39 unresolved sources.
This temperature assumption is to represent other possible temperature distributions that make our SMA observations (Section \ref{sub:obs}) in the Rayleigh-Jeans limit.
The $T_{\mbox{\scriptsize dust,avg}}=$ 100 K face value is not to be considered as a realistic physical value. 

We adopted the DSHARP standard dust opacities  (20\% water ice, 32.91\% astronomical sillicates, 7.43\% trollite, and 39.66\% refractory organics; \citealt{Birnstiel2018ApJ...869L..45B}).
A brief discussion about the dust opacity table is given in Appendix \ref{appendix:diana}.
When evaluating the size-averaged dust opacities, we assumed a power law grain size (a) distribution $n(a) \propto a^s$ with $s=-3.5$ over the minimum and maximum grain sizes ($a_{\mbox{\scriptsize min}}$, $a_{\mbox{\scriptsize max}}$).
The size-averaged dust opacities are insensitive to the exact value of $a_{\mbox{\scriptsize min}}$ which we set to $0.1 \mu$m.
To generally represent the optically thick dust slabs, we adopted a constant dust column density $\Sigma = 10^{3}$ g\,cm$^{-2}$ such that the optical depth $\tau_{\mbox{\scriptsize 230 GHz}}$ will be greater than 50 when $a_{\mbox{\scriptsize max}}$ is in the range of 0.01--10 mm.

We evaluated flux densities of the 39 dust slabs (with different $T_{\mbox{\scriptsize dust,avg}}$ and same $a_{\mbox{\scriptsize max}}$ and $\Sigma$) at the 12 observing frequencies of the present SMA survey based on the radiative transfer solution (Equations 10--20) provided in \citet{Birnstiel2018ApJ...869L..45B}.
The solution includes the effect of dust self-scattering by introducing the scattering opacity $\kappa_\nu^{\mbox{\tiny sca}}$. 
We measured the spectral index of each dust slab by fitting a single power law on the synthesized the 200--420 GHz SED using the {\tt curve\_fit} function in the scipy software package (\citealt{2020SciPy-NMeth}). 

The observed spectral index distribution (Figure \ref{fig:spid_hist}) was affected by the uncertainties of our spectral index measurements (Section \ref{sub:SED}).
To make a fair comparison, we need to incorporate the similar uncertainties in our population synthesis.
This was implemented by randomly re-sampling the spectral index of each of the 39 dust slabs by 1000 times based on the assumed error distribution.
We assumed the error distributions to be Gaussian that the standard deviations are the same with the 1-$\sigma$ spectral index errors we labeled in Figures \ref{fig:SED_0510_resolved}--\ref{fig:SED_0510_1} (Section \ref{sub:SED}).

\begin{figure}
\includegraphics[width=9cm]{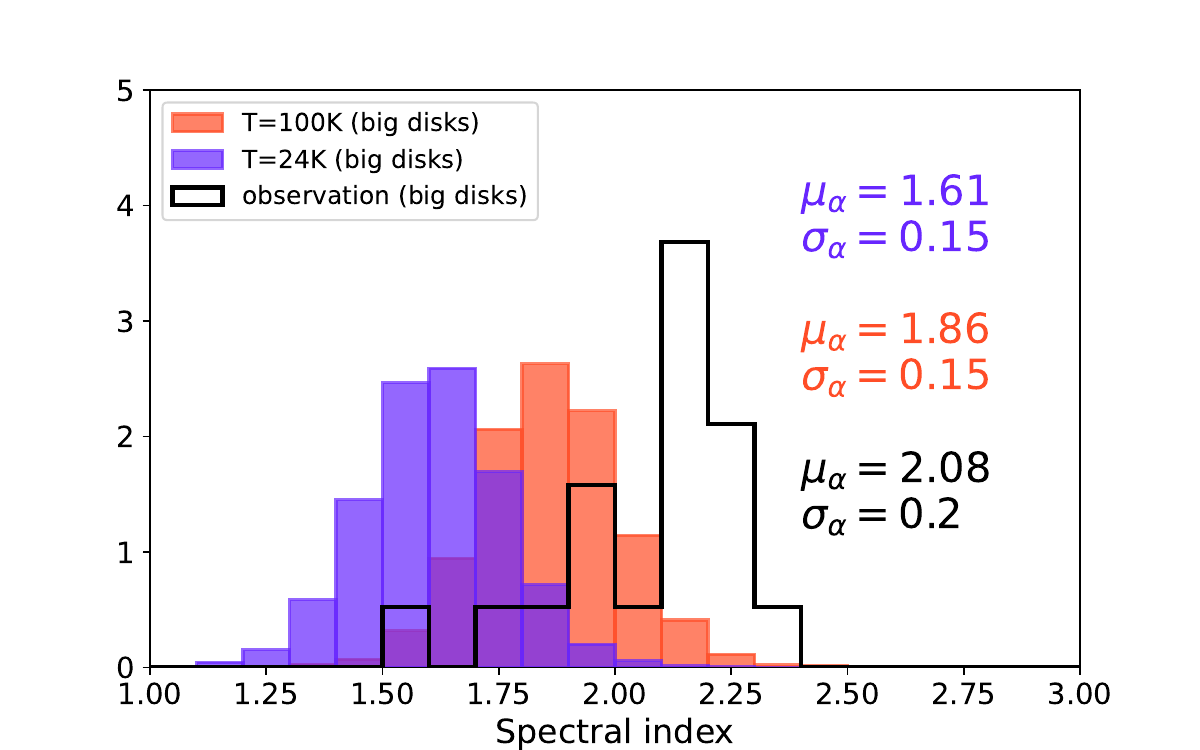}\\
\includegraphics[width=9cm]{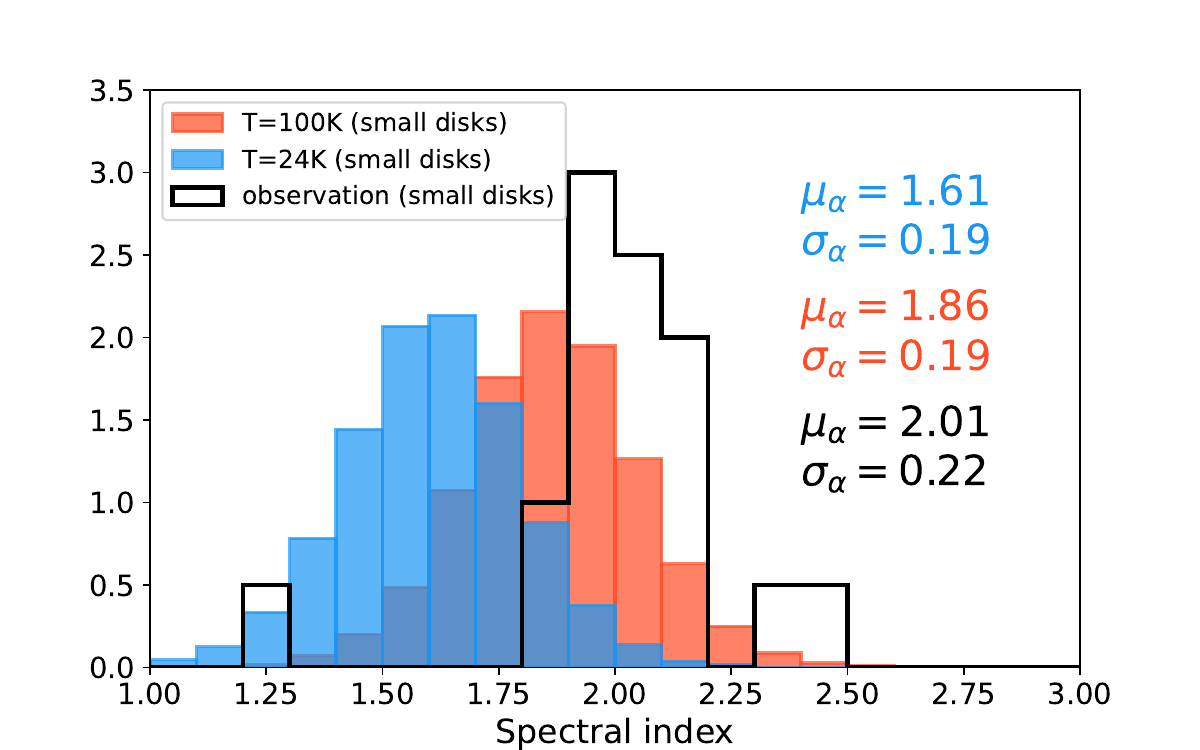}
\caption{
The spectral index distributions ($\alpha_{\mbox{\scriptsize 200-420 GHz}}$) of the larger ($R_{\mbox{\scriptsize disk}}>46$ AU) 19 ({\it top}) and smaller 20 ({\it bottom}) of the 39 unresolved Class~II disks in our SMA survey, which are defined based on the estimates of the disk radii ($R_{\mbox{\scriptsize disk}}$; Section \ref{sub:statistical}). 
These spectral index distributions are compared with those produced from our population synthesis assuming $a_{\mbox{\scriptsize max}} = 0.05$ mm (Section \ref{sub:slab}).
The style of presentation is the same as that of Figure \ref{fig:spid_hist}. 
}
\label{fig:spid_hist_2grp_size}
\end{figure}

\begin{figure*}
    \hspace{-1.8cm}
    \begin{tabular}{ lll }
    \includegraphics[width=6.5cm]{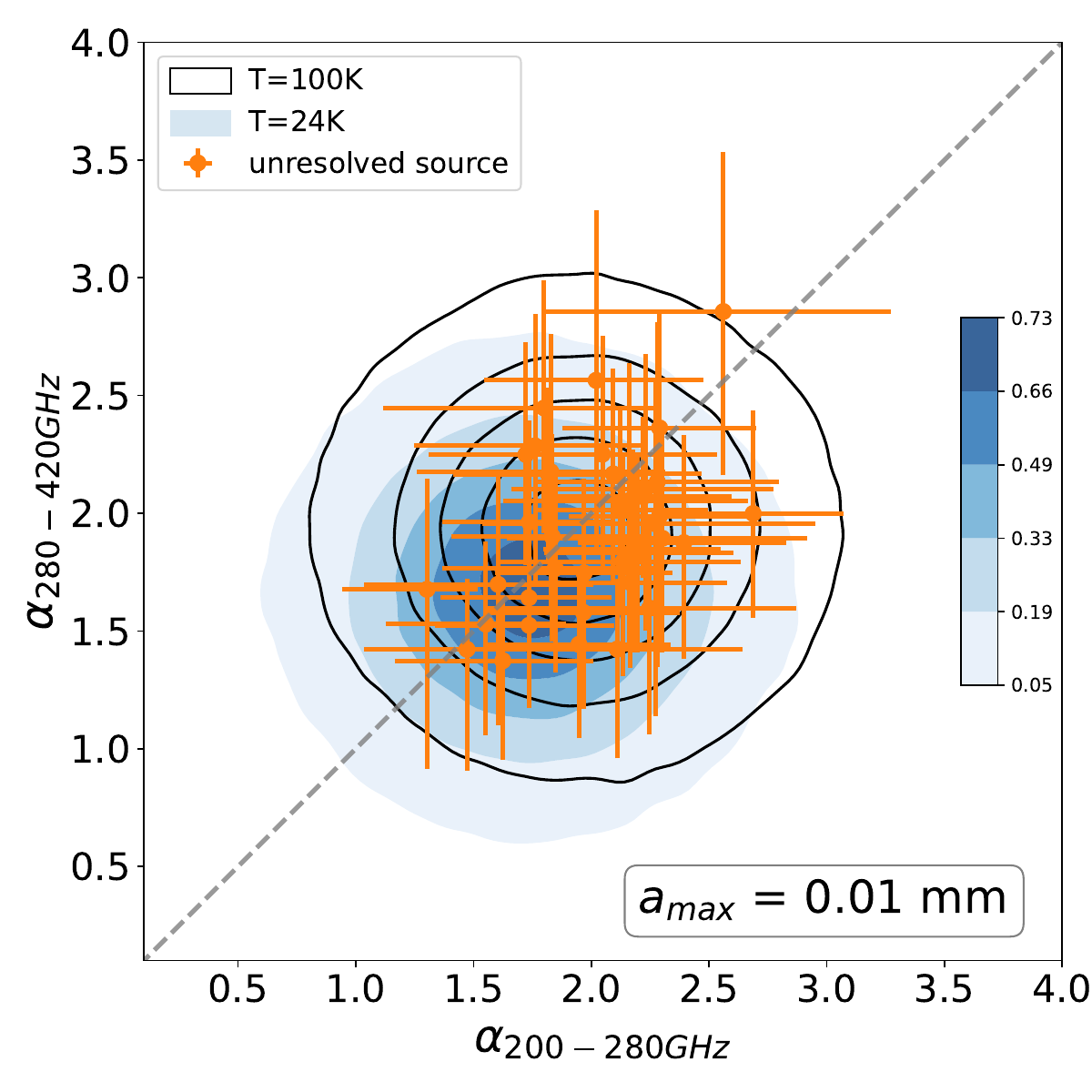}&
    \hspace{-0.5cm}
    \includegraphics[width=6.5cm]{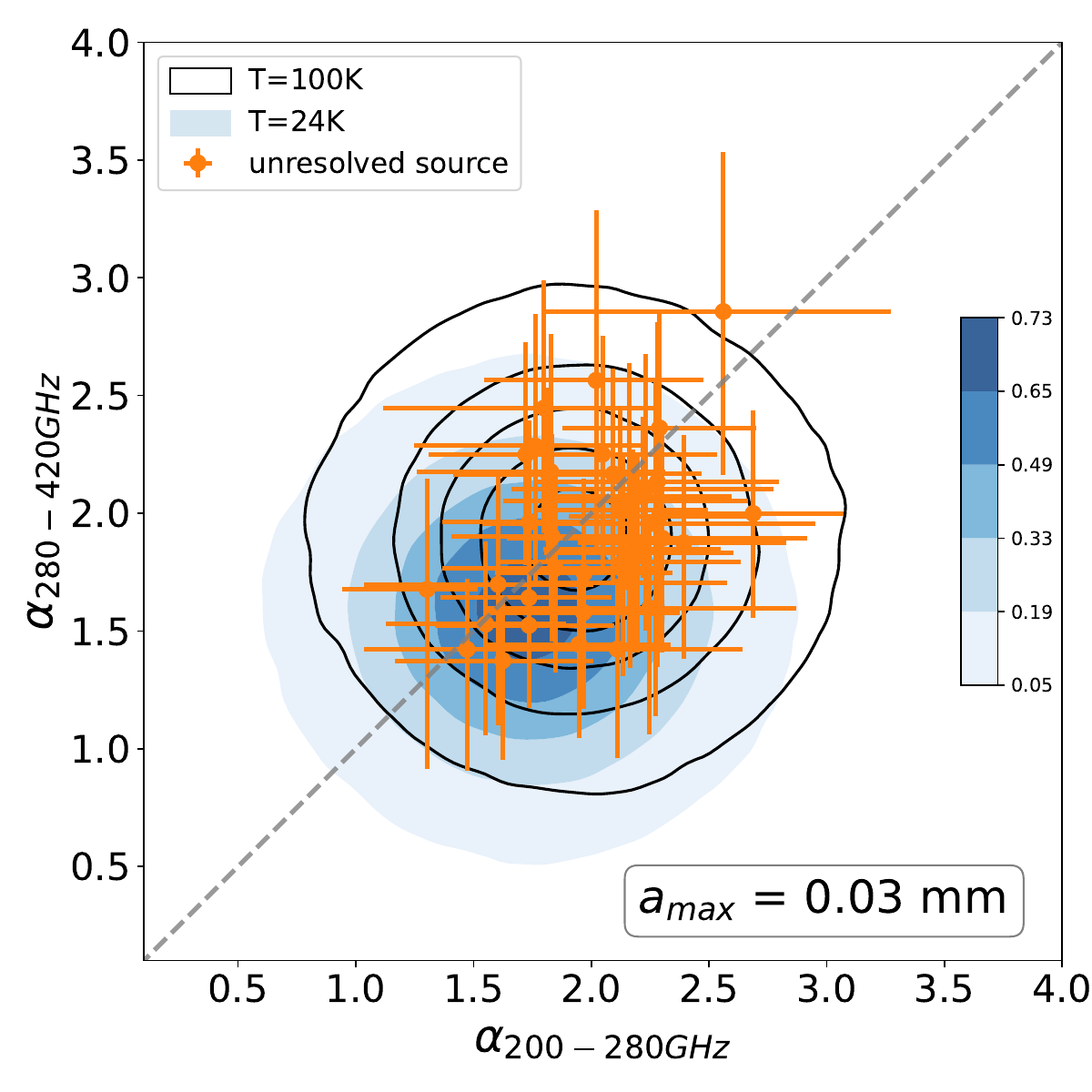}&
    \hspace{-0.5cm}
    \includegraphics[width=6.5cm]{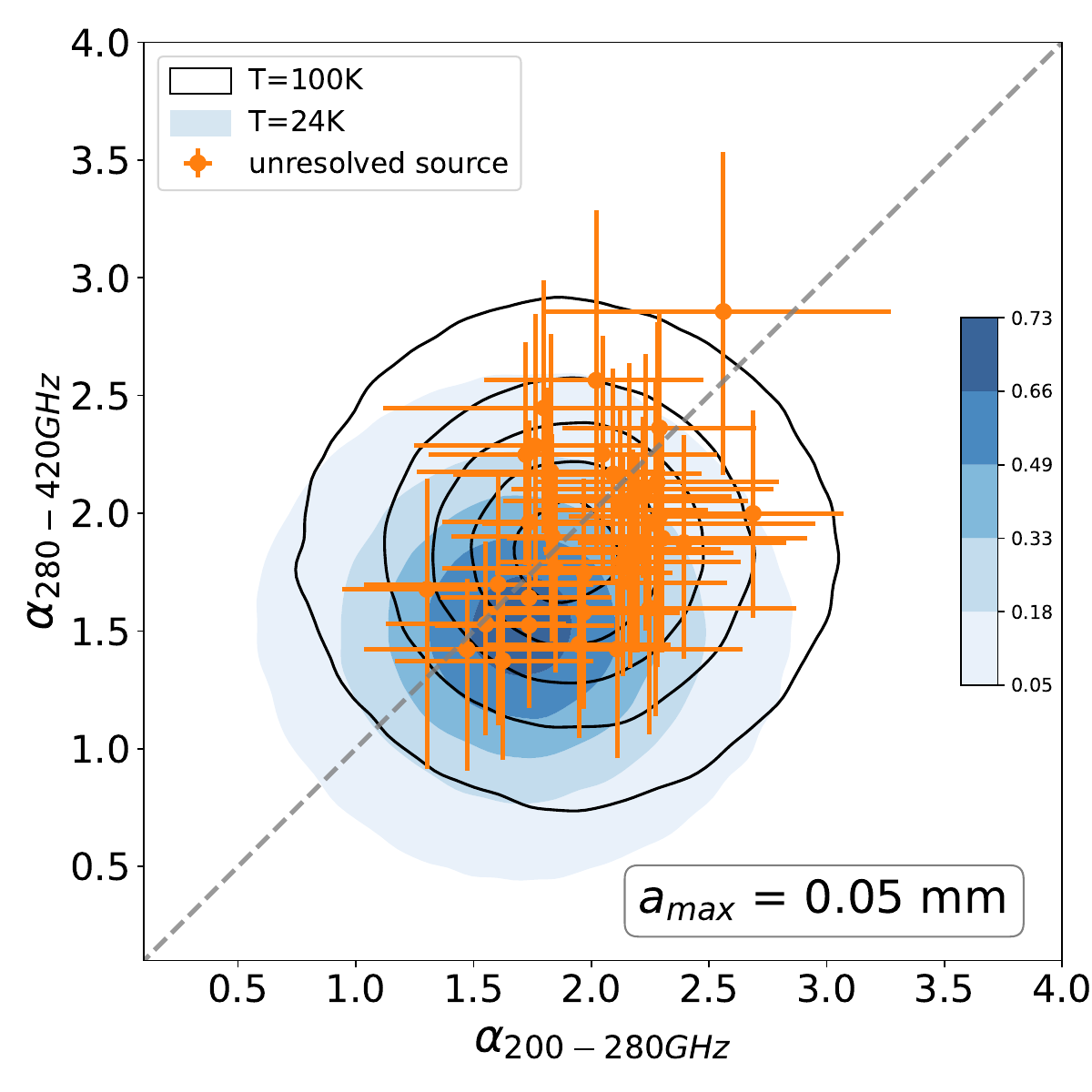}\\
    \includegraphics[width=6.5cm]{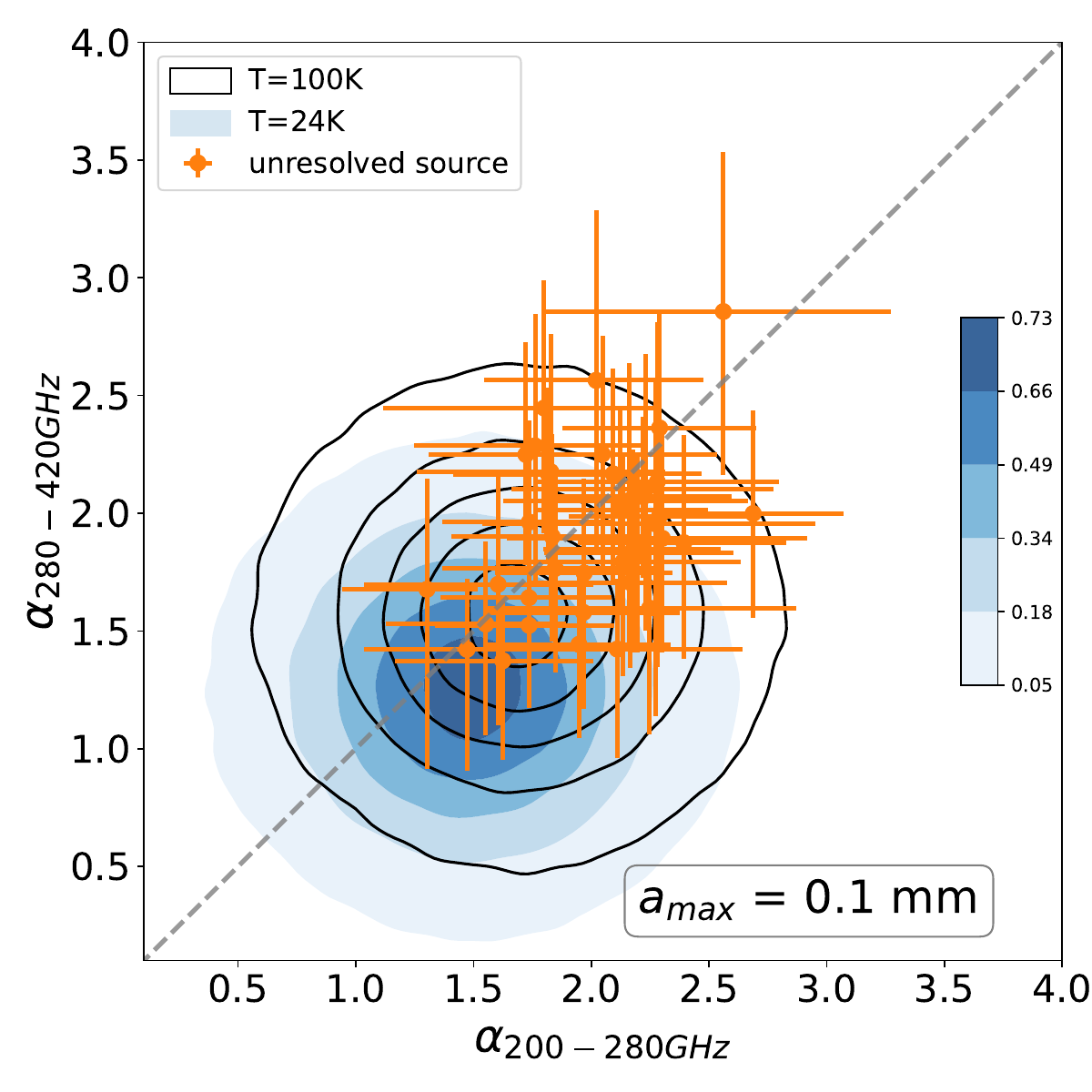}&
    \hspace{-0.5cm}
    \includegraphics[width=6.5cm]{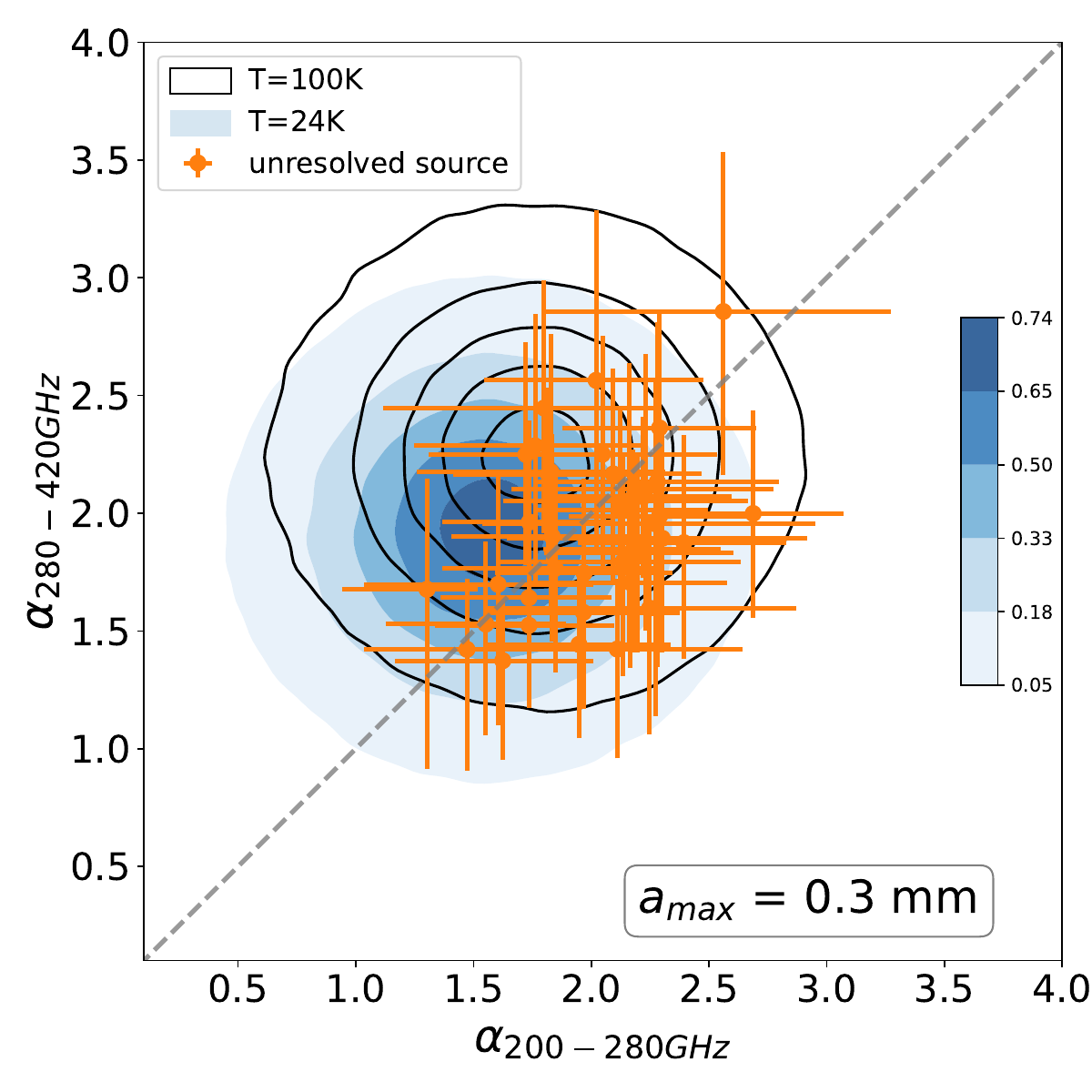}&
    \hspace{-0.5cm}
    \includegraphics[width=6.5cm]{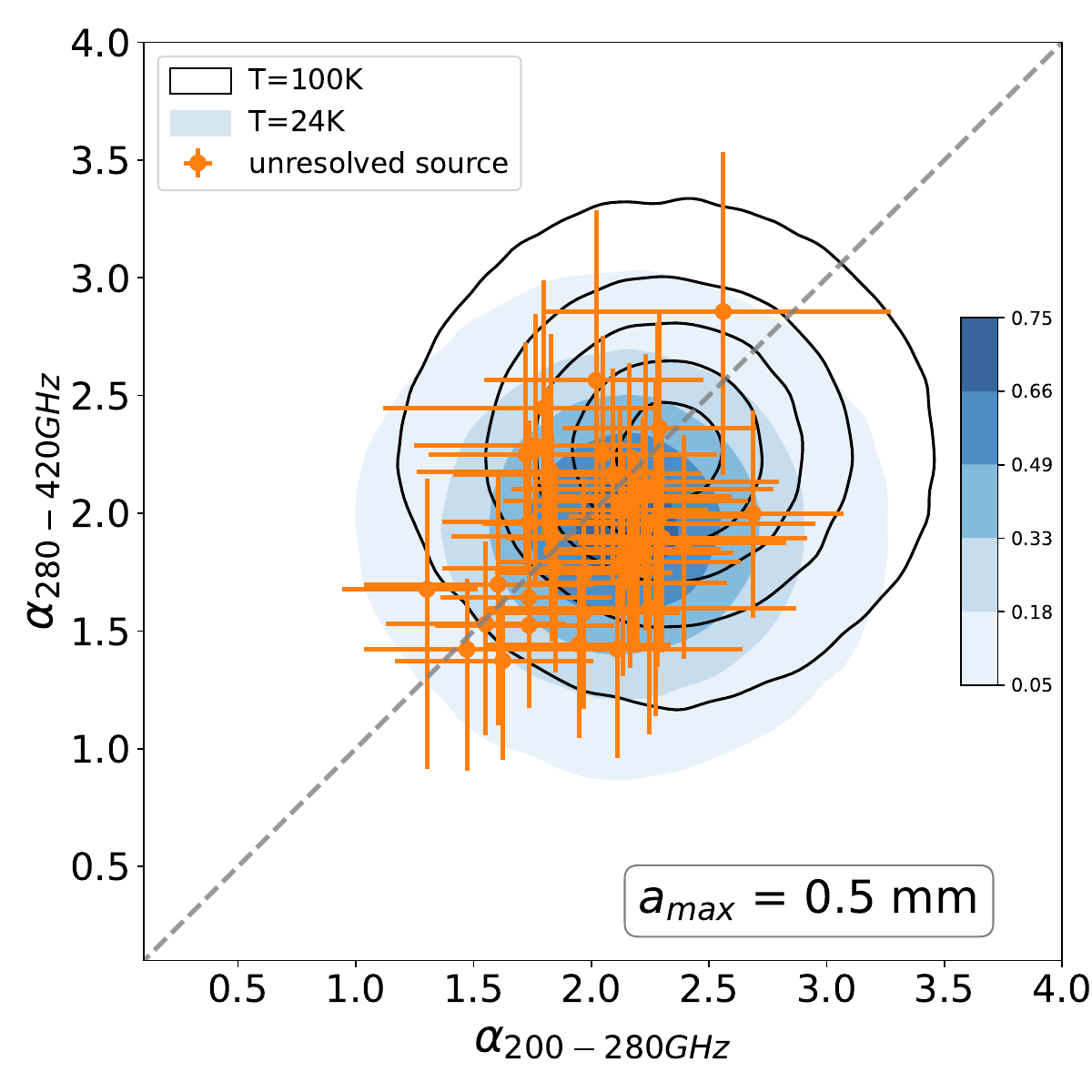}\\
    \includegraphics[width=6.5cm]{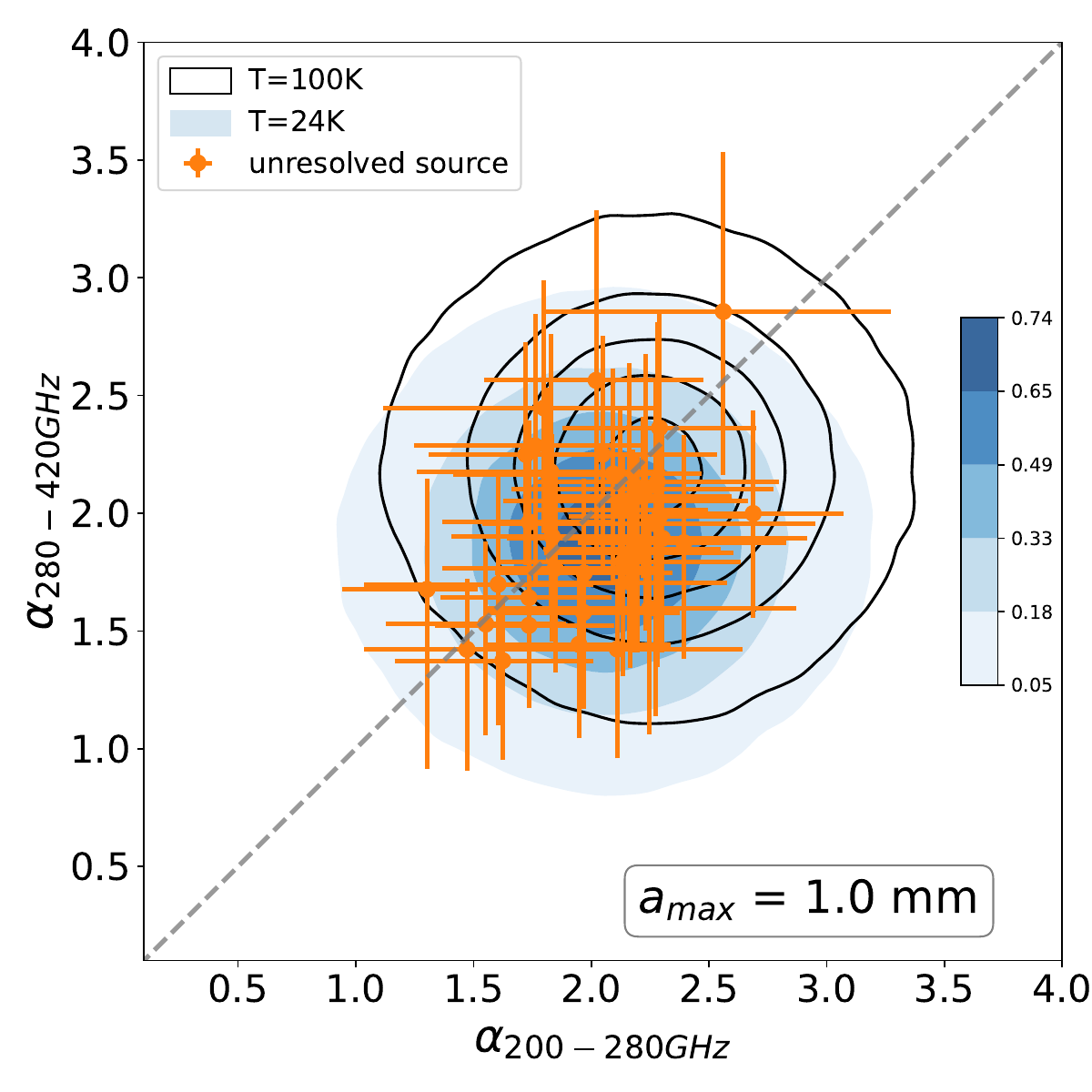}&
    \hspace{-0.5cm}
    \includegraphics[width=6.5cm]{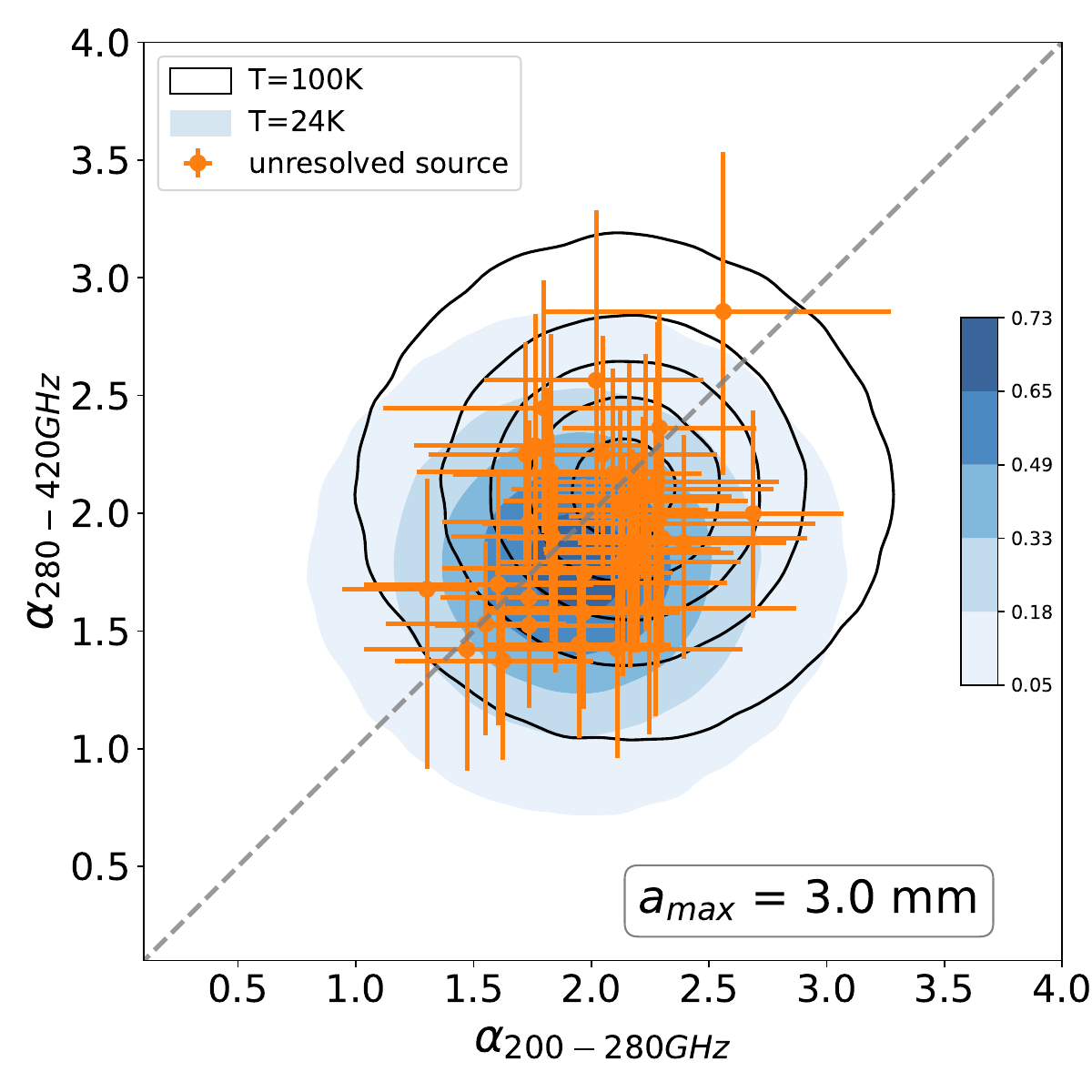}&
    \hspace{-0.5cm}
    \includegraphics[width=6.5cm]{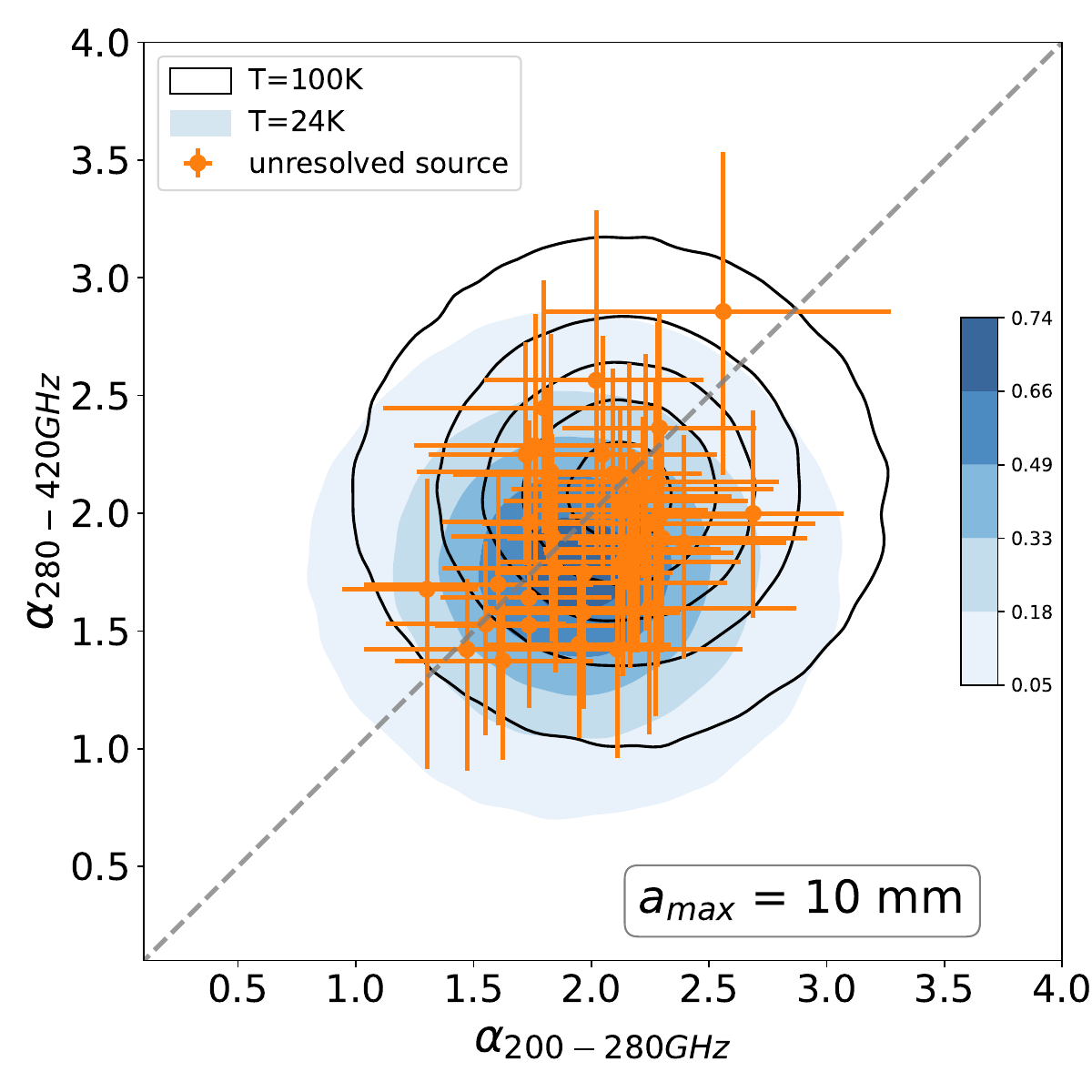}\\
    \end{tabular}
    \caption{Synthesized spectral index ($\alpha_{\mbox{\scriptsize 280-420 GHz}}$ and $\alpha_{\mbox{\scriptsize 200-280 GHz}}$) distributions in 2D histogram for the 39 unresolved sources derived from our population synthesis (Section \ref{sub:slab}) overplotted with the observed  $\alpha_{\mbox{\scriptsize 280-420 GHz}}$ and $\alpha_{\mbox{\scriptsize 200-280 GHz}}$ (symbols; see Figure \ref{fig:337GHz_flux_del_spidx}). The top 3 panels show the synthesized distributions for $a_{\mbox{\scriptsize max}} = 0.01, 0.03, 0.05$ mm. The middle 3 panels show the synthesized distributions for $a_{\mbox{\scriptsize max}} = 0.1, 0.3, 0.5$ mm. The bottom 3 panels show the synthesized distributions for $a_{\mbox{\scriptsize max}} = 1.0, 3.0, 10$ mm. In each panel, the filled blue contour shows the synthesized distribution generated by assuming the $T_{\mbox{\scriptsize dust,avg}}$ to be the median value (24 K) of the distribution in Figure \ref{fig:Tdust_hist}. The black contour shows the synthesized distribution generated by assuming (sub)millimeter flux density is dominated by dust sources in the Rayleigh-Jeans limit (e.g., $T_{\mbox{\scriptsize dust,avg}}=$100 K). The contours in each panel (from innermost to outermost) enclose 10\%, 30\%, 50\%, 70\% and 90\% of the probability mass. The color bar on the right of each panel shows the normalized probability density at each contour level in the filled blue contours. The black contours and filled blue contours are demonstrated in the same way such that they share the same probability density levels and thus each contour level in the two distributions can be compared directly. 
    } 
    \label{fig:spid_2dhist}
\end{figure*}

\subsubsection{Comparison with observations}\label{sub:Popcompar}

Figure \ref{fig:spid_hist} shows the comparison of the observed spectral index distribution with those resulting from our population synthesis. 
We present the cases of $a_{\mbox{\scriptsize max}} = 0.01,  0.03,  0.05,  0.1,  0.3, 0.5,  1.0,  3.0,  10$ mm. 

We found that the population syntheses, under the identical $T_{\mbox{\scriptsize dust,avg}}$ assumptions (i.e., 24 K or 100 K), yielded $\sigma_{\scriptsize \alpha}=$0.17, which is very reasonable considering that mixing little amount of optically thin emission can lift $\sigma_{\scriptsize \alpha}$ to the observed $\sigma_{\scriptsize \alpha}=$0.22.
The observed correlation between $R_{\mbox{\scriptsize disk}}$ and $\alpha_{\mbox{\scriptsize 200--400 GHz}}$ (Figure \ref{fig:disksize_spidx}) indeed can be due to mixing optically thin emission (see Section \ref{sub:qualitative}).
The assumption of $T_{\mbox{\scriptsize dust,avg}}=$100 K yielded systematically higher synthesized spectral indices than the assumption of $T_{\mbox{\scriptsize dust,avg}}=$24 K.

Note that in the population syntheses, we assumed that the dust slabs are extremely optically thick such that the $\alpha_{\mbox{\scriptsize 200-420 GHz}}$ is not sensitive to the assumed dust column density. 
This helps explain the observed, narrow distribution of spectral indices. 
To evaluate the effect of optical depth, we labeled the $\mu_{\scriptsize \alpha}$ of the synthesized distribution when $\tau_{\mbox{\scriptsize 230 GHz}} = 1.0, 3.0, 10$ and 30 in each case of $a_{\mbox{\scriptsize max}}$ (Figure \ref{fig:spid_hist}). 
The results demonstrate that for the observed sample to concentrate to a narrow $\alpha$ range, either all observed sources need to have rather high optical depths (e.g., $\tau\gtrsim$5), or they need to have nearly the same optical depths.
Otherwise, the optical depth distribution will lead to broadening of the observed $\alpha$ distribution.
We disfavored the interpretation with nearly identical optical depths since this requires a coincidence in the large observed sample.
For most of the $a_{\mbox{\scriptsize max}}$ we tried, the synthesized $\alpha$ values converged when we increased  $\tau$ to $\gtrsim$10 values (Figure \ref{fig:spid_hist}).
The exceptions are the cases with $a_{\mbox{\scriptsize max}}$ range of 0.3--1.0 mm, which still show noticeable changes of $\alpha$ when varying $\tau$ in the range of 10--30.
We therefore disfavored the $a_{\mbox{\scriptsize max}}=$0.3--1.0 mm cases since for such disks to concentrate to a narrow $\alpha$ distribution, their $\tau$ need to be $\gtrsim$30, which is not necessarily realistic (Figure \ref{fig:spid_hist}).
On the contrary, the cases of $a_{\mbox{\scriptsize max}} \le 0.1$ mm allow $\tau_{\mbox{\scriptsize 230 GHz}}$ to vary in a wide range from 1.0 to 30 without significantly broaden the synthesized spectral index distribution; and the cases of $a_{\mbox{\scriptsize max}} \ge 3.0$ mm allow $\tau_{\mbox{\scriptsize 230 GHz}}$ in a range of 3.0 to 30. 
The $T_{\mbox{\scriptsize dust,avg}} = 24$ K case allows lower $\tau_{\mbox{\scriptsize 230 GHz}}$ and wider range of $\tau_{\mbox{\scriptsize 230 GHz}}$ than the $T_{\mbox{\scriptsize dust,avg}} = 100$ K case. 

In the case that $T_{\mbox{\scriptsize dust,avg}} = 24$ K (Figure \ref{fig:spid_hist}), the synthesized spectral index distributions (blue) are generally lower than the observed distributions. 
Considering that mixing optically thin dust emission will also increase $\alpha_{\mbox{\scriptsize 200--400 GHz}}$, a realistic choice of $a_{\mbox{\scriptsize max}}$ should yield a synthesized spectral index distribution that is systematically lower than the observed distribution.
From Figure \ref{fig:spid_hist}, the cases of $a_{\mbox{\scriptsize max}} \ge 0.3$ mm appear to be rather realistic when $T_{\mbox{\scriptsize dust,avg}} = 24$ K. 
In the case with $T_{\mbox{\scriptsize dust,avg}} = 100$ K, the synthesized spectral index distributions (Figure \ref{fig:spid_hist}, red) are too high to compare with the observed spectral indices when the $a_{\mbox{\scriptsize max}}$ value is greater than 0.5 mm. 
The assumptions $a_{\mbox{\scriptsize max}} = 0.1$ mm yield too low $\mu_{\scriptsize \alpha}$ to be compatible with the observed $\mu_{\scriptsize \alpha} = 2.04$. 
Therefore, $a_{\mbox{\scriptsize max}} \lesssim 0.1$ mm cases are favorable when $T_{\mbox{\scriptsize dust,avg}} = 100$ K.


\begin{figure}
\vspace{-0.8cm}
\includegraphics[width=8.5cm]{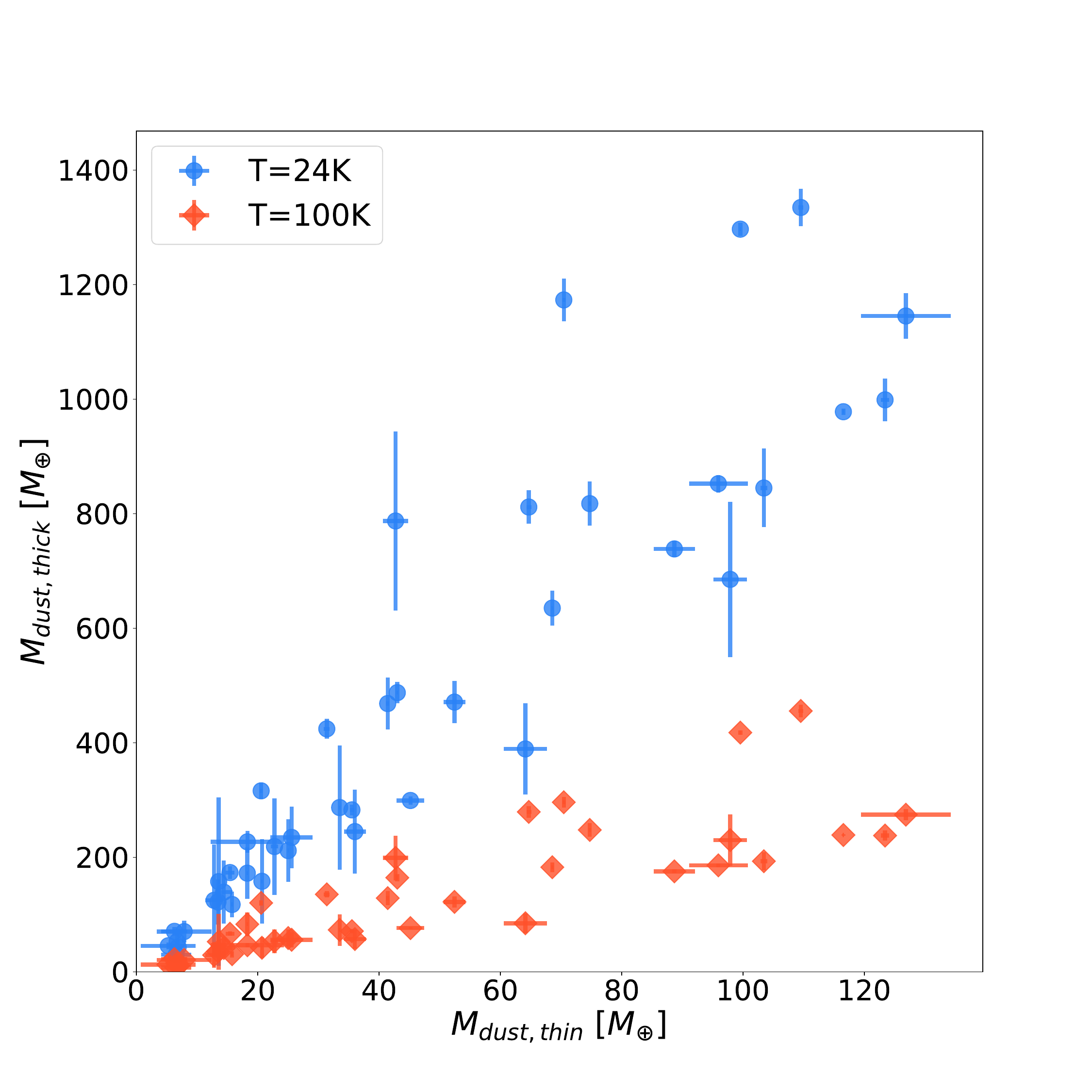}
\includegraphics[width=8.5cm]{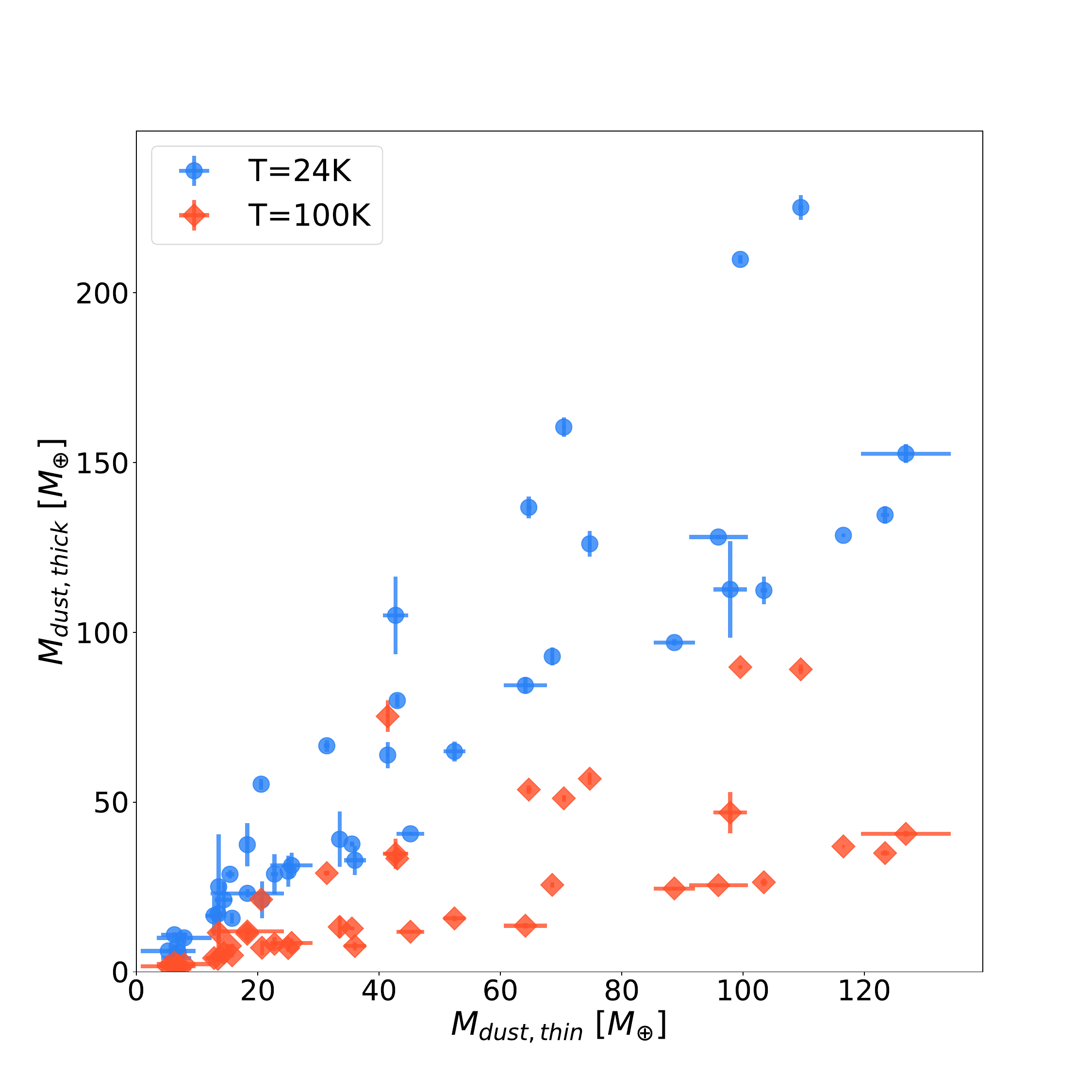}
\caption{ Dust mass lower limit for the 45 sources in the survey assuming (1) $T_{\mbox{\scriptsize dust,avg}} = $24 K and (2) $T_{\mbox{\scriptsize dust,avg}} = $100 K within Rayleigh-Jeans limit. The dust mass ($M_{\mbox{\scriptsize dust, thick}}$) estimated in this work assuming optically thick and considering dust scattering effect is compared with the dust mass ($M_{\mbox{\scriptsize dust, thin}}$) estimated in \citet{Manara2023ASPC..534..539M} assuming optically thin. 
{\it Top} panel: the $a_{\mbox{\scriptsize max}} = $0.05 mm case. {\it Bottom} panel: the $a_{\mbox{\scriptsize max}} = $3.0 mm case. 
We excluded the two sources, CY~Tau and GI~Tau, because their $\alpha_{\mbox{\scriptsize 200--400 GHz}}$ are too low to be produced by the simple physical model (See Appendix \ref{appendix:dustmass}). }
\label{fig:dust_mass_100K}
\end{figure}

For a further check, we evaluated  $\alpha_{\mbox{\scriptsize 200-280 GHz}}$ and $\alpha_{\mbox{\scriptsize 280-420 GHz}}$ from the population syntheses introduced in Section \ref{sub:Popsynthesis}, for the cases that assumes $T_{\mbox{\scriptsize dust,avg}}=$100 K and $T_{\mbox{\scriptsize dust,avg}}=$24 K  for all 39 unresolved sources.
Figure \ref{fig:spid_2dhist} compares the observed $\alpha_{\mbox{\scriptsize 200-280 GHz}}$ and $\alpha_{\mbox{\scriptsize 280-420 GHz}}$ with those resulted from the population synthesis with $a_{\mbox{\scriptsize max}} = 0.01,  0.03,  0.05,  0.1,  0.3,  0.5,  1.0,  3.0,  10$ mm.
Similar to Figure \ref{fig:spid_hist}, the $a_{\mbox{\scriptsize max}} > 0.3$ mm cases are disfavored when $T_{\mbox{\scriptsize dust,avg}}=$100 K as they predict too high $\alpha_{\mbox{\scriptsize 200-280 GHz}}$ and $\alpha_{\mbox{\scriptsize 280-420 GHz}}$.
The $\alpha_{\mbox{\scriptsize 280-420 GHz}}=0.3$ mm case may also be disfavored since even in the case of $T_{\mbox{\scriptsize dust,avg}}=$100 K, it predicts a resolvable offset between  $\alpha_{\mbox{\scriptsize 200-280 GHz}}$ and $\alpha_{\mbox{\scriptsize 280-420 GHz}}$ which is inconsistent with observations.
The $a_{\mbox{\scriptsize max}} = 0.1$ mm case yields $\alpha_{\mbox{\scriptsize 200-280 GHz}}$ and $\alpha_{\mbox{\scriptsize 280-420 GHz}}$ that are a lot lower than the observed spectral indices, which is due to the self-scattering effect introduced in Section \ref{sec:intro} (\citealt{Liu2019ApJ...877L..22L,Zhu2019ApJ...877L..18Z}).

As a summary, based on the analyses presented in this section, the favored $a_{\mbox{\scriptsize max}}$ range for the unresolved disks in our sample is $\lesssim$ 0.1 mm if the $T_{\mbox{\scriptsize dust,avg}}$ values of them are high enough ($T_{\mbox{\scriptsize dust,avg}}=$100 K) to make the 200--400 GHz observing frequency in the Rayleigh-Jeans limit, and if these disks are optically thick.
Otherwise, if all the 39 unresolved sources have very similar $T_{\mbox{\scriptsize dust,avg}}$ (e.g., 24 K) and are optically thick, the favored $a_{\mbox{\scriptsize max}}$ range is 3.0--10 mm.
We cannot rule out that the $a_{\mbox{\scriptsize max}}$ of some of these sources are very close to 0.1 mm (e.g., CY~Tau), or are larger than 0.5 mm.
The two temperature assumptions mentioned above are not strictly mutually exclusive.
For example, if these two temperature assumptions correspond to two disk populations that have smaller and larger radii, the two favored $a_{\mbox{\scriptsize max}}$ranges can coexist.
The $T_{\mbox{\scriptsize dust,avg}} = 100$ K case may correspond to the compact disks that can hardly be resolved in ALMA observation.
The $T_{\mbox{\scriptsize dust,avg}} = 24$ K assumption may be more realistic for some extended disks. 
%
We remark that the favored $a_{\mbox{\scriptsize max}}$ ranges discussed above were out of the assumptions that the disks are optically thick at 200--400 GHz.
The resolved disks in our sample (Figure \ref{fig:337GHz_flux_spidx}), which are spatially rather extended, are not necessarily optically thick.
The $a_{\mbox{\scriptsize max}}$ values of these resolved disks can be $\lesssim$100 $\mu$m in spite that their $T_{\mbox{\scriptsize dust,avg}}$ may be comparably lower than 24 K.

In Figure \ref{fig:spid_hist_2grp_size}, we separated the observed 39 unresolved sources into the larger ($R_{\mbox{\scriptsize disk}}>46$ AU) 19 ones (big disks) and the smaller 20 ones (small disks) based on the estimated $R_{\mbox{\scriptsize disk}}$, and derived the $\alpha_{\mbox{\scriptsize 200--400 GHz}}$ distributions of these two source sub-groups. 
We performed a K-S test on the $\alpha_{\mbox{\scriptsize 200--400 GHz}}$ of the two sub-groups. The small p-value, $p = 0.04$, indicates that the two samples are realizations of distinct distributions. 
We compared these $\alpha_{\mbox{\scriptsize 200--400 GHz}}$ distributions with the population synthesis of the $a_{\mbox{\scriptsize max}} = 0.05$ mm case. 
The observed distribution for big disk have a larger $\mu_{\scriptsize \alpha}$ likely because of the mixing of optically thin emission. 
On the other hand, the two synthesized spectral index distributions show good consistency in the small disk group, and they are in good agreement with the observed distribution. 
Our optically thick dust slab models better describe the small disk population, which is consistent with our interpretation on the observed correlation between $R_{\mbox{\scriptsize disk}}$ and $\alpha_{\mbox{\scriptsize 200--400 GHz}}$. 

\subsection{Physical implication}\label{sub:physics}
We include the discussion about dust mass budget, gravitational instability, dust growth in protoplanetary disks, and planet formation in Sections \ref{subsub:massbudget}, \ref{subsub:GI}, \ref{subsub:growth}, and \ref{subsub:planetformation}, respectively.

\subsubsection{Dust mass budget}\label{subsub:massbudget}
Under the optically thick interpretation, to make the $\alpha_{\mbox{\scriptsize 200-420 GHz}}$ values of the unresolved sources close to 2.0, it requires a very high dust optical depth $\tau\gtrsim$5 at $>$200 GHz frequency.
If the unresolved sources are indeed this optically thick, the previous studies that were based on an optically thin assumption (e.g., \citealt{Andrews_Williams_2005,Andrews_Williams_2007,Ansdell2016ApJ...828...46A,Barenfeld2016ApJ...827..142B,Testi2016A&A...593A.111T,Akeson_2019,Babaian2019AAS...23410503B,Cazzoletti2019A&A...626A..11C,Williams2019ApJ...875L...9W}) could have underestimated optical depths by at least one order of magnitude.
Potentially, this corresponds to underestimates of dust masses by similar factors although it also depends on the assumptions of dust opacities (c.f. Appendix \ref{appendix:diana}).
While some previous studies suggested that the median dust mass of the Class II disks may be $\sim$10 $M_{\oplus}$, the actual dust masses may be generally greater than $\sim$100 $M_{\oplus}$.
If Class II disks indeed generally harbor $\gtrsim$100 $M_{\oplus}$ dust masses, they may still form rocky planets.

We tried to give a lower limit to this underestimation by estimating the lower limit of dust disk mass ($M_{\mbox{\scriptsize dust, thick}}$) assuming adequate optical depth using the dust slab model in Section \ref{sub:slab}. 
Figure \ref{fig:dust_mass_100K} compares the $M_{\mbox{\scriptsize dust, thick}}$ lower limits we estimated of the 47 sources with the previous dust mass estimates that were based on an optically thin assumption ($M_{\mbox{\scriptsize dust, thin}}$; quoted from \citealt{Manara2023ASPC..534..539M}). 
Our derivations of $M_{\mbox{\scriptsize dust, thick}}$ lower limits were based on (1) the assumption of $T_{\mbox{\scriptsize dust,avg}}=24$ K, and (2) the assumption of $T_{\mbox{\scriptsize dust,avg}}=100$ K for all sources. 
The detailed calculations and assumptions for the dust disk mass are presented in Appendix \ref{appendix:dustmass}. 
In Figure \ref{fig:dust_mass_ratio_amax}, we present the lower limits of mass underestimation ratios for each $a_{\mbox{\scriptsize max}}$ assumption by the median value of the calculated $M_{\mbox{\scriptsize dust, thick}}/M_{\mbox{\scriptsize dust, thin}}$ lower limits for the 47 sources.
The favorable $a_{\mbox{\scriptsize max}}$ ranges resulting from the population synthesis (Section \ref{sub:Popcompar}) for the $T_{\mbox{\scriptsize dust,avg}}=24$ K and 100 K cases are presented by the blue and orange shaded area, respectively. 
Within the favorable ranges of $a_{\mbox{\scriptsize max}}$ under the two $T_{\mbox{\scriptsize dust,avg}}$ assumptions, the lower limits of the underestimation ratios of dust masses are in the range of 1.2--2.8.

We note that based on the sophisticated radiative transfer modelling on the multi-wavelength observations, \citet{Ballering2019AJ....157..144B} and \citet{Ribas2020A&A...642A.171R} also suggested that the dust masses in T~Tauri disks might have been systematically underestimated by 1--5 times, which overlaps with our estimates (Figure \ref{fig:dust_mass_ratio_amax}).
We also note that some previous works based on multi-frequency observations towards samples of Class 0/I YSOs also argued that the dust masses have been systematically underestimated while the grain sizes have been systematically overestimated (e.g., \citealt{Li2017ApJ...840...72L,Galvan2018ApJ...868...39G,Xu2022ApJ...934..156X}; see Section 4.1 in the ArXiv-astroph preprint of \citealt{Liu2021ApJ...914...25L} for some comprehensive qualitative discussion).

\begin{figure}
\hspace{-0.5cm}
\includegraphics[width=8.5cm]{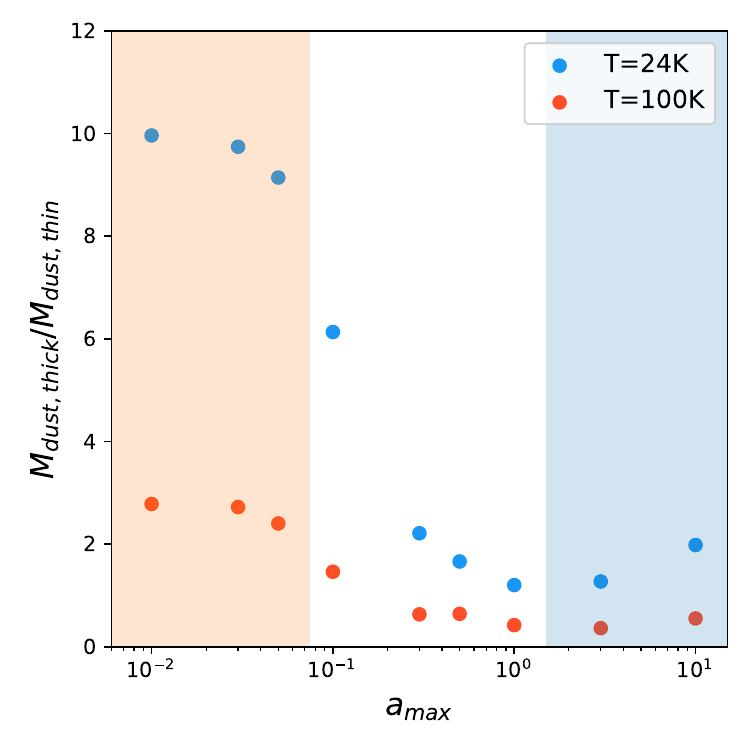} 
\caption{
The lower limits of the underestimation ratio of dust mass as a function of $a_{\mbox{\scriptsize max}}$ (Section \ref{subsub:massbudget}). 
The blue and red points show the median value of the underestimation ratio lower limits for the 47 sources in the survey, which were derived based on the assumptions of (1) $T_{\mbox{\scriptsize dust,avg}}=$24 K and (2) $T_{\mbox{\scriptsize dust,avg}}=$100 K, respectively. The blue and orange shaded area show the preferred maximum grain size ranges in the population synthesis of $\alpha_{\mbox{\scriptsize 200-420 GHz}}$ assuming $T_{\mbox{\scriptsize dust,avg}}=$24 K and $T_{\mbox{\scriptsize dust,avg}}=$100 K, respectively (Section \ref{sub:Popcompar}). }
\label{fig:dust_mass_ratio_amax}
\end{figure}

\subsubsection{Gravitational instability}\label{subsub:GI}

An isolated protoplanetary disk may be gravitationally unstable when its mass exceeds 10\% of the protostellar mass (c.f., \citealt{Shu1987ARA&A..25...23S}).
A gravitationally unstable disk is prone to fragmentation, the formation of azimuthally symmetric (sub)structures and spiral arms, and the triggering of accretion outbursts (e.g., \citealt{Vorobyov2010ApJ...719.1896V,Liu2016SciA....2E0875L,Xu2023ApJ...954..190X}, and references therein).
The high binary fraction may be a consequence of disk gravitational instability (\citealt{Vorobyov2013A&A...552A.129V}); disk fragmentation may also lead to the formation of gas giants (e.g., \citealt{Machida2011ApJ...729...42M}).

A protostar likely had acquired most of its mass during the Class 0/I stage (c.f., \citealt{Evans2009ApJS..181..321E} and references therein).
Immediately after the dispersal of the circumstellar envelope (i.e., the earliest Class II stage), the typical protostellar masses ($M_{*}$) are $\sim$0.5--1 $M_{\odot}$ (e.g., \citealt{Kroupa2002Sci...295...82K}, \citealt{Manara2016A&A...585A.136M} and references therein).
If we consider the protoplanetary disks to be self-regulated by gravitational instability (e.g., \citealt{Xu2022ApJ...934..156X} and references therein), then the typical, critical initial masses of isolated Class II disks maybe 0.05--0.1 $M_{\odot}$ (c.f., \citealt{Galvan2018ApJ...868...39G}).
Assuming a gas-to-dust ratio ($\eta$) of 100, the corresponding initial dust mass is 170--330 $M_{\oplus}$.
This is comparable to the lower limits of dust masses we estimated for some disks (Figure \ref{fig:dust_mass_100K}). 
This comparison may imply that a high fraction of initial dust mass budget is preserved during the Class II stage (more in Section \ref{subsub:growth}).

One might wonder if there is a tension between the dust masses estimates in Figure \ref{fig:dust_mass_100K} and the fact that most high-resolution observations on isolated Class II disks did not exhibit features of gravitational instability (e.g., \citealt{Andrews_2018}; \citealt{Long2018ApJ...869...17L}).
At this moment, this tension is not serious since our selected Class II objects are relatively bright at 0.85--1 mm bands and are not necessarily typical, and (ii) are not necessarily isolated (Section \ref{sub:source}).
The tension can be further alleviated if the values of $\eta$ are lower than 100 in the Class II stage, which may be consistent with the previous measurement of  \citet{Miotello2017A&A...599A.113M}.
However, we point out that the measurements of $\eta$ in \citet{Miotello2017A&A...599A.113M} were based on the assumption that dust emission is optically thin at 0.85--1 mm bands.
Therefore, those measurements of $\eta$ cannot be self-consistently applied in our present study.
The deeper discussion of $\eta$ values is beyond the scope of the present work and deserves future studies.

Lastly, the typically quoted initial disk mass ($\sim$0.1 $M_{\odot}$) may be underestimated because it is the sum of local disk mass ($\pi R^2 \Sigma$) assuming a uniform density profile. 
In fact, these Class II disks can have steep density profiles (e.g. $\Sigma(R) \sim R^{-2}$) such that at each radius the local disk mass remains gravitationally stable. 
In this case, the total disk mass can easily exceed 0.1 $M_{\odot}$ but still be consistent with no gravitational instability. 
The steep density profile could be inherited from the Class 0/I stage with a gravitationally self-regulated initial condition (e.g., \citealt{Xu2023ApJ...954..190X}). 
If Class II disks are possible to harbor a few times 0.1 $M_{\odot}$ mass budget initially, assuming $\eta = 100$, our dust mass estimates of 4--450 $M_{\oplus}$ (Table \ref{tab:Mdust}; Figure \ref{fig:dust_mass_100K}) could be more reasonable as a result of mass transfer and planetesimal formation.


\begin{deluxetable*}{cc|ccc|ccc|c}
\tablecolumns{9}
\tablecaption{The 200-420 GHz spectral index and dust mass estimation for the 47 sources
\label{tab:Mdust}}
\tablehead{ 
\colhead{} & \colhead{} & \multicolumn{3}{c}{$a_{\mbox{\tiny max}}=0.05$ mm, $T_{\mbox{\tiny dust,avg}}=100$ K} & \multicolumn{3}{c}{$a_{\mbox{\tiny max}}=3.0$ mm, $T_{\mbox{\tiny dust,avg}}=24$ K} & \colhead{} \\
\cline{3-5} \cline{6-8}
\colhead{Source} & \colhead{$\alpha_{\mbox{\tiny 200-420 GHz}}$} & \colhead{$\Sigma$} & \colhead{$\tau_{\tiny \mbox{230 GHz}}$} & \colhead{$\mbox{M}_{\scriptsize \mbox{dust, thick}, T=100K}$}  & \colhead{$\Sigma$} & \colhead{$\tau_{\tiny \mbox{230 GHz}}$} & \colhead{$\mbox{M}_{\scriptsize \mbox{dust, thick,}, T=24K}$} & \colhead{$\mbox{M}_{\scriptsize \mbox{dust, thin}}$}  \\
\colhead{} & \colhead{} & \colhead{(g/cm$^2$)} & \colhead{} & \colhead{($M_{\oplus}$)} &  \colhead{(g/cm$^2$)} & \colhead{} & \colhead{($M_{\oplus}$)} & \colhead{($M_{\oplus}$) } 
}
\startdata 
04113+2758A & $1.91^{+0.13}_{-0.13}$ & 9.43 & 4.76 &  $455.54 \pm 11.27$ & 0.61 & 9.27 & $225.13 \pm 3.58$ & $109.51 \pm 0.39$ \\
04113+2758B & $1.91^{+0.13}_{-0.13}$ & 9.43 & 4.76 &  $279.35 \pm 10$ & 0.61 & 9.27 & $136.8 \pm 3.18$ & $64.67 \pm 0.4$ \\
04278+2253 & $2.16^{+0.2}_{-0.19}$ & 4.82 & 2.44 &  $9.96 \pm 8.37$ & 0.33 & 4.94 & $5.6 \pm 2.52$ &  \\
AA Tau & $2.43^{+0.16}_{-0.16}$ & 3.29 & 1.66 &  $57.32 \pm 16.99$ & 0.21 & 3.21 & $32.86 \pm 4.3$ & $36 \pm 1.82$ \\
AB Aur & $3.02^{+0.15}_{-0.14}$ & 1.64 & 0.83 &  $84.88 \pm 17.62$ & 0.04 & 0.62 & $84.37 \pm 2.45$ & $64.11 \pm 3.53$ \\
BP Tau & $2.05^{+0.16}_{-0.15}$ & 6.01 & 3.03 &  $59.05 \pm 14.95$ & 0.4 & 6.06 & $29.67 \pm 4.61$ & $25.01 \pm 0.09$ \\
CIDA-7 & $2.05^{+0.21}_{-0.2}$ & 6.01 & 3.03 &  $20.98 \pm 5.19$ & 0.4 & 6.06 & $10.02 \pm 1.6$ & $7.86 \pm 4.53$ \\
CIDA-9 & $1.9^{+0.18}_{-0.17}$ & 10.02 & 5.06 &  $120.41 \pm 4.92$ & 0.65 & 9.78 & $55.28 \pm 1.57$ & $20.55 \pm 0.23$ \\
CI Tau & $2.4^{+0.13}_{-0.14}$ & 3.41 & 1.72 &  $193.22 \pm 15.83$ & 0.22 & 3.36 & $112.34 \pm 4.09$ & $103.42 \pm 0.59$ \\
CW Tau & $2.23^{+0.14}_{-0.14}$ & 4.35 & 2.2 &  $56.03 \pm 13.06$ & 0.29 & 4.38 & $31.34 \pm 3.73$ & $25.57 \pm 3.47$ \\
CY Tau & $1.54^{+0.14}_{-0.14}$ & 118.18 & 59.65 &   & 3.36 & 50.96 &   & $35.81 \pm 2.67$ \\
DD Tau & $1.95^{+0.2}_{-0.2}$ & 7.89 & 3.98 &  $22.35 \pm 2.22$ & 0.52 & 7.84 & $10.93 \pm 0.71$ & $6.25 \pm 2.16$ \\
DE Tau & $1.97^{+0.18}_{-0.18}$ & 7.3 & 3.69 &  $41.74 \pm 16.36$ & 0.49 & 7.39 & $21.25 \pm 5.25$ & $14.4 \pm 1.45$ \\
DH Tau & $2.07^{+0.18}_{-0.18}$ & 5.77 & 2.91 &  $35.62 \pm 10.35$ & 0.39 & 5.86 & $17.06 \pm 3.21$ & $13.43 \pm 0.06$ \\
DK Tau & $1.87^{+0.18}_{-0.18}$ & 13.68 & 6.91 &  $83.19 \pm 20.93$ & 0.84 & 12.68 & $37.47 \pm 6.4$ & $18.26 \pm 0.12$ \\
DL Tau & $2.25^{+0.13}_{-0.13}$ & 4.12 & 2.08 &  $238.03 \pm 8.86$ & 0.28 & 4.18 & $134.59 \pm 2.54$ & $123.4 \pm 0.68$ \\
DM Tau & $2.08^{+0.14}_{-0.13}$ & 5.65 & 2.85 &  $122.17 \pm 9.49$ & 0.38 & 5.71 & $64.93 \pm 2.92$ & $52.42 \pm 1.83$ \\
DN Tau & $1.8^{+0.13}_{-0.14}$ & 8.48 & 4.28 &  $128.79 \pm 12.08$ & 0.56 & 8.45 & $63.86 \pm 3.88$ & $41.41 \pm 0.29$ \\
DO Tau & $2.01^{+0.14}_{-0.13}$ & 6.6 & 3.33 &  $182.73 \pm 8.58$ & 0.44 & 6.62 & $92.9 \pm 2.69$ & $68.54 \pm 0.15$ \\
DQ Tau & $1.95^{+0.16}_{-0.17}$ & 7.89 & 3.98 &  $247.8 \pm 11.8$ & 0.52 & 7.89 & $126.07 \pm 3.75$ & $74.72 \pm 0.21$ \\
DR Tau & $2.16^{+0.14}_{-0.13}$ & 4.82 & 2.44 &  $296.2 \pm 9.32$ & 0.33 & 4.94 & $160.45 \pm 2.81$ & $70.44 \pm 0.23$ \\
DS Tau & $2.18^{+0.17}_{-0.17}$ & 4.71 & 2.38 &  $30.84 \pm 5.58$ & 0.31 & 4.74 & $15.79 \pm 1.61$ & $15.76 \pm 0.16$ \\
FM Tau & $2.09^{+0.2}_{-0.2}$ & 5.53 & 2.79 &  $12.74 \pm 1.57$ & 0.37 & 5.6 & $6.16 \pm 0.48$ & $5.24 \pm 4.51$ \\
FT Tau & $1.92^{+0.14}_{-0.14}$ & 8.96 & 4.52 &  $164.46 \pm 6.2$ & 0.58 & 8.81 & $79.92 \pm 1.96$ & $42.99 \pm 0.13$ \\
FV Tau & $2.05^{+0.21}_{-0.22}$ & 6.01 & 3.03 &  $14.7 \pm 3.94$ & 0.4 & 6.06 & $7.39 \pm 1.21$ & $6.7 \pm 0.13$ \\
FY Tau & $2.41^{+0.28}_{-0.27}$ & 3.41 & 1.72 &  $7.62 \pm 0.00$ & 0.22 & 3.26 & $4.04 \pm 0.00$ & $6.54 \pm 2.45$ \\
GI Tau & $1.29^{+0.19}_{-0.16}$ & 118.18 & 59.65 &   & 3.36 & 50.96 &   & $8.38 \pm 0.09$ \\
GM Aur & $2.73^{+0.13}_{-0.13}$ & 2.23 & 1.12 &  $185.98 \pm 3.34$ & 0.12 & 1.84 & $128.07 \pm 0.67$ & $95.92 \pm 4.83$ \\
GO Tau & $1.94^{+0.17}_{-0.17}$ & 8.13 & 4.1 &  $135.4 \pm 5.35$ & 0.54 & 8.2 & $66.59 \pm 1.72$ & $31.37 \pm 0.49$ \\
HO Tau & $1.92^{+0.17}_{-0.17}$ & 8.96 & 4.52 &  $52.77 \pm 48.85$ & 0.58 & 8.86 & $25.06 \pm 15.46$ & $13.55 \pm 0.15$ \\
HV Tau & $2.37^{+0.15}_{-0.15}$ & 3.53 & 1.78 &  $29.29 \pm 22.47$ & 0.23 & 3.47 & $16.53 \pm 5.92$ & $12.79 \pm 1.41$ \\
Haro 6-37 & $2.16^{+0.13}_{-0.13}$ & 4.82 & 2.44 &  $199.06 \pm 38.95$ & 0.32 & 4.89 & $105.01 \pm 11.4$ & $42.71 \pm 2.11$ \\
Haro 6-39 & $2.35^{+0.23}_{-0.22}$ & 3.64 & 1.84 &  $10.31 \pm 12.54$ & 0.24 & 3.62 & $6.05 \pm 3.37$ &  \\
IC 2087 IR & $2.23^{+0.13}_{-0.13}$ & 4.35 & 2.2 &  $238.12 \pm 11.87$ & 0.29 & 4.33 & $131.67 \pm 3.4$ &  \\
IP Tau & $2.15^{+0.18}_{-0.19}$ & 4.94 & 2.49 &  $10.87 \pm 3.45$ & 0.33 & 5.04 & $5.73 \pm 1.02$ & $6.87 \pm 0.09$ \\
IQ Tau & $2.15^{+0.14}_{-0.14}$ & 4.94 & 2.49 &  $71.09 \pm 2.27$ & 0.33 & 5.04 & $37.66 \pm 0.67$ & $35.51 \pm 0.36$ \\
LkCa 15 & $2.29^{+0.13}_{-0.13}$ & 4 & 2.02 &  $175.48 \pm 3.43$ & 0.26 & 3.97 & $96.99 \pm 0.93$ & $88.67 \pm 3.44$ \\
RW Aur & $2.2^{+0.17}_{-0.18}$ & 4.59 & 2.32 &  $53.52 \pm 20.69$ & 0.3 & 4.59 & $28.82 \pm 5.89$ & $22.75 \pm 0.59$ \\
RY Tau & $2.2^{+0.13}_{-0.13}$ & 4.59 & 2.32 &  $238.96 \pm 1.39$ & 0.3 & 4.59 & $128.57 \pm 0.4$ & $116.53 \pm 0.11$ \\
SU Aur & $1.73^{+0.25}_{-0.24}$ & 7.19 & 3.63 &  $46.49 \pm 3.92$ & 0.47 & 7.18 & $23.13 \pm 1.22$ & $18.28 \pm 6.03$ \\
T Tau & $1.93^{+0.14}_{-0.13}$ & 8.48 & 4.28 &  $417.82 \pm 3.67$ & 0.56 & 8.45 & $209.84 \pm 1.18$ & $99.54 \pm 0.13$ \\
UX Tau & $2.19^{+0.15}_{-0.15}$ & 4.59 & 2.32 &  $76.45 \pm 1.82$ & 0.31 & 4.64 & $40.67 \pm 0.53$ & $45.16 \pm 2.27$ \\
UY Aur & $1.91^{+0.22}_{-0.23}$ & 9.43 & 4.76 &  $66.6 \pm 3.74$ & 0.61 & 9.27 & $28.75 \pm 1.19$ & $15.41 \pm 0.72$ \\
UZ Tau & $1.92^{+0.13}_{-0.13}$ & 8.96 & 4.52 &  $230.12 \pm 45.06$ & 0.58 & 8.86 & $112.64 \pm 14.26$ & $97.87 \pm 2.75$ \\
V710 Tau & $2.15^{+0.15}_{-0.15}$ & 4.94 & 2.49 &  $73.28 \pm 27.53$ & 0.33 & 4.99 & $39.05 \pm 8.15$ & $33.49 \pm 0.12$ \\
V836 Tau & $2.15^{+0.21}_{-0.21}$ & 4.94 & 2.49 &  $42.11 \pm 18.34$ & 0.33 & 5.04 & $21.17 \pm 5.43$ & $20.68 \pm 0.13$ \\
V892 Tau & $2.23^{+0.13}_{-0.13}$ & 4.35 & 2.2 &  $274.48 \pm 9.65$ & 0.29 & 4.38 & $152.63 \pm 2.76$ & $126.83 \pm 7.41$ \\
\enddata
\end{deluxetable*}

\subsubsection{Dust growth outside of water snow lines}\label{subsub:growth}
We note that in the Class II disks, the water snow lines are typically at $\lesssim$1--2 au radii.
Limited by the angular resolutions, our SMA observations (and most of the existing ALMA observations) preferentially detect the emission of the dust mass reservoirs outside of the water snow lines.
The dust grains that dominate our detected (sub)millimeter flux densities are likely coated with water ices.
In addition, limited by the high optical depth, our observations at (sub)millimeter wavelength only probe the dust properties within the $\tau\sim1$ surface region of the disk. 
In this region, dust grain have relatively uniform properties. 
The disk midplane could have complex structures with varying $a_{\mbox{\scriptsize max}}$ in different radial ranges. 
As the observation migrates to longer wavelengths (e.g. millimeter to centimeter), we would be able to unveil the regions where the dust grain growth is productive in the midplane (\citealt{Liu2021ApJ...923..270L,Hashimoto2022ApJ...941...66H,Hashimoto2023arXiv230811837H,Liu2024arXiv240202900L}). 

Intriguingly, the recent analytic calculation (e.g., \href{https://ui.adsabs.harvard.edu/abs/2015ApJ...812...67K/abstract}{Kimura et al. 2015}) and the latest laboratory experiments for dust stickiness (e.g., \href{https://ui.adsabs.harvard.edu/abs/2019ApJ...874...60S/abstract}{Steinpilz et al. 2019}, \href{https://ui.adsabs.harvard.edu/abs/2019ApJ...873...58M/abstract}{Musiolik \& Wurm 2019}, \href{https://ui.adsabs.harvard.edu/abs/2021A\%26A...652A.106P/abstract}{Pillich et al. 2021}) suggested that the water-ice-coated dust grains are a lot less sticky than the grains made of dry silicate or refractory organics.
The fragmentation velocity of water-ice-coated dust grains may be $\sim$1 m\,s$^{-1}$, which is a lot lower than how it is considered in some theoretically studies of dust growth (e.g., \citealt{Drazkowska2017A&A...608A..92D,Vorobyov2019A&A...627A.154V}; see also the discussion in \citealt{Jiang2023arXiv231107775J}).
In this case, the bouncing/fragmentation barriers will limit $a_{\mbox{\scriptsize max}}$ to $\sim$100 $\mu m$ in the relatively turbulent disks, and to a few mm in the disks with low turbulence (\citealt{Zsom2010A&A...513A..57Z,Birnstiel2016SSRv..205...41B,Jiang2023arXiv231107775J}).
Since small grains undergo slow radial drift and little planet formation, the dust mass budget would deplete slowly, in agreement to the relatively high dust masses we found. 

In our present work, we found that CY~Tau, SU~Aur and GI~Tau in our sample are consistent with anomalously low ($<2.0$) $\alpha_{\mbox{\scriptsize 200-420 GHz}}$. 
Since our selected samples are the relatively bright Class II disks in the Taurus-Auriga region (Figure \ref{fig:totalClassII}), if the correlation between  $R_{\mbox{\scriptsize disk}}$ and $\alpha_{\mbox{\scriptsize 200-420 GHz}}$ can be extended to smaller $R_{\mbox{\scriptsize disk}}$, in light of the millimeter size-luminosity relation (\citealt{Tripathi_2017,Andrews_2018}), we may expect to confirm more anomalously low spectral indices if the errors in $\alpha_{\mbox{\scriptsize 200-420 GHz}}$ can be suppressed by new, more sensitive observations towards the spatially compact, fainter sources.
The $\alpha_{\mbox{\scriptsize 200-420 GHz}}<2.0$ may be interpreted by the aforementioned dust self-scattering effect in these optically thick disks which indicates  $a_{\mbox{\scriptsize max}}\sim$100 $\mu$m (\citealt{Liu2019ApJ...877L..22L}, \citealt{Zhu2019ApJ...877L..18Z}). 
With the interpretations of less sticky ice-coated dust grains and bouncing/fragmentation barrier, we can explain the observed low  $a_{\mbox{\scriptsize max}}$ values outside of the water snowline. 
The high spectral indices detected from the spatially resolved, extended disks (e.g., AB~Aur; Figure \ref{fig:disksize_spidx}) can be coherently interpreted with $\lesssim$100 $\mu$m grain sizes outside of the water snowline (\citealt{Testi2014prpl.conf..339T}).
If the 200--400 GHz SMA observing frequencies are in the Rayleigh-Jeans limits of the 39 unresolved disks in our sample, the favored $a_{\mbox{\scriptsize max}}\lesssim$0.1 $\mu$m range (Section \ref{sub:Popcompar}) is consistent with the scenario of less sticky water-ice coated dust.
According to the estimates of \citet{Jiang2023arXiv231107775J}, the few millimeter $a_{\mbox{\scriptsize max}}$ favored by the $T_{\mbox{\scriptsize dust,avg}}=$24 K population syntheses (Section \ref{sub:Popcompar}) is  consistent with the low fragmentation velocity $v_{\mbox{\scriptsize frag}}\sim$1 m\,s$^{-1}$.

Another possible origin of a $<2$ spectral index could be a vertical temperature gradient, which can be produced by internal viscous heating in an optically thick disk. Internal viscous heating has been suggested as a dominant heating mechanism for younger disks and/or at small radii (e.g., \citealt{Liu2019ApJ...884...97L,Liu2021ApJ...914...25L,Zamponi2021MNRAS.508.2583Z,Xu2022ApJ...934..156X,Xu2023ApJ...954..190X}); it is less clear whether it still plays a significant role in the older disks in our sample.


\subsubsection{Implications to the formation of planets and dusty rings}\label{subsub:planetformation}

\citet{Hasen2012ApJ...751..158H} argued that, to form the observed hot Neptune and super-Earth systems, it may require 50--100 $M_{\oplus}$ of solid material to be interior to the 1 au radius. 
In light of this, under our optically thick interpretation, the observed Class II disks appear to possess adequate dust mass budget (see top panels of Figure \ref{fig:dust_mass_100K}) to form terrestrial planets. 
In other words, the issue that T~Tauri disks may not posses sufficient dust masses for forming planets (e.g., \citealt{Manara2018A&A...618L...3M}) may be fictitious.
Terrestrial planet formation may be postponed to later than the Class II stage, and may not naturally occur in the regions outside of water snow lines that can be easily spatially resolved by the ALMA observations.
The questions are where, when, and how the $\lesssim$100 $\mu$m sized dust grains can be converted to $>$1 km sized planetesimals and planets? 

The recent laboratory experiment suggested that the water-ice free dust in the very hot ($T > 500$ K) inner disk (\citealt{Kimura2015ApJ...812...67K,Steinpilz2019ApJ...874...60S,Pillich2021A&A...652A.106P}) may be sticky.
The high fragmentation velocity in such high temperature environments may be prone to the formation of planetesimals and terrestrial planets. 
Large grain size or steep radial variation of grain size have been observed in a few young or massive disks (\citealt{Liu2021ApJ...923..270L,Yamamuro2023ApJ...949...29Y,Houge2024MNRAS.527.9668H,Xu2023ApJ...954..190X}). 
The rocky planets formed in the high temperature environments will likely be deficient in water and carbon, which may be similar to the planets in the inner Solar system (\citealt{Li2021SciA....7.3632L}).

Outside of water snow lines, dust growth and planetesimal formation, in general, may not be as efficient as what used to be considered.
Forming gas giants via accreting icy pebbles (e.g., \citealt{Lambrechts2012A&A...544A..32L}) may not be trivial.
This may partly explain why gas giants are rare.
Gas giants may need to form in some exceptional environments, for examples, the dust traps that have accumulated large amounts of masses.
Traps of $\sim$0.1--1 mm sized dust may initial appear like rings at $\lesssim$1 mm wavelengths (e.g., \citealt{Okuzumi2019ApJ...878..132O,Jiang2023arXiv231107775J}).
When the gas-to-dust ratios become sufficiently low (e.g., $\sim$1) in the dust traps, the dynamic instabilities may promote dust growth and the formation of dense dust substructures. 

The recent observations on the ZZ~Tau~IRS, CIDA1, and DM~Tau disks (\citealt{Hashimoto2021ApJ...911....5H,Hashimoto2022ApJ...941...66H,Liu2024arXiv240202900L}) resolved dusty rings at $\lesssim$1 mm wavelengths; but resolved clumpy dusty structures that harbor $a_{\mbox{\scriptsize max}}\gtrsim$1 mm grains at $\gtrsim$7 mm wavelengths.
These results may be interpreted by the more efficient dust growth in some localized dust vortices (e.g., \citealt{Huang2020ApJ...893...89H,Li2020ApJ...892L..19L}), which may facilitate the formation of ice giants or the rocky cores of gas giants.
Whether or not such localized clumpy dusty structures are common outside of the water snow lines remain uncertain, which needs to be address by future high angular resolution surveys at $>$7 mm wavelengths.

In some relatively massive and gravitational unstable disks, gas giants can also form via disk gravitational fragmentation (e.g., \citealt{Machida2011ApJ...729...42M,Vorobyov2013A&A...552A.129V}).
Given our relatively high mass estimates (Section \ref{subsub:massbudget}, \ref{subsub:GI}), this is not improbable.

The challenge of forming gas giant cores may also be overcome by converting a small fraction of dust mass into larger pebbles (e.g., \citealt{Windmark2012A&A...540A..73W}).  
For example, \citealt{Xu2023ApJ...946...94X} found that $\sim$50 $M_{\oplus}$ of 1--10 cm pebble may form during early disk evolution while most dust 
mass remains below the $<$mm fragmentation barrier. 
This allows cores to form from large pebbles while typical grain size (which dominates the opacity) remains small, which is in line with our observations.



\section{Conclusion}\label{sec:conclusion}
We present the SMA 200--420 GHz survey towards a sample of 47 Class II sources in the Taurus-Auriga region. 
Except for the undetected source, CoKu Tau 1, we obtained 12 independent samples of the 200--420 GHz SEDs on the 47 sources.

We fit the 200--420 GHz SED by power law using the MCMC method and derived the $\alpha_{\mbox{\scriptsize 200-420 GHz}}$ for each source. 
The 7 marginally resolved disks, GM~Aur, DL~Tau, CI~Tau, AB~Aur, DM~Tau, AA~Tau and GO~Tau, consistently show $\alpha_{\mbox{\scriptsize 200-420 GHz}}>2.0$, except for GO Tau. 
The other 40 unresolved sources show $\alpha_{\mbox{\scriptsize 200-420 GHz}}\sim2.0$. 
GI Tau, CY Tau and SU Aur show relatively low $\alpha_{\mbox{\scriptsize 200-420 GHz}}<2.0$ of 1.29, 1.54 and 1.73. 

With an improved survey and data reduction strategy, we robustly confirmed the low ($\sim$2) spectral index of the 47 Taurus sources reported in \citet{Andrews_Williams_2005}. 
Moreover, with a dense sampling (at 10--40 GHz intervals) of SED, we did not detect spectral index variation over the frequency range of 200--420 GHz.

The $\alpha_{\mbox{\scriptsize 200-420 GHz}}$ of the 39 unresolved sources form an extremely narrow distribution of $2.04 \pm 0.22$ and show no dependence on the $F_{\mbox{\scriptsize 337 GHz}}$. 
The 7 resolved (diameter > 250 AU) disks have larger $\alpha_{\mbox{\scriptsize 200-420 GHz}}$ compared to the distribution of the unresolved sources. 
The effective radius ($R_{\mbox{\scriptsize 95\%}}$ or $R_{\mbox{\scriptsize 90\%}}$) of 46 disks are positively correlated to the $\alpha_{\mbox{\scriptsize 200-420 GHz}}$ with a correlation coefficient of $\hat{\rho} = 0.67 ^{+0.12} _{-0.15}$. 

The statistical results suggest that the (sub)millimeter luminosity is dominated by optically thick structures in the Class II disk. 
There is a caveat that the spectral index and the dust properties are varying across the disk (e.g. \citealt{Guidi2022A&A...664A.137G}) in multi-wavelength observations. 
The globally-averaged spectral indices detected in this survey can be reproduced with the optically thick contribution in the total flux densities and reveal the dust properties within the $\tau\sim1$ surface region. 
Future high-resolution studies at longer wavelengths may help resolve the different properties in the optically thin and optically thick disk structures. 

Measurements of dust masses based on (sub)millimeter observations and an optically thin assumption therefore were likely underestimated.
The dust mass budget in Class II disks may be sufficient to feed later planet formation. 
The estimation of the lower limits of dust masses based on optically thick dust slab model yields a range of $M_{\mbox{\scriptsize dust, thick}}=$ 4--450 $M_{\tiny \bigoplus}$. 
It implies that optically thin assumption has resulted in at least $1.2-2.8$ times underestimation of dust mass in the Class II disks. 

We performed population synthesis using simple physical models of optically thick, uniform dust slabs with two $T_{\mbox{\scriptsize dust,avg}}$ assumptions, 24 K and 100 K, for the 39 unresolved sources. 
From the comparison of the synthesized spectral index distributions with different $T_{\mbox{\scriptsize dust,avg}}$ assumptions and the observed $2.04 \pm 0.22$ population, we found that the $T_{\mbox{\scriptsize dust,avg}}$ of the 39 unresolved sources are either narrowly distributed around a low temperature (e.g., $\sim$ 24 K) or high enough to be within Rayleigh-Jeans limit. 
The observed spectral index distribution could be described by the optically thick dust slabs with $a_{\mbox{\scriptsize max}}\lesssim0.1$ mm for the 39 unresolved sources when $T_{\mbox{\scriptsize dust,avg}} = 100$ K and with $a_{\mbox{\scriptsize max}}\gtrsim3.0$ mm when $T_{\mbox{\scriptsize dust,avg}} = 24$ K. 

Three of the disks, CY Tau, SU Aur and GI Tau, in our sample are consistent with anomalously low ($<2.0$) $\alpha_{\mbox{\scriptsize 200-420 GHz}}$. 
This implies the dust self-scattering effect could be prominent in these optically thick disks and points to a $a_{\mbox{\scriptsize max}}\sim100 \mu m$.

Limited by the angular resolutions of our SMA observations, our observations preferentially trace the emission of the water-ice-coated dust grains residing outside of the 150--170 K water snow line.
Therefore, we infer that the coagulation of the water-ice-coated dust grains might be generally limited by the inward migration barrier and bouncing/fragmentation barrier.
This is in line with the recent analytical calculations and laboratory experiments that suggested that water-ice-coated dust grains are a lot less sticky than how they used to be considered; instead, those experiments indicated that the water-ice free dust inside of the water snow line are stickier can thus can coagulate more efficiently. 

Our implication of the dust mass budget in the Class II disks and the stickiness of dust grains have an impact on the current understanding of planet-formation, which deserve follow-up studies. 

\begin{acknowledgments}
We thank the anonymous referee for the constructive suggestions which helped improve the quality of our paper. We thank Dr. Tilman Birnstiel who provided insightful discussion about the potential issues of the DSHARP opacity table.
The Submillimeter Array is a joint project between the Smithsonian Astrophysical Observatory and the Academia Sinica Institute of Astronomy and Astrophysics, and is funded by the Smithsonian Institution and the Academia Sinica (\citealt{Ho2004ApJ...616L...1H}).
H.B.L. and C.Y.C. are supported by the National Science and Technology Council (NSTC) of Taiwan (Grant Nos. 111-2112-M-110-022-MY3).
Support for F.L. was provided by NASA through the NASA Hubble Fellowship grant \#HST-HF2-51512.001-A awarded by the Space Telescope Science Institute, which is operated by the Association of Universities for Research in Astronomy, Incorporated, under NASA contract NAS5-26555. 
\end{acknowledgments}

%

\vspace{5mm}
\facilities{JVLA, ALMA, SMA}


\software{
          astropy \citep{2013A&A...558A..33A},  
          Numpy \citep{VanDerWalt2011}, 
          MIR IDL \citep{Qi2003cdsf.conf..393Q},
          Miriad \citep{Sault1995ASPC...77..433S},
          emcee \citep{Foreman-Mackey2013PASP},
          corner \citep{corner},
          PyAstronomy \citep{pya}
          }



\appendix

\begin{figure*}[ht]
    \hspace{-1.8cm}
    \begin{tabular}{ lll }
    \includegraphics[width=7cm]{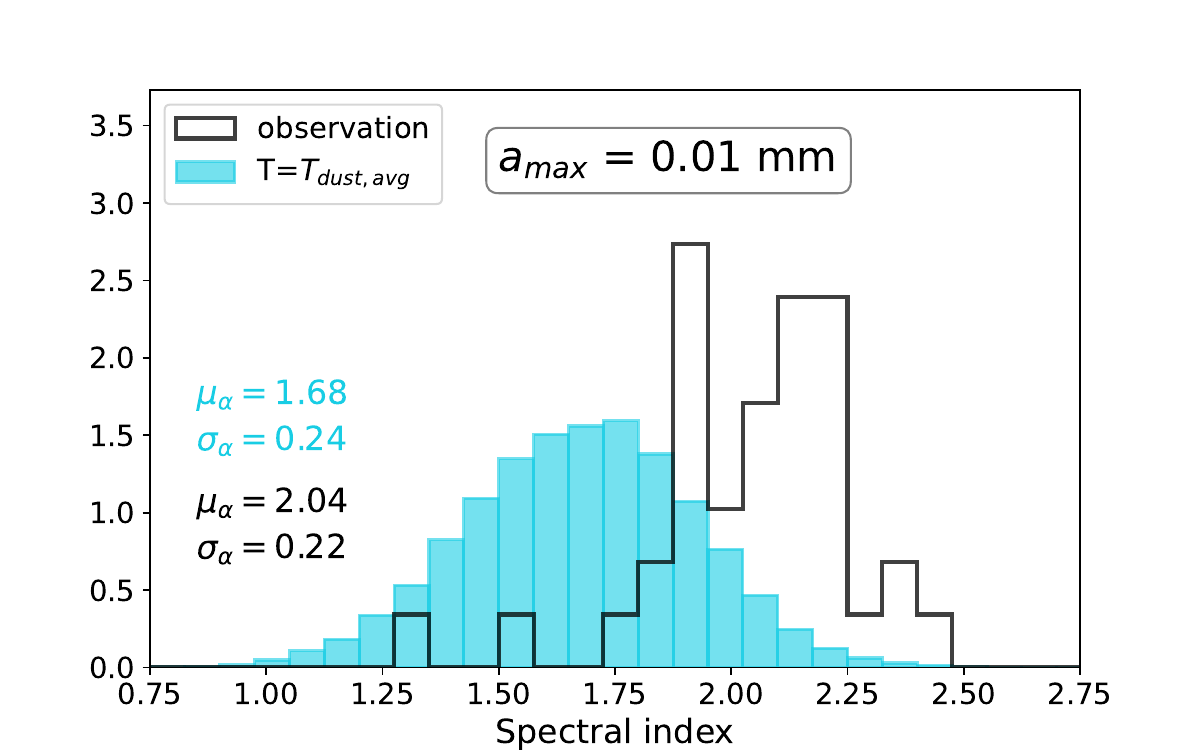}&
    \hspace{-1.1cm}
    \includegraphics[width=7cm]{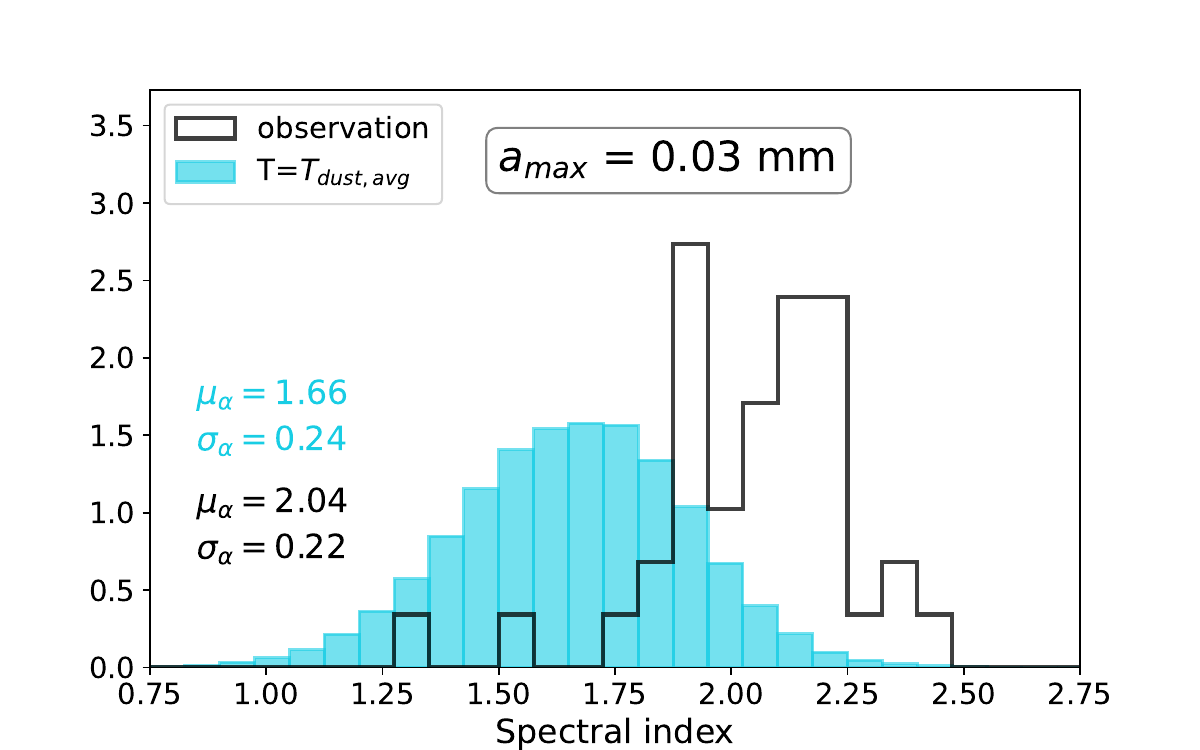}&
    \hspace{-1.1cm}
    \includegraphics[width=7cm]{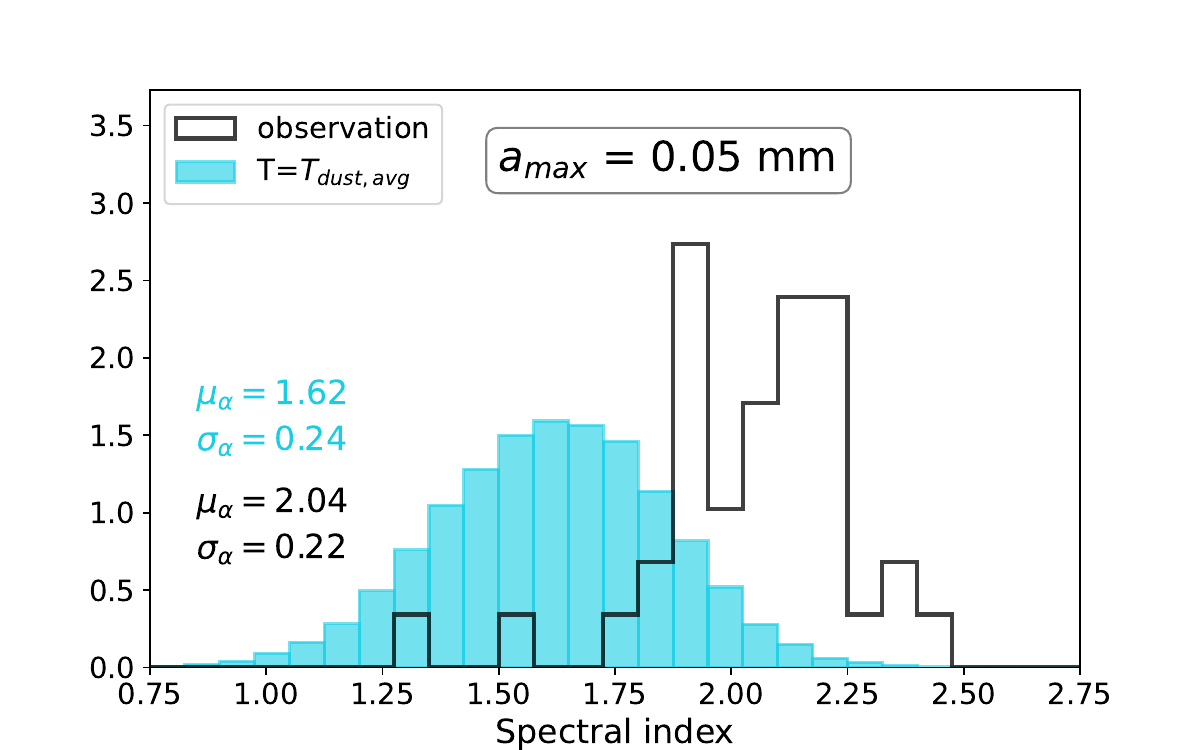}\\
    \includegraphics[width=7cm]{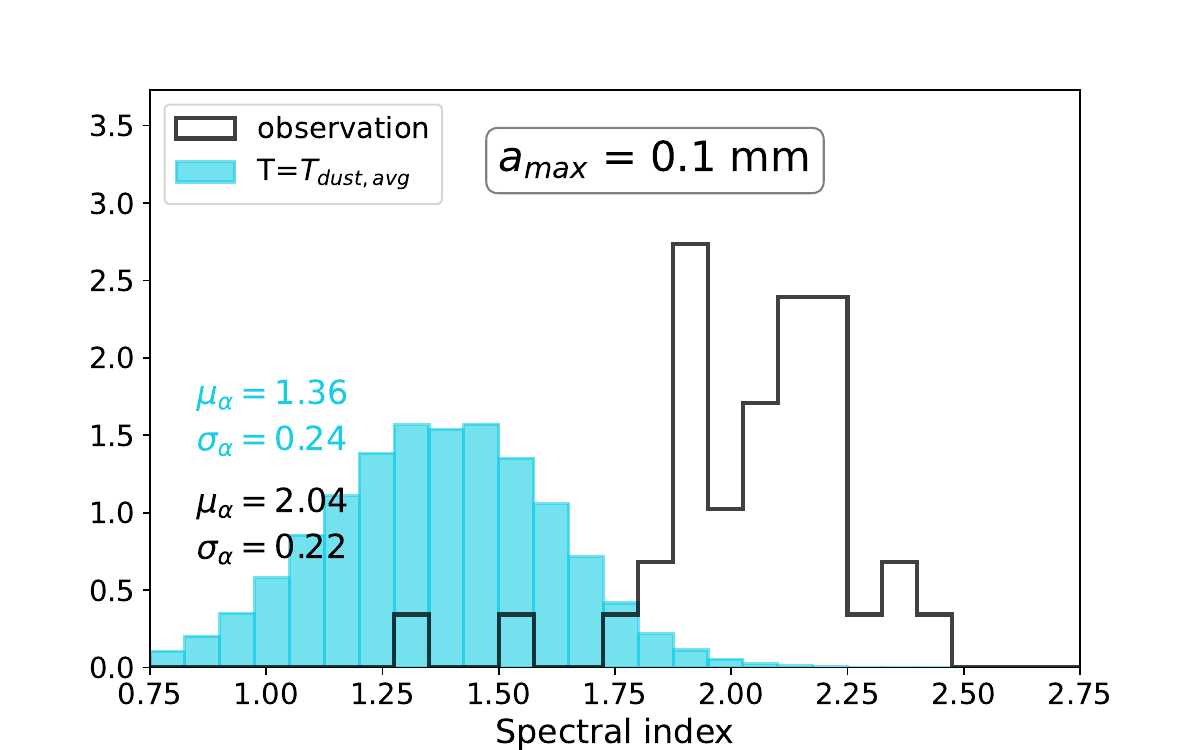}&
    \hspace{-1.1cm}
    \includegraphics[width=7cm]{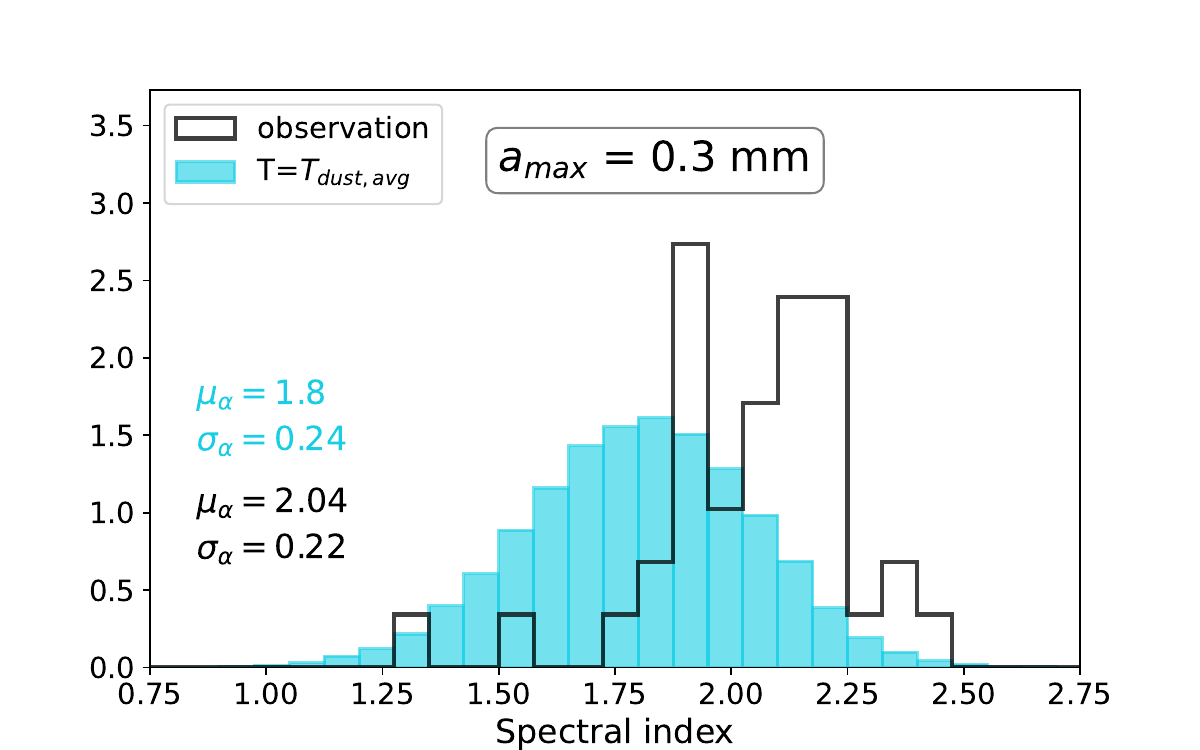}&
    \hspace{-1.1cm}
    \includegraphics[width=7cm]{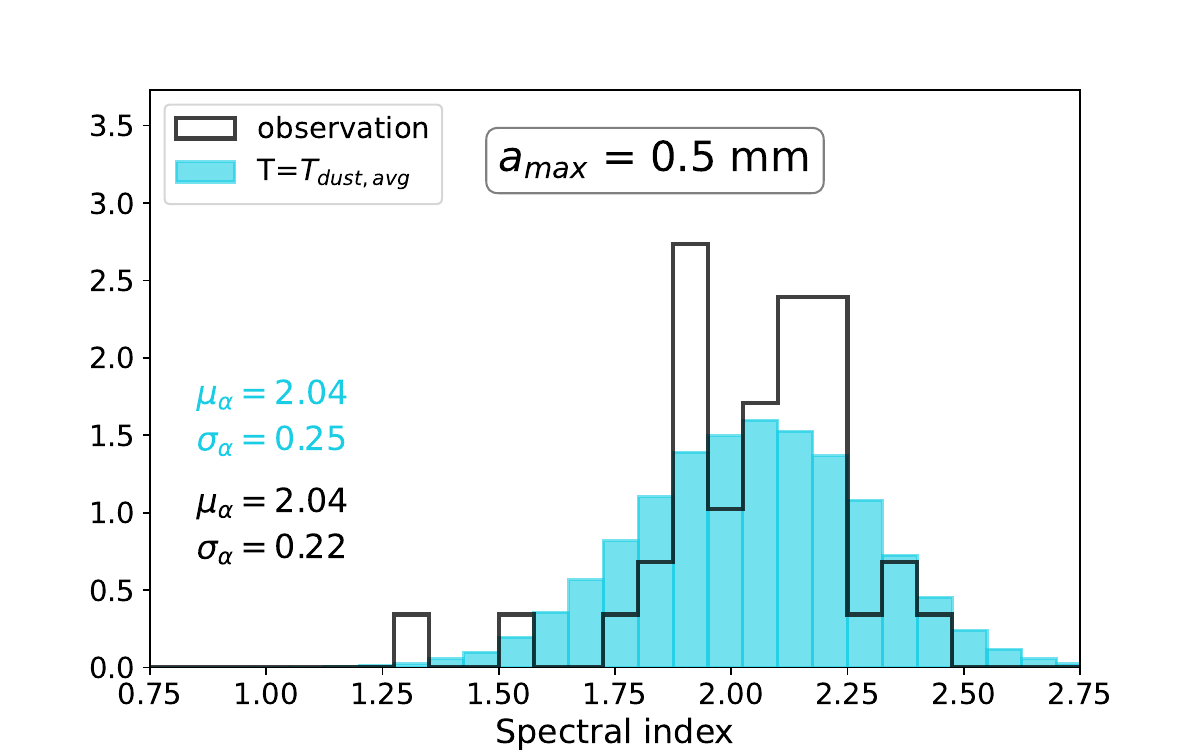}\\
    \includegraphics[width=7cm]{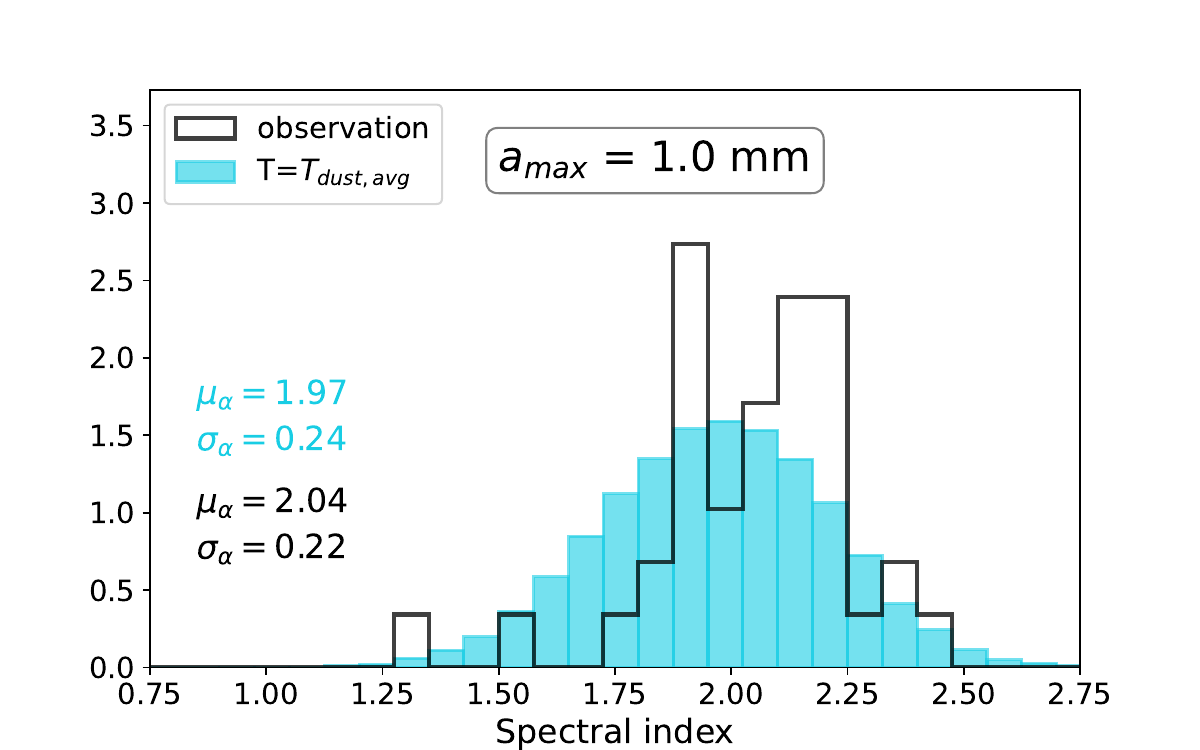}&
    \hspace{-1.1cm}
    \includegraphics[width=7cm]{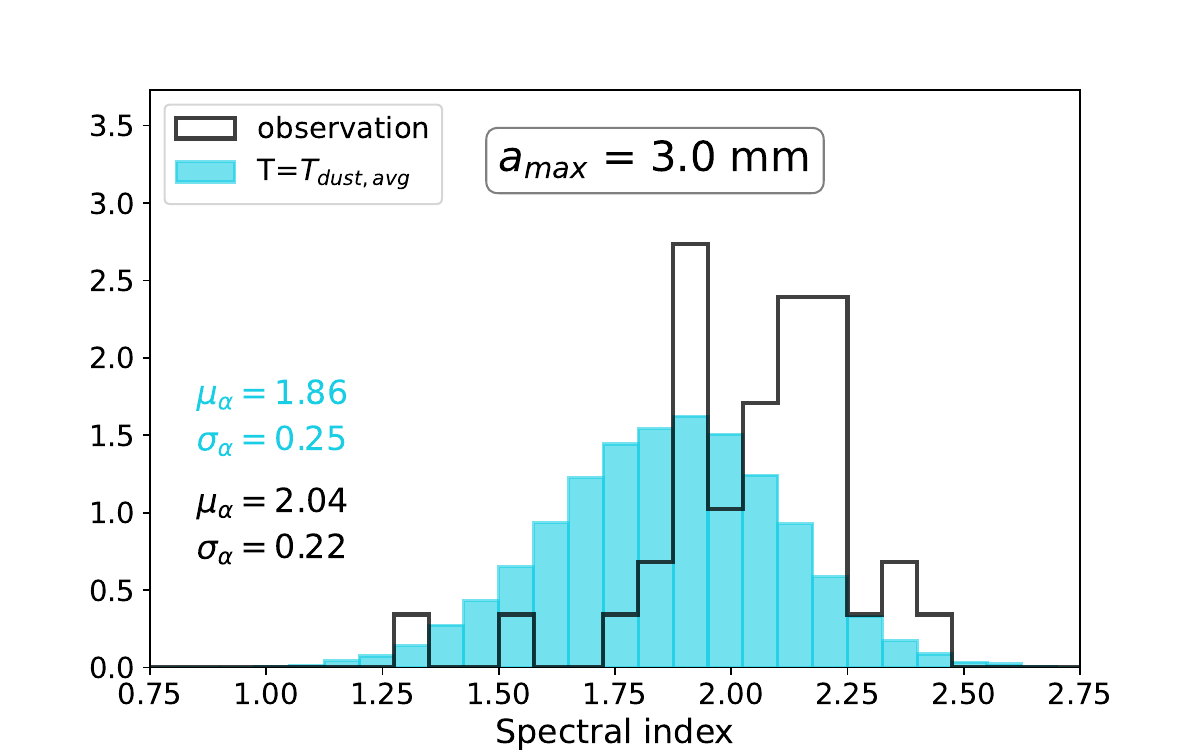}&
    \hspace{-1.1cm}
    \includegraphics[width=7cm]{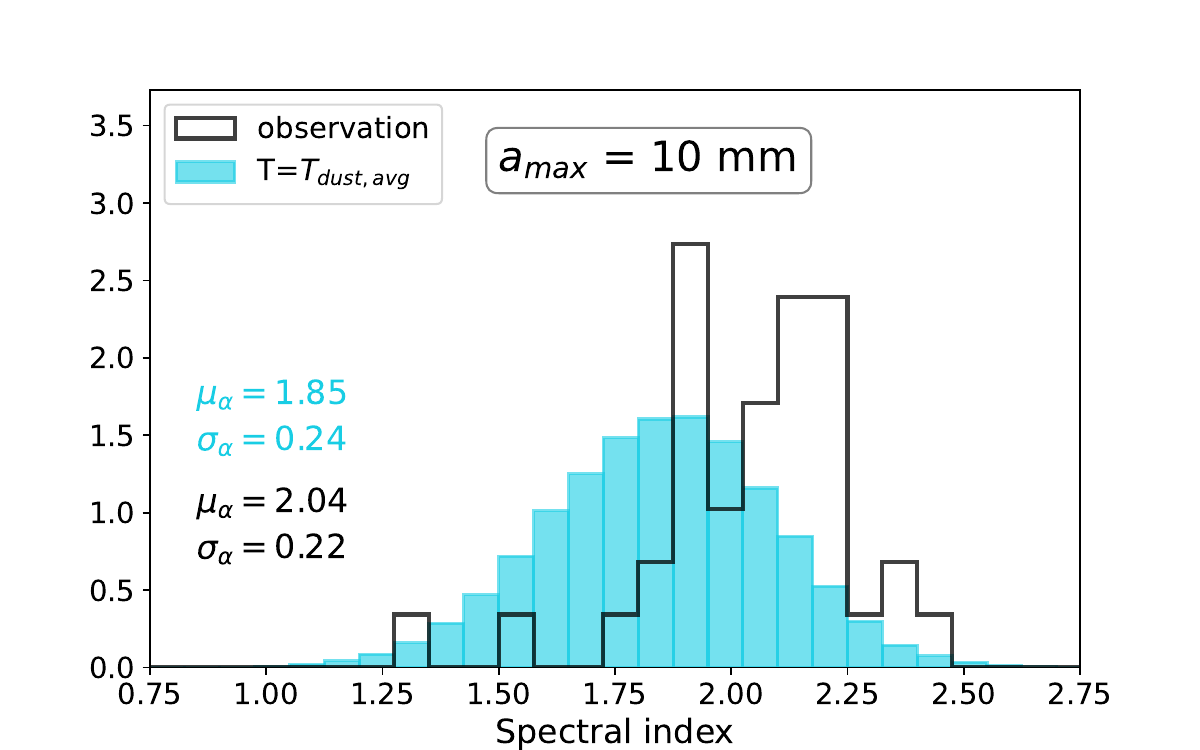}\\
    \end{tabular}
    \caption{ Synthesized spectral index distribution ($\alpha_{\mbox{\scriptsize 200-420 GHz}}$) derived from the dust slab model overplotted with the spectral index distribution from the observation for the 39 unresolved sources. The arrangement of panels for each $a_{\mbox{\scriptsize max}}$ is the same as that in Figure \ref{fig:spid_hist}. In each panel, the cyan histogram shows the synthesized distribution generated by assuming the $T_{\mbox{\scriptsize dust,avg}}$ distribution in Figure \ref{fig:Tdust_hist}. The black histogram shows the spectral index distribution in the observational results. The derived $\mu_{\scriptsize \alpha}$ and $\sigma_{\scriptsize \alpha}$ of the synthesized distributions and observational results are shown in the left and have the same color with the histograms. }
    \label{fig:spid_hist_Tavg}
\end{figure*}

\section{Population synthesis using estimated averaged dust temperature}\label{appendix:popsynthesis}

Figure \ref{fig:spid_hist_Tavg} shows the comparison of the observed spectral index distribution with those from our population synthesis assuming the $T_{\mbox{\scriptsize dust,avg}}$ distribution in Figure \ref{fig:Tdust_hist}. 
We present the cases of $a_{\mbox{\scriptsize max}} = 0.01,  0.03,  0.05,  0.1,  0.3, 0.5,  1.0,  3.0,  10$ mm. 
It appears that for all considered cases of $a_{\mbox{\scriptsize max}}$, the population synthesis (Section \ref{sub:Popsynthesis}) using this temperature assumption yielded too broad spectral index distributions to be consistent with the observed ones.
While $\sigma_{\scriptsize \alpha}$ of the synthesized spectral index distributions (based on the method in Section \ref{sub:Popsynthesis} is in general $\sim$0.24-0.25, the $\sigma_{\scriptsize \alpha}$ of the observed spectral index distribution is only 0.22.
Furthermore, when we excluded the two outliers with exceptionally low spectral indices, CY~Tau and GI~Tau, from the observation distribution, we obtained a smaller standard deviation of $\sigma_{\scriptsize \alpha} = 0.16$. 
This enhances the discrepancy between the observed $\sigma_{\scriptsize \alpha}$ and the population synthesis results.

Physically, $\sigma_{\scriptsize \alpha}$ can be enlarged due to the following reasons: (1) $a_{\mbox{\scriptsize max}}$ are distributed over a broad range, (2) $T_{\mbox{\scriptsize dust}}$ are distributed over a broad range, and the observing frequencies are not necessarily in the Rayleigh-Jeans limits, and (3) mixing optically thin dust emission.
Since our population synthesis assumed the same $a_{\mbox{\scriptsize max}}$ for all 39 unresolved sources and the extremely optically thick limit, the explanations for the high $\sigma_{\scriptsize \alpha}$ in the synthesized spectral index distributions in Figure \ref{fig:spid_hist_Tavg} is either the recipes in Section \ref{sub:Popsynthesis} underestimated the $\mbox{T}_{\mbox{\scriptsize dust, avg}}$ for some of the observed Class~II disks, or the recipes over-estimated the dispersion of $T_{\mbox{\scriptsize dust,avg}}$.
The underestimates of $\mbox{T}_{\mbox{\scriptsize dust, avg}}$ may be due to overestimating disk radii (e.g., due to the limit angular resolutions of the existing observations), or due to the inaccurate or over-simplified assumptions for the power law index of the radial temperature profile (see the discussion in Section \ref{sub:qualitative}).
The smaller standard deviation of observed distribution than that of the model also indicates that the distributions for the physical parameter, such as $\mbox{T}_{\mbox{\scriptsize dust, avg}}$ or $a_{\mbox{\scriptsize max}}$, should be rather uniform since any discrepancy will result in broadening of $\alpha$ distribution.  

\section{Dust mass estimation}\label{appendix:dustmass}
We estimated the dust mass for the detected 47 sources based on the simple physical model in Section \ref{sub:slab}, in which the Class II disk is approximated as a dust slab with uniform dust column density, temperature and $a_{\mbox{\scriptsize max}}$. 
We provide two versions of estimates that were based on different assumptions of averaged dust temperatures $\mbox{T}_{\mbox{\scriptsize dust, avg}}$.
The first assumed that $\mbox{T}_{\mbox{\scriptsize dust, avg}}=100$ K, such that our observations for all sources are in the Rayleigh-Jeans.
The second adopted the median value of the temperature (24 K) estimates for the individual sources introduced in Section \ref{sub:Popsynthesis} (Table \ref{tab:Tdust}).
We adopted the DSHARP dust opacities  (\citealt{Birnstiel2018ApJ...869L..45B}) and a power law grain size distribution with power s$=-3.5$ over an interval of ($a_{\mbox{\scriptsize min}},a_{\mbox{\scriptsize max}}$; c.f. Section \ref{sub:Popsynthesis}). 
The rationale of our choice of dust opacity is discussed in Appendix \ref{appendix:diana}.

To estimate the lower limits of dust mass underestimation ratio, everytime we assumed a common $a_{\mbox{\scriptsize max}}$ for all 47 disks. 
Following the assumptions of $a_{\mbox{\scriptsize max}}$ in population synthesis (Section \ref{sub:Popsynthesis}), we calculated the dust mass in the cases of $a_{\mbox{\scriptsize max}} = 0.01,  0.03,  0.05,  0.1,  0.3,  0.5,  1.0,  3.0,  10$ mm. 
We evaluated the synthesized $\alpha_{\mbox{\scriptsize 200-420 GHz}}$ from the synthesized 200--420 GHz SED using the same fitting method in Section \ref{sub:Popsynthesis}. 
We increased the optical depth from $\tau_{\mbox{\scriptsize 230 GHz}} =$ 0 to $\tau_{\mbox{\scriptsize 230 GHz}}=$ 50 (extremely optically thick case) and determined over which the synthesized $\alpha_{\mbox{\scriptsize 200-420 GHz}}$ becomes lower than the observed $\alpha_{\mbox{\scriptsize 200-420 GHz}}$. 
We then assume this critical dust column density (corresponding to the critical optical depth) to estimate the dust mass lower limit ($M_{\mbox{\scriptsize dust, thick}}$) in the optically thick assumption.
The observed $\alpha_{\mbox{\scriptsize 200-420 GHz}}$, estimated dust column density ($\Sigma$) and optical depth ($\tau_{\mbox{\scriptsize 230 GHz}}$) are listed in the second, third/sixth and fourth/seventh columns in Table \ref{tab:Mdust}, respectively. 
When calculating the critical $\tau_{\mbox{\scriptsize 230 GHz}}$, we found that the $\alpha_{\mbox{\scriptsize 200-420 GHz}}$ of some sources could not be produced by the model even though we have increased the $\tau_{\mbox{\scriptsize 230 GHz}}$ to the maximum value of 50. 
In this situation, we allow the synthesized $\alpha_{\mbox{\scriptsize 200-420 GHz}}$ to be 1-$\sigma$ observed spectral index error larger than the observed spectral index and then recalculated the critical $\tau_{\mbox{\scriptsize 230 GHz}}$. 

\begin{figure}
\includegraphics[width=8.5cm]{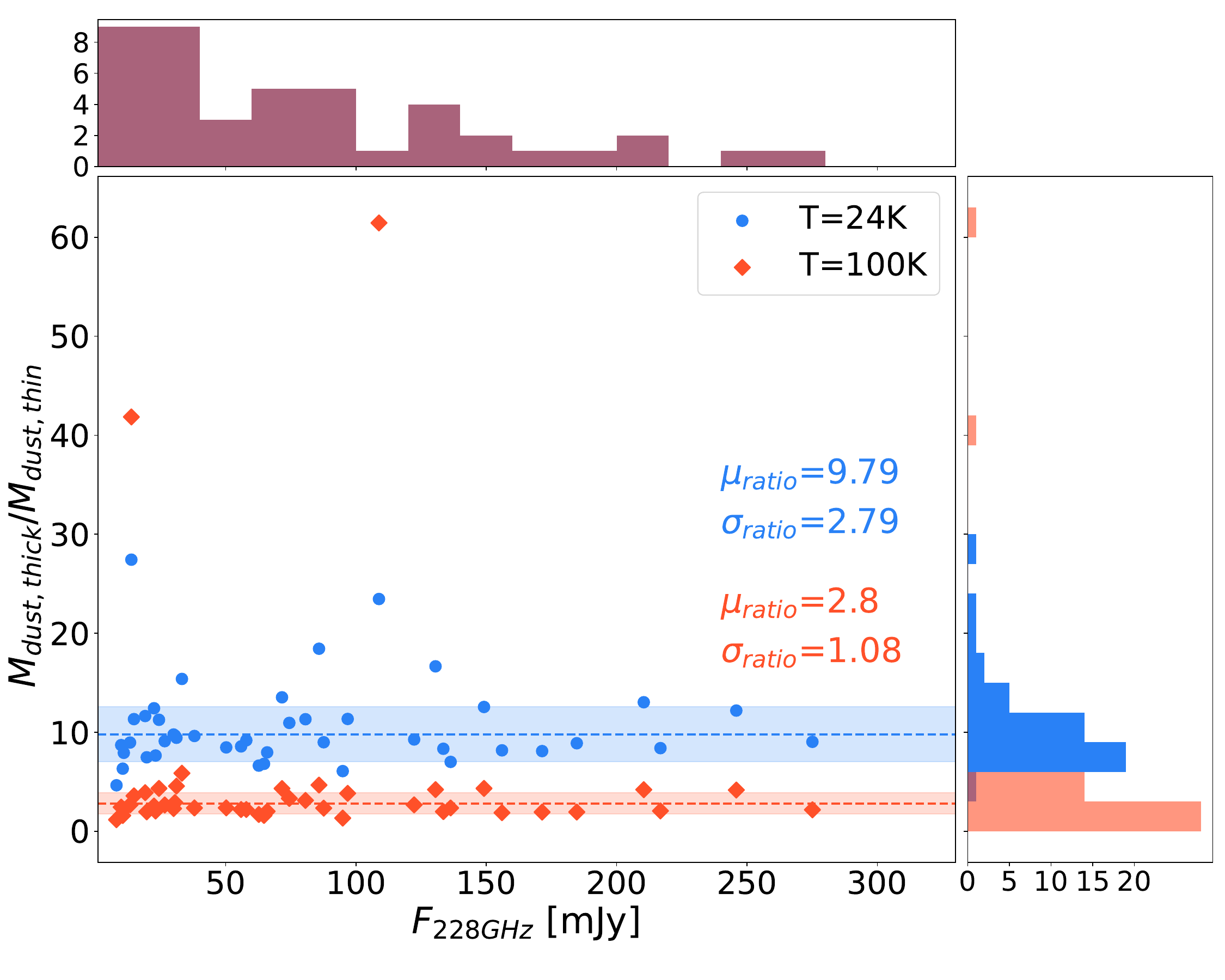} \\
\includegraphics[width=8.5cm]{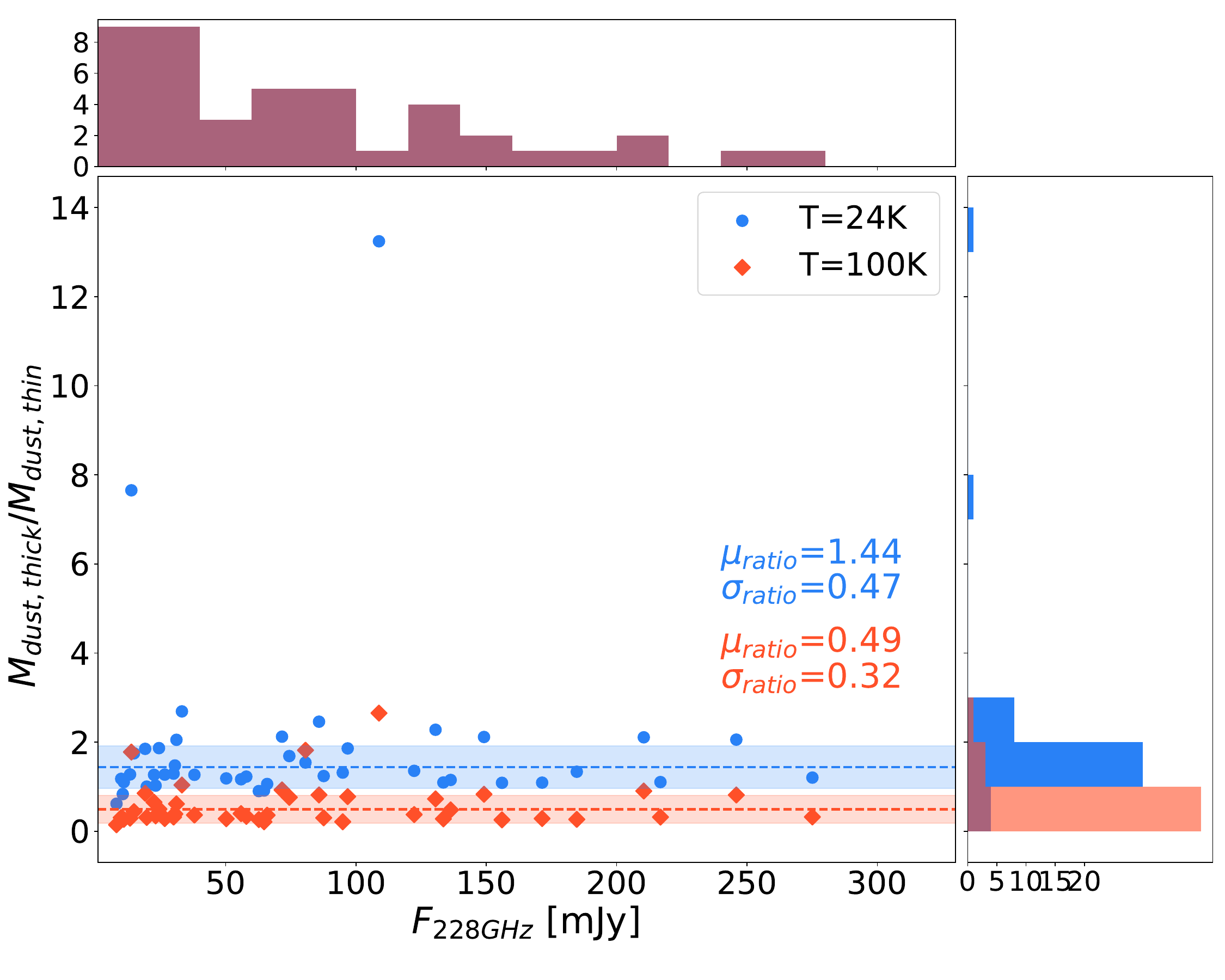} \\
\caption{
Dust mass underestimation ratio for the 47 sources in the survey assuming (1) $T_{\mbox{\scriptsize dust,avg}}=$ 100 K within Rayleigh-Jeans limit and (2) $T_{\mbox{\scriptsize dust,avg}}=$ 24 K, the median value of $T_{\mbox{\scriptsize dust,avg}}$ in Table \ref{tab:Tdust}. {\it Top} panel shows the estimated dust mass ($M_{\mbox{\scriptsize dust, thick}}$) in this work assuming $a_{\mbox{\scriptsize max}}=0.05$ mm. {\it Bottom} panel shows the estimated dust mass ($M_{\mbox{\scriptsize dust, thick}}$) assuming $a_{\mbox{\scriptsize max}}=3.0$ mm. The dashed line and the shaded area are the mean and standard deviation of the ratio, respectively. The outlying points, CY~Tau and GI~Tau, have been excluded when we calculated the mean and standard deviation. }
\label{fig:dust_mass_ratio}
\end{figure}

We estimated the solid angle ($\Omega$) for the emission from the observed 228 GHz flux density ($F_{\mbox{\scriptsize 228 GHz}}$) based on the same dust slab model with same assumptions of dust column density, temperature and $a_{\mbox{\scriptsize max}}$. 
We increased the solid angle from 0 to a circular region with 100 AU radius and determined over which the calculated $F_{\mbox{\scriptsize 228 GHz}}$ becomes higher than the observed value. 
We then took the solid angle to estimate the total area of the emission in dust mass estimation.
With the estimated $\Sigma$ and $\Omega$ and the quoted distance, we calculated the $M_{\mbox{\scriptsize dust, thick}}$ for the 47 disks. 
The estimated dust mass lower limit ($M_{\mbox{\scriptsize dust, thick}}$) and the previous measurements ($M_{\mbox{\scriptsize dust, thin}}$) quoted from \citet{Manara2023ASPC..534..539M} are listed in Table \ref{tab:Mdust}. 
In the table, we only present cases of $a_{\mbox{\scriptsize max}} = 0.05$ mm under $\mbox{T}_{\mbox{\scriptsize dust, avg}}=100$ K assumption and cases of $a_{\mbox{\scriptsize max}} = 3.0$ mm under $\mbox{T}_{\mbox{\scriptsize dust, avg}}=24$ K assumption, which are within the preferred ranges in the results of the population synthesis.

Figure \ref{fig:dust_mass_ratio} shows the lower limits of the underestimation ratio ($M_{\mbox{\scriptsize dust, thick}}/M_{\mbox{\scriptsize dust, thin}}$) of the 47 sources in the survey under the assumption of (1) $T_{\mbox{\scriptsize dust,avg}}= 24$ K, and (2) $T_{\mbox{\scriptsize dust,avg}}= 100$ K. 
The top panel shows the case of $a_{\mbox{\scriptsize max}}=0.05$ mm, and the bottom panel shows the case of $a_{\mbox{\scriptsize max}}=3.0$ mm. 
In case (1), the lower limits of underestimation ratios may be regarded as less conservative estimations, given that the values $T_{\mbox{\scriptsize dust,avg}}$ (Table \ref{tab:Tdust}) are likely underestimated (c.f. Section \ref{sub:Popcompar}); In case (2), the assumed, high $T_{\mbox{\scriptsize dust,avg}}$ provide more conservative estimations of mass underestimation ratios. 

In case (1), the two sources, CY Tau and GI Tau, have too low spectral indices to be reproduced by the model. 
Although we have considered the 1-$\sigma$ observed spectral index error and increased the $\tau_{\mbox{\scriptsize 230 GHz}}$ to the maximum value of 50, the spectral index cannot be as low as 1.67 and 1.47, respectively. 
We inferred that in these two sources, the $a_{\mbox{\scriptsize max}}$ may have grown to a size very close to 0.1 mm, so the spectral index cannot be produced with other $a_{\mbox{\scriptsize max}}$ mm assumptions. 
This condition caused the two outlying underestimation ratios of dust mass in the bottom panel of Figure \ref{fig:dust_mass_ratio}.
In case (2), 3 sources have low spectral indices that cannot be produced with the assumed $a_{\mbox{\scriptsize max}}$. 
However, the dust column density required to make $\tau_{\mbox{\scriptsize 230 GHz}} =$ 50 is smaller than that in case (1), and therefore the outliers are not obvious in the bottom panel of Figure \ref{fig:dust_mass_ratio}. 
To avoid our statistics being biased by these failed mass calculations, we take the median value of the underestimation ratio of all the 47 sources as a representative underestimation ratio for each $a_{\mbox{\scriptsize max}}$ case. 


\section{Dust mass opacities}\label{appendix:diana}
Our discussion was mainly based on the default DSHARP opacity table (\citealt{Birnstiel2018ApJ...869L..45B}; Section \ref{sub:Popsynthesis}, \ref{sub:Popcompar}, \ref{sub:physics}).
In this case, the absorption opacity at $\sim$250 GHz frequency ($\lambda\sim$1.2 mm) is approximately in the range of $\sim$0.5--2 cm$^{2}$\,g$^{-1}$ when $a_{\mbox{\scriptsize max}}$ is in the range of 0.1--1 mm.
This choice of opacity table made it straightforward to compare with the results of the DSHARP project (see \citealt{Andrews_2018} for a summary) and the other recent works that also adopted this opacity table.
Many demographic surveys towards Class~II disks also adopted the compatible values of absorption opacities (e.g., \citealt{Andrews_Williams_2005,Andrews_Williams_2007,Carpenter2014ApJ...787...42C,Ansdell2016ApJ...828...46A,Barenfeld2016ApJ...827..142B,Testi2016A&A...593A.111T,Akeson_2019,Babaian2019AAS...23410503B,Cazzoletti2019A&A...626A..11C,Williams2019ApJ...875L...9W,Testi2022A&A...663A..98T}) in spite that dust scattering opacity was not considered in their dust mass estimates.

We acknowledge that there are other publicly available opacity tables that also have been widely adopted in the researches of protoplanetary disks.
Their assumptions of dust compositions can be different from that of the DSHARP opacity table. 
For our discussion in this section, we categorize the dust opacity tables into (1) opaque tables, and (2) other tables, based on their assumptions of the dominant form of interstellar carbon (more below).

The opacity tables in the opaque table category assumed that the the dominant form of interstellar carbon is graphite or amorphous carbon; in the latter cases, they quoted the absorption opacities of the specific "BE amorphous carbon" sample (\citealt{Preibisch1993A&A...279..577P}) or "ACAR amorphous carbon" sample (\citealt{Mennella1998ApJ...496.1058M}).
Examples of opacity tables in this category include the DIANA opacity table (\citealt{Woitke2016A&A...586A.103W}) and the absorption opacities used/evaluated in \citet{Ricci2010A&A...512A..15R} and \citet{Draine1984ApJ...285...89D}, etc.

The opacity tables in the other category either assumed that the dominant form of interstellar carbon is amorphous carbon and quoted the absorption opacities of amorphous carbon from \citet{Jager1998A&A...332..291J}, or assumed that the dominant form of interstellar carbon is refractory organics (e.g., \citealt{Henning1996A&A...311..291H}).
The DSHARP opacity table (\citealt{Birnstiel2018ApJ...869L..45B}) we adopted for the present analyses is in this category.

\begin{figure}
\includegraphics[width=8.5cm]{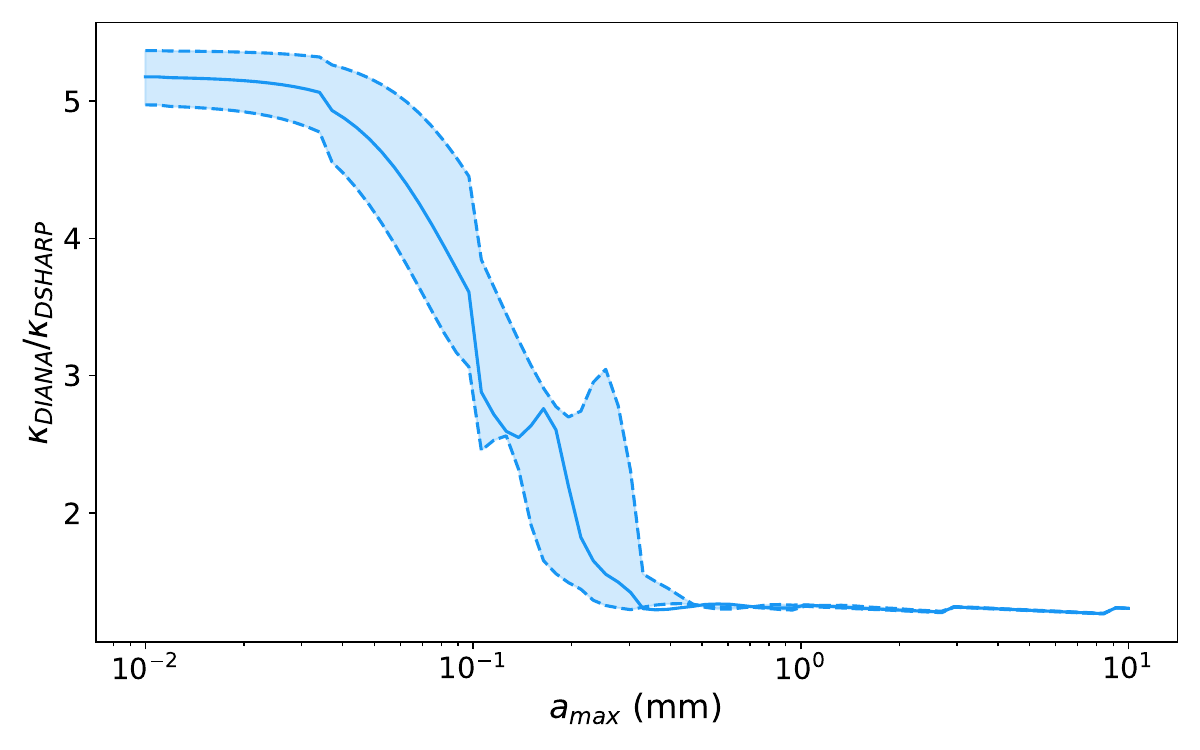} 
\vspace{-1.5cm}
\caption{
Ratios between the DIANA absorption opacity ($\kappa_{\mbox{\tiny DIANA}}$) and the DSHARP absorption opacity ($\kappa_{\mbox{\tiny DSHARP}}$).
}
\label{fig:opacity_ratio}
\end{figure}

At (sub)millimeter wavelengths, the $\kappa^{\mbox{\scriptsize abs}}$ of graphite and the BE and ACAR amorphous carbon samples are several orders of magnitude higher than the $\kappa^{\mbox{\scriptsize abs}}$ of the other carbonaceous materials. 
This is because that graphite and the BE and ACAR amorphous carbon samples carry free charges. 
As a consequence, at (sub)millimeter wavelengths, for a specific choice of  $a_{\mbox{\scriptsize max}}$, the absorption opacities $\kappa^{\mbox{\scriptsize abs}}$ of the tables in the opaque table category are considerably higher than the $\kappa^{\mbox{\scriptsize abs}}$ of those in the other category (c.f., \citealt{Birnstiel2018ApJ...869L..45B,Zormpas2022A&A...661A..66Z}).
For example, Figure \ref{fig:opacity_ratio} shows a comparison between $\kappa^{\mbox{\scriptsize abs}}_{\mbox{\tiny DIANA}}$ and $\kappa^{\mbox{\scriptsize abs}}_{\mbox{\tiny DSHARP}}$.

Naively, we might expect those who adopted an opacity table in the opaque table category to yield systematically lower dust mass estimates than those who adopted an opacity table in the other category.
Practically, the choices of opacity table interact with the SEDs fittings in a more complicated way.
How swapping opacity table changes the resulting $a_{\mbox{\scriptsize max}}$ and $M_{\mbox{\scriptsize dust}}$ derivations cannot be predicted with simple calculations, in particular, when the scattering opacity is taken into consideration.

We are not aware of any strong experimental or observational evidence that favors the opacity tables in one category over the other.
We note that based on an observational point of view, \citet{Draine2006ApJ...636.1114D} pointed out that the dominant form of interstellar carbon is unlikely to be graphite or carbonaceous material that is similar to the BE and ACAR amorphous carbon samples.
Therefore, we presently do not adopt an opacity table in the opaque table category.

\begin{figure*}[]
    \hspace{-0.8cm}
    \begin{tabular}{c}
    \includegraphics[width=18cm]{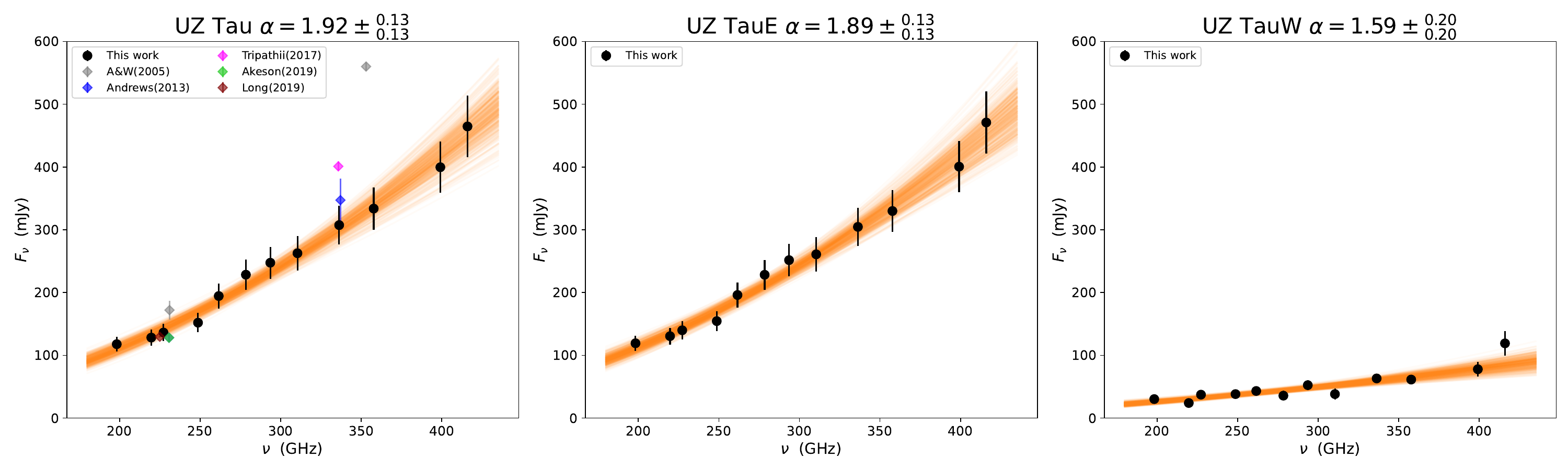} 
    \end{tabular}
    \caption{ Two flux measurement results of UZ~Tau using two model assumptions in visibility fitting: (1) a point-source model, (2) two point-source components model. {\it Left} panel: The flux densities of UZ~Tau when we treated it as an unresolved source and fit it with model (1). This panel is identical to the UZ~Tau panel in Figure \ref{fig:SED_0510_0}. {\it Middle \& Right} panels: The individual flux densities of UZ Tau E and UZ Tau W when we fit them as unresolved binary fit them with model (2). The presentation style of 200--420 GHz SED is the same as that of Figure \ref{fig:SED_0510_resolved}. 
    }
    \label{fig:SED_UZTau}
\end{figure*}

The size-averaged dust opacities also depend on the assumptions of grain size distributions $n(a)$.
In the fittings of observed (sub)millimeter SEDs, the typical assumption is $n(a)\propto a^{s}$ in between the minimum and maximum grain sizes $a_{\mbox{\scriptsize min}}$ and $a_{\mbox{\scriptsize max}}$ and is zero elsewhere (c.f., \citealt{Hildebrand1983QJRAS..24..267H}). 
The most common choice of the power-law index is $s=-3.5$, which is adopted in the present work. 
It was motivated by the assumption that the dust grain growth is limited by the fragmentation barrier (c.f., \citealt{Birnstiel2016SSRv..205...41B} and references therein).
When dust grain growth is limited by the inward drift barrier instead of the fragmentation barrier, the value of $s$ may be close to $-2.5$ (\citealt{Birnstiel2016SSRv..205...41B}), which corresponds to a top heavier size distribution function.
Whether dust grain growth in the protoplanetary disks is mainly limited by the fragmentation barrier or inward drift barrier, remains uncertain. 
When fitting (sub)millimeter dust SEDs, it may be anticipated that assuming $s=-2.5$ instead of $s=-3.5$ will yield slightly lower $a_{\mbox{\scriptsize max}}$ values. 
This is because the $s=-2.5$ case weights the contribution from the large dust grains more.
In this sense, the $a_{\mbox{\scriptsize max}}$ value suggested in our present work may be regarded as an upper limit. 
In practice, how the choice of $s$ biases the resulting $a_{\mbox{\scriptsize max}}$ values also cannot be predicted with simple calculations.
Nevertheless, in our experiences, when the observed $\alpha$ at (sub)millimeter bands is close to 2.0, assuming a compact dust morphology and the dust composition used in the DSHARP dust opacity table, the assumptions of $s=-2.5$ and $s=-3.5$ yield very similar $a_{\mbox{\scriptsize max}}$ values (Aso et al. submitted)

Finally, dust opacities also depend on the morphology of dust grains. 
For a specific $a_{\mbox{\scriptsize max}}$, the values of $\kappa^{\mbox{\scriptsize abs}}$ may become smaller if we assume a porous dust morphology (see Figure 11 of \citealt{Birnstiel2018ApJ...869L..45B}).
The DSHARP opacity table assumes a compact dust morphology.
Following the discussion of \citet{Tazaki2019ApJ...885...52T}, we think it is fair to assume that the dust grains are not highly porous.

In general, the derivations of $a_{\mbox{\scriptsize max}}$ and $M_{\mbox{\scriptsize dust}}$ are uncertain due to the uncertainties in the dust opacities.
The detailed comparison of dust opacity tables is beyond the scope of the present work.

\section{Class II Taurus disks}\label{appendix:source}
\subsection{Excluded flux density measurements}
As we mentioned in the end of Section \ref{sub:selcal}, we have excluded the visibility data of some faint sources at certain frequency band due to possible residual phase error. 
We excluded the flux measurements at the frequency bands if both the visibility amplitude shows negative values at some {\it uv}-ranges and the clean image shows no or diffused signal at the phase center. 
For CoKu Tau 1, we excluded the flux densities at 199, 228, 248, 262, 278, 294, 310, 337, 399.5 and 415.5 GHz. 
For FM Tau, we excluded the 278 GHz flux density.
For FY Tau, RW Aur and V836 Tau, we excluded their 357 GHz flux densities. 
For SU Aur and UY Aur, we excluded their 399.5 GHz flux densities. 
For GI Tau, we excluded the 415.5 GHz flux density. 


\begin{deluxetable}{c|cccc}
\tabletypesize{\footnotesize}
\tablecolumns{5}
\vspace{0.8cm}
\tablecaption{DQ Tau flux densities at individual epoch  \label{tab:DQ_Tau_flux}}
\tablehead{\colhead{Obs. date} & \colhead{$F_{\mbox{\tiny 199 GHz}}$} & \colhead{$F_{\mbox{\tiny 219 GHz}}$} & \colhead{$F_{\mbox{\tiny 228 GHz}}$} & \colhead{$F_{\mbox{\tiny 248 GHz}}$} \\
\colhead{} & \colhead{(mJy)} & \colhead{(mJy)} & \colhead{(mJy)} & \colhead{(mJy)}}
\startdata 
2021 10 18 & $123.1 \pm 12.8$ & $146.1 \pm 15.1$ & $128.2 \pm 13.3$ & $135.5 \pm 14.1$\\
2021 11 26 & $60.1 \pm 6.5$ & $72.5 \pm 7.7$ & $74.3 \pm 7.8$ & $82.6 \pm 8.7$ \\
\hline \noalign {\medskip}
Obs. date & $F_{\mbox{\tiny 262 GHz}}$ & $F_{\mbox{\tiny 278 GHz}}$ & $F_{\mbox{\tiny 294 GHz}}$ & $F_{\mbox{\tiny 310 GHz}}$ \\
\hline \noalign {\smallskip}
2021 11 20 & $ 112.4\pm 11.9$ & $88.5 \pm 10.7$ & $132.4 \pm 14.3$ & $134.2 \pm 14.7$\\
\hline \noalign {\medskip}
Obs. date & $F_{\mbox{\tiny 337 GHz}}$ & $F_{\mbox{\tiny 357 GHz}}$ & $F_{\mbox{\tiny 399.5 GHz}}$ & $F_{\mbox{\tiny 415.5 GHz}}$ \\
\hline \noalign {\smallskip}
2021 09 03 & $151.5 \pm 16.4$ & $174.0 \pm 19.9$ & $175.4 \pm 22.9$ & $189.2 \pm 32.1$\\
2021 09 25 & $144.0 \pm 15.2$ & $154.1 \pm 17.3$ & $190.7 \pm 24.6$ & $230.8 \pm 36.1$\\
2021 11 22 & $146.5 \pm 15.5$ & $176.0 \pm 19.4$ & $232.5 \pm 29.1$ & $296.6 \pm 40.7$\\
\enddata
\end{deluxetable}

\subsection{Variable source DQ Tau}\label{appendix:dqtau}
DQ Tau is a spectroscopic binary surrounded by a circumbinary disk. We observed varying flux densities between 230 GHz-1 track taken on October 18 and 230 GHz-2 track taken on November 26. 
The flux densities decreased by a factor of 2 from the former date to the latter date. 
\cite{Salter_2008} and \cite{Getman_2022} reported flares from DQ Tau. 

\subsection{Quadruple system UZ Tau}\label{appendix:uztau}
UZ Tau is a quadruple system that consists of two components, UZ Tau E and UZ Tau W. 
UZ Tau E is a spectroscopic binary. 
UZ Tau W is resolved as a binary with a $\sim 0.5 ''$ separation in ALMA observations.
The East and West components have a separation of $\sim 3 ''$, which can be resolved in our observations.

We tried fitting the UZ Tau visibility using two model assumptions in Section \ref{sub:flux}: (1) a point-source (2) two point-source components. 
In case (1), UZ Tau was viewed as an unresolved source, and we set the center of the point-source model at UZ Tau E. 
In case (2), the two components were fitted simultaneously by two point sources, and the two point sources were centered at UZ Tau E and UZ Tau W, respectively. 
Figure \ref{fig:SED_UZTau} compares the flux density measurements derived from the two model assumptions in visibility fitting. 
We found that UZ Tau E dominates the visibility fitting in case (1), which makes the results of UZ Tau in the left panel similar to that of UZ Tau E in the middle panel. 
From the right panel, we found UZ Tau W contributes little flux densities, and we could not robustly constrain the flux densities of UZ Tau W in the two components fitting.

\begin{figure}[]
    \hspace{-0.8cm}
    \begin{tabular}{c}
    \includegraphics[width=9.5cm]{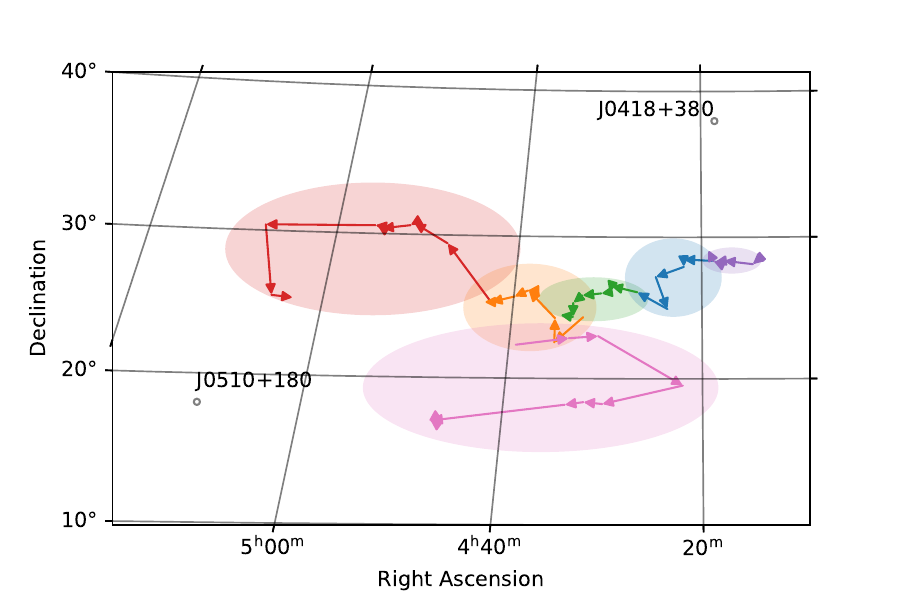} \\
    \includegraphics[width=9.5cm]{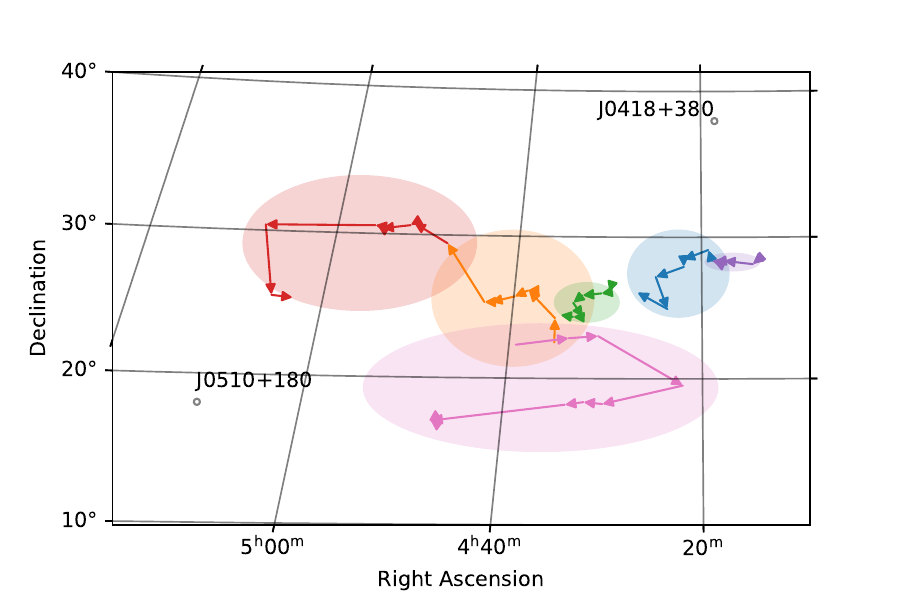}
    \end{tabular}
    \caption{
    The 47 selected Class II sources on the sky, which are divided into six groups (color coded). {\it Top :--} grouping for tracks 230 GHz-1, 230 GHz-2, 270 GHz-3. {\it Bottom :--} grouping for tracks 400 GHz-4, 400 GHz-5, 400 GHz-6. The arrows indicate the slewing paths in each observing group. The purple and blue groups are calibrated by J0418+380. The other groups are calibrated by J0510+180.
    }
    \label{fig:target_group}
\end{figure}

\begin{figure*}
    \begin{center}
    \includegraphics[width=14cm]{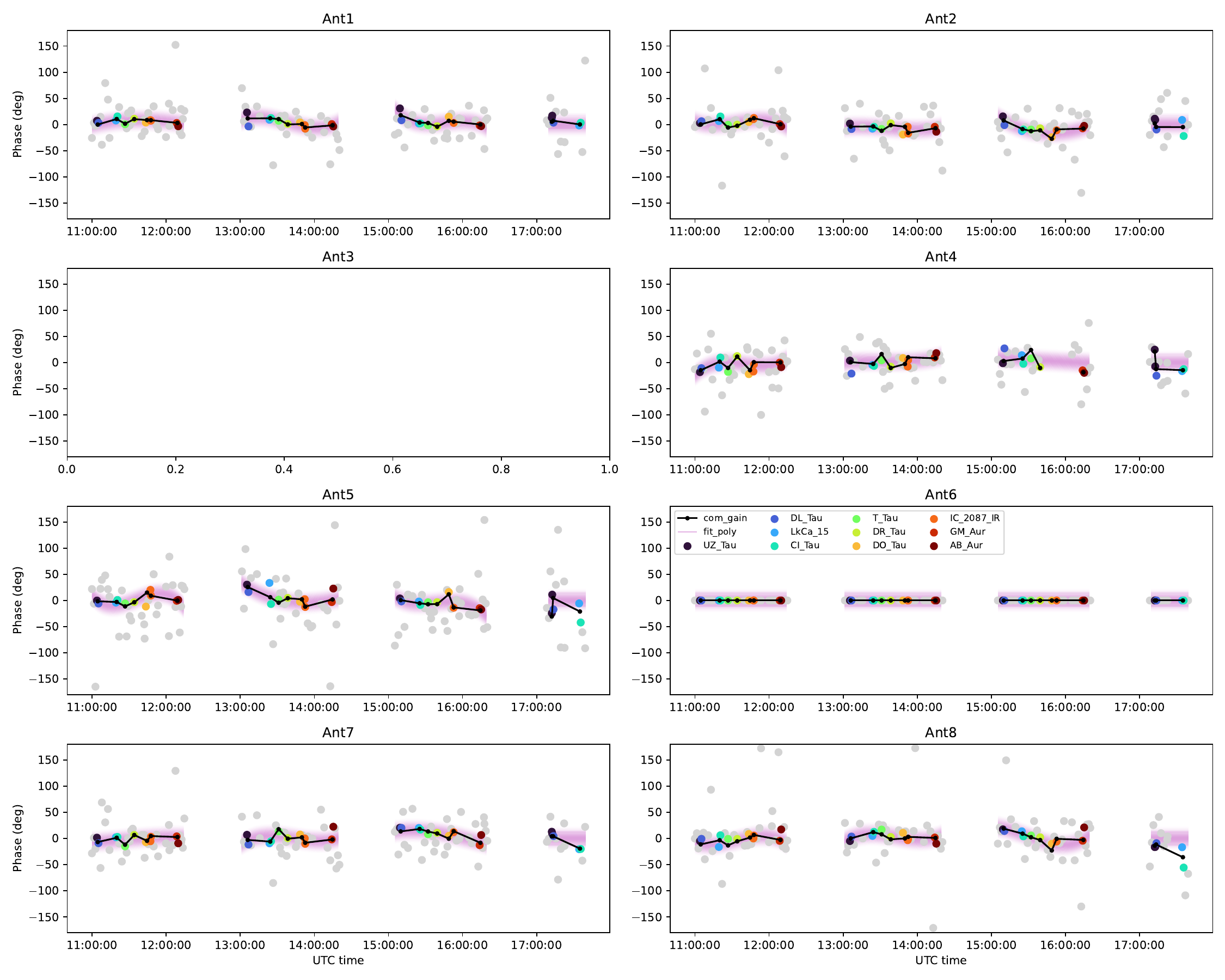}
    \end{center}
    \caption{
     The gain phase self-calibration solutions of track 400 GHz-5 at 337 GHz. In this figure, we show the solutions derived from the 34 sources that were gain calibrated by the quasar J0510+180. Each panel shows the antenna-based gain phase self-calibration solutions for an antenna; the reference antenna was Antenna 6, that the phases were defined to be zero. Dots present the solutions that were derived independently from each of the 34 sources: colorful dots are those derived from the 10 self-calibratable bright sources, while gray dots are those derived from the remaining 24 sources. The black line segments present the multi-pointing gain phase self-calibration solutions in each 12 minutes calibration time cycle (Section \ref{sub:selcal}). The pink line segments are the time-interpolated solutions that were derived from the polynomial fittings to the multi-pointing solutions. 
    }
    \label{fig:selcal_sol} 
\end{figure*}

\subsection{ALMA observations}
\subsubsection*{AA Tau}
AA Tau has a single disk with multiple ring structures and an inclination angle of $59^{\circ}.1$ (\citealt{Loomis2017ApJ...840...23L}). 
They presented the high-resolution ALMA 0.87 and 1.3 mm continuum observation and resolved three rings at 48.7, 94.9 and 142.6 AU disk radius. 

\subsubsection*{AB Aur}
AB Aur is a Herbig Ae star with spectral type A1.0 and hosts a large transition disk with $R_{\mbox{\scriptsize out}}\sim193$ AU (\citealt{Tang_2014}) at 1.3 mm. 
The ALMA continuum observations at 1.3 mm showed a dust ring which locates at 120 AU (\citealt{Tang_2017}), and the CO(J=2-1) map revealed two spiral arms in the cavity inside the dust ring. 

\subsubsection*{CW Tau}
CW Tau has a compact disk with $R_{\mbox{\scriptsize out}}\sim$48 AU (\citealt{Ueda_2022}) at 1.3 mm. 
The ALMA Band 4 and 6 images revealed one gap at $\sim$20 AU (\citealt{Ueda_2022}). 
They obtained the $\alpha_{\mbox{\scriptsize 0.75-1.3 mm}}$ $=2.0 \pm 0.24$ and $\alpha_{\mbox{\scriptsize 2.17-3.56 mm}}$ $=3.7 \pm 0.29$ from Band 4, 6, 7 and 8 observations and estimated a $M_{\mbox{\scriptsize dust}}$ $\sim 250 M_{\oplus}$. 

\subsubsection*{DM Tau}
DM Tau hosts a transition disk with multiple ring structures revealed by 1.3 mm ALMA observations. 
\citet{Hashimoto2021ApJ...911....5H} identified an inner ring at radius 3 AU, a outer ring at 20 AU which encloses the inner cavity, and the other two rings at 90 AU and 110 AU in the extended outer structure. 

\subsubsection*{GM Aur}
GM Aur hosts a transition disk with multiple ring structures revealed by 1.1 and 2.1 mm ALMA observations (\citealt{Huang2020ApJ...891...48H}). 
In the 1.1 mm continuum, they identified two rings at 40 AU and 84 AU which encloses the inner cavity, and another ring at 168 AU in the extended outer structure. 
They obtained the $\alpha_{\mbox{\scriptsize 1.1-2.1 mm}}$ $=2.7 \pm 0.2$ from Band 4 and 6 observations. 

\subsubsection*{LkCa 15}
LkCa 15 has a single disk with multiple ring structures and a inclination angle of $50^{\circ}.2$ (\citealt{Long2022ApJ...937L...1L}). 
They presented the high-resolution ALMA 0.87 and 1.3 mm continuum observation and resolved three rings at 42, 69 and 101 AU disk radius. 

\subsubsection*{UY Aur}
UY Aur is a binary system exhibiting a circumbinary disk. 
We quoted their separation from the ALMA 0.87 and 1.3 mm observations in \cite{Akeson_2014}.
\cite{Tang_2014} found extremely low $\alpha_{\mbox{\scriptsize 0.85 mm-6 cm}}$ $\sim 1.62 $ for both sources.

\section{Observing loop and target source group}\label{appendix:group}
Figure \ref{fig:target_group} shows the 6 target groups in two groupings for $<310$ GHz tracks and $>310$ GHz tracks. The arrows indicate the slewing path in each observing group. 

\section{Issues in the observations}\label{appendix:obs_issue}
When inspecting the data taken in 400 GHz-5 track on September 3, we found the amplitude of gain
calibrator J0510+180 dropped significantly in 400 GHz receiver probably due to its spectral index. 
The poor SNRs at $\sim$400 GHz make the phase of J0510+180 scatter over a larger range than that of 3C84. 
Therefore, we included 3C84 in the gain phase and amplitude calibration to enhance the SNRs of calibrators. 
In 400 GHz-4 and 400-GHz-6, we added 3C84 scan after every J0510+180 scan to solve the poor SNRs problem. 

In 270 GHz-3 track, on November 20, our target sources were close to the Moon during observation, while the gain calibrators are relatively far from it. 
Some of the targets are blocked by the Moon in certain UTC hours, and it is possible that the sidelobe of the antennas might response to the flux from the Moon.
We decided the limit angular separation of 1 degree from the Moon in which the data should be excluded and flagged out CI Tau scan around UTC time 12:00. 

\section{Gain phase self-calibration solutions}\label{appendix:selcal}
Figure \ref{fig:selcal_sol} shows the gain phase self-calibration solutions of track 400 GHz-5 at 337 GHz. The solutions were derived based on the procedure described in Section \ref{sub:selcal}. 

\section{Image profiles}\label{appendix:image}
\begin{figure*}[]
    \centering
    \includegraphics[width=15cm]{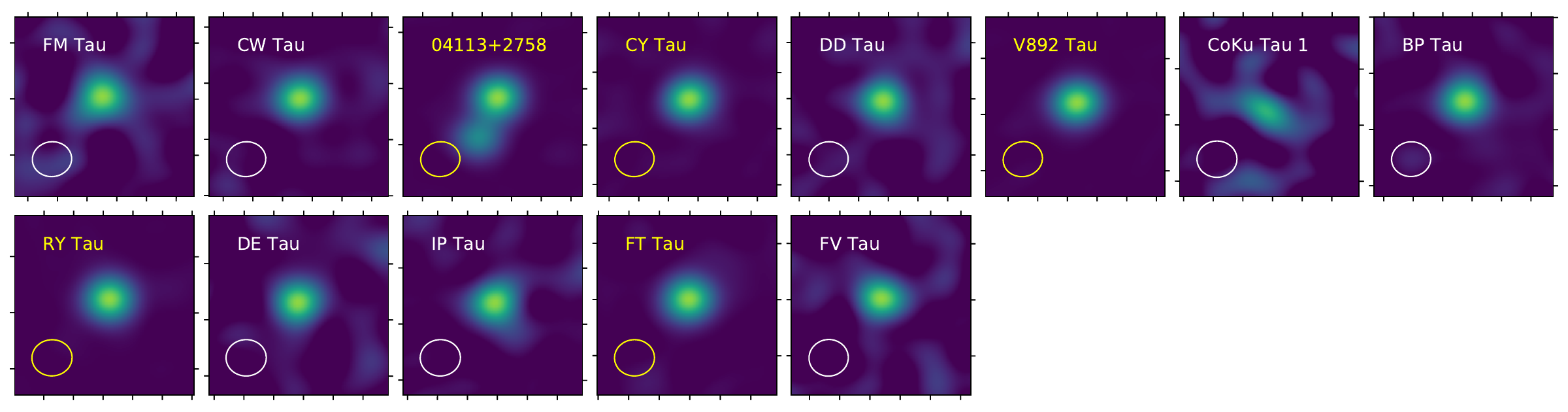}
    \includegraphics[width=15cm]{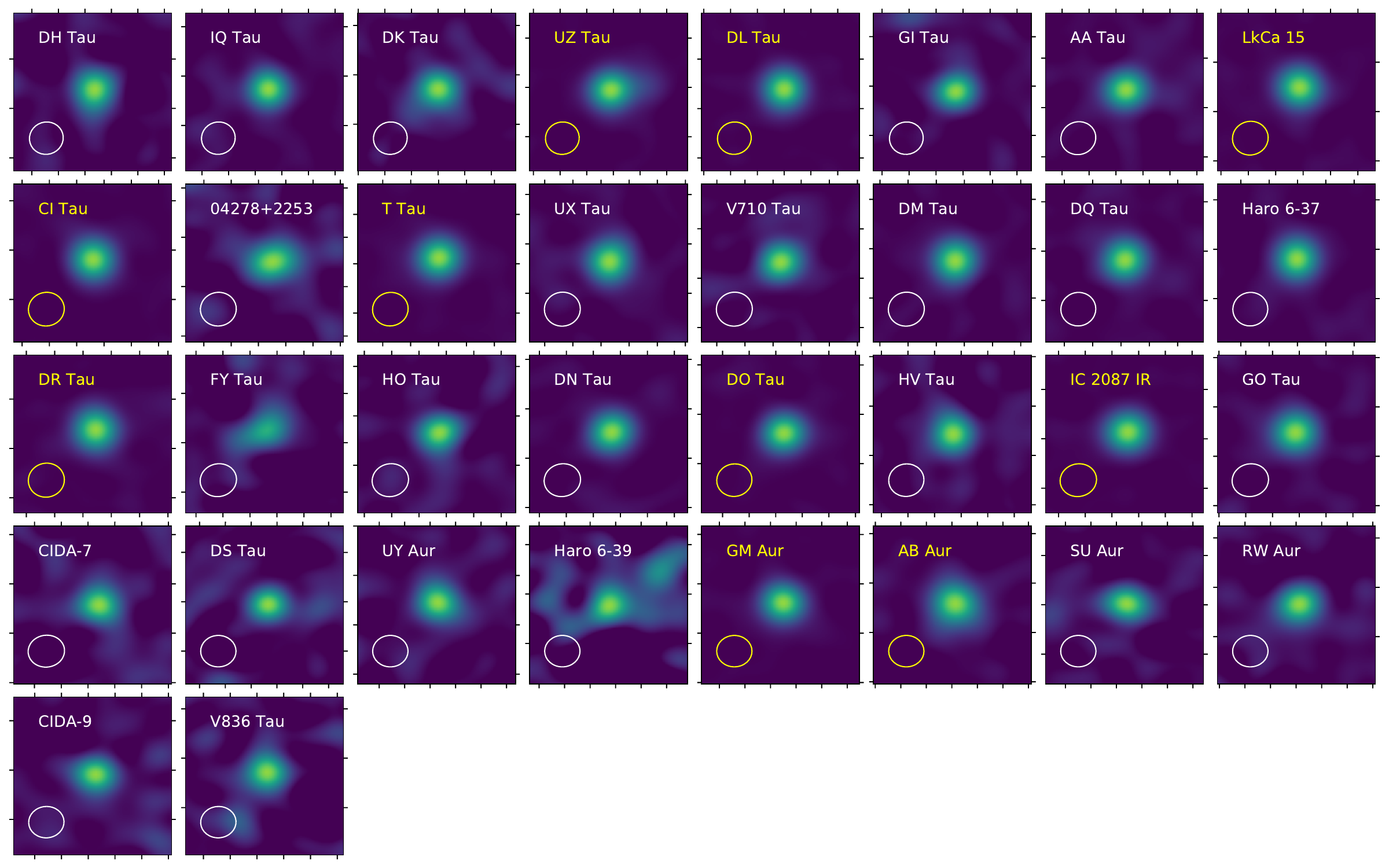}
    \caption{
    Similar to Figure \ref{fig:347GHz_image} but for the observations at 209 GHz.
    }
    \label{fig:209GHz_image}
\end{figure*}

\begin{figure*}[]
    \centering
    \includegraphics[width=15cm]{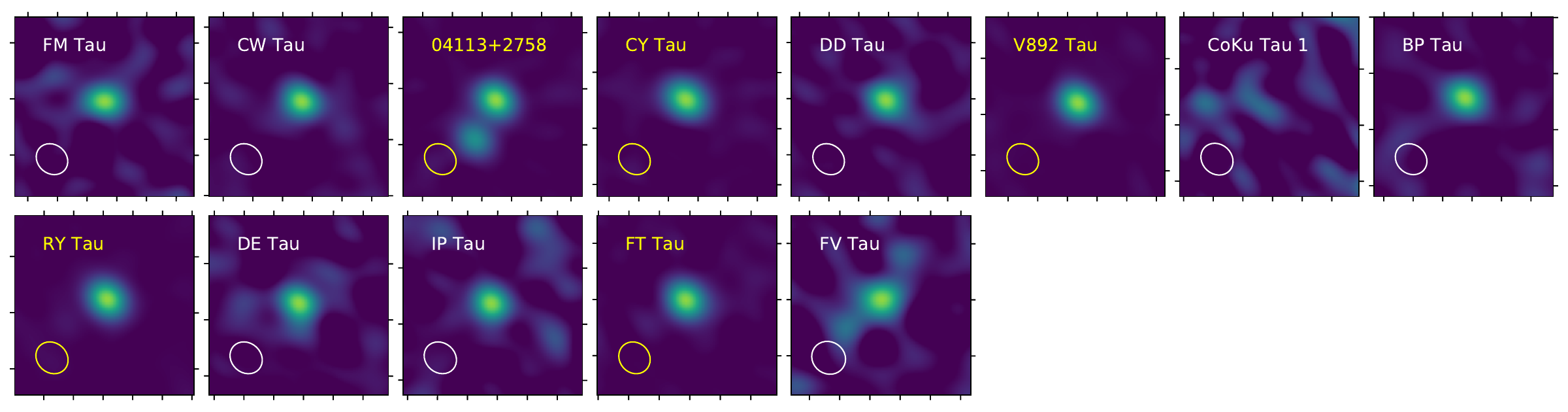}
    \includegraphics[width=15cm]{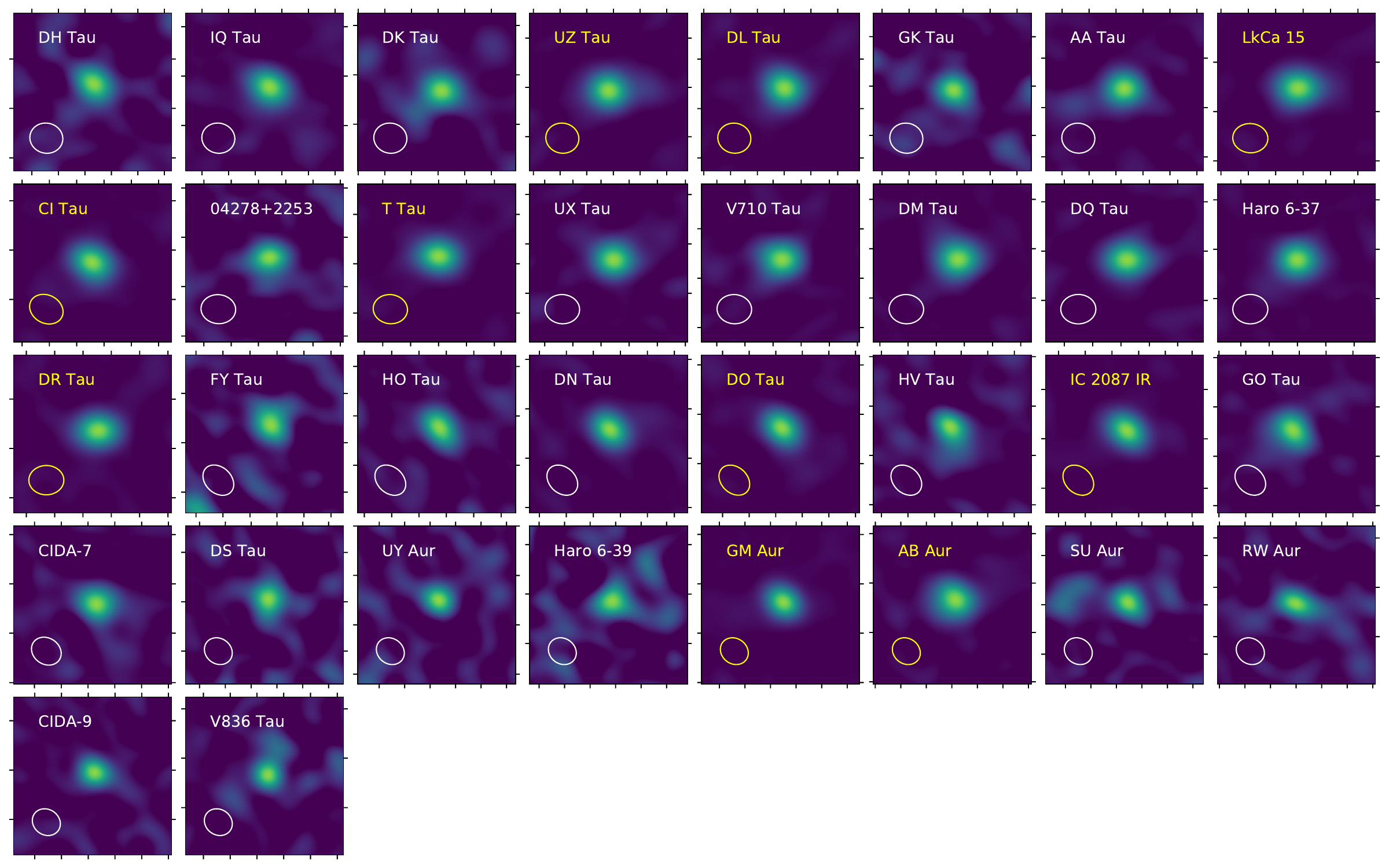}
    \caption{
    Similar to Figure \ref{fig:347GHz_image} but for the observations at 270 GHz.
    }
    \label{fig:270GHz_image}
\end{figure*}

\begin{figure*}[]
    \centering
    \includegraphics[width=15cm]{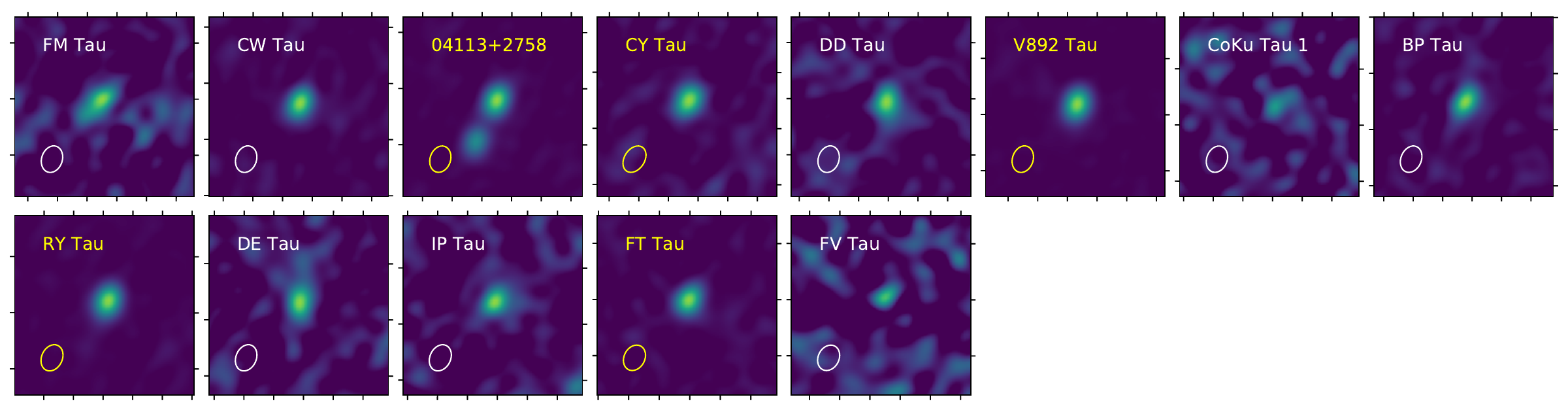}
    \includegraphics[width=15cm]{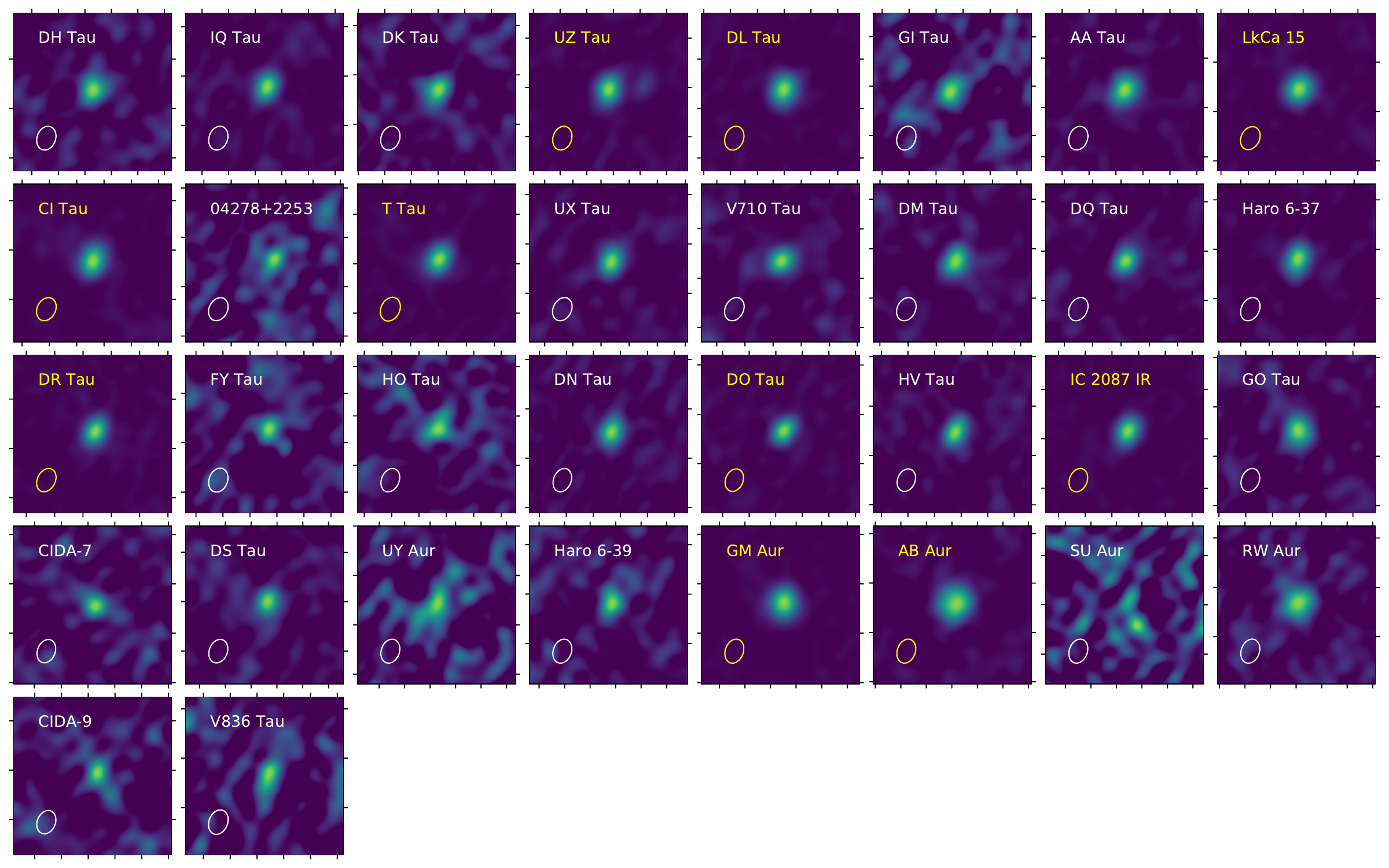}
    \caption{
    Similar to Figure \ref{fig:347GHz_image} but for the observations at 407.5 GHz.
    }
    \label{fig:407.5GHz_image}
\end{figure*}

Figure \ref{fig:209GHz_image}--\ref{fig:407.5GHz_image} show the 209 GHz, 270 GHz and 407.5 GHz continuum images of the 47 Class II sources. 
Table \ref{tab:beam_size_rms} summarizes the achieved synthesized beams and rms noises at the representative frequencies for each selected target sources.


\bibliography{main}{}
\bibliographystyle{aasjournal}

\begin{deluxetable*}{ccccccccccccccc}
\tabletypesize{\scriptsize}
\tablecolumns{15}
\tablewidth{0pt}
\tablecaption{Imaging profiles\label{tab:beam_size_rms}}
\tablehead{ 
\colhead{Source} & \colhead{RA} & \colhead{Dec} & \colhead{$\theta_{\mbox{\tiny 209 GHz}}$} & \colhead{$\theta_{\mbox{\tiny 238 GHz}}$} & \colhead{$\theta_{\mbox{\tiny 270 GHz}}$} & \colhead{$\theta_{\mbox{\tiny 302 GHz}}$} & \colhead{$\theta_{\mbox{\tiny 347 GHz}}$} & \colhead{$\theta_{\mbox{\tiny 407.5 GHz}}$} & \colhead{$\sigma_{\mbox{\tiny 209 GHz}}$} & \colhead{$\sigma_{\mbox{\tiny 238 GHz}}$} & \colhead{$\sigma_{\mbox{\tiny 270 GHz}}$} & \colhead{$\sigma_{\mbox{\tiny 302 GHz}}$} & \colhead{$\sigma_{\mbox{\tiny 347 GHz}}$} & \colhead{$\sigma_{\mbox{\tiny 407.5 GHz}}$} \\
\colhead{} & \colhead{(hh:mm:ss)} & \colhead{(dd$^{\circ}$mm$'$ss$''$)} & \colhead{(arcsec)} & \colhead{(arcsec)} & \colhead{(arcsec)} & \colhead{(arcsec)} & \colhead{(arcsec)} & \colhead{(arcsec)} & \colhead{(mJy)} & \colhead{(mJy)} & \colhead{(mJy)} & \colhead{(mJy)} & \colhead{(mJy)} & \colhead{(mJy)} 
}
\startdata 
04113+2758 & 04:14:26.31 & $+28^{\circ}06'1\farcs8$ & $3.5\times3.1$ & $3.0\times2.9$ & $3.0\times2.6$ & $2.3\times2.1$ & $2.9\times2.4$ & $3.1\times2.6$ & 7.0 & 8.5 & 14.8 & 17.9 & 16.8 & 35.7\\
04278+2253 & 04:30:50.3 & $+23^{\circ}00'8\farcs29$ & $3.6\times3.4$ & $3.2\times3.1$ & $3.5\times2.9$ & $3.0\times2.3$ & $2.5\times2.3$ & $2.4\times1.9$ & 0.9 & 1.0 & 1.6 & 2.0 & 2.1 & 5.3\\
AA Tau & 04:34:55.42 & $+24^{\circ}28'52\farcs59$ & $3.6\times3.3$ & $3.2\times3.0$ & $3.4\times3.0$ & $2.9\times2.3$ & $2.5\times2.2$ & $2.5\times1.9$ & 3.0 & 3.2 & 6.0 & 7.5 & 5.4 & 12.7\\
AB Aur & 04:55:45.84 & $+30^{\circ}33'3\farcs71$ & $3.6\times3.2$ & $3.0\times3.0$ & $3.0\times2.6$ & $2.3\times2.1$ & $2.6\times2.2$ & $2.5\times1.8$ & 2.2 & 2.9 & 5.4 & 6.8 & 6.9 & 16.7\\
BP Tau & 04:19:15.84 & $+29^{\circ}06'26\farcs2$ & $3.5\times3.1$ & $3.0\times2.9$ & $3.0\times2.6$ & $2.3\times2.1$ & $2.9\times2.4$ & $3.2\times2.6$ & 2.4 & 2.7 & 4.8 & 5.6 & 7.4 & 20.1\\
CIDA-7 & 04:42:21.05 & $+25^{\circ}20'34\farcs05$ & $3.7\times3.2$ & $3.2\times3.0$ & $3.2\times2.7$ & $2.5\times2.2$ & $2.5\times2.3$ & $2.5\times1.8$ & 0.9 & 0.9 & 1.9 & 2.1 & 2.4 & 6.3\\
CIDA-9 & 05:05:22.83 & $+25^{\circ}31'30\farcs46$ & $3.6\times3.2$ & $3.1\times3.0$ & $3.0\times2.6$ & $2.3\times2.1$ & $2.5\times2.3$ & $2.4\times1.8$ & 1.8 & 2.1 & 4.3 & 4.4 & 4.7 & 13.1\\
CI Tau & 04:33:52.02 & $+22^{\circ}50'29\farcs68$ & $3.6\times3.4$ & $3.3\times3.1$ & $3.5\times2.8$ & $2.8\times2.4$ & $2.5\times2.3$ & $2.4\times1.9$ & 3.0 & 3.8 & 9.2 & 10.8 & 8.3 & 17.6\\
CW Tau & 04:14:17.01 & $+28^{\circ}10'57\farcs18$ & $3.5\times3.1$ & $3.0\times2.9$ & $3.0\times2.6$ & $2.3\times2.1$ & $2.9\times2.4$ & $3.1\times2.6$ & 2.4 & 2.8 & 5.1 & 5.7 & 6.2 & 16.4\\
CY Tau & 04:17:33.74 & $+28^{\circ}20'46\farcs23$ & $3.5\times3.1$ & $3.0\times2.9$ & $3.0\times2.6$ & $2.3\times2.1$ & $2.8\times2.4$ & $3.1\times2.5$ & 3.0 & 3.2 & 6.2 & 6.7 & 9.0 & 27.2\\
CoKu Tau 1 & 04:18:51.47 & $+28^{\circ}20'26\farcs55$ & $3.6\times3.2$ & $3.2\times3.0$ & $3.0\times2.6$ & $2.3\times2.1$ & $2.5\times2.2$ & $2.4\times1.8$ & 1.2 & 1.2 & 0.0 & 1.9 & 4.2 & 8.9\\
DD Tau & 04:18:31.15 & $+28^{\circ}16'28\farcs05$ & $3.5\times3.1$ & $3.0\times2.9$ & $3.0\times2.6$ & $2.3\times2.1$ & $2.9\times2.4$ & $3.1\times2.6$ & 1.1 & 1.3 & 2.2 & 2.4 & 4.0 & 12.1\\
DE Tau & 04:21:55.65 & $+27^{\circ}55'5\farcs63$ & $3.6\times3.2$ & $3.2\times3.0$ & $3.1\times2.6$ & $2.3\times2.1$ & $2.9\times2.4$ & $3.1\times2.6$ & 2.1 & 2.6 & 4.1 & 4.7 & 7.4 & 18.1\\
DH Tau & 04:29:41.56 & $+26^{\circ}32'57\farcs62$ & $3.4\times3.3$ & $3.1\times3.0$ & $3.4\times3.1$ & $2.9\times2.3$ & $2.5\times2.2$ & $2.5\times1.9$ & 1.9 & 2.0 & 4.1 & 4.4 & 3.2 & 8.2\\
DK Tau & 04:30:44.26 & $+26^{\circ}01'24\farcs25$ & $3.4\times3.3$ & $3.1\times3.0$ & $3.4\times3.1$ & $2.9\times2.3$ & $2.5\times2.2$ & $2.5\times1.9$ & 2.4 & 2.7 & 5.3 & 6.1 & 4.7 & 11.7\\
DL Tau & 04:33:39.09 & $+25^{\circ}20'37\farcs62$ & $3.4\times3.3$ & $3.1\times3.0$ & $3.4\times3.0$ & $2.9\times2.3$ & $2.5\times2.2$ & $2.5\times1.9$ & 3.6 & 4.3 & 9.9 & 12.8 & 7.6 & 15.7\\
DM Tau & 04:33:48.75 & $+18^{\circ}10'9\farcs44$ & $3.7\times3.4$ & $3.3\times3.1$ & $3.5\times3.0$ & $3.0\times2.3$ & $2.5\times2.3$ & $2.5\times1.8$ & 2.6 & 3.2 & 6.8 & 7.3 & 6.6 & 14.5\\
DN Tau & 04:35:27.39 & $+24^{\circ}14'58\farcs38$ & $3.7\times3.3$ & $3.2\times3.1$ & $3.5\times2.6$ & $3.0\times2.3$ & $2.5\times2.3$ & $2.5\times1.8$ & 2.4 & 2.9 & 6.1 & 7.1 & 5.6 & 14.3\\
DO Tau & 04:38:28.6 & $+26^{\circ}10'48\farcs92$ & $3.6\times3.3$ & $3.2\times3.0$ & $3.5\times2.6$ & $3.0\times2.3$ & $2.5\times2.2$ & $2.4\times1.8$ & 2.5 & 2.9 & 7.0 & 8.3 & 5.9 & 14.8\\
DQ Tau & 04:46:53.06 & $+16^{\circ}59'59\farcs72$ & $3.5\times3.4$ & $3.2\times3.0$ & $3.6\times3.0$ & $3.1\times2.3$ & $2.5\times2.2$ & $2.5\times1.8$ & 2.8 & 2.9 & 6.2 & 6.6 & 5.3 & 11.9\\
DR Tau & 04:47:6.22 & $+16^{\circ}58'42\farcs44$ & $3.7\times3.4$ & $3.3\times3.1$ & $3.6\times3.0$ & $3.1\times2.2$ & $2.5\times2.2$ & $2.5\times1.8$ & 3.2 & 3.4 & 7.1 & 8.3 & 6.4 & 14.3\\
DS Tau & 04:47:48.61 & $+29^{\circ}25'10\farcs71$ & $3.6\times3.2$ & $3.0\times3.0$ & $3.0\times2.6$ & $2.3\times2.1$ & $2.6\times2.3$ & $2.5\times1.8$ & 1.7 & 2.0 & 3.9 & 4.5 & 2.4 & 6.4\\
FM Tau & 04:14:13.59 & $+28^{\circ}12'48\farcs56$ & $3.5\times3.1$ & $3.0\times2.9$ & $3.0\times2.6$ & $2.3\times2.1$ & $2.9\times2.4$ & $3.1\times2.6$ & 1.0 & 1.1 & 1.8 & 1.9 & 2.9 & 7.5\\
FT Tau & 04:23:39.2 & $+24^{\circ}56'13\farcs75$ & $3.6\times3.2$ & $3.1\times3.0$ & $3.0\times2.7$ & $2.4\times2.1$ & $2.8\times2.5$ & $3.1\times2.6$ & 2.6 & 2.9 & 5.7 & 6.9 & 7.1 & 18.5\\
FV Tau & 04:26:53.52 & $+26^{\circ}06'53\farcs53$ & $3.5\times3.3$ & $3.1\times3.0$ & $3.1\times2.8$ & $2.6\times2.2$ & $2.8\times2.5$ & $3.1\times2.6$ & 1.2 & 1.3 & 2.6 & 2.9 & 4.3 & 10.8\\
FY Tau & 04:32:30.59 & $+24^{\circ}19'56\farcs68$ & $3.7\times3.3$ & $3.2\times3.1$ & $3.5\times2.6$ & $3.0\times2.3$ & $2.5\times2.2$ & $2.5\times1.9$ & 1.2 & 1.3 & 2.3 & 2.9 & 2.7 & 6.8\\
GI Tau & 04:33:34.06 & $+24^{\circ}21'16\farcs56$ & $3.4\times3.3$ & $3.1\times3.0$ & $3.4\times3.0$ & $2.9\times2.3$ & $2.5\times2.2$ & $2.5\times1.9$ & 1.0 & 1.1 & 2.2 & 2.5 & 1.9 & 5.1\\
GM Aur & 04:55:10.99 & $+30^{\circ}21'58\farcs83$ & $3.6\times3.2$ & $3.0\times3.0$ & $3.0\times2.6$ & $2.3\times2.1$ & $2.6\times2.2$ & $2.5\times1.8$ & 3.1 & 3.8 & 8.0 & 10.2 & 8.6 & 20.0\\
GO Tau & 04:43:3.08 & $+25^{\circ}20'18\farcs29$ & $3.7\times3.3$ & $3.2\times3.1$ & $3.5\times2.6$ & $3.0\times2.3$ & $2.5\times2.3$ & $2.4\times1.8$ & 2.4 & 2.7 & 5.8 & 6.3 & 5.6 & 14.7\\
HO Tau & 04:35:20.23 & $+22^{\circ}32'14\farcs24$ & $3.7\times3.3$ & $3.2\times3.1$ & $3.5\times2.6$ & $3.0\times2.3$ & $2.6\times2.3$ & $2.5\times1.8$ & 1.1 & 1.3 & 2.4 & 3.0 & 2.9 & 7.4\\
HV Tau & 04:38:35.51 & $+26^{\circ}10'41\farcs1$ & $3.6\times3.3$ & $3.2\times3.0$ & $3.5\times2.6$ & $3.0\times2.3$ & $2.5\times2.2$ & $2.4\times1.8$ & 2.0 & 2.2 & 4.9 & 6.2 & 3.0 & 8.0\\
Haro 6-37 & 04:46:59.09 & $+17^{\circ}02'39\farcs56$ & $3.5\times3.4$ & $3.2\times3.0$ & $3.6\times3.0$ & $3.1\times2.3$ & $2.5\times2.2$ & $2.5\times1.8$ & 2.4 & 2.8 & 5.6 & 7.1 & 5.6 & 13.6\\
Haro 6-39 & 04:52:9.72 & $+30^{\circ}37'44\farcs99$ & $3.6\times3.2$ & $3.0\times3.0$ & $3.0\times2.6$ & $2.3\times2.1$ & $2.6\times2.3$ & $2.5\times1.8$ & 1.5 & 1.7 & 2.8 & 3.4 & 1.9 & 5.0\\
IC 2087 IR & 04:39:55.75 & $+25^{\circ}45'1\farcs48$ & $3.7\times3.3$ & $3.2\times3.1$ & $3.5\times2.6$ & $3.0\times2.3$ & $2.5\times2.3$ & $2.4\times1.8$ & 3.7 & 4.5 & 9.9 & 12.0 & 9.1 & 21.2\\
IP Tau & 04:24:57.08 & $+27^{\circ}11'56\farcs08$ & $3.6\times3.2$ & $3.2\times3.0$ & $3.1\times2.6$ & $2.3\times2.1$ & $2.8\times2.4$ & $3.1\times2.6$ & 1.1 & 1.3 & 2.2 & 2.5 & 3.2 & 8.6\\
IQ Tau & 04:29:51.57 & $+26^{\circ}06'44\farcs25$ & $3.4\times3.3$ & $3.1\times3.0$ & $3.4\times3.1$ & $2.9\times2.3$ & $2.5\times2.2$ & $2.5\times1.9$ & 2.7 & 3.0 & 7.0 & 7.3 & 5.3 & 12.2\\
LkCa 15 & 04:39:17.81 & $+22^{\circ}21'2\farcs95$ & $3.6\times3.4$ & $3.3\times3.1$ & $3.6\times3.0$ & $3.1\times2.3$ & $2.5\times2.3$ & $2.5\times1.9$ & 3.1 & 3.7 & 7.7 & 9.5 & 8.0 & 17.1\\
RW Aur & 05:07:49.57 & $+30^{\circ}24'4\farcs61$ & $3.6\times3.2$ & $3.0\times3.0$ & $3.0\times2.6$ & $2.3\times2.1$ & $2.6\times2.2$ & $2.5\times1.8$ & 2.0 & 2.4 & 4.9 & 5.1 & 4.9 & 12.6\\
RY Tau & 04:21:57.42 & $+28^{\circ}26'34\farcs93$ & $3.6\times3.2$ & $3.2\times3.0$ & $3.1\times2.6$ & $2.3\times2.1$ & $2.9\times2.4$ & $3.1\times2.6$ & 3.6 & 4.6 & 8.4 & 11.0 & 12.4 & 31.9\\
SU Aur & 04:55:59.41 & $+30^{\circ}34'1\farcs01$ & $3.6\times3.2$ & $3.0\times3.0$ & $3.0\times2.6$ & $2.3\times2.1$ & $2.6\times2.2$ & $2.5\times1.8$ & 1.7 & 2.1 & 3.7 & 4.2 & 4.6 & 12.0\\
T Tau & 04:21:59.45 & $+19^{\circ}32'5\farcs94$ & $3.6\times3.4$ & $3.2\times3.0$ & $3.5\times2.9$ & $3.0\times2.3$ & $2.5\times2.4$ & $2.5\times1.9$ & 4.0 & 4.7 & 9.2 & 11.5 & 8.8 & 18.1\\
UX Tau & 04:30:4.01 & $+18^{\circ}13'49\farcs04$ & $3.6\times3.4$ & $3.3\times3.1$ & $3.5\times3.0$ & $3.0\times2.3$ & $2.5\times2.3$ & $2.5\times1.8$ & 2.5 & 2.8 & 5.3 & 6.3 & 5.6 & 14.3\\
UY Aur & 04:51:47.38 & $+30^{\circ}47'12\farcs65$ & $3.6\times3.2$ & $3.0\times3.0$ & $3.0\times2.6$ & $2.3\times2.1$ & $2.6\times2.3$ & $2.5\times1.8$ & 1.8 & 2.0 & 4.2 & 4.6 & 4.5 & 11.7\\
UZ Tau & 04:32:43.08 & $+25^{\circ}52'30\farcs48$ & $3.4\times3.3$ & $3.1\times3.0$ & $3.4\times3.0$ & $2.9\times2.3$ & $2.5\times2.2$ & $2.5\times1.9$ & 3.7 & 4.3 & 9.2 & 10.5 & 6.5 & 14.4\\
V710 Tau & 04:31:57.8 & $+18^{\circ}21'37\farcs45$ & $3.7\times3.4$ & $3.3\times3.1$ & $3.5\times3.0$ & $3.0\times2.3$ & $2.5\times2.3$ & $2.5\times1.8$ & 2.4 & 2.7 & 5.6 & 6.3 & 5.6 & 15.3\\
V836 Tau & 05:03:6.62 & $+25^{\circ}23'19\farcs09$ & $3.6\times3.2$ & $3.1\times3.0$ & $3.0\times2.6$ & $2.3\times2.1$ & $2.6\times2.3$ & $2.6\times1.9$ & 2.0 & 2.3 & 4.4 & 5.2 & 4.9 & 13.4\\
V892 Tau & 04:18:40.61 & $+28^{\circ}19'14\farcs99$ & $3.5\times3.1$ & $3.1\times2.9$ & $3.0\times2.6$ & $2.3\times2.1$ & $2.9\times2.4$ & $3.1\times2.6$ & 4.4 & 5.4 & 10.9 & 14.0 & 13.3 & 32.6\\
\enddata
\end{deluxetable*}

\begin{longrotatetable}
\begin{deluxetable*}{ccccccccccccc}
\tabletypesize{\scriptsize}
\tablecolumns{13}
\tablewidth{0pt}
\tablecaption{Flux densities\label{tab:flux}}
\tablehead{ 
\colhead{Source} & \colhead{$F_{\mbox{\tiny 199 GHz}}$} & \colhead{$F_{\mbox{\tiny 219 GHz}}$} & \colhead{$F_{\mbox{\tiny 228 GHz}}$} & \colhead{$F_{\mbox{\tiny 248 GHz}}$} & \colhead{$F_{\mbox{\tiny 262 GHz}}$} & \colhead{$F_{\mbox{\tiny 278 GHz}}$} & \colhead{$F_{\mbox{\tiny 294 GHz}}$} & \colhead{$F_{\mbox{\tiny 310 GHz}}$} & \colhead{$F_{\mbox{\tiny 337 GHz}}$} & \colhead{$F_{\mbox{\tiny 357 GHz}}$} & \colhead{$F_{\mbox{\tiny 399.5 GHz}}$} & \colhead{$F_{\mbox{\tiny 415.5 GHz}}$} \\
\colhead{} & \colhead{(mJy)} & \colhead{(mJy)} & \colhead{(mJy)} & \colhead{(mJy)} & \colhead{(mJy)} & \colhead{(mJy)} & \colhead{(mJy)} & \colhead{(mJy)} & \colhead{(mJy)} & \colhead{(mJy)} & \colhead{(mJy)} & \colhead{(mJy)}
}
\startdata 
04113+2758A & $196.8\pm19.8$ & $239.2\pm24.0$ & $245.9\pm24.7$ & $289.1\pm29.0$ & $344.8\pm34.7$ & $380.3\pm38.4$ & $418.2\pm42.1$ & $482.8\pm48.6$ & $519.2\pm52.0$ & $591.2\pm59.3$ & $739.8\pm74.6$ & $812.8\pm83.0$ \\
04113+2758B & $108.2\pm11.0$ & $136.9\pm13.9$ & $149.1\pm15.1$ & $172.1\pm17.4$ & $203.7\pm20.7$ & $227.6\pm23.4$ & $244.5\pm24.9$ & $266.2\pm27.2$ & $310.9\pm31.2$ & $336.5\pm34.0$ & $443.6\pm45.3$ & $453.7\pm48.3$ \\
04278+2253 & $9.5\pm1.2$ & $8.7\pm1.3$ & $10.6\pm1.3$ & $15.4\pm1.8$ & $11.0\pm1.9$ & $17.8\pm2.9$ & $18.3\pm2.7$ & $19.7\pm3.0$ & $26.2\pm3.0$ & $30.4\pm3.8$ & $35.4\pm5.3$ & $40.4\pm8.0$ \\
AA Tau & $58.3\pm6.8$ & $71.7\pm8.1$ & $64.7\pm7.7$ & $83.0\pm9.4$ & $94.9\pm10.7$ & $110.0\pm17.4$ & $121.1\pm15.5$ & $151.5\pm19.6$ & $186.4\pm19.3$ & $205.9\pm22.0$ & $281.9\pm32.5$ & $345.0\pm43.6$ \\
AB Aur & $68.3\pm8.0$ & $83.1\pm9.4$ & $94.9\pm10.4$ & $127.8\pm13.8$ & $148.1\pm17.2$ & $178.2\pm22.7$ & $217.0\pm25.6$ & $244.8\pm27.7$ & $322.9\pm32.9$ & $378.6\pm39.1$ & $581.3\pm62.1$ & $561.4\pm66.5$ \\
BP Tau & $36.8\pm4.3$ & $46.6\pm5.2$ & $50.2\pm5.5$ & $55.3\pm6.1$ & $67.3\pm7.7$ & $66.8\pm8.7$ & $92.5\pm10.4$ & $100.6\pm11.5$ & $108.9\pm11.3$ & $101.4\pm11.2$ & $168.3\pm19.1$ & $194.7\pm25.1$ \\
CIDA-7 & $8.4\pm1.2$ & $9.9\pm1.3$ & $13.3\pm1.6$ & $13.5\pm1.6$ & $17.9\pm2.4$ & $16.8\pm2.8$ & $20.7\pm2.9$ & $22.6\pm3.2$ & $19.4\pm2.5$ & $27.3\pm3.7$ & $45.6\pm6.8$ & $56.6\pm10.5$ \\
CIDA-9 & $27.3\pm3.2$ & $31.9\pm3.7$ & $33.2\pm3.8$ & $41.7\pm4.6$ & $37.2\pm4.9$ & $46.9\pm6.6$ & $57.3\pm7.0$ & $64.4\pm8.1$ & $72.4\pm7.9$ & $79.6\pm9.4$ & $90.1\pm13.4$ & $123.6\pm20.9$ \\
CI Tau & $114.0\pm11.8$ & $142.1\pm14.7$ & $156.0\pm16.0$ & $198.0\pm20.2$ & $237.2\pm25.1$ & $260.5\pm28.7$ & $303.9\pm32.2$ & $320.9\pm34.1$ & $390.8\pm39.4$ & $485.1\pm49.4$ & $605.3\pm62.6$ & $679.4\pm73.4$ \\
CW Tau & $45.1\pm5.0$ & $57.4\pm6.2$ & $57.9\pm6.2$ & $70.8\pm7.5$ & $82.4\pm9.1$ & $94.9\pm10.9$ & $123.3\pm13.2$ & $133.5\pm14.5$ & $145.4\pm14.7$ & $170.1\pm17.4$ & $211.4\pm22.3$ & $220.1\pm25.4$ \\
CY Tau & $95.5\pm9.8$ & $106.0\pm10.9$ & $108.8\pm11.1$ & $128.7\pm13.1$ & $161.8\pm16.6$ & $160.8\pm16.9$ & $171.8\pm17.8$ & $188.8\pm19.7$ & $205.4\pm20.9$ & $222.8\pm23.0$ & $269.1\pm29.6$ & $302.7\pm36.9$ \\
CoKu Tau 1 &  & $5.8\pm1.3$ &  &  &  &  &  &  &  & $16.5\pm3.3$ &  &  \\
DD Tau & $14.1\pm1.8$ & $15.2\pm1.9$ & $14.8\pm1.9$ & $20.2\pm2.4$ & $22.3\pm2.9$ & $28.8\pm3.9$ & $31.8\pm4.0$ & $33.4\pm4.3$ & $28.6\pm3.3$ & $35.8\pm4.5$ & $69.9\pm8.8$ & $56.5\pm11.0$ \\
DE Tau & $25.0\pm3.2$ & $24.9\pm3.4$ & $30.5\pm3.8$ & $39.0\pm4.6$ & $41.9\pm5.6$ & $44.4\pm6.9$ & $42.3\pm6.2$ & $66.1\pm8.5$ & $59.0\pm6.6$ & $76.3\pm9.0$ & $95.6\pm13.1$ & $104.8\pm18.9$ \\
DH Tau & $20.0\pm2.6$ & $28.4\pm3.3$ & $26.6\pm3.1$ & $28.2\pm3.4$ & $24.2\pm4.1$ & $53.1\pm7.2$ & $38.2\pm5.7$ & $46.8\pm6.7$ & $63.0\pm6.6$ & $66.3\pm7.4$ & $86.6\pm10.6$ & $87.3\pm14.2$ \\
DK Tau & $29.7\pm3.7$ & $33.5\pm4.2$ & $31.1\pm3.9$ & $38.6\pm4.7$ & $52.2\pm6.9$ & $48.0\pm8.2$ & $51.3\pm7.9$ & $56.7\pm9.0$ & $77.0\pm8.3$ & $81.9\pm9.4$ & $107.7\pm13.9$ & $95.4\pm18.3$ \\
DL Tau & $138.2\pm14.4$ & $176.3\pm18.2$ & $171.4\pm17.6$ & $211.7\pm21.6$ & $254.4\pm26.5$ & $291.5\pm31.0$ & $344.9\pm36.1$ & $409.6\pm42.6$ & $431.7\pm43.4$ & $519.7\pm52.6$ & $622.0\pm63.8$ & $723.1\pm76.4$ \\
DM Tau & $71.6\pm7.9$ & $90.0\pm9.6$ & $87.6\pm9.2$ & $116.1\pm12.4$ & $155.1\pm17.1$ & $171.7\pm20.0$ & $191.3\pm21.2$ & $202.7\pm22.8$ & $227.8\pm23.5$ & $266.8\pm28.4$ & $297.9\pm34.0$ & $297.7\pm41.9$ \\
DN Tau & $63.3\pm6.7$ & $78.8\pm8.2$ & $80.6\pm8.3$ & $91.9\pm9.5$ & $124.9\pm13.2$ & $116.3\pm13.1$ & $147.5\pm15.7$ & $164.7\pm17.6$ & $176.0\pm17.9$ & $188.8\pm19.6$ & $230.6\pm25.3$ & $220.5\pm28.1$ \\
DO Tau & $98.0\pm10.0$ & $127.1\pm12.9$ & $122.3\pm12.4$ & $138.0\pm14.0$ & $160.4\pm16.6$ & $183.9\pm19.3$ & $231.1\pm23.8$ & $243.1\pm25.1$ & $272.0\pm27.4$ & $293.7\pm29.8$ & $389.8\pm40.2$ & $452.9\pm48.3$ \\
DQ Tau & $60.1\pm6.5$ & $72.5\pm7.7$ & $74.4\pm7.9$ & $82.6\pm8.7$ & $112.3\pm11.9$ & $88.4\pm10.6$ & $132.4\pm14.3$ & $134.1\pm14.8$ & $146.4\pm15.6$ & $176.1\pm19.4$ & $232.6\pm29.2$ & $296.6\pm40.8$ \\
DR Tau & $104.3\pm10.7$ & $136.1\pm13.8$ & $130.5\pm13.2$ & $159.5\pm16.1$ & $200.4\pm20.4$ & $208.2\pm21.6$ & $243.5\pm24.9$ & $256.1\pm26.3$ & $307.4\pm30.9$ & $361.1\pm36.4$ & $473.5\pm48.2$ & $533.4\pm55.6$ \\
DS Tau & $11.9\pm2.0$ & $20.0\pm2.7$ & $19.7\pm2.7$ & $23.2\pm3.0$ & $28.3\pm4.4$ & $23.7\pm5.4$ & $33.7\pm5.4$ & $39.5\pm6.2$ & $43.2\pm4.6$ & $48.6\pm5.4$ & $68.3\pm8.3$ & $69.4\pm10.5$ \\
FM Tau & $7.5\pm1.1$ & $9.9\pm1.3$ & $9.9\pm1.3$ & $10.7\pm1.4$ & $18.0\pm2.3$ &  & $15.6\pm2.4$ & $20.3\pm2.9$ & $20.8\pm2.4$ & $22.3\pm3.0$ & $38.9\pm5.4$ & $41.4\pm7.9$ \\
FT Tau & $83.8\pm8.6$ & $88.2\pm9.1$ & $96.8\pm9.9$ & $112.3\pm11.5$ & $128.6\pm13.4$ & $151.9\pm16.3$ & $155.9\pm16.4$ & $176.0\pm18.6$ & $195.3\pm19.8$ & $238.6\pm24.3$ & $274.2\pm28.9$ & $377.3\pm41.2$ \\
FV Tau & $9.4\pm1.4$ & $9.9\pm1.6$ & $10.9\pm1.6$ & $11.5\pm1.7$ & $18.9\pm2.9$ & $19.6\pm3.7$ & $23.2\pm3.6$ & $19.1\pm3.7$ & $26.1\pm3.2$ & $33.7\pm4.4$ & $29.2\pm6.1$ & $37.7\pm10.4$ \\
FY Tau & $7.7\pm1.3$ & $9.7\pm1.5$ & $8.1\pm1.3$ & $10.3\pm1.6$ & $13.4\pm2.5$ & $18.2\pm3.5$ & $20.5\pm3.4$ & $31.6\pm4.5$ & $19.4\pm2.6$ &  & $33.8\pm6.2$ & $57.4\pm10.8$ \\
GI Tau & $12.0\pm1.5$ & $15.8\pm1.8$ & $13.8\pm1.7$ & $15.8\pm1.9$ & $15.5\pm2.3$ & $22.3\pm3.4$ & $25.4\pm3.4$ & $24.2\pm3.6$ & $20.7\pm2.4$ & $24.4\pm3.1$ & $35.9\pm5.1$ &  \\
GM Aur & $128.8\pm13.3$ & $163.2\pm16.7$ & $184.7\pm18.8$ & $228.1\pm23.1$ & $280.2\pm28.7$ & $315.9\pm32.9$ & $415.1\pm42.6$ & $487.7\pm49.9$ & $525.7\pm52.8$ & $601.9\pm60.7$ & $833.3\pm84.8$ & $1000.7\pm104.0$ \\
GO Tau & $49.1\pm6.1$ & $66.1\pm7.8$ & $71.6\pm8.2$ & $82.5\pm9.3$ & $95.0\pm12.4$ & $117.3\pm16.1$ & $103.9\pm13.2$ & $98.7\pm13.1$ & $161.9\pm17.0$ & $149.3\pm17.1$ & $203.3\pm26.5$ & $281.0\pm44.9$ \\
HO Tau & $15.5\pm1.9$ & $16.6\pm2.0$ & $19.1\pm2.2$ & $20.5\pm2.4$ & $24.9\pm3.2$ & $25.0\pm3.9$ & $40.8\pm4.9$ & $32.0\pm4.6$ & $44.4\pm4.8$ & $45.5\pm5.3$ & $49.3\pm7.5$ & $61.2\pm11.4$ \\
HV Tau & $28.4\pm3.4$ & $33.8\pm3.9$ & $30.0\pm3.6$ & $42.2\pm4.7$ & $54.4\pm6.8$ & $52.6\pm8.0$ & $73.4\pm9.2$ & $84.3\pm10.5$ & $88.9\pm9.1$ & $98.5\pm10.3$ & $133.8\pm14.5$ & $150.3\pm17.9$ \\
Haro 6-37 & $67.2\pm7.0$ & $92.7\pm9.5$ & $85.8\pm8.8$ & $101.1\pm10.3$ & $135.1\pm14.1$ & $142.3\pm15.4$ & $172.4\pm18.0$ & $184.7\pm19.4$ & $224.2\pm22.6$ & $249.3\pm25.4$ & $303.2\pm31.7$ & $321.8\pm35.8$ \\
Haro 6-39 & $7.9\pm1.6$ & $10.2\pm1.8$ & $12.8\pm2.0$ & $12.0\pm2.0$ & $17.6\pm3.3$ & $20.5\pm4.4$ & $14.9\pm3.7$ & $20.3\pm4.5$ & $25.2\pm2.8$ & $34.8\pm4.0$ & $39.8\pm5.6$ & $52.9\pm8.4$ \\
IC 2087 IR & $153.9\pm15.5$ & $196.1\pm19.7$ & $201.5\pm20.3$ & $247.2\pm24.8$ & $287.1\pm29.0$ & $312.6\pm31.8$ & $389.9\pm39.4$ & $426.0\pm43.0$ & $503.4\pm50.4$ & $562.5\pm56.5$ & $742.4\pm74.9$ & $781.5\pm80.1$ \\
IP Tau & $10.3\pm1.5$ & $14.4\pm1.9$ & $10.5\pm1.5$ & $13.0\pm1.8$ & $18.7\pm2.6$ & $25.3\pm3.7$ & $24.1\pm3.3$ & $23.6\pm3.6$ & $29.7\pm3.3$ & $36.4\pm4.2$ & $42.6\pm5.7$ & $43.3\pm8.3$ \\
IQ Tau & $52.1\pm5.7$ & $62.4\pm6.7$ & $65.9\pm7.0$ & $75.0\pm7.9$ & $97.6\pm10.8$ & $122.5\pm14.0$ & $129.8\pm14.2$ & $129.7\pm14.6$ & $157.8\pm16.0$ & $160.6\pm16.8$ & $239.6\pm25.5$ & $255.9\pm29.8$ \\
LkCa 15 & $97.7\pm10.0$ & $125.2\pm12.7$ & $133.5\pm13.5$ & $169.0\pm17.1$ & $204.0\pm20.8$ & $241.9\pm24.9$ & $265.3\pm27.0$ & $297.3\pm30.4$ & $346.7\pm34.8$ & $381.6\pm38.5$ & $508.6\pm51.8$ & $517.9\pm54.5$ \\
RW Aur & $27.3\pm3.3$ & $33.0\pm3.9$ & $38.0\pm4.3$ & $36.1\pm4.3$ & $47.0\pm6.1$ & $55.0\pm7.8$ & $67.7\pm8.2$ & $82.9\pm10.0$ & $81.6\pm8.7$ &  & $109.2\pm14.3$ & $152.6\pm22.1$ \\
RY Tau & $168.9\pm17.0$ & $205.3\pm20.6$ & $216.8\pm21.8$ & $260.5\pm26.2$ & $316.8\pm31.9$ & $364.8\pm36.9$ & $407.3\pm41.0$ & $476.0\pm47.9$ & $525.1\pm52.6$ & $584.2\pm58.6$ & $758.0\pm76.3$ & $857.2\pm87.1$ \\
SU Aur & $23.7\pm2.9$ & $22.0\pm2.8$ & $22.5\pm2.9$ & $29.2\pm3.5$ & $31.8\pm4.6$ & $24.2\pm5.4$ & $47.6\pm6.3$ & $47.3\pm6.9$ & $50.4\pm5.8$ & $44.1\pm6.6$ &  & $113.9\pm20.1$ \\
T Tau & $165.2\pm16.7$ & $205.6\pm20.7$ & $210.4\pm21.1$ & $239.7\pm24.1$ & $278.4\pm28.1$ & $296.7\pm30.2$ & $350.5\pm35.4$ & $366.9\pm37.2$ & $431.1\pm43.2$ & $502.6\pm50.5$ & $637.1\pm64.4$ & $715.3\pm73.5$ \\
UX Tau & $55.3\pm6.0$ & $59.4\pm6.4$ & $62.7\pm6.6$ & $76.2\pm8.0$ & $92.5\pm10.1$ & $90.1\pm10.8$ & $117.1\pm12.8$ & $124.9\pm13.8$ & $158.3\pm16.2$ & $172.7\pm18.1$ & $238.2\pm25.9$ & $251.1\pm30.5$ \\
UY Aur & $21.8\pm2.7$ & $28.9\pm3.4$ & $24.4\pm3.0$ & $22.1\pm3.0$ & $34.1\pm4.7$ & $43.3\pm6.4$ & $48.4\pm6.4$ & $48.0\pm6.8$ & $51.2\pm5.9$ & $62.6\pm7.8$ &  & $89.0\pm18.4$ \\
UZ Tau & $117.7\pm12.0$ & $128.3\pm13.1$ & $136.3\pm13.8$ & $152.0\pm15.4$ & $194.5\pm20.0$ & $228.4\pm23.8$ & $247.5\pm25.4$ & $262.6\pm27.1$ & $307.4\pm30.9$ & $333.8\pm33.7$ & $399.6\pm40.9$ & $464.6\pm48.9$ \\
V710 Tau & $41.3\pm4.7$ & $53.9\pm5.9$ & $55.9\pm6.0$ & $73.6\pm7.8$ & $85.5\pm9.4$ & $84.4\pm10.4$ & $88.9\pm10.3$ & $108.8\pm12.5$ & $126.6\pm13.1$ & $144.8\pm15.5$ & $194.6\pm21.9$ & $233.1\pm29.0$ \\
V836 Tau & $17.2\pm2.5$ & $23.6\pm3.1$ & $23.1\pm3.1$ & $26.7\pm3.5$ & $32.3\pm5.0$ & $36.7\pm6.6$ & $51.3\pm7.0$ & $55.7\pm7.9$ & $55.6\pm6.5$ &  & $66.4\pm12.6$ & $94.3\pm21.9$ \\
V892 Tau & $206.7\pm20.8$ & $252.7\pm25.4$ & $275.1\pm27.6$ & $322.8\pm32.4$ & $391.9\pm39.4$ & $434.8\pm43.8$ & $515.9\pm51.8$ & $569.8\pm57.3$ & $643.9\pm64.5$ & $733.2\pm73.5$ & $972.6\pm97.7$ & $1075.7\pm108.8$ \\
\enddata
\end{deluxetable*}
\end{longrotatetable}



\end{CJK}
\end{document}